\documentclass[11pt,a4paper]{bookandy}
\usepackage{amssymb}
\usepackage{picins}
\usepackage[includehead,nofoot,height=21.6cm,width=13.5cm,top=3cm,centering]{geometry}
\usepackage{fancyhdr}
\usepackage{thesisandy}
\usepackage{cite}
\input epsf
\begin{document}





\thispagestyle{empty}
\begin{figure}[ht]
\begin{center}
      \leavevmode
      \epsfxsize=105mm
      \epsfbox{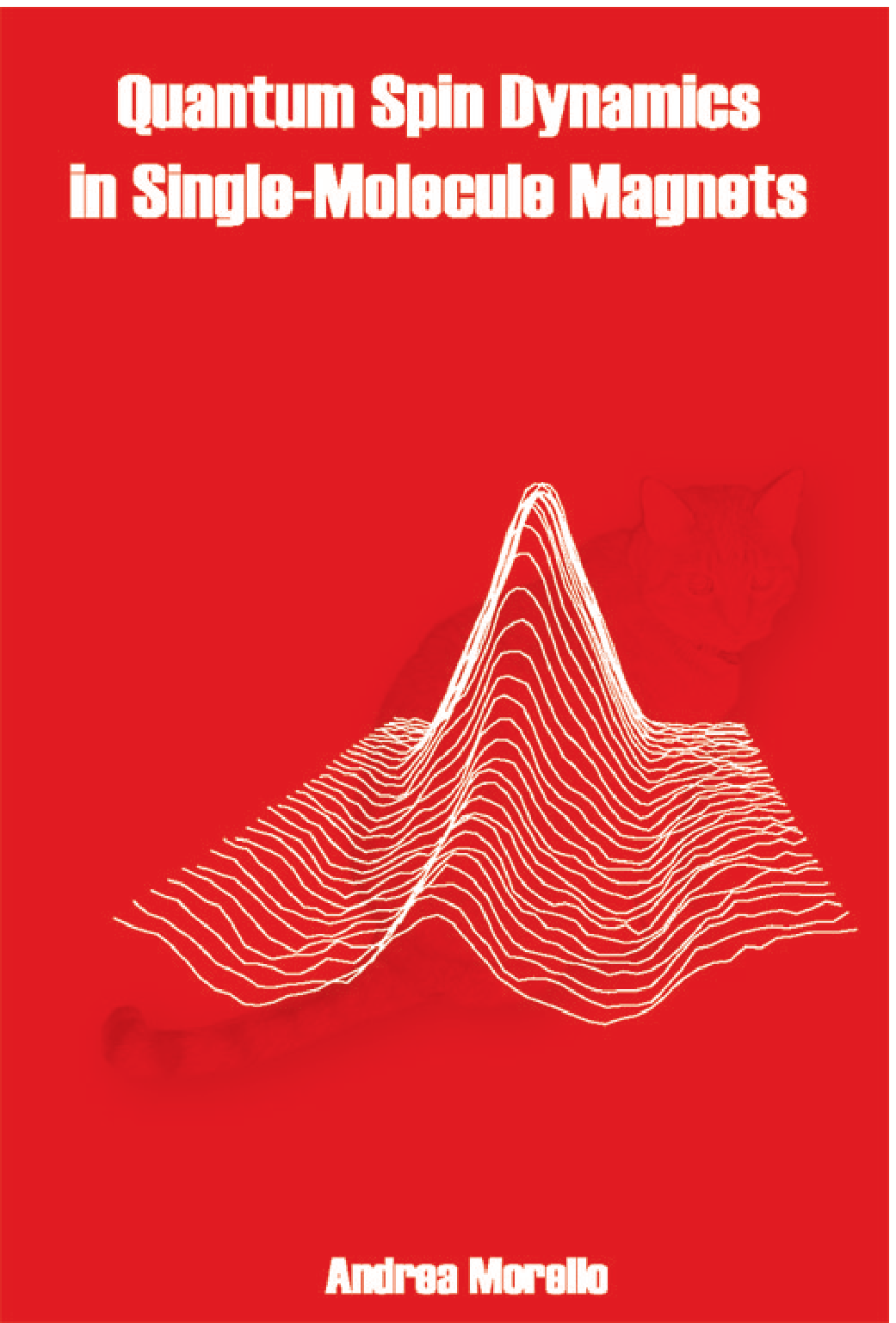}
\end{center}
\end{figure}

\pagebreak

\thispagestyle{empty}

\ \\

\pagebreak

\begin{center}
\thispagestyle{empty}
\Huge{Quantum spin dynamics in
single-molecule magnets}

\vspace{2cm}

\Large{Proefschrift}

\vspace{2cm}

\large{ter verkrijging van\\ de graad van Doctor aan de
Universiteit Leiden,\\ op gezag van de Rector Magnificus Dr. D. D.
Breimer,\\ hoogleraar in de faculteit der Wiskunde en\\
Natuurwetenschappen en die der Geneeskunde,\\ volgens besluit van
het College voor Promoties\\ te verdedigen op donderdag 18 maart 2004\\
te klokke 15:15 uur.}

\vspace{1cm}

door

\vspace{1cm}

\Large{Andrea Morello}

\vspace{3mm}

\large{geboren te Pinerolo (Itali\"{e}) in 1972}
\end{center}

\pagebreak

\thispagestyle{empty}

\noindent \setlength{\parindent}{0mm} \textbf{Promotiecommissie}\\

\setlength{\parindent}{0mm}
\begin{tabular}[t]{ll}
Promotor: & \medskip Prof.\ dr.\ L. J. de Jongh \\
Referent: &
\medskip Prof.\ dr.\ G. Aeppli (University College London, Groot Britanni\"{e})\\
Overige leden: & Dr.\ H. B. Brom \\ & Prof.\ dr.\ G. Frossati
\\ & Prof.\ dr. P. H. Kes \\ & Prof.\ dr.\ M. Orrit \\ & Prof.\ dr.\ J. Zaanen \\

\end{tabular}

\vspace{10cm}
ISBN: 90-77595-20-1\\
Printed by Optima Grafische Communicatie, Rotterdam: www.ogc.nl\\
\ \\
This work is part of the research program of the Stiching voor
Fundamenteel Onderzoek der Materie (FOM), which is supported by
the Nederlandse Organisatie voor Wetenschappelijk Onderzoek
(NWO).\\

\rule{50mm}{0.1pt}

\begin{small}
Cover: a sequence of nuclear spin echo traces obtained at
increasing time intervals after the inversion of the $^{55}$Mn
nuclear polarization in Mn$_{12}$-ac. The recovery of the
equilibrium polarization is driven by quantum tunneling of the
cluster spin.
\end{small}

\newpage

\thispagestyle{empty}

\ \\

\vspace*{3cm}

\hspace*{8cm}Said the straight man\\
\hspace*{8cm}to the late man:\\
\hspace*{8cm}``Where have you been?''\\
\hspace*{8cm}``I've been here and\\
\hspace*{8cm}I've been there and\\
\hspace*{8cm}I've been \ldots in between.''\\ \\
\hspace*{8cm}\textit{King Crimson,}\\
\hspace*{8cm}\textit{``I talk to the wind'', 1969.}\\

\pagebreak

\setlength{\parindent}{5mm} \thispagestyle{empty}
\pagenumbering{Roman} \setcounter{page}{1} \tableofcontents

\def\baselinestretch{1}

\chapter{Introduction}
\setcounter{page}{1}\pagenumbering{arabic}

Physicists, chemists and biologists have worked since decades on
the study of systems of increasing complexity, just as much as
engineers and technologists have been trying to miniaturize the
devices they design for practical applications. It seems that
these two directions have now come very close to a joining point
that happens to be located at the nanometer-scale. Accordingly,
the name ``Nanoscience'' has been given to the resulting broad
field of research. The rate at which laboratories and research
groups are being converted to activities in nanoscience, may raise
some fears that it will become a fashion-driven hype. On the other
hand, it is undeniable that nanoscience is in fact nothing else
than the natural evolution of last century's science, with a solid
history of interests behind it. The subject of this thesis, that
could be summarized as the quantum dynamics of magnetic molecules,
is a good example of multidisciplinary topic that combines
fundamental and practical aspects of the research at the nanometer
scale.

The field of molecular magnetism has its roots in the interest of
chemists towards the synthesis of large molecules, and the
assembly of macroscopic amounts of them in regular structures.
Along this route, it became clear that the synthesis of molecules
having magnetic ions as constituents could give rise to structures
where each molecule can be seen as a single-domain magnetic
particle \cite{gatteschi93S}, often called ``cluster'',
coordinated by a shell of organic ligands. Importantly, such
molecules are stoichiometric chemical compounds that can be packed
in a crystalline structure, where the identical magnetic units
interact only weakly with each other. In this way, each molecule
can be treated in first approximation as a single nanometer-scale
magnet, from where the name ``Single Molecule Magnet'' (SMM), and
is characterized by a large total spin $\vec{S}$ that arises from
the combination of the atomic electron spins. The possibility of
combining magnetic ions and organic ligands in the most diverse
ways allows to tune the physical properties of these systems to
obtain a wide range of magnetic behaviors.

One of the most essential physical properties of SMMs is their
magnetic anisotropy, meaning that it may be energetically
favorable for the magnetic moment of each molecule to align along
a certain axis. At temperatures much lower than the anisotropy
energy\footnote{Throughout this thesis we shall always express the
energies in temperature units, i.e. divided by the Boltzmann
constant $k_B$.}, the molecular spin is effectively ``frozen'' in
a certain direction along the anisotropy axis, giving rise to
single-molecule magnetic hysteresis \cite{sessoli93N}. This
discovery in a molecular cluster containing a core of 12 manganese
ions (Mn$_{12}$-ac) opened the perspective of using SMMs as the
ultimate magnetic memory units \cite{joachim00N}.

At the same time, people interested in the fundamental aspects of
quantum mechanics at the macroscopic scale (the ``Schr\"{o}dinger
cat'' problem \cite{schrodinger35}) realized that SMMs could be
candidates for the observation of quantum phenomena at the
macromolecular level, in particular quantum tunneling of the
magnetic moment \cite{caldeira81PRL,chudnovsky88PRL}. This has
been indeed recently achieved by observing regular steps in the
magnetic hysteresis loops of Mn$_{12}$-ac
\cite{friedman96PRL,thomas96N}, which occur when spin states on
opposite sides of the anisotropy barrier have the same energy, so
that the cluster spin may invert its direction by resonant quantum
tunneling. Again, this discovery is important for both practical
and fundamental reasons. On the one hand, it makes clear that a
memory unit based on systems as small as a single molecule would
be useless if we cannot avoid the memory being self-erased by
quantum tunneling. On the other hand, it suggests the possibility
of using SMMs as a test ground for theories concerning the
transition from quantum to classical physics
\cite{zurek91PT,leggett02JPCM}, in particular as far as the role
of the environment is concerned.

One of the great advantages of SMMs for fundamental research on
quantum mechanics at the large scale is precisely that the
influence of the environmental degrees of freedom on the giant
molecular spin can be accurately calculated, thanks to the
knowledge of the crystal structure and the magnetic couplings
between the \emph{localized} moments. Furthermore, the
experimental investigations can profit of all the best known
techniques for solid-state physics. Since the pure quantum
behavior in SMMs is typically achieved only in the subkelvin
temperature range, this research requires the use of outstanding
low-temperature facilities for magnetic measurements. In chapter
II we describe the working principles and the guidelines for the
design of several ultra-low temperature setups that are
particularly suitable for experiments on quantum magnetism.

As mentioned above, the magnetic properties of SMMs are determined
in the first place at the chemistry level, but an essential aspect
is that it is also possible, given a certain SMM, to tune its
quantum mechanical behavior ``in situ'' by applying an external
magnetic field. In particular, one can increase the tunneling rate
of $\vec{S}$ by applying a field perpendicular to the anisotropy
axis of the molecule \cite{luis00PRL}. This can be pushed to such
an extent that we may expect the possibility to observe coherent
quantum oscillations of the magnetic moment trough the anisotropy
barrier \cite{stamp03CM}, a phenomenon that has no classical
analog. In this case, SMMs would become suitable qubits for
quantum computing \cite{bennett00N}, with the interesting feature
that the operating frequency can be tuned locally by the simple
application of a magnetic field.

A large part of this thesis is dedicated to the study of the
interaction between SMMs and their environment, in particular the
nuclear spins. This interaction is crucial from many points of
view. In the zero- or low-field regime, the tunneling probability
is very small, in fact so small that, until recently, there were
serious doubts about the observability of quantum tunneling of the
magnetic moment in SMMs. This is due the fact that, in order to
tunnel, the total electron spin states on opposite sides of the
anisotropy barrier must be in resonance within an energy $\Delta$
that can be as small as $\sim 10^{-11}$ K in Mn$_{12}$-ac. Any
\emph{static} perturbation larger than that, would destroy the
resonance condition and make tunneling impossible. Prokof'ev and
Stamp \cite{prokof'ev96JLTP,prokof'ev00RPP} succeeded in
explaining why tunneling was actually observed in the experiments,
by noticing that the coupling of the cluster spin with the nuclear
spins in its surroundings, although several order of magnitude
larger than $\Delta$, is \emph{dynamic}. The electron spin energy
levels are therefore fluctuating in time with respect to each
other, which gives rise to an effective ``tunneling window'' that
depends on the strength of the coupling with the nuclei, as was
subsequently demonstrated by experiments with isotopically
substituted samples \cite{wernsdorfer00PRL}. In the opposite
regime, where the fast quantum oscillations of $\vec{S}$ can be
obtained by applying a strong perpendicular field, the nuclear
spins play the role of an intrinsic source of decoherence.
Understanding this phenomenon is obviously essential for any
attempt to use a magnetic qubit for quantum computing, since
nuclear spins are an unavoidable presence in practically any
material. The decoherence due to nuclear spins is indeed important
also for systems like the flux qubit \cite{chiorescu03S}. Chapter
III of this thesis is dedicated to a review of the theoretical
aspects of the quantum behavior of SMMs, from the basic notions to
the details of the Prokof'ev-Stamp theory.

Despite its relevance, the nuclear spin dynamics in SMMs at
ultra-low temperature remained an essentially unexplored field. In
chapter IV we report a thorough investigation of the dynamics of
$^{55}$Mn nuclei in Mn$_{12}$-ac. Our results uncover many basic
aspects, some of which confirm the validity of the Prokof'ev-Stamp
theory, while others are totally new and have never been
considered in any theoretical treatment of the coupled system of
``quantum spin + nuclei''. For instance, we demonstrate that the
quantum tunneling of the electron spin is a mechanism capable of
producing nuclear relaxation at an unexpectedly high rate. We also
show that nuclei in different molecules are coupled with each
other, an essential ingredient for the creation of a dynamic
tunneling window. The analysis of the nuclear spin dynamics in
large perpendicular fields confirms earlier results obtained by
specific heat experiments \cite{luis00PRL} concerning the increase
of the tunneling rate. Finally, we consider for the first time the
issue of the nuclear spin temperature in the presence of
\emph{temperature-independent} quantum tunneling fluctuations.
Surprisingly, the nuclear spin temperature is found to remain in
equilibrium with the lattice temperature down to $T \simeq 20$ mK!
We discuss therefore the need for an extension of the
Prokof'ev-Stamp theory to take into account inelastic tunneling
events.

In chapter V we present our research on the low-temperature
magnetic properties of a rather peculiar SMM, containing Mn$_6$
molecular clusters that are characterized by a very small magnetic
anisotropy because of the highly symmetric structure. We find that
the electron spin-lattice relaxation remains fast down to
millikelvin temperatures, which allows the observation of a
long-range magnetic ordering of the cluster spins, due only to the
mutual dipolar interactions. The nature of this ordered state is
already interesting in itself, but even more so when compared to
the situation in the highly anisotropic SMMs, where the freezing
of the cluster spins has made the investigation of the
thermodynamic ground state impossible so far. Experiments in
disordered rare-earth spin systems have shown that the application
of a transverse field greatly facilitates the system in reaching
its ground state \cite{brooke99S}, a strategy that seems promising
for anisotropic SMMs as well. Because of the absence of
anisotropy, and therefore of quantum tunneling, the nuclear spin
dynamics in Mn$_6$ offers also an interesting comparison with the
case of Mn$_{12}$-ac.

As may have become clear already, this thesis is focused uniquely
on the fundamental aspects of the quantum spin dynamics of SMMs.
The attempts to find useful applications for molecular magnets
have not stopped meanwhile, and we think that the recent progress
in patterning \cite{cavallini03NL} and surface deposition
\cite{cornia03AC} of Mn$_{12}$-ac molecules may represent an
actual breakthrough for both the magnetic storage and the quantum
computing purposes. Wherever this progress will lead to, we hope
that the work presented here will help the scientific community to
gain a deeper understanding of how the quantum phenomena at the
molecular scale are influenced by the interaction with the
environment.

\def\baselinestretch{1}

\chapter{Experimental techniques}

This chapter begins with a survey of the principles and the design
of the setups used for the experiments reported in chapters IV and
V. In addition, we discuss the design and construction of two
ultra low-$T$ setups for static magnetic measurements, a SQUID
magnetometer and a torquemeter, which are not involved in the
research presented in this thesis, but constitute a new and
significant extension to the experimental facilities of our
laboratory: \S\ref{sec:squid} and \S\ref{sec:torque} are therefore
meant as future reference for the newcomers in our research group,
or anybody interested in building ultra low-$T$ magnetometers.

\section{Dilution refrigerator} \label{sec:dilution}

The heart of of our ultra low-$T$ setup is a $^3$He/$^4$He
dilution refrigerator. The working principle (see e.g.
\cite{pobell} for details) is based on the fact that, even at
$T=0$, a liquid mixture of $^3$He and $^4$He does not separate
completely: the Van der Waals force between a $^3$He and a $^4$He
atom is larger than between two $^3$He, thus a certain fraction of
$^3$He atoms will spontaneously dilute into the $^4$He phase,
until the Fermi energy of $^3$He in the dilute phase equals the
difference in binding energies between $^3$He-$^3$He and
$^3$He-$^4$He. If one distils some $^3$He out of the dilute phase,
more $^3$He atoms will migrate from the pure $^3$He phase into the
dilute one to reestablish the equilibrium. The increase of entropy
associated with this process corresponds to the absorption of heat
by the $^3$He/$^4$He mixture, i.e. an effective cooling power. The
cooling can be made continuous if the $^3$He extracted from the
dilute phase is recondensed and injected on the side of the pure
$^3$He phase, taking care that the incoming $^3$He stream is
properly cooled by exchanging heat with the cold $^3$He atoms that
flow through the dilute phase.

\begin{figure}[p]
\begin{center}
      \leavevmode
      \epsfxsize=120mm
      \epsfbox{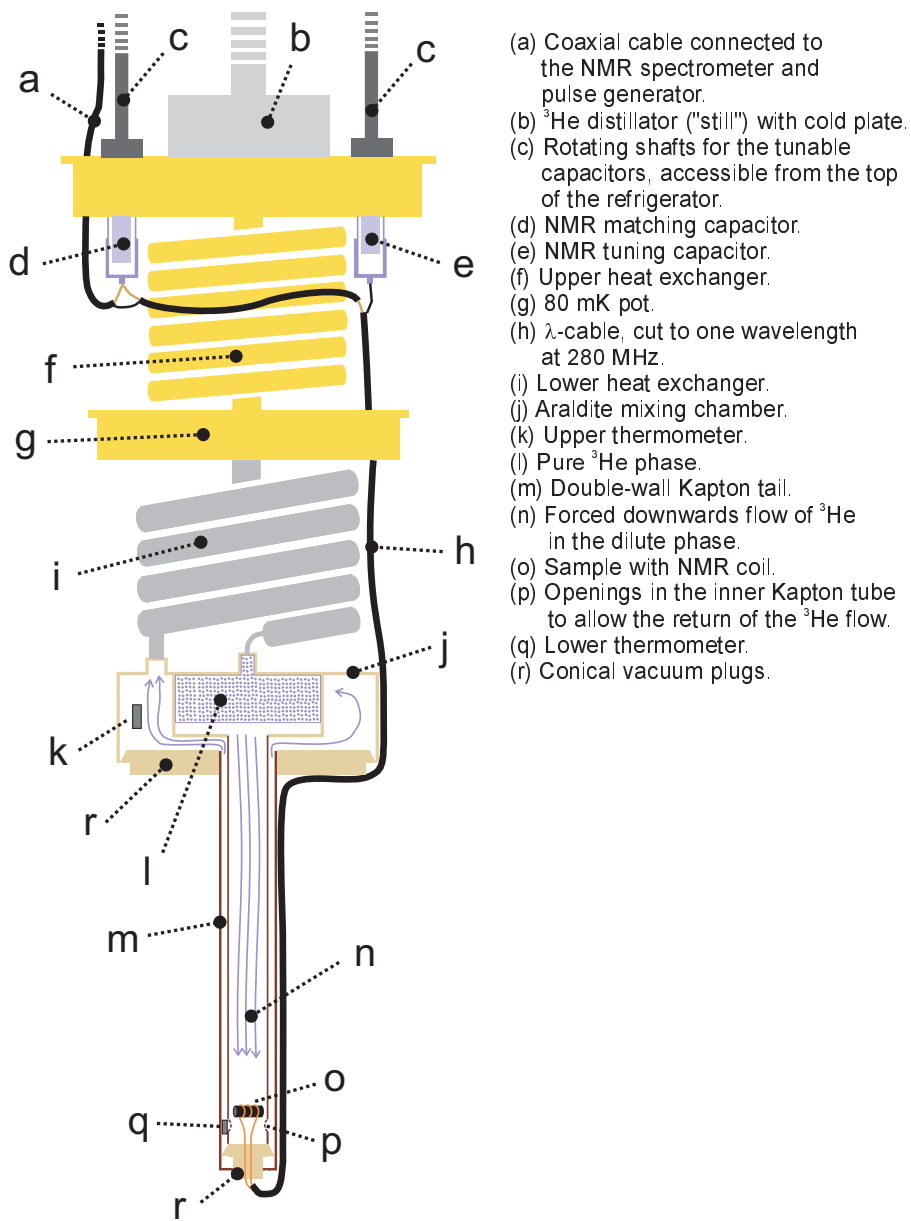}
\end{center}
\caption{\label{dilution} Scheme of the low-$T$ part of the
dilution refrigerator with NMR circuitry. For clarity we omit the
vacuum can and the radiation shields anchored at the still and at
the 80 mK pot. Only the narrow tail of the refrigerator is
inserted in the bore of a 9 T superconducting magnet (not shown).}
\end{figure}

Our system is based on a Leiden Cryogenics MNK126-400ROF dilution
refrigerator, fitted with a plastic mixing chamber that allows the
sample to be thermalized directly by the $^3$He flow, specially
designed for our purposes in collaboration with G. Frossati. A
scheme of the low-temperature part of the refrigerator is shown in
Fig. \ref{dilution}, together with the NMR circuitry (see
\S\ref{NMRcircuit}). The mixing chamber consists of two concentric
tubes, obtained by rolling a Kapton foil coated with Stycast 1266
epoxy. The tops of each tube are glued into concentric Araldite
pots: the inner pot receives the downwards flow of condensed
$^3$He and, a few millimeters below the inlet, the phase
separation between pure $^3$He phase and dilute $^3$He/$^4$He
phase takes place. The circulation of $^3$He is then forced
downwards along the inner Kapton tube, which has openings on the
bottom side to allow the return of the $^3$He stream through the
thin space in between the tubes. Both the bottom of the Kapton
tail and the outer pot are closed by conical Araldite plugs
smeared with Apiezon N grease. The Kapton tail is about 35 cm long
and is surrounded by two silver-coated brass radiation shields,
one anchored at the 80 mK pot, the other at the still. The whole
low-$T$ part of the refrigerator is closed by a vacuum can, which
has itself a thin brass tail to surround the Kapton part of the
mixing chamber, and is inserted into the bore of an Oxford
Instruments 9 Tesla NbTi superconducting magnet. In this way, only
Kapton and brass cylinders (plus the lowest Araldite plug) are
placed in the high-field region, whereas all other
high-conductivity metal parts (heat exchangers, 80 mK pot, $^3$He
distillator, etc.) are outside the magnet bore and subject only to
a small stray field. This design was intended to minimize eddy
currents heating while moving the refrigerator through the pick-up
coil of the SQUID magnetometer (\S\ref{sec:squid}), while
thermalizing the sample directly by the contact with the $^3$He
flow. As it turned out, the excellent thermalization obtained is
this way is also essential for the success of the NMR experiments
described in chapter IV, whose significance depends crucially on
the efficient cooling of the nuclear spins (\S\ref{spinT}).

The temperature inside the mixing chamber is monitored by
measuring with a Picowatt AVS-47 bridge the resistance of two
Speer carbon thermometers, one in the outer top Araldite pot, the
other at the bottom of the Kapton tail, just beside the sample. We
have verified that the temperature is very uniform along the whole
chamber (except in the presence of sudden heat pulses): even at
the lowest $T$, the mismatch between the measured values is
typically $\lesssim 0.5$ mK. A Leiden Cryogenics Triple Current
Source is used to apply heating currents to a manganin wire,
anti-inductively wound around a copper joint just above the $^3$He
inlet in the mixing chamber. In this way we can heat the incoming
$^3$He stream and uniformly increase the mixing chamber
temperature.

For the $^3$He circulation we employ an oil-free pumping system,
consisting of a Roots booster pump (Edwards EH500) with a pumping
speed of 500 m$^3$/h, backed by two 10 m$^3$/h dry scroll pumps
(Edwards XDS10). The main pumping line is a $\varnothing$ 100 mm
solid tube, fixed at one side with a flexible rubber joint that
allows for an inclination of a few degrees, and connected to the
head of the fridge by a ``T'' piece with two extra rubber bellows,
to reduce the vibrations transmitted by the pumping system. In
this configuration, the system reaches a base temperature that can
be as low as 9 mK; with the extra wiring for NMR experiments, the
base temperature is $\sim 12$ mK, and the practical operating
temperature while applying $rf$-pulses is 15 - 20 mK (see
\S\ref{sec:thermaleff}). The typical $^3$He circulation rate at
the base temperature is $\dot{n} \sim 350$ $\mu$mol/s, and the
cooling power at 100 mK is $\dot{Q} \sim 150$ $\mu$W. When the
Kapton tail is replaced by a flat plug, $\dot{Q}$ can be increased
up to 700 $\mu$W @ 100 mK and $\dot{n} \sim 1200$ $\mu$mol/s by
applying extra heat to the still; with the tail in place it's more
difficult to increase the circulation rate, and $\dot{Q}$ @ 100 mK
hardly exceeds 250 $\mu$W.

\section{Nuclear Magnetic Resonance}

The NMR technique is applied since more than 50 years
\cite{purcell46PR} to widely different research fields; in this
thesis, we are interested in the use of nuclear spins as local
probes for magnetic fluctuations in solid-state systems, but also
in the dynamics of the nuclei themselves, which appears to play an
essential role in the quantum behavior of single-molecule magnets.

\subsection{Basics of pulse NMR} \label{NMRbasics}

The Hamiltonian of a nuclear spin $\vec{I}$ placed in a magnetic
field $\vec{B}_z$ applied along a certain axis $\vec{z}$ is:
\begin{eqnarray}
\mathcal{H} = \hbar \gamma \vec{I} \cdot \vec{B}_z,
\label{nuclearhamiltonian}
\end{eqnarray}
where $\gamma$ is the gyromagnetic ratio, such that $\mu = \hbar
\gamma I$ is the magnetic moment of the nucleus. The eigenstates
$| m \rangle$ are the $2I+1$ projections of the spin along
$\vec{z}$, having energies $E_m$, $m = -I \ldots I$, separated
from each other by the Zeeman splitting $E_{m+1}-E_m = \hbar
\gamma B_z$, which corresponds to the Larmor frequencies $\omega =
\gamma B_z$ for the classical precession. Transitions between
adjacent Zeeman levels can be produced by introducing a
time-dependent field $\vec{B}_{rf}(t)$, provided that the matrix
elements $\langle m| B_{rf} |m+1 \rangle \neq 0$, i.e.
$\vec{B}_{rf}$ is not parallel to $\vec{z}$. Considering for
simplicity a spin $I=1/2$ and taking $\vec{B}_{rf}(t) = B_{rf}
\vec{x} \cos(\omega_N t)$, the expectation value of the $z$
component of the magnetic moment will oscillate in time at a
frequency $\omega_{\mathrm{Rabi}} = \gamma B_{rf}$ (Rabi
oscillations):
\begin{eqnarray}
\langle \mu_z (t) \rangle = \langle \mu_z (0) \rangle
\cos(\omega_{\mathrm{Rabi}} t).
\end{eqnarray}
In a classical picture, this means that, if $\vec{\mu}(0) = \hbar
\gamma I \vec{z}$, after a time $t_{\pi/2} = \pi/(2 \gamma
B_{rf})$ the magnetic moment has been turned 90$^{\circ}$ away
from the $\vec{z}$ axis and lies in the $xy$ plane. For this
reason, the application of such an alternating field of frequency
$\omega_N$ for a time $t_{\pi/2}$ is called ``$\pi/2$-pulse''. The
rotation angle can in fact take any value, by choosing the
appropriate duration and strength of $B_{rf}$. After a
$\pi/2$-pulse the state of the system is $|\psi\rangle =
1/\sqrt{2}(|+1/2\rangle + |-1/2\rangle)$, and the time evolution
caused by the Hamiltonian (\ref{nuclearhamiltonian}) is equivalent
to the classical Larmor precession of the magnetic moment within
the $xy$ plane, which produces a rotating magnetic field that can
be detected via the electromotive force induced in a pick-up coil
with axis $\parallel \vec{x}$. In practice the same coil is used
to produce $B_{rf}$ and to detect the Larmor precession.

Since we always work with macroscopic ensembles of spins, it is
convenient to describe the system in terms of its density matrix
$\rho$. If the state of each spin in the system is described as
$|\psi\rangle = \sum_m a_m |m\rangle$, then the matrix elements
$\rho_{m,m'}$ are the average over the sample of $a_m a_{m'}^*$.
The diagonal elements of $\rho$ represent the populations of the
levels, and the non-diagonal elements account for the correlation
between different spins. Under thermal equilibrium at a
temperature $T$, the non-diagonal terms of $\rho$ are zero, and
the diagonal terms obey:
\begin{eqnarray}
\frac{\rho_{m,m}}{\rho_{m+1,m+1}} = \exp \left(-\frac{\hbar \gamma
B_z}{k_B T} \right), \label{rhomm}
\end{eqnarray}
\noindent which corresponds to a longitudinal ($\parallel B_z$)
nuclear magnetization, $M_z(T)$, according to the Curie law:
\begin{eqnarray}
M_z(T) = N \hbar^2 \gamma^2 \frac{I(I+1)B_z}{3k_B T},
\end{eqnarray}
\noindent where $N$ is the number of spins per unit volume.

By means of \textit{rf}-pulses as described above, it is possible
to perturb the system and change its density matrix. For instance,
the effect of a $\pi$-pulse on an ensemble of spins $1/2$ in
thermal equilibrium is to invert the diagonal elements of $\rho$,
i.e. the nuclear magnetization. After the inversion, the thermal
equilibrium can be reestablished in the presence of perturbations
that induce transitions between the Zeeman levels, and of a
reservoir that can absorb the heat released by the nuclear spins.
The time constant for the recovery of the equilibrium values of
the diagonal elements of $\rho$ is $T_1$, the nuclear spin-lattice
relaxation time.

Far from equilibrium, it is possible to define a nuclear spin
temperature $T_{\mathrm{nucl}}$ different from the lattice $T$,
provided that the ratio of the populations formally obeys a
relation like (\ref{rhomm}) with $T_{\mathrm{nucl}}$ replacing
$T$, and that $\rho_{m,m'} = 0$ for $m \neq m'$. For example, a
$\pi$-pulse on a spin 1/2 produces a negative spin temperature.
$T_1$ can be then interpreted as timescale for the reequilibration
of the spin- and lattice temperatures.

After a $\pi/2$ pulse, the diagonal elements of $\rho$ are
identical and nonzero non-diagonal elements appear. This means
that the spins are coherently precessing, the correlation between
the single Larmor precessions decaying with a time constant $T_2
\leq T_1$, which is also the decay time for the non-diagonal
elements of $\rho$.

\begin{figure}[t]
\begin{center}
      \leavevmode
      \epsfxsize=120mm
      \epsfbox{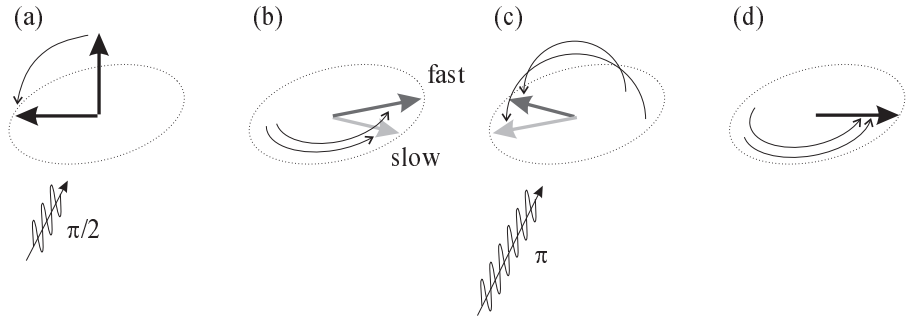}
\end{center}
\caption{\label{echo} Sketch of the spin-echo sequence: (a)
$\pi/2$-pulse; (b) Larmor precession and dephasing of the slow
(light gray) and fast (dark gray) spins; (c) $\pi$-pulse; (d)
refocusing of the precessions and formation of the echo.}
\end{figure}

In a real experiment, the signal produced by the Larmor precession
after a $\pi/2$-pulse, called ``free induction decay'', actually
decays on a time $T_2^* \leq T_2$ because of static field
inhomogeneity, i.e. the fact that spins in different spatial
positions may have different $B_z$, thus different $\omega_N$.
This static dephasing can be recovered by means of spin echo. By
applying a $\pi$-pulse at a time $\tau > T_2^*$ after the
$\pi/2$-pulse, each spin is rotated 180$^{\circ}$ about the axis
of the coil. In this way, the spins that are precessing ``too
slow'' find themselves ahead of the fast ones. At time $2\tau$ all
spins are again precessing in phase (Fig. \ref{echo}), giving rise
to an echo of coherent precession whose amplitude depends on the
effective correlation $\propto \exp(-2\tau/T_2)$ at that instant.
We can therefore measure $T_2$ by applying a $\pi/2 - \pi$ pulse
sequence with increasing values of $\tau$, and observing the
$\tau-$dependence of the echo amplitude.

\begin{figure}[t]
\begin{center}
      \leavevmode
      \epsfxsize=120mm
      \epsfbox{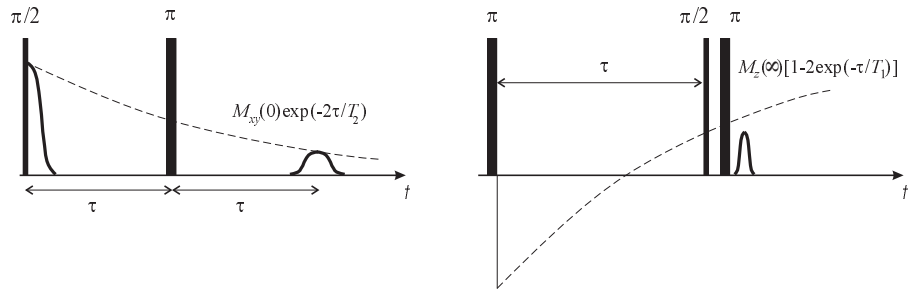}
\end{center}
\caption{\label{measureT1T2} Left: $T_2$ measurement by echo
decay. Right: $T_1$ measurement by inversion recovery.}
\end{figure}

As regards $T_1$, we remark that the coil is sensitive only to the
projection of the Larmor precession along its axis, i.e. $\perp
\vec{z}$. The effect of a $\pi/2$-pulse, or of an echo sequence,
is indeed to project the $z$-component of the nuclear
magnetization into the $xy$ plane and make it measurable. For
instance, an echo sequence after a $\pi$-pulse yields a signal
which is proportional to the initial magnetization, but with
opposite phase as compared to the signal without $\pi$-pulse. By
introducing a waiting time $\tau$ between the $\pi$-pulse and the
echo sequence, one can access the time evolution of the
$z$-component of the magnetization, i.e. the recovery of the
equilibrium state and thereby the time constant $T_1$. A sketch of
the pulse sequences used to measure $T_1$ and $T_2$ is shown in
Fig. \ref{measureT1T2}.

\subsection{Instruments and circuitry} \label{NMRcircuit}

\begin{figure}[p]
\begin{center}
      \leavevmode
      \epsfxsize=120mm
      \epsfbox{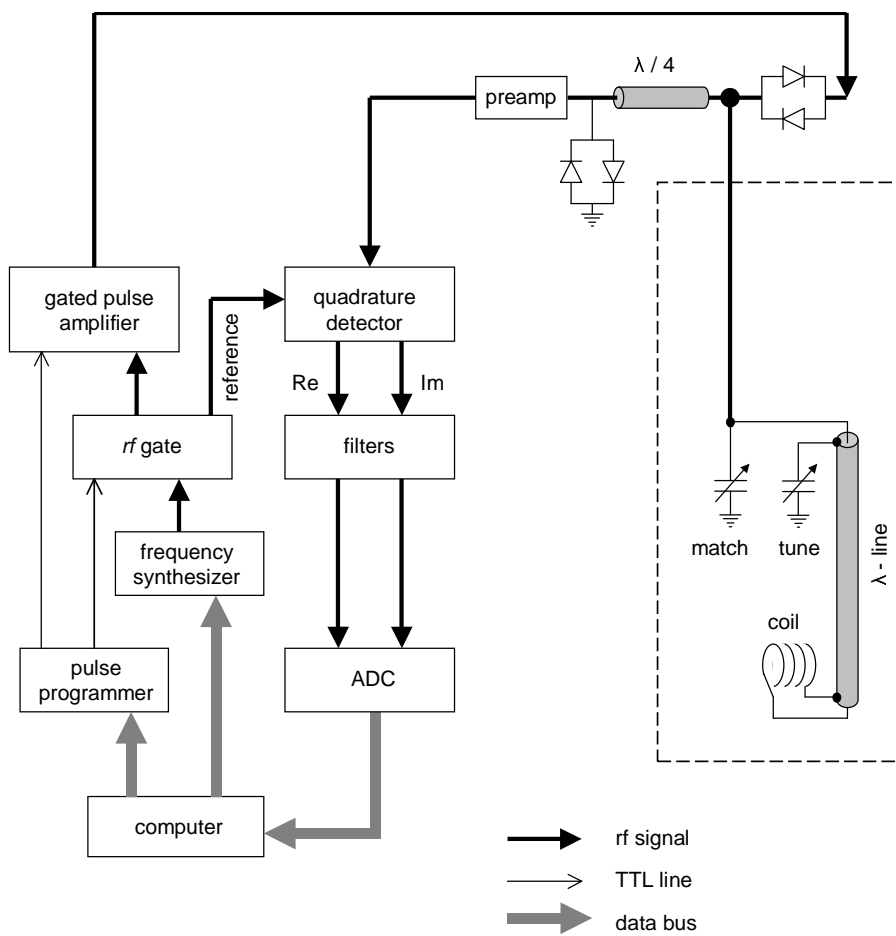}
\end{center}
\caption{\label{NMRsetup} Scheme of the NMR setup. The parts
enclosed in the dashed box are inserted in the dilution
refrigerator.}
\end{figure}

\begin{figure}[p]
\begin{center}
      \leavevmode
      \epsfxsize=125mm
      \epsfbox{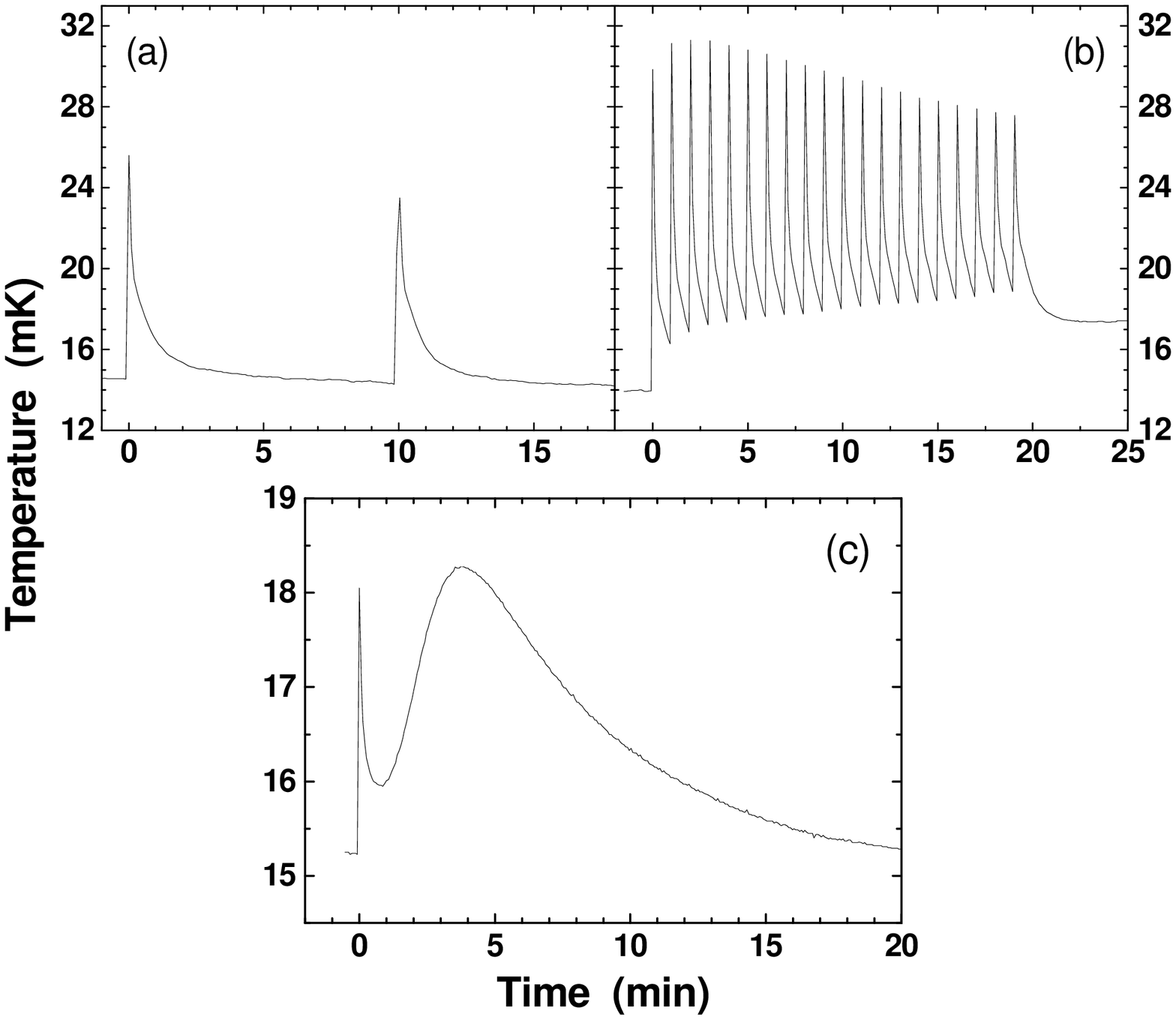}
\end{center}
\caption{\label{thermaleff} Time evolution of the mixing chamber
temperature while applying 12 $\mu$s - 24 $\mu$s spin-echo pulse
sequences. (a) Response of the thermometer at the bottom of the
Kapton tail, next to the sample, after two spin-echo sequences at
10 min intervals. (b) Like (a), after 20 sequences at 1 min
intervals. (c) Thermometer at the top of the mixing chamber, close
to the $^3$He outlet, after a single spin-echo sequence.}
\end{figure}

A scheme of the NMR setup used for the present work is shown in
Fig. \ref{NMRsetup}. The non-commercial electronic devices have
been developed by R. Hulstman, and a computer running a Pascal
management program written by J. Witteveen \cite{witteveenT}
controls the generation of the NMR pulses and the detection of the
signal. The pulse sequences are obtained by programming a
home-made device, which opens the gate of the pulse amplifier
(Kalmus 166UP) at the specified times: one typically waits $5$
$\mu$s before the full power of the amplifier (typ. $\sim 100$ W)
is available, and closes the gate as soon as the power is no
longer necessary, to avoid leakage currents to the NMR coil. The
\textit{rf}-signal to be amplified is generated by the frequency
synthesizer (Farnell SSG1000), but is first handled by a
\textit{rf}-gate, driven by the pulse programmer. The purpose of
the \textit{rf}-gate is to produce pulses of the desired length
and phase. For instance, to measure a spin echo we first produce a
$\pi/2 - \pi$ sequence with all the pulses in phase with the
reference signal, and record the result; then we repeat the
sequence but now with a $180^{\circ}$ phase shift on the
$\pi/2$-pulse, and we subtract the measured signal from the
previous one. In this way, the ``true'' NMR signal from the second
sequence, which has opposite sign with respect to the first
because of the phase shift in the $\pi/2$-pulse, is actually added
to the first one, whereas we cancel out the offset of the
spectrometer and the ringing (i.e. the fictitious signal that
appears because of the electrical resonance of the circuit)
produced by the $\pi/2$-pulse \cite{fritschijT}. The amplified
pulses are then fed to the resonant circuit.

The circuitry we used includes two tunable cylindrical teflon
capacitors, mounted at the still of the dilution refrigerator (see
Fig. \ref{dilution}) and driven by rotatable shafts. The capacitor
in parallel to the \textit{rf}-line is used to match the impedance
of the circuit to $50$ $\Omega$, whereas the one in series to the
NMR coil tunes the frequency of the resonator. The peculiarity of
our circuit is that, because of the distance ($\sim 50$ cm)
separating the capacitors from the NMR coil in the mixing chamber,
we had to introduce a $\lambda$-cable between coil and tuning
capacitor. At the frequency where the cable is precisely one
wavelength, the circuit is indeed identical to the standard
concentrated circuit. Away from the $\lambda$-frequency, the cable
adds extra inductance or capacitance. More importantly, since the
$\lambda$-cable is a low-conductivity thin brass coax for low-$T$
applications, the quality factor of the resonator (which includes
the cable!) is drastically reduced. Although this affects the
sensitivity of the circuit, it also broadens the accessible
frequency range without need to retune the capacitors. We found
that, cutting the cable for one wavelength at $\sim 280$ MHz, the
circuit is usable between (at least) 220 and 320 MHz. Once the
desired frequency range is chosen and the $\lambda$-cable is cut
accordingly, the NMR coil must be constructed by first mounting it
just near the tuning capacitor, i.e. making a ``standard''
concentrated circuit. If this circuit resonates in the same range
as the $\lambda$-frequency, the coil can be safely moved into the
mixing chamber.

The NMR signal induced in the coil is first passed though a
low-noise DOTY preamplifier placed at the top of the cryostat,
then detected by a phase-sensitive quadrature detector, and
finally stored in the memory buffer of a flash A/D converter. The
results of the positive and negative pulse sequences are
subtracted and dumped in the measurement data file. The diodes in
parallel to the preamplifier and the $\lambda/4$-cable, together
with the diodes in series to the power line, are used to separate
the small NMR signal from the high-power pulses. For the pulse
amplifier, the diodes in series are a short circuit, whereas the
$\lambda/4$-cable is an open circuit since it is short-circuited
at the opposite end by the diodes in parallel. In this way all the
power goes in the resonator and does not damage the preamplifier.
Conversely, the series diodes are an open circuit for the very
small signal picked up by the NMR coil, which is therefore
entirely conveyed towards the preamplifier.

\subsection{Heating effects} \label{sec:thermaleff}

Applying NMR pulses of typically 100 W for several microseconds to
a copper coil at $T \simeq 15$ mK is potentially very harmful for
the thermal stability of the sample and of the $^3$He/$^4$He
mixture in which it is immersed. As discussed in
\S\ref{sec:dilution}, our setup is specially designed to
circumvent this problem. In Fig. \ref{thermaleff} we show three
examples of the magnitude of the heating effects that arise when
applying spin-echo NMR pulse sequences consisting of a 12 $\mu$s
$\pi/2$-pulse followed by a 24 $\mu$s $\pi$-pulse after 45 $\mu$s.
Panels (a) and (b) show the temperatures as measured by the
thermometer placed at the bottom of the Kapton tail, just next to
the NMR coil and the sample, whereas the temperature in panel (c)
is taken at the top of the mixing chamber in the outer part of the
wide Araldite pot, i.e. 35 cm past the sample in the $^3$He path.
In both cases, a very sharp peak in $T$ is visible at the instant
when the pulses are applied, which is simply due to the direct
(and instantaneous) heating of the thermometers by the
electromagnetic field produced by the coil, thus it does not
reflect a ``real'' increase in the temperature of the $^3$He bath.
The situation is particularly clear in panel (c), where the
radiative heating spike is quickly recovered, and only after a few
minutes the wave of warmer $^3$He atoms comes along. By comparing
(a) and (b) with (c), it is also clear that the height of the
radiative spike is lower in the upper thermometer, which is indeed
farther away from the coil. Panel (a) demonstrates that, with an
interval of 10 minutes between the sequences, the temperature of
the sample can be kept very stable at $\simeq 15$ mK. Even with
just 1 minute waiting time (b), the increase of the baseline of
the $^3$He bath temperature does not exceed 5 mK!

\section{\textit{ac} - susceptometer} \label{acsusc}

\begin{figure}[t]
\begin{center}
      \leavevmode
      \epsfxsize=120mm
      \epsfbox{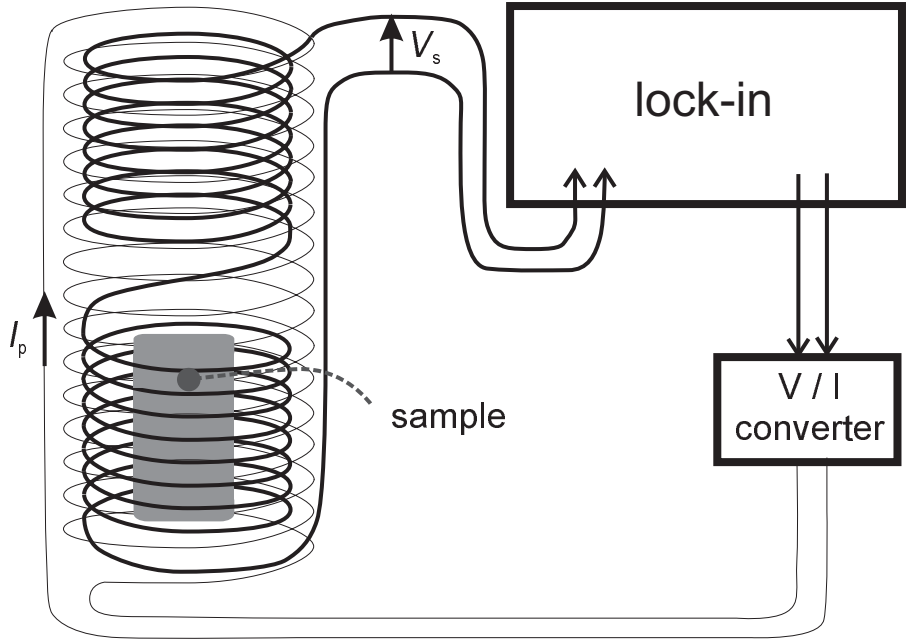}
\end{center}
\caption{\label{suscept} Schematic drawing of the
\textit{ac}-susceptometer.}
\end{figure}

For the frequency-dependent \textit{ac}-susceptibility
measurements reported in chapter V we used a simple and compact
home-made susceptometer based on the mutual inductance principle.
A sketch of the instrument is shown in Fig. \ref{suscept}. A
primary coil, consisting of $250$ turns of $\varnothing$ $100$
$\mu$m NbTi superconducting wire is wound on top of two secondary
coils, each consisting of 600 turns of $\varnothing$ $40$ $\mu$m
copper wire, and connected in series with opposite polarity. When
an alternating current $I_p(t) = I_p \cos(\omega t)$ is passed
through the primary coil, it induces a voltage $V_s = V_1 - V_2$
in the secondary, where $V_1$ and $V_2$ are the voltages at each
section of the secondary coil, and the minus sign is due to the
opposite winding directions:
\begin{eqnarray}
V_s = \omega I_p (\mathcal{M}_1 - \mathcal{M}_2) \cos(\omega t +
\pi/2).
\end{eqnarray}
The above relation holds if there is no current flowing in the
secondary coils, e.g. when they are connected to an ideal
voltmeter. The mutual inductances are given by $\mathcal{M}_i = (1
+ \chi_i)f$, where ($i= 1,2$) and$f = f(N_p,N_s,A_p,A_s,\ldots)$
is a factor that depends on the geometry, the area and the number
of turns of the primary and secondary coils, and $\chi_i$ is the
magnetic susceptibility of the material filling each secondary
coil. By leaving coil 2 empty ($\chi_2 = 0$) and filling coil 1
with the sample to be investigated, we obtain:
\begin{eqnarray}
V_s = \omega I_p \cos(\omega t + \pi/2)  \chi.
\end{eqnarray}
The susceptibility of the sample is therefore easily measured by
connecting the secondary coil to a lock-in amplifier (Stanford
SR830); the phase-sensitive detection allows to discern both the
real and the imaginary part of the complex susceptibility $\chi =
\chi' + i \chi"$.

To apply the primary current $I_p$ we either introduced a 500
k$\Omega$ resistor in series to the 5 V excitation circuit of the
lock-in, or we used a home-built V/I converter that allows to have
more current at low frequencies, since the induced voltage is
$\propto \omega$. Using the V/I converter we may easily measure
down to frequencies $\sim 1$ Hz, whereas above $\sim 1$ kHz it is
advisable to simply use the 500 k$\Omega$ series resistor, which
eliminates the slight phase rotations that the V/I converter
starts to introduce at high frequencies. With a typical value of
$I_p \sim 50$ $\mu$A, the alternating field produced at the sample
is $\mu_0 H_p \sim 1$ $\mu$T.

The susceptometer is small enough to fit vertically in the tail of
the dilution refrigerator, or even horizontally in the outer part
of the upper Araldite pot (cf. Fig. \ref{dilution}), for
measurements without static external field.

\section{SQUID magnetometer} \label{sec:squid}

The SQUID magnetometer described in this section is designed to
allow high-sensitivity measurements on powder samples down to $T
\sim 10$ mK. For example, metallic nanoclusters show spectacular
quantum-size effects in their thermodynamic properties
\cite{volokitin96N} at very low temperatures, but their magnetic
susceptibility is very small in this regime. Furthermore, most of
those materials are highly air-sensitive, thus it is very
convenient to introduce and keep the sample in a sealed glass
tube, which is also ideal to perform preliminary measurements at
$T>1.5$ K in commercial SQUID magnetometers. One of the
requirements for the design discussed here is indeed the
compatibility with the sealed sample holders used in other
experiments, without any need to further manipulate the sample.
The really challenging part of the design consists in moving the
sample through the pick-up coils while it is at $T\sim 10$ mK
inside the mixing chamber of the dilution refrigerator. For
samples with such extremely small magnetic signals such a measure
is needed because the two components of the pick-up coil will
never perfectly compensate one another, leaving an empty-coil
signal that could wash out the signal of the sample to be
measured. As we shall discuss below, this has required a system
that moves the \emph{whole} dilution refrigerator insert.

\subsection{Working principle}

A SQUID (Superconduting QUantum Interference Device) is basically
an ultra-sensitive flux-to-voltage converter, that exploits the
peculiar quantum properties of closed superconduting circuits. To
understand its working principle, we recall that the
superconducting state can be described as the condensation of
paired electrons (Cooper pairs) into an ordered phase
characterized by a complex order parameter:
\begin{eqnarray}
\psi = |\psi(\vec{r})|e^{i\varphi(\vec{r})},
\label{orderparameter}
\end{eqnarray}
where $|\psi(\vec{r})|^2$ is the density of Cooper pairs and
$\varphi(\vec{r}) = \vec{p}\cdot \vec{r} / \hbar$ is the phase. In
the presence of a magnetic field $\vec{B} = \vec{\nabla} \times
\vec{\mathcal{A}}$, the generalized momentum is expressed as
$\vec{p} = 2m\vec{v} - 2e\vec{\mathcal{A}}$, since the Cooper
pairs have mass $2m$ and charge $2e$ ($e = -1.6 \times 10^{-19}$
C).

The current density $\vec{j}$ can be obtained by introducing
(\ref{orderparameter}) in the standard quantum mechanical
expression:
\begin{eqnarray}
\vec{j} = -i \hbar \frac{e}{m}\left(\psi^{*} \vec{\nabla} \psi -
\psi \vec{\nabla} \psi^{*} \right) - \frac{4e^2}{m}  \psi^{*} \psi
\vec{\mathcal{A}},\\
\vec{j} = \left( \frac{2e \hbar}{m} \vec{\nabla} \varphi -
\frac{4e^2}{m} \vec{\mathcal{A}} \right) |\psi|^2.
\end{eqnarray}
The current in a superconductor can flow only within a surface
layer of thickness comparable to the London penetration depth,
since inside the bulk $\vec{j}=0$. By considering a
superconducting ring, the integral $\oint \vec{j} \cdot
\mathrm{d}\vec{l}$ along a closed path deep inside the bulk is
thus obviously zero, i.e.:
\begin{eqnarray}
\oint \left( \frac{2e \hbar}{m} \vec{\nabla} \varphi -
\frac{4e^2}{m} \vec{\mathcal{A}} \right) \cdot \mathrm{d}\vec{l} =
0; \label{intj}
\end{eqnarray}
recalling the Stokes theorem, $\oint \vec{\mathcal{A}} \cdot
\mathrm{d} \vec{l} = \int\hspace{-1.5mm}\int \vec{\nabla} \times
\vec{\mathcal{A}} \cdot \mathrm{d}\vec{S} =
\int\hspace{-1.5mm}\int \vec{B} \cdot \mathrm{d}\vec{S} = \Phi$
yields the magnetic flux enclosed by the ring. Furthermore, since
the order parameter must be single-valued, the total phase
accumulated along a closed path must be an integer multiple of
$2\pi$:
\begin{eqnarray}
\oint \vec{\nabla} \varphi \cdot \mathrm{d}\vec{l} = 2\pi n.
\label{intphi}
\end{eqnarray}
Combining (\ref{intj}) and (\ref{intphi}) yields the quantization
of the magnetic flux in a superconducting ring:
\begin{eqnarray}
\Phi = n \frac{h}{2e} = n\Phi_0,
\end{eqnarray}
where $\Phi_0 = h/2e = 2.07 \times 10^{-15}$ Wb is the flux
quantum. This means also that, since the externally applied flux
$\Phi_{\mathrm{ext}}$ is a classical variable and can be changed
smoothly, the quantization of the flux may require a screening
current $I_s$ to flow in the ring to fulfill the condition:
\begin{eqnarray}
\Phi = \Phi_{\mathrm{ext}} + L I_s = n\Phi_0,
\end{eqnarray}
where $L$ is the self-inductance of the ring.

\begin{figure}[t]
\begin{center}
      \leavevmode
      \epsfxsize=90mm
      \epsfbox{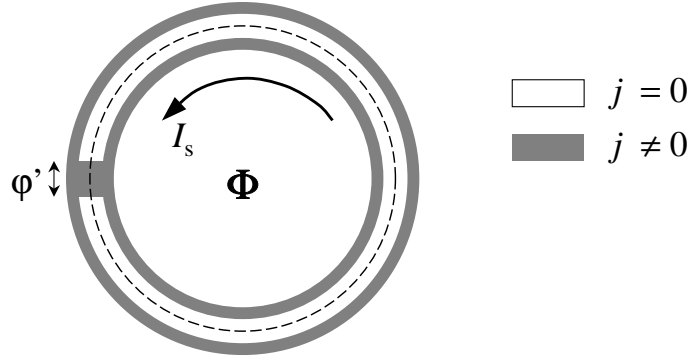}
\end{center}
\caption{\label{squidring} Sketch of superconducting ring
interrupted by a weak link. The dashed line is the integration
path.}
\end{figure}

A SQUID is obtained by interrupting a superconducting ring with
one (\textit{rf}-SQUID) or two (\textit{dc}-SQUID) Josephson
junctions, i.e. weak links where the superconductivity is locally
suppressed but the Cooper pairs can cross by quantum tunneling.
The essential feature added by a weak link is that it is no longer
possible to find a closed path through the ring such that
$\vec{j}=0$ everywhere, since $\vec{j} \neq 0$ through the weak
link (Fig \ref{squidring}). This adds an extra contribution
$\varphi'$ to the phase of the order parameter:
\begin{eqnarray}
\varphi'= \frac{m}{e\hbar |\psi|^2} \int_{w.l.} \vec{j} \cdot
\mathrm{d}\vec{l},
\end{eqnarray}
where the integral is along the weak link, supposed to be so thin
that the magnetic flux in negligible, and $|\psi|^2$ in the limit
of small currents can be taken equal to the bulk value. The
condition of single-valued order parameter becomes:
\begin{eqnarray}
\varphi' + \frac{2e}{\hbar} \int \hspace{-2.5mm} \int \vec{B}
\cdot \mathrm{d}\vec{S} = \varphi'  + 2\pi \frac{\Phi}{\Phi_0} =
\varphi' + 2\pi \frac{\Phi_{\mathrm{ext}} + LI_s}{\Phi_0} =  2\pi
n.
\end{eqnarray}
It follows that the screening current $I_s(\Phi)$ must be a
periodic function of $\Phi_{\mathrm{ext}}/\Phi_0$:
\begin{eqnarray}
I_s = I_c \sin(2\pi \Phi/\Phi_0) = I_c \sin[2\pi(n - \Phi/\Phi_0)]
= I_c \sin \varphi', \label{joseph1}
\end{eqnarray}
which is the famous Josephson phase-current relation. $I_c$ is the
Josephson critical current, i.e. the maximum supercurrent that can
pass through the link without dissipation.

By applying an external flux which varies linearly in time, a
constant voltage $V = - \mathrm{d}\Phi_{\mathrm{ext}}/\mathrm{d}t$
develops across the weak link. Substituting in Eq. (\ref{joseph1})
yields:
\begin{eqnarray}
I_s = I_c \sin \left( 2\pi \frac{Vt + LI_s}{\Phi_0} \right).
\label{IsvsV}
\end{eqnarray}
This implies that, when a constant voltage is applied to the
Josephson junction, an alternating supercurrent circulates in the
ring with frequency $\nu = 2eV/h$ proportional to the applied
voltage. Combining (\ref{IsvsV}) and (\ref{joseph1}) we find that
a constant voltage produces a linear increase of $\varphi'$ with
time, i.e. the voltage is proportional to the time derivative of
the phase:
\begin{eqnarray}
V = \frac{\hbar}{2e} \frac{\mathrm{d}\varphi}{\mathrm{d}t} .
\label{joseph2}
\end{eqnarray}

\begin{figure}[t]
\begin{center}
      \leavevmode
      \epsfxsize=120mm
      \epsfbox{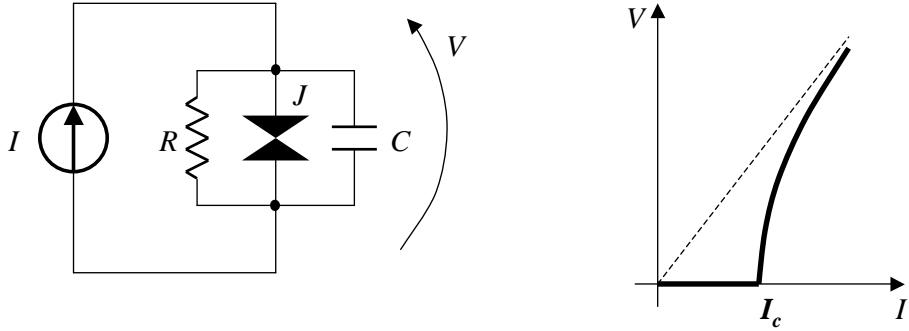}
\end{center}
\caption{\label{RSJ} Model of current-biased Josephson junction
(J) with dissipation (R) and capacitance (C), and the relative
$V-I$ characteristic.}
\end{figure}

This relation is useful to obtain the $V-I$ characteristic of a
current-biased Josephson junction. When $I>I_c$, the current
through the weak link causes dissipation, which can be accounted
for by a resistance $R$ in parallel to the ideal Josephson
junction described by (\ref{joseph1}). Moreover, a realistic
junction is formed by two superconducting pads separated by a very
thin barrier, so that we have to add a capacitance $C$ to the
model. We may then write:
\begin{eqnarray}
I = I_c \sin \varphi' + \frac{V}{R} +
C\frac{\mathrm{d}V}{\mathrm{d}t} = I_c \sin \varphi' +
\frac{\hbar}{2eR} \frac{\mathrm{d}\varphi'}{\mathrm{d}t} +
\frac{\hbar C}{2e} \frac{\mathrm{d}^2 \varphi'}{\mathrm{d}t^2},
\end{eqnarray}
which leads to a $V/I$ curve of the type shown in Fig. \ref{RSJ}.

\begin{figure}[t]
\begin{center}
      \leavevmode
      \epsfxsize=110mm
      \epsfbox{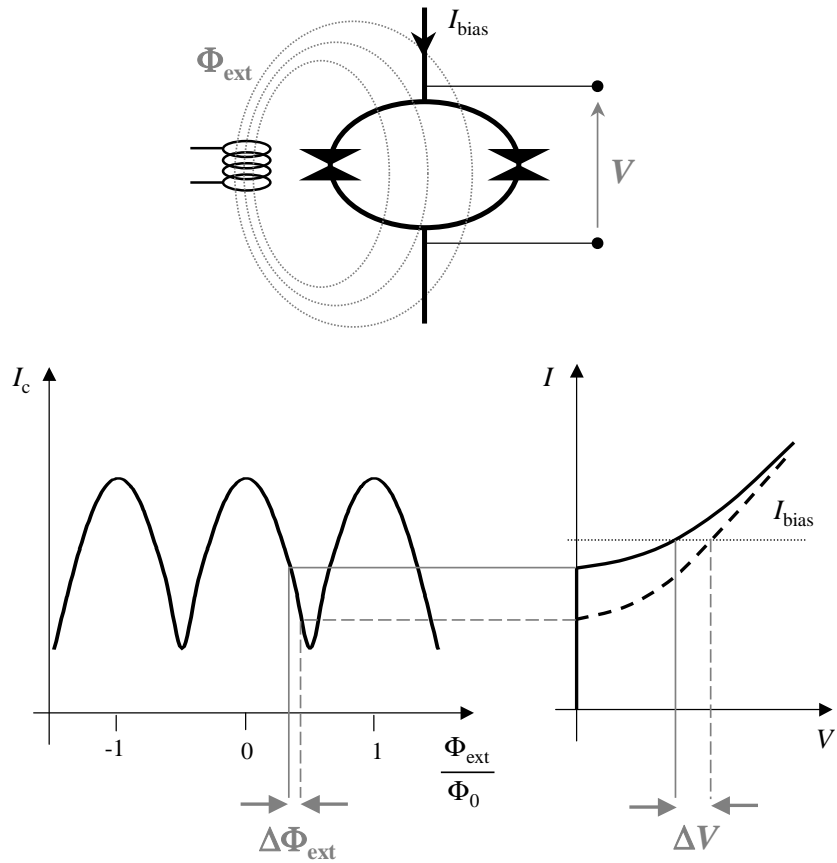}
\end{center}
\caption{\label{VvsPhi} Working principle of a \textit{dc}-SQUID
as flux-to-voltage converter.}
\end{figure}

A \textit{dc}-SQUID, like the one used in our setup, is operated
by biasing the junctions $J_1$ and $J_2$ with a current
$I_{\mathrm{bias}}>I_c$, such that a voltage develops across them
(Fig. \ref{VvsPhi}). The essential feature is that the voltage can
be modulated by an applied flux, since the critical current $I_c$
also depends on $\Phi$. Indeed, after some calculation one finds
that the screening current $I_s$ is related to the critical
current $I_c$ of the junctions (assumed to be identical) by:
\begin{eqnarray}
I_s = \frac{I_c}{2} \left[ \sin \varphi_1' - \sin \left(
\varphi_1' - 2\pi \frac{\Phi_{\mathrm{ext}} + LI_s}{\Phi_0}
\right) \right]. \label{IsDC}
\end{eqnarray}
$I_c$ is obtained by maximizing $I_s$, yielding the periodic form
of $I_c(\Phi_{\mathrm{ext}})$ shown in Fig. \ref{VvsPhi}. By
properly choosing the bias conditions, the \textit{dc}-SQUID
operates therefore as a flux-to-voltage converter, where the
external flux can be applied by injecting a current in the input
coil. Notice that small a fraction of a flux quantum can produce
voltage changes of the order of millivolts! In the practice a
\textit{dc}-SQUID is used in feedback mode by employing a
so-called Flux-Lock Loop (FLL), i.e. adding a feedback coil that
produces a compensating flux such that $V = const$. This increases
the accuracy and the dynamic range of the measurement, and allows
to implement noise-reducing detection schemes. For more details on
SQUID sensors, see \cite{lounasmaaB,gallopB}.

To use a \textit{dc}-SQUID in an actual magnetometer, it is still
necessary\footnote{Except in a rather radical design like the
``microSQUID'' \cite{wernsdorfer01ACC}} to produce a current
proportional to the magnetic moment of the sample to be measured.
Such a current can be injected in the input coil to produce a flux
that is coupled to the SQUID ring by the mutual inductance
$\mathcal{M}$. Typically, the input current is obtained by
constructing a closed superconducting circuit which includes the
SQUID's input coil on one side, and terminates with a pick-up coil
on the other side. Once the circuit has been cooled down below the
superconducting critical temperature, the enclosed flux is
constant. Any change in the magnetic permeability of the circuit,
possibly due to the sample, will result in a screening current
that, while keeping the total flux constant, produces the required
flux in the input coil. In our case, the change in permeability of
the pick-up circuit is obtained by vertically moving the
\emph{whole} dilution refrigerator (\S\ref{verticalmov}), whose
tail, that contains the sample, is inserted in the pick-up coil.

\begin{figure}[t]
\begin{center}
      \leavevmode
      \epsfxsize=100mm
      \epsfbox{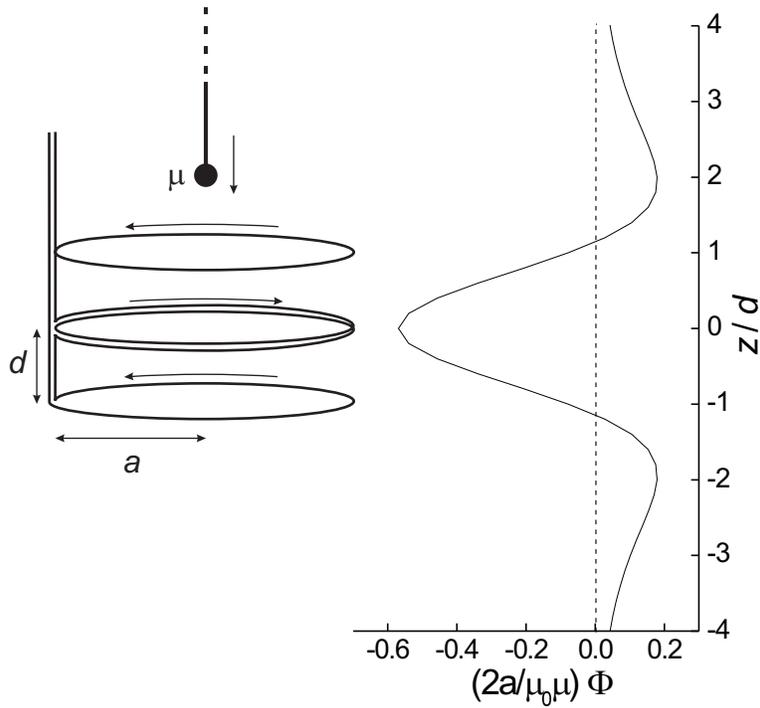}
\end{center}
\caption{\label{gradiometer} A second-order gradiometer in
Helmoltz geometry ($a = 2d$) and the magnetic flux induced by a
dipole moving along $z$. Notice that the side peaks in $\Phi(z)$
do not coincide with the positions of the external coils.}
\end{figure}

Because of the high sensitivity, it is essential to make sure that
no sources of flux other than the sample may couple with the
SQUID. This can be done by employing a gradiometer, which in its
simplest form consists of two coils wound in opposite direction,
so that the flux produced by any uniform magnetic field cancels
out, and only the gradient of $B$ can be detected (first-order
gradiometer). A further improvement is the second-order
gradiometer, shown in Fig. \ref{gradiometer}, which is obtained by
inserting a coil with $2N$ windings between two coils with N turns
each, wound opposite to the central one. This design eliminates
also the effects of linear field gradients. Furthermore, according
to the reciprocity principle \cite{mallinson66JAP}, the flux
$\Phi$ produced in a coil of arbitrary geometry by a magnetic
moment $\mu$ at position $\vec{r}$ is related to the field
$\vec{B}(\vec{r})$ produced by the same coil in $\vec{r}$ when
carrying a current $I$ such that:
\begin{eqnarray}
\vec{B}\cdot \vec{\mu} = \Phi I. \label{recip}
\end{eqnarray}
The field produced by a single-loop coil is equivalent to the
field of a magnetic dipole, whereas a first-order gradiometer is a
magnetic quadrupole, and a second-order is an octupole, so that
the fields they produce vary in space like $r^{-3}$, $r^{-4}$ and
$r^{-5}$, respectively. From Eq.(\ref{recip}) it is clear that a
second-order gradiometer is the least sensitive to magnetic fields
produced outside of the coil system.

The magnetic moment of the sample can be detected by moving it
through the pick-up coil. The enclosed flux is easily obtained
from the flux induced by a dipole with magnetic moment $\vec{\mu}
\parallel \vec{z}$ at a position $z$ along the axis in a loop of radius
$a$ placed at $z=z_0$:
\begin{eqnarray}
\Phi_{\mathrm{loop}}(z) = \frac{\mu_0}{2} f(z-z_0) \mu,\\
f(z-z_0) =  \frac{a^2}{[a^2 + (z-z_0)^2]^{3/2}} = \frac{1}{a}
\left[1 + \frac{z_0^2}{a^2}\left(\frac{z}{z_0}-1\right)^2
\right]^{-3/2}. \label{fdiz}
\end{eqnarray}
If the upper and lower coils of the gradiometer are placed at
$z=d$ and $z=-d$, respectively, then the total picked-up flux is:
\begin{eqnarray}
\Phi_{\mathrm{pu}}(z) = N\frac{\mu_0}{2} [f(z-d) - 2f(z) + f(z+d)]
\mu.   \label{Phipu}
\end{eqnarray}
Typically one adopts the Helmoltz geometry, $a = 2d$, resulting in
a $\Phi(z)$ as shown in Fig. \ref{gradiometer}\footnote{The factor
$a$ in the $x$-axis scale of Fig. \ref{gradiometer} is obtained by
using the second form of Eq. (\ref{fdiz}).}.

The picked-up flux is related to the screening current in the
circuit, $I_s$, by:
\begin{eqnarray}
\Phi_{\mathrm{pu}} = (L_{\mathrm{pu}} + L_{\mathrm{leads}} +
L_{\mathrm{in}})I_s,
\end{eqnarray}
where $L_{\mathrm{pu}}$, $L_{\mathrm{leads}}$ and
$L_{\mathrm{in}}$ are the inductances of the pick-up coil, the
leads and the SQUID input coil, respectively. The flux at the
SQUID sensor $\Phi_{\mathrm{SQUID}}$ is thus given by:
\begin{eqnarray}
\Phi_{\mathrm{SQUID}} = \mathcal{M}I_s = f_{\mathrm{tr}} \Phi_{\mathrm{pu}}, \nonumber\\
f_{\mathrm{tr}} = \frac{\mathcal{M}}{L_{\mathrm{pu}} +
L_{\mathrm{leads}} + L_{\mathrm{in}}}, \label{phisquid}
\end{eqnarray}
where $f_{\mathrm{tr}}$ is the flux-transfer ratio
\cite{claassen75JAP}.

\subsection{Design and construction of the pick-up circuit}
\label{SQUIDconstr}

The circuitry for our SQUID magnetometer is, with a few additions,
based on the principles discussed above. The niobium
\textit{dc}-SQUID sensor is part of a Conductus LTS iMAG system,
which includes a FLL circuitry to be placed just outside the
cryostat, and is connected to the SQUID controller by a hybrid
optical-electric cable. To couple a magnetic signal to the SQUID
we constructed a second-order gradiometer by winding a
$\varnothing$ $100$ $\mu$m NbTi wire on a brass coil-holder, to be
inserted in the bore of the Oxford Instruments 9 T NbTi
superconducting magnet. The gradiometer coils have $\varnothing$
$36$ mm ($a=18$ mm) and $d=9$ mm, due to the $\varnothing$ $34$ mm
vacuum can of the refrigerator that contains the sample. The leads
of the pick-up circuit, which are tightly twisted and shielded by
a Nb capillary, are screwed onto the input pads of the SQUID to
obtain a closed superconducting circuit. The input inductance of
the SQUID sensor is $L_{\mathrm{in}} = 600$ nH; since
$L_{\mathrm{leads}} \sim 200$ nH and, given the dimensions,
$L_{\mathrm{pu}} \sim L_{\mathrm{in}} + L_{\mathrm{leads}}$
already with $N=1$, it follows from Eq. (\ref{phisquid}) that the
most convenient choice of windings for the
gradiometer\footnote{Recall that $\Phi_{\mathrm{pu}} \propto N$
but $L_{\mathrm{pu}} \propto N^2$.} is 1-2-1. The mutual
inductance between SQUID ring and input coil is $\mathcal{M} = 10$
nH, which means that $f_{\mathrm{tr}} \sim 1/200$.

The FLL included in the Conductus electronics is able to
compensate for $500$ $\Phi_0$ at the SQUID ring, which means that
a maximum flux $\Phi_{\mathrm{pu}}^{\mathrm{(max)}} =
500/f_{\mathrm{tr}} \sim 10^5$ $\Phi_0$ can be picked up by the
gradiometer without saturating the system. The maximum magnetic
moment $\mu^{\mathrm{(max)}}$ is therefore [cf. Eq. (\ref{Phipu})
and Fig. \ref{gradiometer}]:
\begin{eqnarray}
\mu^{\mathrm{(max)}} \approx \frac{2a}{0.6 \mu_0}
\Phi_{\mathrm{pu}}^{\mathrm{(max)}} \sim 10^{18}\mu_B \sim 10^{-2}
\mathrm{\ emu}.
\end{eqnarray}
In order to extend the dynamic range, we have built an extra flux
transformer on the pick-up circuit, which allows to introduce a
magnetic flux from the outside to compensate for the flux induced
by the sample. By using the SQUID as a null-meter, there is the
extra advantage that no current circulates in the pick-up circuit,
thus the field at the sample is precisely the field produced by
the magnet, without the extra field that would be produced by a
current in the gradiometer. The flux transformer consists of 16
turns of NbTi wire, wound on top of 1 loop of the pick-up wire and
shielded by a closed lead box. In the same box we glued the
pick-up leads on a 100 $\Omega$ chip resistor, which is used to
locally heat the circuit above the superconducting $T_c$ and
eliminate the trapped flux. The flux transformer is fed by a
Keithley 220 current source, whereas the heater is operated by one
of the three current sources in the Leiden Cryogenics Triple
Current Source used for the refrigerator; both sources are
controlled by a computer program (see \S\ref{automation})

Despite the efforts to isolate the system from external
vibrations, the powerful pumps of the dilution refrigerator still
provide a non-negligible mechanical noise. In particular, the
Roots pump has a vibration spectrum with a lowest peak at $\sim
25$ Hz. To prevent the possible flux changes induced by such
vibrations when coupled to a magnetic field, we added an extra
low-pass filter on the pick-up circuit \cite{meschkeT} in the form
of a 17 mm long $\varnothing$ $0.3$ mm copper wire in parallel
with the pick-up leads. At $T=4.2$ K such a wire has a resistance
$R \simeq 11$ $\mu \Omega$, which together with the input
inductance $L_{\mathrm{in}} = 600$ nH of the SQUID yields a cutoff
frequency
\begin{eqnarray}
f_{l.p.} = \frac{1}{2\pi} \frac{R}{L_{\mathrm{in}}} \simeq 3
\mathrm{\ Hz}.
\end{eqnarray}
The filter is contained in a separate Pb-shielded box inserted
between the flux transformer and the SQUID. The pick-up leads and
the copper wire are contacted via Nb pads.

\begin{figure}[p]
\begin{center}
      \leavevmode
      \epsfxsize=120mm
      \epsfbox{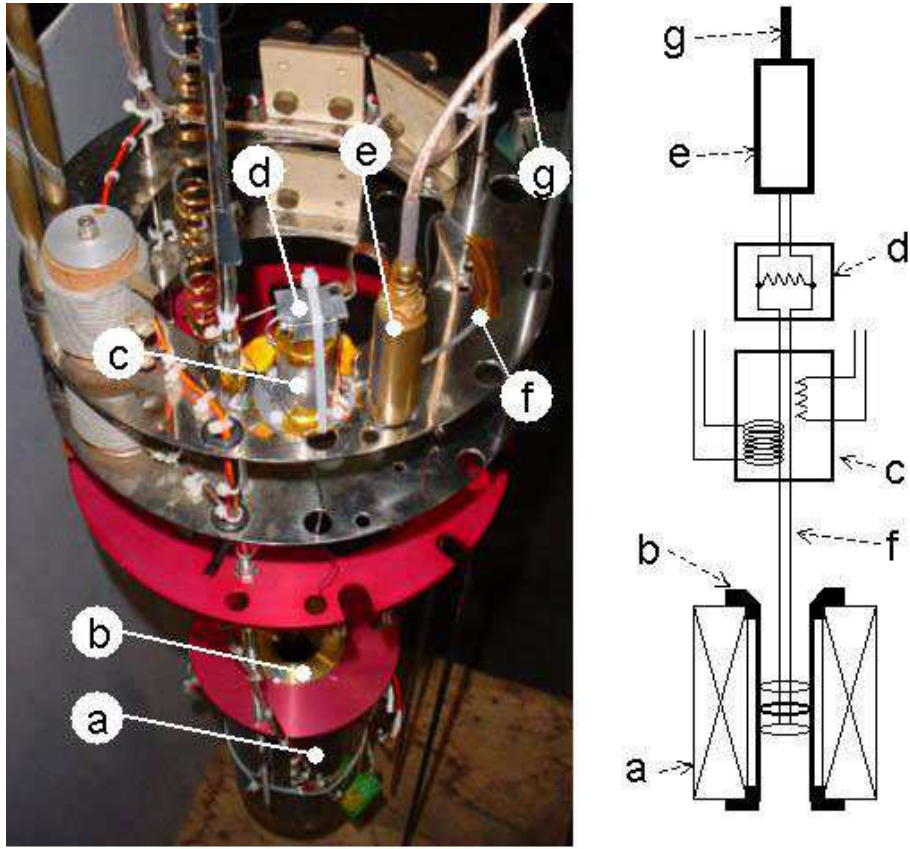}
\end{center}
\caption{\label{SQUIDcirc} Picture and scheme of the SQUID
circuitry. (a) Superconducting magnet. (b) Coil holder. (c) Flux
transformer and heater in lead shield. (d) 3 Hz lead shielded
low-pass filter. (e) SQUID sensor. (f) Pick-up leads with Nb
capillary shield. (g) SQUID cryocable.}
\end{figure}

A picture of the SQUID circuitry is shown in Fig. \ref{SQUIDcirc}.
The SQUID, the filter and the flux transformer are mounted on a
radiation shield of the magnet hanging. The whole system is
inserted in a 90 liter helium cryostat. The magnet hanging is
stabilized by triangular phosphor-bronze springs (not shown in Fig
\ref{SQUIDcirc}) that press against the inner walls of the
cryostat.

\subsection{Vertical movement} \label{verticalmov}

The movement of the sample through the gradiometer is obtained by
lifting the whole dilution refrigerator. For this purpose, the
refrigerator is fixed on a movable flange, while the top of the
cryostat is closed by a rubber bellow. On the flange we screwed
the nuts of three recirculating balls screws (SKF SN3 $20\times
5\mathrm{R}$), which allow a very smooth displacement of the nut
by turning the screw\footnote{The friction between nut and screw
is so low that, by placing the screw vertically and leaving the
nut free, the nut would start to turn and fall off the screw just
by gravity!}. The base of each screw is mounted on ball bearings
and is fitted with a gearwheel. The gearwheels are connected by a
toothed belt (Brecoflex 16 T5 / 1400) driven by a three-phase AC
servomotor (SEW DFY71S B TH 2.5 Nm) that can exert a torque up to
2.5 Nm. In this way, the rotation of the servomotor is converted
into the vertical movement of the dilution refrigerator insert.
The movement is so smooth that the consequent vibrations are
hardly perceptible and do not exceed the vibrations due to the
$^3$He pumping system.

The motor is driven by a servo-regulator (SEW MDS60A0015-503-4-00)
that can be controlled by a computer. For safety reasons an
electromagnetic brake is fitted, that blocks the motor in case of
power failure. In addition, a set of switches is mounted along one
of the pillars that support the screws: when the flange reaches
the highest or the lowest allowed position, the switches force the
motor to brake independently of the software instructions. The
maximum allowed vertical displacement is 10 cm.

\begin{figure}[p]
\begin{center}
      \leavevmode
      \epsfxsize=120mm
      \epsfbox{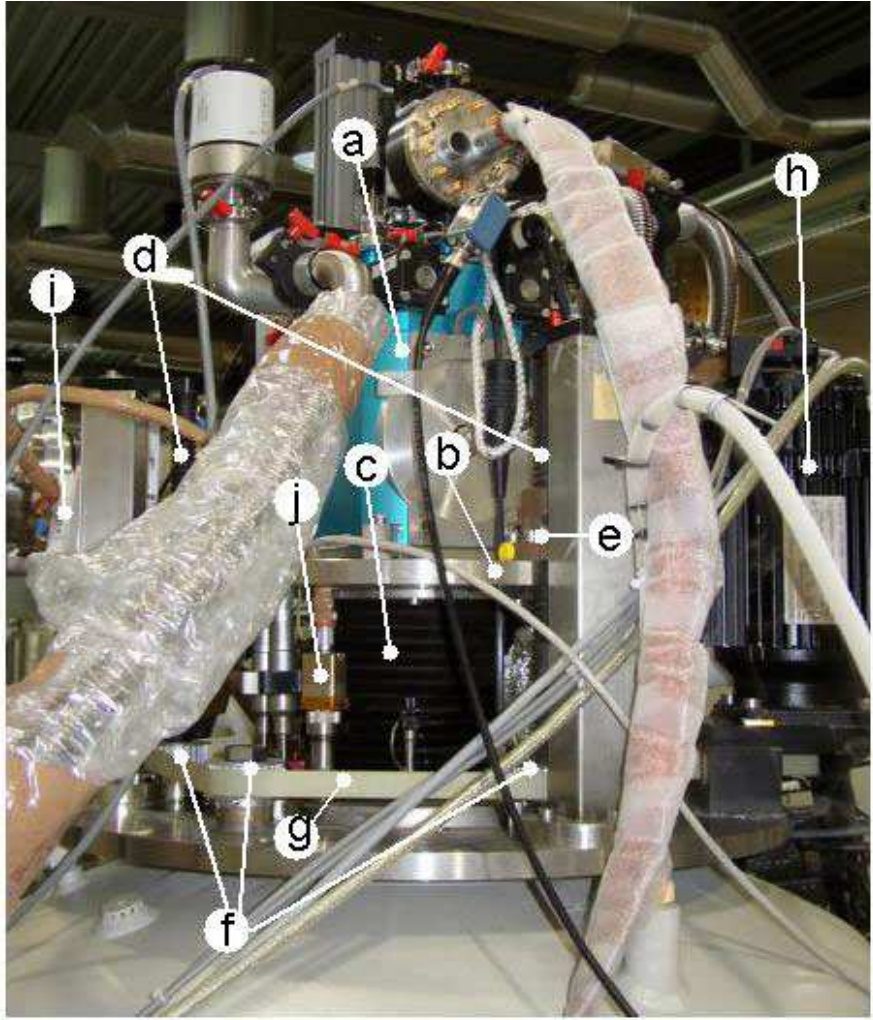}
\end{center}
\caption{\label{setuptop} Picture of the top of the cryostat, with
the elements for the vertical movement. (a) Head of the dilution
refrigerator. (b) Movable flange. (c) Rubber bellow. (d) Screws.
(e) Nut with recirculating balls. (f) Gearwheels. (g) Belt. (h)
Servomotor. (i) SQUID FLL electronics. (j) Feedthrough to the
SQUID cryocable.}
\end{figure}

A picture of the top of the cryostat with the vertical movement
elements is shown in Fig. \ref{setuptop}.

\subsection{Grounding and shielding}

The shielding of the SQUID sensor and electronics is of course a
crucial issue for the successful design of a SQUID-based
magnetometer. As mentioned in \S\ref{SQUIDconstr}, the
low-temperature parts are shielded by superconducting Pb boxes or
Nb capillaries. Strange as it may sound, the electronics outside
the cryostat posed in fact many more problems. The reason is that
the Conductus iMAG system uses beautiful hybrid optical-electric
cables for the communication between the SQUID controller and the
FLL electronics on top of the cryostat, but in the cable that
connects the FLL to the SQUID sensor, the ground is used as return
line for the signals! This means that \emph{any ground loop}
involving the SQUID electronics will \emph{completely destroy} the
functionality of the system. Obviously, all the outer metallic
parts in the system (e.g. the case of the SQUID sensor, the vacuum
feedthrough for the cryocable, etc.) are connected to the same
electrical ground, including the GPIB communication terminals.

The best way to avoid ground loops in the SQUID system would be to
ground the SQUID sensor and its cable at the cryostat (and take
care that it remains a very clean ground), and connect the
controller to an isolation transformer \cite{vleemingT}. This
method is not practicable when communication with the computer via
the GPIB bus is needed, and the same computer is connected to
another instrument that requires a common ground with the cryostat
(this is the case for the Picowatt AVS-47 resistance bridge). The
only choice is therefore to ground the controller at its power
cord and \emph{float the whole SQUID circuitry}, all the way to
the SQUID sensor inside the cryostat and the shields of the
pick-up circuit (which must be connected to the SQUID ground).
This is already a rather cumbersome operation, but it's not yet
sufficient. We found out that, in this configuration, the FLL
electronics and the room-temperature cables around it are not
enough shielded from the electromagnetic interference\footnote{We
obviously took care that the motor does not touch the cryostat
ground.} generated by the motor during the vertical movement of
the dilution refrigerator. This sort of interference does not
annihilate the functionality of the SQUID like a ground loop would
do, nor does it simply induce an increase of the instrumental
noise: the effect of bad shielding is that the SQUID system
behaves as if it were connected to a non-superconduting, inductive
circuit! It obviously took some time before we realized this,
since the inductive behavior of the pick-up system is one of the
most expectable failures, which can be due to any weakening of the
superconductivity in the circuit, for instance because of a bad
contact on the SQUID input pads. The full functionality of the
magnetometer was reached by enclosing the FLL and the
room-temperature SQUID cables into an extra copper shield,
grounded at the cryostat but separated from the SQUID ground.

\subsection{Automation} \label{automation}

The operation of the SQUID magnetometer can be programmed in a
completely automatic way by a Delphi software developed by W. G.
J. Angenent \cite{angenentT}, that integrates the SQUID
controller, the magnet power supply, the servomotor controller,
the thermometers resistance bridge and the current sources for the
flux transformer and the pick-up heater.

The software includes a proportional controller that drives the
Keithley 220 current source in order to null the SQUID voltage, in
case the user chooses the measure in feedback mode rather than
just recording the SQUID voltage. Another option is between
continuous or step-by-step vertical movement, the latter being
more convenient for the feedback measuring mode. Furthermore, the
software already contains routines for fitting the SQUID voltage
or the feedback current and extract the magnetization. By
programming a sequence of, for instance, several measurements at
constant temperature and increasing fields, one can obtain a whole
magnetization curve without need of any intervention of the user.

\begin{figure}[t]
\begin{center}
      \leavevmode
      \epsfxsize=120mm
      \epsfbox{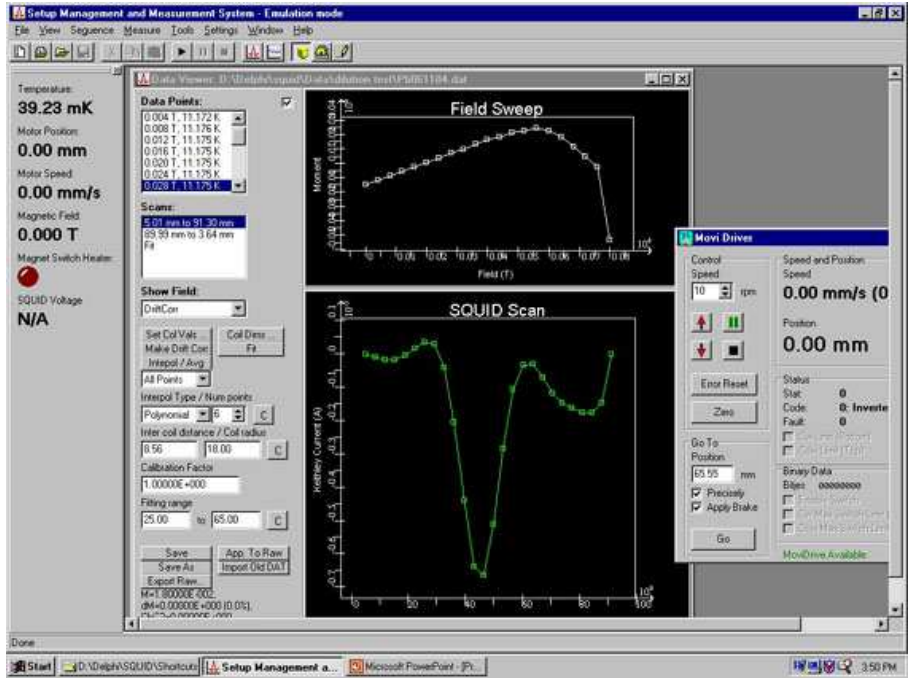}
\end{center}
\caption{\label{screenshot} A screenshot of the user interface of
the SQUID magnetometer management program.}
\end{figure}

A very user-friendly Windows interface is supplied, by which the
user can set all the parameters of the measurement, control each
single instrument, program measurement sequences and analyze the
data. A screenshot of the user interface is shown in Fig.
\ref{screenshot}.

\section{Torque magnetometer} \label{sec:torque}

A torque magnetometer (torquemeter, in brief) is the ideal
instrument to complement the SQUID on the high-field side, since
its sensitivity grows linearly with the field, and it suffers none
of the limitations due to the critical field of the
superconducting parts of a SQUID magnetometer. We constructed a
torquemeter to be installed in a different cryostat, fitted with a
18 T Oxford Instruments Nb$_3$Sn superconducting magnet and a
smaller dilution refrigerator (Oxford Kelvinox 25), which was used
previously for specific heat measurements \cite{mettesT}. Again we
took care of preserving the compatibility with other setups,
namely the calorimeter: the samples used for specific heat
experiments can be directly recycled for torque magnetometery.

\subsection{Principles of cantilever magnetometery}

A sample with magnetic moment $\vec{\mu}$ placed in a magnetic
field $\vec{B}$ experiences a torque $\vec{\mathcal{T}}$ given by:
\begin{eqnarray}
\vec{\mathcal{T}} = \vec{\mu} \times \vec{B}.
\end{eqnarray}
The torque is therefore nonzero only if $\vec{\mu}$ is not
parallel to $\vec{B}$, which can happen if the sample has an
intrinsic magnetic anisotropy and the anisotropy axis is not
aligned with the field, or in a single-crystalline isotropic
material, provided the crystal itself has a certain shape
anisotropy. For a powder isotropic sample, or a non-oriented
powder anisotropic sample, $\mathcal{T}=0$. Still, such a sample
can experience a force $\vec{F} = \vec{\nabla}(\vec{\mu} \cdot
\vec{B}) $ if the magnetic field is not uniform. For instance, the
magnetic field along the axis ($\vec{z}$) of a magnet has, to a
very good approximation, a quadratic dependence on the distance
from the field center:
\begin{eqnarray}
B(z) = B_0 (1-\mathcal{G}z^2), \label{Bvsz}
\end{eqnarray}
where $\mathcal{G}$ is a constant that only depends on the
geometry of the magnet coils. The magnetic force on a powder
sample is therefore $\vec{F} = (\mathrm{d}B/\mathrm{d}z)\vec{\mu}
= - 2\mathcal{G}zB_0 \vec{\mu}$. If the sample is mounted on a
torsion cantilever at a distance $r$ from the rotation axis, it
exerts a torque:
\begin{eqnarray}
\vec{\mathcal{T}} = - (2\mathcal{G}zB_0) \vec{r} \times \vec{\mu},
\end{eqnarray}
which produces a displacement of the extremity of the
cantilever\footnote{In reality it is a rotation, that can be
approximated as a displacement of the extremity when the rotation
angle is very small.} $\Delta z = \mathcal{T}/\mathcal{E}$, where
$\mathcal{E}$ is an elastic constant. If $\Delta z \ll z$, then
$2\mathcal{G}zr/\mathcal{E} = \mathcal{K}$ is just a constant that
depends on the cantilever and its position along the
magnet\footnote{Notice that $\mathcal{K}$ has a sign:
$\mathcal{K}>0$ above the field center, and $\mathcal{K}<0$
below.}. The magnetic moment can therefore be extracted from a
measurement of the displacement $\Delta z$:
\begin{eqnarray}
\Delta z = -\mathcal{K} B_0 \mu. \label{deltaZ}
\end{eqnarray}

\begin{figure}[t]
\begin{center}
      \leavevmode
      \epsfxsize=100mm
      \epsfbox{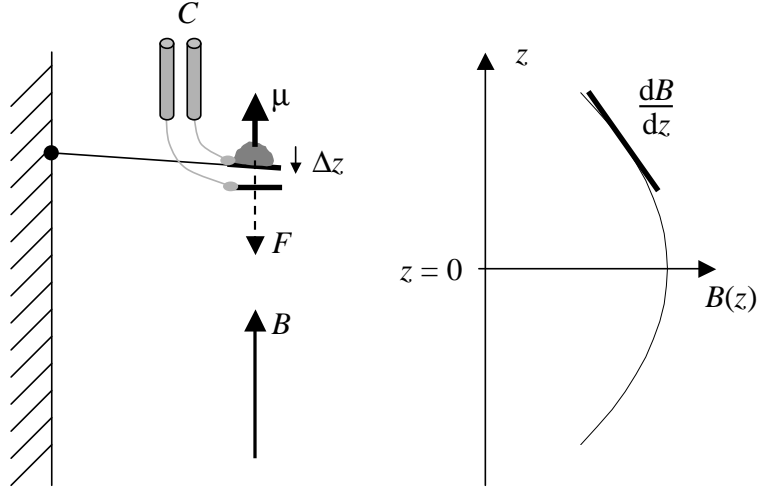}
\end{center}
\caption{\label{torquescheme} Scheme of a typical configuration
for capacitive detection of the magnetic force on a paramagnetic
isotropic sample in a field gradient. The magnitude of the field
inhomogeneity is largely exaggerated.}
\end{figure}

The deflection of the cantilever can be measured in several ways,
including piezo-resistive \cite{willemin98JAP} and optical
\cite{schaapman02APL} methods, but a simplest effective option is
to shape the extremity of the cantilever as a platelet, and
measure the changes in the capacitance formed by the platelet
facing a suitable counterelectrode
\cite{brooks87RSI,qvarford92RSI,rossel98RSI}. Calling $C_0 =
\epsilon_0 A/d_0$ the capacitance of the system at rest, where $A$
is the area of the platelet and $d_0$ the distance at rest from
the counterelectrode, the displacement $\Delta z$ produced by the
magnetic force translates into a change in the capacitance:
\begin{eqnarray}
\Delta C = C_0 \left( -\frac{\Delta z}{d_0} + \left( \frac{\Delta
z}{d_0} \right)^2 + \ldots \right). \label{deltaC}
\end{eqnarray}
For a paramagnetic sample and a cantilever placed above the field
center, Eq. (\ref{deltaZ}) simply implies that the sample is
pushed down towards the field maximum, thus if the
counterelectrode is below the cantilever, $C$ increases with
field. In particular, as long as $\Delta z \ll d_0$ so that we can
retain only the linear term in (\ref{deltaC}), we may write:
\begin{eqnarray}
\Delta C = \mathcal{K} \frac{C_0}{d_0} B \mu. \label{deltaCfin}
\end{eqnarray}
The scheme of a typical measurement configuration is shown in Fig.
\ref{torquescheme}.

Notice that, at constant $\mu$, the capacitance change (which sets
the instrumental sensitivity) grows linearly with field; this is
why torque magnetometery is essentially a high-field technique.
The magnetic moment is typically obtained by performing
field-sweep measurements of the unbalance of a capacitance bridge,
then dividing the results by the applied field. Obviously, one
should discard the data around zero field, since even the tiniest
noise would produce a diverging effect. It is not advisable to
perform temperature sweeps at constant field, since the
$T$-dependence of $C_0$ and $d_0$ would affect the results.

\subsection{Design and construction}

From the previous discussion, it is clear that there are several
possibilities to increase the sensitivity of a torquemeter: (i)
use a soft cantilever (small $\mathcal{E}$); (ii) increase the
distance $r$ of the sample from the rotation axis; (iii) build a
capacitor with large area and very small distance between the
plates; (iv) place the torquemeter far from the field center. The
fourth option must be taken with care, and in our case $z$ is
determined by the position of the three experimental slots
available in the refrigerator \cite{volokitinT}: one slot is
precisely at the center, the other two are $\sim 25$ mm above and
below. The longitudinal inhomogeneity factor [cf. Eq.
(\ref{Bvsz})] of the Nb$_3$Sn magnet used here is $\mathcal{G}
\simeq 2.8 \cdot 10^{-5}$ T/mm$^2$, which means that a torquemeter
in the upper slot experiences a field gradient
$\mathrm{d}B/\mathrm{d}z \simeq 0.035B_0$ T/mm. Contrary to the
setup described in \S\ref{sec:dilution}, in this refrigerator the
sample is in vacuum, thus it must be thermalized by the cantilever
itself.

\begin{figure}[p]
\begin{center}
      \leavevmode
      \epsfxsize=120mm
      \epsfbox{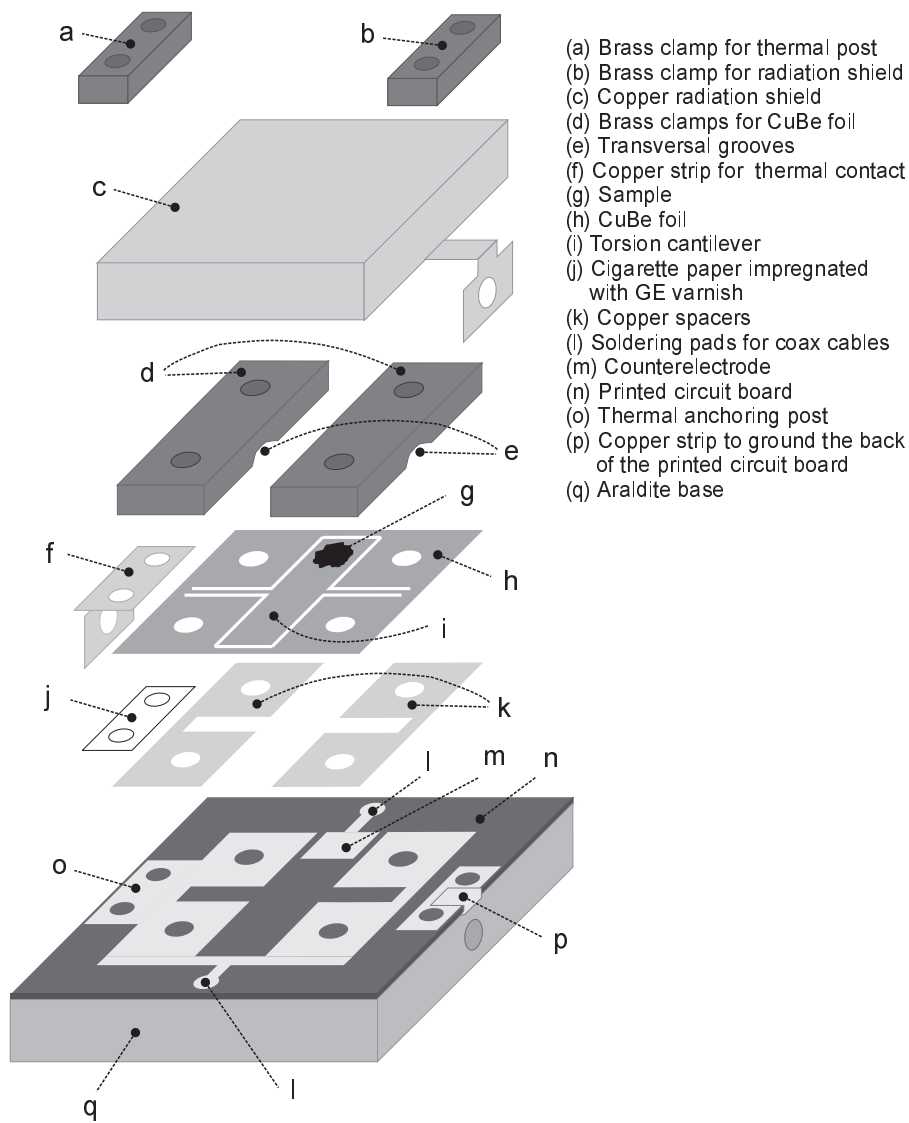}
\end{center}
\caption{\label{torquemeter} Drawing of the disassembled torque
magnetometer.}
\end{figure}

The system, shown in Fig. \ref{torquemeter}, lies on an Araldite
base, screwed laterally between two wide copper slabs that
constitute the cold finger connected to the mixing chamber (not
shown). The torquemeter is designed as a symmetric torsion
balance, to avoid the possible contribution of the platelet to the
magnetic force. The $20 \times 4$ mm central strip is held in
position by two thin arms, which constitute the elastic twistable
element. This cantilever is obtained by spark erosion on a 50
$\mu$m thick copper-beryllium (CuBe) foil, subsequently annealed
at 315 $^{\circ}$C for two hours in order to eliminate the
mechanical stress and obtain a perfectly flat foil. The sample,
mixed with Apiezon N grease for thermal contact, is placed at one
extremity of the balance, which constitutes the upper plate of the
capacitor. The lower plate is obtained from a square copper pad on
a printed circuit board. The CuBe foil is held by two brass clamps
screwed onto the base, with the interposition of 25 $\mu$m thick
copper spacers. The clamps have transversal grooves in
correspondence with the torsion arms of the cantilever. To avoid
the premature touch of the cantilever on the counterelectrode, we
typically put two spacers on top of each other. Pads with the same
shape as the spacers are drawn on the printed circuit board; their
function is to thermalize the system and to contact electrically
the CuBe foil to a coax cable soldered on the board. Since the
cantilever constitutes one of the plates of the capacitor and may
not be connected to the electrical ground, a L-shaped copper strip
is screwed on top of a thermal anchoring post, obtained from teh
printed circuit board. A cigarette paper impregnated with GE
varnish is inserted between the copper strip and the anchoring
post, to provide good thermal contact while maintaining electrical
insulation. The copper strip is then screwed laterally between the
Araldite base and the cold finger. To shield the sample from
thermal radiation and electromagnetic noise we cover the system
with a copper box, coated with thin Kapton tape, screwed laterally
to the cold finger and therefore also connected to the electrical
ground. To complete the shielding, a small copper strip connects
the front to the back side of the printed circuit board, which is
completely copper-plated.

Two coaxial cables are soldered to copper pads on the printed
circuit board, one connected to the counterelectrode, the other to
the cantilever. We can therefore measure the capacitance between
cantilever and counterelectrode by connecting the coaxials to a
General Radio 1615-A capacitance bridge. A Stanford SR830 lock-in
amplifier provides the 5 V, $\sim 1$ kHz excitation, and measures
the unbalance of the bridge. The measurements are controlled by a
LabView program developed by W. G. J. Angenent. Before starting a
field-sweep, the temperature is set to the desired value, then the
voltage due to the unbalance of the capacitance bridge at
zero-field is accurately measured, in order to subtract it from
the data afterwards. While sweeping the magnetic field, typically
at a rate $\mathrm{d}B/\mathrm{d}t = 0.05 - 0.1$ T/min, the
lock-in voltage $V_L$ is measured and divided by $B$ after
subtraction of the zero-field offset. This yields the sample
magnetization $M(B) \propto \mu(B)$, since $V_L \propto \Delta C$
[cf. Eq. (\ref{deltaCfin})]. The exact conversion factor between
the lock-in voltage and the magnetization is very difficult to
obtain in a reliable and reproducible way\footnote{It depends for
instance on the precise position of the sample on the platelet
($r$), the distance at rest between the capacitor plates ($d_0$),
etc.}; we shall therefore express $M$ in the ``electrical units''
$\mu$V/T.

\subsection{Performance}

An example of the performance of our torquemeter is given by the
experiment on a $\sim 0.25$ mg Cerium Magnesium Nitrate (CMN)
sample, which is an ideal Curie paramagnet down to at least $T
\sim 10$ mK.

\begin{figure}[t]
\begin{center}
      \leavevmode
      \epsfxsize=130mm
      \epsfbox{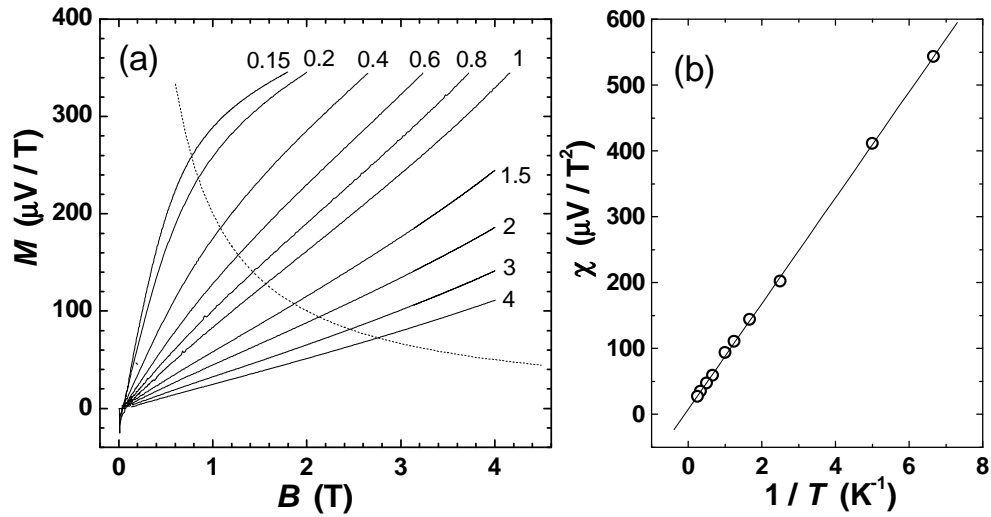}
\end{center}
\caption{\label{torqueCMN} (a) Magnetization of a 0.25 mg CMN
sample, at the indicated temperatures (in Kelvin). Dashed line:
(emprical) threshold of the linear response region. (b) Static
susceptibility as a function of the inverse temperature, fitted by
the Curie law.}
\end{figure}

The measured magnetization, shown in Fig. \ref{torqueCMN}(a),
exhibits the expected Brillouin behavior up to a certain
threshold, where the non-linear terms in (\ref{deltaC}) start to
be important. The condition of linear response $\Delta C \propto
\Delta z$ is obeyed below a certain maximum displacement, i.e. a
horizontal line in the $V_L - B$ plane, which translates into a
hyperbola in the $M - B$ plane.

By fitting the initial linear part of $M(B)$ at different
temperatures, we obtain the static susceptibility $\chi(T)$. As
shown in Fig. \ref{torqueCMN}(b), the measured susceptibility
obeys indeed the Curie law, $\chi \propto 1/T$, and confirms the
proper thermalization of the sample.

Knowing that the magnetization of CMN is $\simeq 3.3$ emu/g at $B
= 1$ T and $T = 1$ K (as deduced from Fig. 14 in \cite{paulsenB}),
we can estimate that the linear response region is roughly $\mu
\cdot B < 2 \cdot 10^{-3}$ emu$\cdot$T. The instrumental noise,
typically $\sim 10$ nV, translates into an equivalent noise of
$\sim 10^{-7}$ emu in the magnetic moment at $B=1$ T. By
increasing the field to e.g. 10 T, another order of magnitude is
gained in the equivalent magnetic noise, but the maximum
measurable signal is also 10 times smaller. Therefore, the
appropriate amount of sample must be chosen with care, as a
function of the range of $B$ and $\mu$ of interest.

\def\baselinestretch{1}
\chapter{Theoretical aspects of molecular magnetism}

This chapter contains a selection of theoretical approaches
necessary to understand the physics of molecular magnets and their
interaction with the environment. After an introduction to the
relevance of this subject for the extrapolation of quantum
mechanics to the large scale, we proceed to a ``top-down''
physical description, starting from the model for a single giant
spin in its crystalline environment, then illustrating the effect
of a magnetic field and the coupling to nuclear spins and the
mutual coupling via dipolar fields. We then discuss the standard
theory of the nuclear spin dynamics, mainly to show that such a
treatment is no longer adequate in the presence of macroscopic
quantum tunneling: the Prokof'ev-Stamp theory is the necessary
approach to a unified description of a giant quantum spin coupled
to a spin bath. One of the most crucial outcomes of the present
work is that even the Prokof'ev-Stamp theory needs an extension in
order to account for our experimental results presented in chapter
IV. We shall therefore come back to this issue in an extra
theoretical section at the end of that chapter.

\section{Macroscopic quantum phenomena}

Besides the perspective of possible applications (magnetic
recording media, spin qubits, etc.), single-molecule magnets are
very attractive systems to study the observability of quantum
phenomena at the macroscopic scale. The motivation behind this
interest dates back to the formulation of the Schr\"{o}dinger cat
paradox \cite{schrodinger35} and the so-called ``measurement
problem'' in quantum mechanics. In fact, the ``weak point'' of
quantum mechanics in its present formulation is the rather
artificial distinction between the (microscopic) quantum system
and the (macroscopic) measurement apparatus, which is supposed to
obey the laws of classical physics, and whose action forces the
wave function of the quantum system to be projected (collapse)
onto one of its eigenstates. Nothing is specified about where the
border between a microscopic (quantum) and a macroscopic
(classical) system lies, neither how the collapse of the wave
function actually takes place. The proposals to solve these
problems range from the most pragmatic idea of decoherence
\cite{zurek91PT}, all the way to theories that effectively add
extra postulates to the standard quantum mechanics
\cite{ghirardi86PRD,pearle89PRA}. Before one even attempts to use
the experimental observations on SMMs to elucidate the fundamental
problems mentioned above, it is essential to quantify the degree
of ``Schr\"{o}dinger cattiness'' of the quantum states being
observed. Here we briefly review the approach of Leggett
\cite{leggett80SPTP,leggett02JPCM}, who usefully introduces two
parameters to quantify the macroscopicity of a quantum
superposition of states: the ``extensive difference'' and the
``disconnectivity''.

\subsection{Extensive difference}

Let us consider a system in a quantum superposition of states,
i.e. whose wave function $|\psi\rangle$ can be expressed as a
linear combination of two wave functions:
\begin{eqnarray}
|\psi\rangle = a|\psi_a\rangle + b|\psi_b\rangle,
\end{eqnarray}
with the assumption that $|\psi_a\rangle$ and $|\psi_b\rangle$
represent states of the system which are by some reasonable
criterion ``classically different''. We may therefore characterize
the two branches of the superposition by some extensive quantities
$\{i\}$ (e.g. charge, magnetic moment, position of the center of
mass, etc.) which should be considerably different in the states
$|\psi_a\rangle$ and $|\psi_b\rangle$. Next, we define $\Lambda_i$
as the difference between the values of the extensive quantity $i$
in the two branches of the superposition, divided by a ``typical
value'' of $i$ at the atomic scale (e.g. 1 electron charge, 1 Bohr
magneton, 1 \AA ngstrom, etc.). The ``extensive difference''
$\Lambda$ of the quantum superposition is then the maximum value
assumed among the $\Lambda_i$'s.

This sounds at first as a very natural definition of macroscopic
distinctness, but in fact it's easy to find examples where
$\Lambda$ is a very large number, although the system is not what
one would like to call ``macroscopic''. Think of a neutron passing
through an interferometer with arms e.g. 10 cm apart: one finds
$\Lambda \sim 10^9$, although this is clearly not the sort of
examples we are looking for as a challenge to the interpretations
of quantum mechanics.

\subsection{Disconnectivity}

Leggett introduces therefore another parameter, $\mathcal{D}$,
called ``disconnectivity''. The precise definition can be found in
Ref. \cite{leggett80SPTP}, but for our purpose it is sufficient to
say that $\mathcal{D}$ is the number of particles that have
substantially different behavior in the states $|\psi_a\rangle$
and $|\psi_b\rangle$. More precisely, $\mathcal{D}$ is the number
of particle correlations that must be measured in order to
distinguish the coherent superposition $|\psi\rangle$ from a
statistical mixture of $|\psi_a\rangle$ and $|\psi_b\rangle$. For
instance, a system of $N$ identical particles prepared in a state
like
\begin{eqnarray}
|\psi\rangle = a|\psi_a\rangle^N + b|\psi_b\rangle^N
\end{eqnarray}
has $\mathcal{D} = N$, since $N$ particles are simultaneously
superimposed, whereas a state
\begin{eqnarray}
|\psi\rangle = (a|\psi_a\rangle + b|\psi_b\rangle)^N
\label{psisuper}
\end{eqnarray}
has $\mathcal{D} = 1$, being just the product of one-particle
superpositions. The above-mentioned neutron in a diffractometer
has obviously $\mathcal{D} = 1$, thereby reestablishing the fact
that it does not constitute a ``true'' macroscopic quantum state.
An interesting example is the one of a superconductor: it's easy
to realize that $\mathcal{D} = 2$, since the standard BCS wave
function is a product of two-particle correlations. The ``charge
qubit'' \cite{nakamura99N}, which is one of the most exciting
developments in solid-state quantum devices of the last few years,
again has $\mathcal{D} = 2$ since it is based on the superposition
of a Cooper pair (2 particles) being inside or outside a
superconducting island; the fact that the device itself has a size
$\sim 1$ $\mu$m does not automatically make it relevant for the
problem of quantum mechanics at the large scale!

The observation of high-$\mathcal{D}$ quantum superpositions has
been realized only very recently: for instance, by diffraction of
C$_{60}$ molecules \cite{arndt99N} a value of $\mathcal{D} = 1048$
has been achieved. Even higher values have been reached by the
``flux-qubit'' \cite{chiorescu03S,friedman00N}, which features the
superposition of counter-rotating supercurrents in a SQUID loop:
in this case all the current-carrying Cooper pairs (i.e. those
within the London penetration depth from the surface) are behaving
differently in the two branches, yielding $\mathcal{D} > 10^6$ !

\subsection{Macroscopicity of a single-molecule magnet}

We can now apply the concepts defined above to the case of a SMM.
In general, we will be interested in phenomena that arise from the
quantum superposition of two different projections of the total
spin along the $z$ axis . For example, for Mn$_{12}$-ac the total
spin $S=10$ can give rise to a total magnetic moment of 20 $\mu_B$
pointing along $+\vec{z}$ or $-\vec{z}$. A superposition of such
two states would be characterized by an extensive difference
$\Lambda = 40$. As for the disconnectivity, we anticipate (see
\S\ref{paramMn12}) that the total spin of the molecule is the
result of the ferrimagnetic coupling of 8 Mn$^{3+}$ ions with spin
$s=2$, i.e. 4 electron spins per ion, and 4 Mn$^{4+}$ ions with
$s=3/2$ (3 electrons). In total, the number of electrons having
different states when $\vec{S}$ is along $+\vec{z}$ or $-\vec{z}$
is $\mathcal{D} = 4 \times 8 + 3 \times 4 = 44$. Although it
doesn't reach the macroscopicity of a fullerene molecule or a
SQUID qubit, a SMM is therefore substantially more macroscopic
that an atomic-size quantum system.

\section{Effective Spin Hamiltonian}

The magnetic properties of the single-molecule magnets
investigated in this thesis are determined in the first place by
the net electron spin of each single ion, arising from the partial
filling of $3d$ shells. Because of the strong crystal field
effects, the electron angular momentum is quenched, so that no
orbital contribution needs to be included. Furthermore, the
electron spins within a cluster are magnetically coupled by
superexchange interactions, typically due to the orbital overlap
through oxygen bridges. In this way, one obtains a magnetic ground
state which can be characterized by a high value of the total
cluster spin $S$, like in the case of Mn$_{12}$-ac and Mn$_6$, and
is separated from other spin states by an energy determined by the
strength of the intracluster superexchange couplings. As long as
the temperature is kept much lower than the energy separation
between the ground and the first excited total spin state, it is
very convenient to adopt the effective spin Hamiltonian
description, which means that the cluster is treated as an object
characterized by just the number of energy levels compatible with
the total spin ground state, i.e. $2S+1$. The resulting energy
levels can be further split by the effect of crystal field
anisotropy or couplings with magnetic fields. A typical effective
spin Hamiltonian to describe the crystal field anisotropy is the
following:
\begin{eqnarray}
\mathcal{H}_{\mathrm{CFA}}=-DS_z^2 - B S_z^4 + E(S_x^2 - S_y^2) +
C(S_+^4 + S_-^4). \label{hamiltonianCFA}
\end{eqnarray}
\noindent In the next subsections we shall discuss which
predictions about the physical behavior of a single-molecule
magnet can be made on basis of the above Hamiltonian.

\subsection{Superparamagnetic blocking}

The terms $-DS_z^2$ and $-B S_z^4$ (with $|B| \ll |D|$, typically)
represent uniaxial anisotropies: if $D>0$, then $z$ is the easy
axis of magnetization. In manganese-based clusters the uniaxial
anisotropy typically arises from the Jahn-Teller distortion of the
coordination octahedra at Mn$^{3+}$ sites; the total cluster
anisotropy is then obtained as the vector sum of the single-ion
anisotropies \cite{benciniB}. Notice that by considering only
these two terms, $\mathcal{H}_{\mathrm{CFA}}$ commutes with the
$S_z$ operator, thus the eigenstates $|m\rangle$ of $S_z$ are also
exact eigenstates of $\mathcal{H}_{\mathrm{CFA}}$. The energy
levels scheme, as shown in the left panel of Fig.
\ref{eigenstates}, can be regarded as a set of doublets of
degenerate states, $|+m\rangle$ and $|-m\rangle$, with energies
$E_{+m}=E_{-m}$, each state being localized on the left or right
side of the anisotropy barrier of height $U$. At temperatures
comparable with the barrier height, thermally activated
transitions from one side to the other are very fast, and the
molecule behaves as a high-spin paramagnet. Since the high spin
results from several ions behaving collectively as a single-domain
particle, the system is called superparamagnet \cite{bean59JAP}.
The relaxation rate by thermal activation, $\tau^{-1}$, depends on
temperature according to the Arrhenius law:
\begin{eqnarray}
\tau^{-1} = \tau_0^{-1} \exp(-U/k_B T).
\end{eqnarray}
At sufficiently low temperatures, $\tau^{-1}$ may become
exceedingly long and give rise to hysteresis in the magnetization
loops \cite{sessoli93N} and the appearance of a
frequency-dependent dissipation peak in the
\textit{ac}-susceptibility \cite{caneschi91JACS}. The temperature,
$T_B$, at which the cluster spin can no longer follow an external
driving field is called blocking temperature, and obviously
depends on the timescale relevant for the experiment in question.

\subsection{Spin quantum tunneling}

\begin{figure}[t]
\begin{center}
      \leavevmode
      \epsfxsize=100mm
      \epsfbox{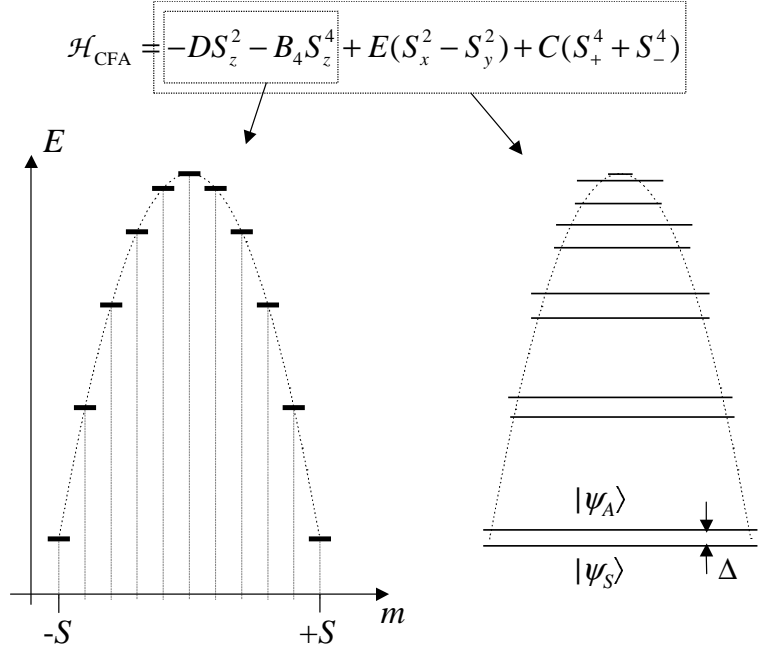}
\end{center}
\caption{\label{eigenstates} Sketch of the energy levels scheme
for the Hamiltonian (\ref{hamiltonianCFA}) when considering only
the diagonal terms (left) or including the non-diagonal ones
(right). The magnitude of the tunneling splittings is largely
exaggerated and even not to relative scale.}
\end{figure}

The quantum aspects of SMMs are encountered by considering
non-diagonal anisotropy terms like $E(S_x^2 - S_y^2)$, which
describes a hard-axis anisotropy caused by a rhombic distortion,
and $C(S_+^4 + S_-^4)$ which is the lowest-order non-diagonal term
allowed by tetragonal symmetry. These terms do not commute with
$S_z$, so the eigenstates $|\psi\rangle$ of the complete
$\mathcal{H}_{\mathrm{CFA}}$ are no longer a set of doublets of
localized states (Fig. \ref{eigenstates}). Expressing
$|\psi\rangle$ in the basis of the eigenstates $|m\rangle$ of
$S_z$:
\begin{eqnarray}
|\psi\rangle = \sum_m c_m |m\rangle, & & m= -S \ldots +S,
\end{eqnarray}

\noindent one finds that each doublet now consists of a state
$|\psi _S\rangle$ with symmetric coefficients, $c_{+m} = c_{-m}$,
and an antisymmetric state $|\psi_A\rangle $ with $c_{+m} =
-c_{-m}$, separated by an energy gap $\Delta = E_A - E_S$. In
particular for the ground doublet, one finds to a very good
approximation:
\begin{eqnarray}
|\psi_S\rangle &=& \frac{1}{\sqrt{2}} (|+S\rangle + |-S\rangle) \nonumber \\
|\psi_A\rangle &=& \frac{1}{\sqrt{2}} (|+S\rangle - |-S\rangle).
\end{eqnarray}
\noindent This means that a state localized on one side of the
barrier, e.g. $|+S\rangle$, must now be expressed as a
superposition of the actual eigenstates:
\begin{eqnarray}
|+S\rangle = \frac{1}{\sqrt{2}} (|\psi _S\rangle +
|\psi_A\rangle).
\end{eqnarray}
It is clear from basic quantum mechanics \cite{cohen} that if one
would prepare the system at $t=0$ in such a state,
$|\psi(t=0)\rangle = |+S\rangle$, the time evolution should obey
the Schr\"odinger equation
\begin{eqnarray}
|\psi(t)\rangle = \frac{1}{\sqrt{2}} e^{-i\frac{E_S+E_A}{2
\hbar}t}(|\psi _S\rangle e^{+i\frac{\Delta}{2 \hbar}t} +
|\psi_A\rangle e^{-i\frac{\Delta}{2 \hbar}t}). \label{tunnelosc}
\end{eqnarray}
This is equivalent to having coherent oscillations of the cluster
spin between the states $|+S\rangle$ and $|-S\rangle$ with
frequency $\omega_T = \Delta / \hbar$, which implies that the spin
is tunneling back and forth through the anisotropy barrier. For
this reason the energy gap $\Delta$ is called tunneling splitting.

\subsection{Coupling with external fields: coherent and incoherent
tunneling}\label{coupling}

It must be stressed that, in real SMMs, the tunneling splitting
(particularly the one of the ground doublet) is a very small
quantity when produced by non-diagonal anisotropy terms only. As
we shall discuss below, the coupling with environmental degrees of
freedom is many orders of magnitude larger than $\Delta$, which
basically eliminates the possibility to observe coherent tunneling
oscillations as described in Eq. (\ref{tunnelosc}). One of the
great advantages of single-molecules magnets as systems to study
macroscopic quantum effects, is that the parameters of the spin
Hamiltonian, and thereby the whole physical properties, can be
easily manipulated by applying an external magnetic field
$\vec{B}$. This introduces an extra term in the effective spin
Hamiltonian:
\begin{eqnarray}
\mathcal{H} = \mathcal{H}_{CFA} - g \mu_B \vec{S} \cdot \vec{B}.
\label{hamiltonianB}
\end{eqnarray}
By choosing the direction of $\vec{B}$ with respect to the
anisotropy axis $\vec{z}$, we have the freedom to either introduce
extra non-diagonal terms ($\vec{B} \perp \vec{z}$) that enhance
the quantum behavior, or to destroy the symmetry of the energy
level scheme ($\vec{B}
\parallel \vec{z}$) and thereby reduce the coupling between states on opposite
sides of the anisotropy barrier.

\begin{figure}[t]
\begin{center}
      \leavevmode
      \epsfxsize=100mm
      \epsfbox{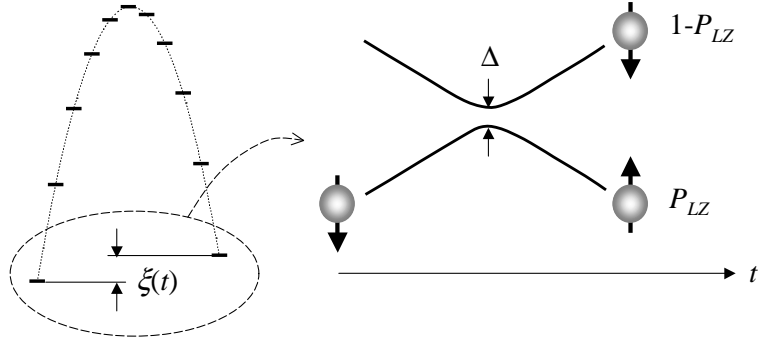}
\end{center}
\caption{\label{FigLZ} If the bias $\xi(t)$ crosses the tunneling
resonance, the cluster spin may tunnel through the anisotropy
barrier with a probability given by Eq. (\ref{PLZ}). The drawing
on the right represents the time evolution of a doublet of
electron spin levels while crossing the resonance.}
\end{figure}

Very important for the present discussion is the case where a
longitudinal magnetic field $B_z (t)$ fluctuates in time and
produces a bias, $\xi(t) = 2 g \mu_B B_z (t) S$, on the cluster
spin doublets. If $\xi(t)$ spans a range much larger than $\Delta$
but crosses several times through the tunneling resonance, one
finds that at each crossing there is a nonzero probability for the
cluster spin to tunnel through the barrier, but there is no
correlation between subsequent tunneling events; therefore this is
called ``incoherent tunneling''. The probability $P_{\mathrm{LZ}}$
of a single tunneling event upon crossing through the tunneling
resonance can be calculated with the Landau-Zener formula
\cite{zener32}:
\begin{eqnarray}
P_{\mathrm{LZ}} = 1 - \exp\left(\frac{- \pi \Delta ^2}{2 \hbar
\mathrm{d} \xi (t)/ \mathrm{d} t}\right). \label{PLZ}
\end{eqnarray}
A sketch of a Landau-Zener transition is given in Fig.
\ref{FigLZ}. This phenomenon has been successfully exploited as a
way to extract the tunneling splitting \cite{wernsdorfer99S,
wernsdorfer00JAP} by sweeping an external magnetic field through
the tunneling resonance. As will be discussed below, the
Landau-Zener formalism can also be used to describe the coupling
between a single cluster spin and its environment, since very
often the effect of environmental fluctuations can be treated as
an effective magnetic field that couples to the cluster spin.

\begin{figure}[t]
\begin{center}
      \leavevmode
      \epsfxsize=100mm
      \epsfbox{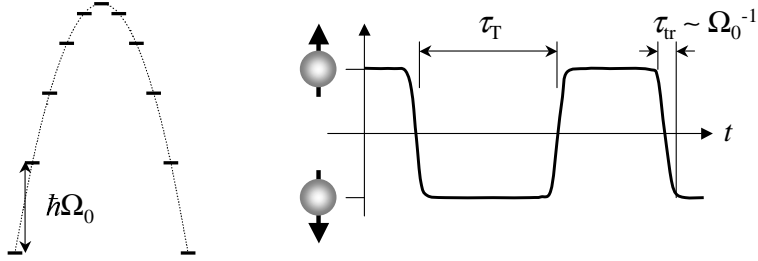}
\end{center}
\caption{\label{FigtauTR} Illustration of the relationship between
tunneling traversal time $\tau_{\mathrm{tr}}$, bounce frequency
$\Omega_0$ and tunneling interval $\tau_T$ for incoherent
tunneling of the cluster spin.}
\end{figure}

In the presence of incoherent tunneling fluctuations, it is
essential to distinguish between the average time interval
$\tau_T$ between subsequent uncorrelated tunneling events, and the
so-called tunneling traversal time $\tau_{\mathrm{tr}}$, which is
in our case the timescale for the reversal of the cluster spin. A
detailed analysis of the tunneling traversal time constitutes a
topic on itself \cite{maccoll32PR,mullen89PRL,hauge89RMP}, but for
our purposes it is sufficient to estimate the order of magnitude
of $\tau_{\mathrm{tr}}$. In the general framework of the
semiclassical instanton technique
\cite{chudnovsky88PRL,chudnovskyB}, $\tau_{\mathrm{tr}}$ is
related to the inverse of the ``bounce frequency'' or ``attempt
frequency'' $\Omega_0$, i.e. the frequency of the small
oscillations at the bottom of each potential well. In SMMs $\hbar
\Omega_0$ is of the order of the energy separation between the two
lowest electron spin doublets, as shown in Fig. \ref{FigtauTR}.
For instance in Mn$_{12}$-ac we have $\hbar \Omega_0 \sim 10$ K,
thus $\tau_{\mathrm{tr}} \sim \Omega_0^{-1} \sim 10^{-12}$ s. This
means that the tunneling events take place in a virtually
instantaneous way, as compared to the timescale of all other
relevant phenomena.

\begin{figure}[t]
\begin{center}
      \leavevmode
      \epsfxsize=90mm
      \epsfbox{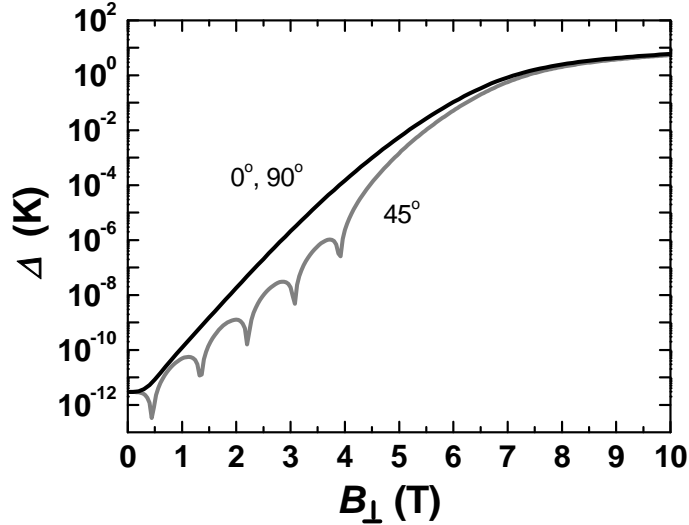}
\end{center}
\caption{\label{FigDeltavsBx} Perpendicular field dependence of
the tunneling splitting of the ground doublet in Mn$_{12}$-ac, as
obtained by numerical diagonalization of the Hamiltonian
(\ref{hamiltonianMn12}). The two lines refer to $\vec{B}_{\perp}$
parallel to the $x$ or $y$ crystallographic axes, or oriented at
$45^{\circ}$ between them.}
\end{figure}

In order to push the system into a regime where coherent tunneling
may take place, it is necessary to increase $\Delta$ above the
strength of the coupling with environmental degrees of freedom.
This can be done by applying a magnetic field $B_{\perp}$
perpendicular to the anisotropy axis. Although sophisticated
calculations of $\Delta(B_{\perp})$ have been proposed
\cite{enz86JPC,hemmen86PB}, it is often more practical to obtain
$\Delta(B_{\perp})$ directly from the numerical diagonalization of
the spin Hamiltonian. An example of the resulting
$\Delta(B_{\perp})$ in the case of Mn$_{12}$-ac is shown in Fig.
\ref{FigDeltavsBx}. A simplified expression for
$\Delta(B_{\perp})$, based on the Hamiltonian $\mathcal{H} =
-DS_z^2 - g \mu_B \vec{B_{\perp}} \cdot \vec{S}$, has been
obtained in the limit $g \mu_B B_{\perp} \ll DS$ and $S \gg 1$ by
Korenblit and Shender \cite{korenblit78SPJETP}:
\begin{eqnarray}
\Delta(B_{\perp}) \approx \frac{4}{\sqrt{\pi}} DS^{3/2} \left(
\frac{e g \mu_B B_{\perp}}{4DS} \right)^{2S}, \label{DeltavsBx}
\end{eqnarray}
where $e$ is the base of the natural logarithm. The more general
expression for the splitting $\Delta_m$ of any doublet of levels
$|+m\rangle$, $|-m\rangle$ is \cite{garanin91JPA}:
\begin{eqnarray}
\Delta_m(B_{\perp}) \approx \frac{2D}{[(2m-1)!]^2}
\frac{(S+m)!}{(S-m)!} \left( \frac{g \mu_B B_{\perp}}{2D}
\right)^{2m}. \label{DeltamvsBx}
\end{eqnarray}

It clearly appears that the simple application of a perpendicular
field has an enormous influence on $\Delta$, which allows to study
the quantum dynamics of SMMs in different regimes.

From Eq. (\ref{DeltamvsBx}) it is also clear that, since $g \mu_B
B_{\perp} \ll DS$, the tunneling splitting increases drastically
(exponentially, in fact) when higher excited doublets (i.e. with
lower $|m|$) are considered\footnote{For the case of a SMM in zero
external field, the role of $B_{\perp}$ in (\ref{DeltamvsBx}) is
taken by the non-diagonal crystal-field anisotropy terms and by
the transverse component of the dipolar field from neighboring
molecules. By condensing such effects into an ``equivalent
$B_{\perp}$'' of constant value, it is easy to see that $\Delta_m
\propto f(m) \cdot (const.)^{2m}$}. Therefore, in the presence of
thermal excitations, tunneling can more easily proceed through
excited doublets than through the ground state. This remains true
in zero external field, although the tunneling splitting is
produced by non-diagonal anisotropy terms. The details of the
mechanism of thermally-assisted tunneling have generated a vast
literature
\cite{hartmann96IJMPB,luis97PRB,garanin97PRB,fort98PRL,leuenberger00PRB},
since a large majority of experiments on SMMs have been carried
out in the temperature regime $T > 1.5$ K, where tunneling through
excited states is essential. The transition to pure ground state
tunneling happens indeed at temperatures that, at a first sight,
appear surprisingly low. The point is that, although the Boltzmann
factors for the population of excited doublets are exponentially
small, $\Delta_m$ grows exponentially upon excitations to higher
doublets (\ref{DeltamvsBx}), so the competition between the two
exponents cannot be treated in a trivial way \cite{stamp04CP}.
Since the present work is focused on the ultra-low temperature
regime, the topic of thermally assisted tunneling will be touched
upon only marginally.

\subsection{Parameters for Mn$_{12}$-ac} \label{paramMn12}

\begin{figure}[t]
\begin{center}
      \leavevmode
      \epsfxsize=130mm
      \epsfbox{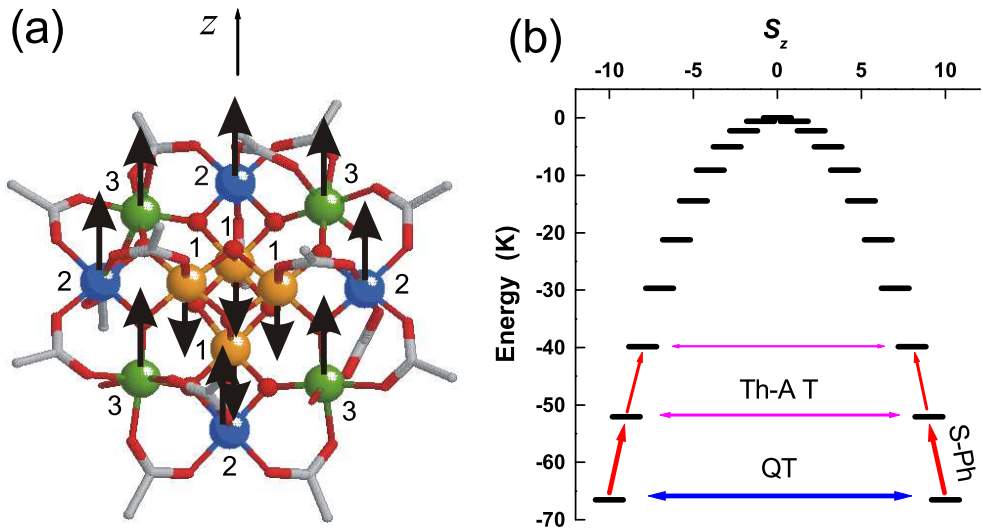}
\end{center}
\caption{\label{structure} (a) Structure of the Mn$_{12}$-ac
cluster, with the labelling of the three inequivalent Mn sites as
described in the text. (b) Energy level scheme for the electron
spin as obtained from the Hamiltonian (\ref{hamiltonianMn12}),
retaining only the terms diagonal is $S_z$. The non-diagonal terms
allow transitions between states on opposite sides of the
anisotropy barrier by means of quantum tunneling (QT). In the
presence of intrawell transitions induced by thermal excitation,
thermally assisted quantum tunneling (Th-A T) between excited
doublets can also take place .}
\end{figure}

The structure of the
[Mn$_{12}$O$_{12}$(O$_{2}$CMe)$_{16}$(H$_{2}$O)$_{4}$]
(Mn$_{12}$-ac) cluster \cite{lis80AC} (Fig. \ref{structure})
contains a core of 4 Mn$^{4+}$ ions with electron spin $s = 3/2$,
which we shall denote as Mn$^{(1)}$, and 8 Mn$^{3+}$ ions ($s =
2$) on two inequivalent crystallographic sites, Mn$^{(2)}$ and
Mn$^{(3)}$. The electron spins are coupled by mutual superexchange
interactions, the strongest being the antiferromagnetic
interaction between Mn$^{(1)}$ and  Mn$^{(2)}$
\cite{sessoli93JACS}. The molecules crystallize in a tetragonal
structure with lattice parameters $a=b=17.319$ \AA and $c=12.388$
\AA. Numerical studies have shown the influence of the magnitude
and sign of the exchange constants on the precise energy level
structure of the cluster \cite{raghu01PRB}, and an attempt has
been made to verify the assignment of the exchange interactions by
a spin-wave analysis of the nuclear spin-lattice relaxation rate
in the thermally activated regime \cite{yamamoto02PRL}. The ground
state of the molecule has a total electron spin $S=10$ and is
separated from the first excited manifold $S=9$ by $\sim 58$ K,
according to neutron scattering experiments \cite{hennion97PRB}.
For the temperature range of interest in the present work ($T<2$
K), we may therefore describe the electron spin of the cluster by
means of the spin Hamiltonian in the $S=10$ manifold:
\begin{eqnarray}
\mathcal{H}=-DS_z^2 - BS_z^4 + C(S_+^4 + S_-^4) + \mu_B \vec{B}
\cdot \mathbf{g} \cdot \vec{S}. \label{hamiltonianMn12}
\end{eqnarray}
We adopt the parameter values $D = 0.548$ K, $B = 1.17\times
10^{-3}$ K and $C = 2.2\times 10^{-5}$ K as obtained by neutron
scattering data \cite{mirebeau99PRL}, and for the $\mathbf{g}$
tensor the values $g_{\parallel} = 1.93$ and $g_{\perp} = 1.96$
from high-frequency EPR\footnote{The anisotropy parameters
obtained by EPR seem to depend on the magnetic field range used in
the experiment \cite{barra97PRB,hill98PRL}, whereas neutron
scattering is a zero-field experiment, and does not require
assumptions on the $\mathbf{g}$ tensor in order to fit the data.}
\cite{barra97PRB}. The uniaxial anisotropy terms $-DS_z^2$ and $-
BS_z^4$ can be attributed to the single-ion anisotropy of the
Mn$^{3+}$ ions\cite{barra97PRB}, which is due to the crystal field
effects resulting in the Jahn-Teller distortions of the
coordination octahedra, where the elongation axes are
approximately parallel to the $\hat{c}$-axis of the crystal. The
energy levels scheme is then a series of doublets of degenerate
states which are separated by a barrier with a total height $DS^2
+ BS^4 \simeq 66.6$ K. The non-diagonal anisotropy term $C(S_+^4 +
S_-^4)$ arises from the fourfold $S_4$ point symmetry of the
molecule, and is the lowest order term that would allow the
quantum tunneling of the electron spin between states $S_z = \pm
m$. An external magnetic field $B_z$ parallel to the anisotropy
axis would introduce an extra term $g_{\parallel} \mu_B B_z S_z$
in the Hamiltonian, thereby destroying the symmetry of the energy
level scheme; at the particular values of field $B_z^{mm'} =
[n/(g_{\parallel} \mu_B)][D + B(m^2 + m'^2)]$, with $n$ an integer
$-S \leq n \leq S$, the states $S_z = m$ on one side of the
barrier come in resonance with the states $S_z = m' = n - m$ on
the opposite side. In the presence of nondiagonal terms in the
spin Hamiltonian, one then expects the occurrence of quantum
tunneling of the magnetization, as indeed observed in the
experiments \cite{thomas96N}. The selection rules imposed by the
term $C(S_+^4 + S_-^4)$ imply that resonant tunneling should be
observed only for $n$ a multiple of 4, in contrast with the
observation that steps of comparable height appear in the magnetic
hysteresis loops for all values of $n$. There is now solid
experimental evidence \cite{hill03PRL,delbarco03PRL} for the
prediction \cite{cornia02PRL} that a disorder in the acetic acid
of crystallization is present and gives rise to six different
isomers of Mn$_{12}$ cluster, four of which have symmetry lower
than tetragonal and therefore possess non-diagonal anisotropy
terms that would allow tunneling transitions for any even value of
$n$. Furthermore, to obtain tunneling resonances also at odd
values of $n$, it is necessary to introduce in
(\ref{hamiltonianMn12}) the effect of dipolar fields originating
from neighboring molecules and the hyperfine couplings with the
nuclear spins, as will be described below.

A peculiarity of the Mn$_{12}$-ac system is the presence in every
real sample of fast-relaxing molecules (FRMs) \cite{aubin97CC},
i.e. clusters characterized by a lower anisotropy barrier and a
much faster relaxation rate, as observed for instance by
\textit{ac}-susceptibility \cite{evangelisti99SSC} and
magnetization measurements \cite{wernsdorfer99EPL}. It has been
recognized that such FRMs originate from Jahn-Teller
isomerism\cite{sun99CC}, i.e. the presence in the molecule of one
or two Mn$^{3+}$ sites where the elongated Jahn-Teller axis points
in a direction roughly perpendicular instead of parallel to the
crystalline $\hat{c}$-axis. This results in the reduction of the
anisotropy barrier to 35 or 15 K in case of one or two flipped
Jahn-Teller axes, respectively\cite{wernsdorferU}, and presumably
in an increase of the non-diagonal terms in the spin Hamiltonian
as well. Furthermore, the anisotropy axis $z$ of the whole
molecules no longer coincides with the crystallographic
$\hat{c}$-axis, but deviates e.g. by $\sim 10^{\circ}$ in the
molecules with 35 K barrier\cite{wernsdorfer99EPL}. The
Jahn-Teller isomerism is very different from the above-mentioned
effect of disorder in solvent molecules, and produces much more
important effects. As shall become clear in chapter IV, the
presence of the FRMs is essential for the interpretation of our
experimental data.

\subsection{Parameters for Mn$_{6}$}  \label{paramMn6}

\begin{figure}[t]
\begin{center}
      \leavevmode
      \epsfxsize=130mm
      \epsfbox{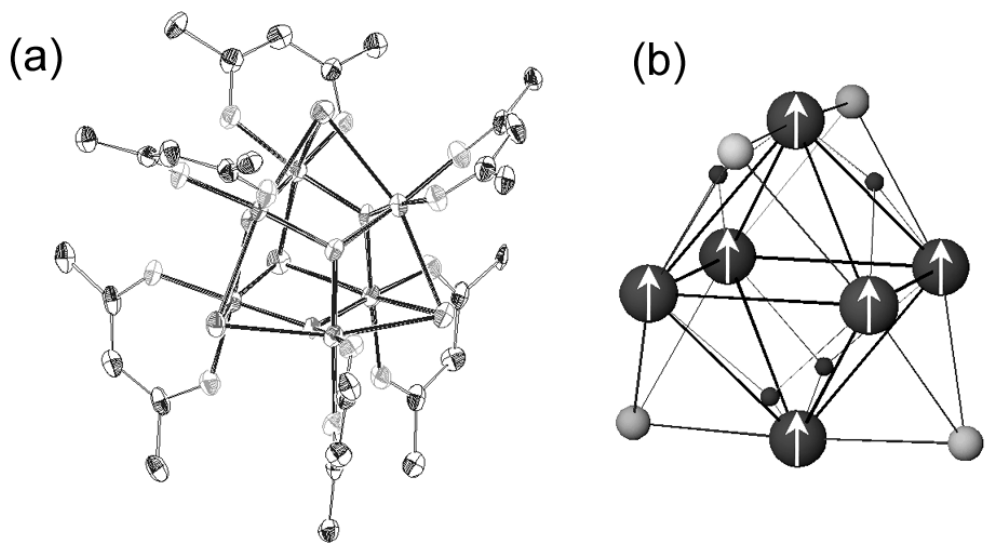}
\end{center}
\caption{\label{structureMn6} (a) Crystal structure of
Mn$_{6}$O$_{4}$Br$_{4}$(Et$_{2}$dbm)$_{6}$ \cite{aromi99JACS}. (b)
Detail of the symmetric octahedral core, containing six
ferromagnetically coupled Mn$^{3+}$ ions, yielding a total spin $S
= 12$ for this molecular superparamagnetic particle.}
\end{figure}

The molecular core of Mn$_{6}$O$_{4}$Br$_{4}$(Et$_{2}$dbm)$_{6}$,
hereafter abbreviated as Mn$_{6}$ \cite{aromi99JACS}, is a highly
symmetric octahedron of Mn$^{3+}$ ions (with spin $s=2$) that are
ferromagnetically coupled via strong intra-cluster superexchange
interactions (Fig. \ref{structureMn6}). Accordingly, the ground
state is a $S=12$ multiplet and the energy of the nearest excited
state is approximately $150$ K higher. The unit cell is
monoclinic, with space group Pc and contains $4$
molecules\footnote{We have used the crystallographic data for a
related complex Mn$_{6}$O$_{4}$Cl$_{4}$(Et$_{2}$dbm)$_{6}$
\cite{aromi99JACS}, in which the Br$^{-}$ ions are replaced by
Cl$^{-}$.} bound together only by Van der Waals forces.
Inter-cluster superexchange is therefore negligible and the only
relevant interaction interaction between cluster spins is the
dipolar one.

To determine the spin Hamiltonian of the $S=12$ manifold, we
notice that the anisotropy of the single Mn$^{3+}$ ions may be
very strong due to Jahn-Teller effects. On the other hand, due to
the highly symmetric octahedral structure of the cluster, the
addition of the single-ion anisotropies may lead to a vanishingly
small net cluster anisotropy, no matter how large is the
zero-field splitting of the constituting atoms \cite{benciniB}.
Indeed, the analysis of magnetization data on basis of the simple
Hamiltonian:
\begin{eqnarray}
\mathcal{H} = -DS_{z}^{2} - g\mu_{B}\vec{B}\cdot\vec{S}.
\label{hamiltonianMn6}
\end{eqnarray}
indicated an upper limit $|D| \lesssim 0.01 K$ for the total
cluster anisotropy \cite{aromi99JACS}, with $g \simeq 1.94 -
1.98$. An independent estimate of $D$ was obtained by
high-frequency ESR data taken in the range $95 - 380$ GHz by J.
Krzystek (NHMFL Tallahassee, USA) \cite{morello03PRL}. Owing to
the combination of very small $D$ and large $S$, as well as the
presence of a signal at $g = 2.00$ arising from a minute amount of
Mn$^{2+}$ impurity (often seen in ESR of Mn$^{3+}$ compounds), the
interpretation of the spectra was not fully conclusive.
Nevertheless, signals with a clearly visible structure on the
low-field end of the spectra could be obtained. It could be
identified as fine structure originating from zero-field splitting
(ZFS), since it was independent of field and frequency.
Simulations of the spectra performed using Eq.
(\ref{hamiltonianMn6}) agree well with the experiment taking
$|D|/k_{B}\sim 0.03$ or $\sim 0.05$ K, depending on the sign of
$D$ (which could not be unequivocally determined). Although a
smaller rhombic component could be present, the data do not
justify a more elaborate fitting.

The Mn$_6$ cluster constitutes therefore a practical example of
how the anisotropy parameters of high-spin molecules can be tuned
by simply modifying the crystal structure, while maintaining the
same constituents as other clusters having very different
properties (cf. Mn$_{12}$-ac in \S\ref{paramMn12}).

\section{Hyperfine couplings} \label{sec:hyperfine}

The coupling between electron and nuclear spins in a single ion
can be generally expressed in the form:
\begin{eqnarray}
\mathcal{H}_{\mathrm{hyp}} = \vec{I} \cdot \mathbf{A} \cdot
\vec{s} = - \gamma_N \hbar \vec{I} \cdot \vec{B}_{\mathrm{hyp}}
\label{Bhypgen}
\end{eqnarray}
where $\vec{I}$ is the nuclear spin, $\vec{s}$ is the electron
spin of the single ion, $\mathbf{A}$ is the hyperfine coupling
tensor, $\gamma_N$ is the nuclear gyromagnetic ratio and
$\vec{B}_{\mathrm{hyp}}$ is the total hyperfine field at the
nucleus. The hyperfine tensor can contain several contributions
depending on the particular nuclear site under consideration. For
instance, $\mathbf{A}$ for the $^{55}$Mn nuclei in a Mn$^{4+}$ ion
is a diagonal tensor: the only contribution comes from the the
isotropic Fermi contact term, arising from the core polarization
of inner $s$ electrons caused by 3$d$ electron spins. For the
$^1$H nuclei, located at the ligands of the molecules, the main
contribution is instead arising from the dipolar field produced by
the electron spins. The resulting $B_{\mathrm{hyp}}$ is then
typically much smaller than for $^{55}$Mn.

\subsection{Hyperfine fields on the $^{55}$Mn nuclei in
Mn$_{12}$-ac}   \label{hyperfineMn12}

\begin{figure}[t]
\begin{center}
      \leavevmode
      \epsfxsize=100mm
      \epsfbox{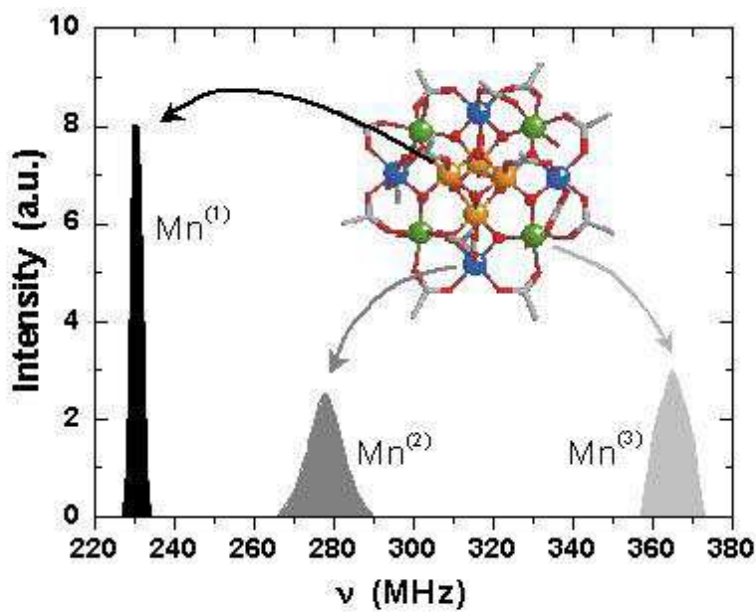}
\end{center}
\caption{\label{Mnlines} $^{55}$Mn NMR spectra in zero field, with
assignment of each line to the corresponding site in the molecule
(adapted from \cite{kubo02PRB}).}
\end{figure}

Thanks to the high anisotropy barrier of Mn$_{12}$-ac, it is
possible to detect the Larmor precession of $^{55}$Mn nuclear
spins even in the absence of an external field, by exploiting the
local hyperfine field from the electron spins which, below the
blocking temperature $T_{\mathrm{B}} \sim 3$ K of the
superparamagnetic clusters, is static on the timescale of an NMR
experiment. Goto \textit{et al.}\cite{goto00PHB} were the first to
find that the $^{55}$Mn NMR spectrum in zero field consists of
three quadrupolar-split lines centered around  230, 280 and 365
MHz. The hyperfine fields in each of the three inequivalent Mn
sites are directly obtained as $B_{\mathrm{hyp}} = 2\pi\nu /
\gamma_N$, where $\gamma_N / 2\pi = 10.57$ MHz/T is the
gyromagnetic ratio of the $^{55}$Mn nuclei, yielding
$B_{\mathrm{hyp}}^{(1)} = 21.8$ T, $B_{\mathrm{hyp}}^{(2)} = 26.5$
T and $B_{\mathrm{hyp}}^{(3)} = 34.5$ T. Subsequently, a detailed
analysis of the hyperfine couplings has been reported by Kubo
\textit{et al.}\cite{kubo02PRB}, who were able to assign these
three NMR lines to specific Mn sites in the molecule. The line
centered at $\nu^{(1)} \simeq 230$ MHz corresponds to the nuclei
in Mn$^{4+}$ ions, and is characterized by a relatively small
quadrupolar splitting $\Delta\nu^{(1)}_Q \simeq 0.72$ MHz. The
direction of $\vec{B}_{\mathrm{hyp}}$ is antiparallel to the
Mn$^{4+}$ electron spin, thus it coincides with the direction of
the total spin $S=10$ of the molecule. In the Mn$^{3+}$ ions, the
hyperfine interaction contains also a dipolar term, which makes
the coupling tensor anisotropic. Furthermore,
$\vec{B}_{\mathrm{hyp}}$ is not exactly antiparallel to the
electron spin, but there's a small canting angle\cite{kubo02PRB}.
Accounting for the effects of a dipolar contribution to the
hyperfine field in Mn$^{3+}$, it was then possible to assign the
line at $\nu^{(2)} \simeq 280$ MHz to the Mn$^{(2)}$ site and the
line $\nu^{(3)} \simeq 365$ MHz to the Mn$^{(3)}$
site\cite{kubo02PRB} (cf. Fig. \ref{Mnlines}). The quadrupolar
splittings of these lines are $\Delta\nu^{(2)}_Q \simeq 4.3$ MHz
and $\Delta\nu^{(3)}_Q \simeq 2.9$ MHz

The structure mentioned above is reflected in the effect on the
NMR frequencies of an external field $\vec{B}_z$ applied along the
anisotropy axis. The total field at the nuclear site becomes
$\vec{B}_{\mathrm{tot}} = \vec{B}_{\mathrm{hyp}} + \vec{B}_z$; in
particular for the Mn$^{(1)}$ site (but approximately also for the
other sites), since $\vec{B}_{\mathrm{hyp}}$ is parallel to
$\vec{B}_z$, $\nu(B_z)= \gamma_N (B_{\mathrm{hyp}} + B_z) = \nu(0)
+ \gamma_N B_z$ exhibits a slope\cite{furukawa01PRB,kubo02PRB}
which is given by the gyromagnetic ratio of $^{55}$Mn.

\subsection{NMR spectra in perpendicular field}
\label{sec:spectraBx}

\begin{figure}[t]
\begin{center}
      \leavevmode
      \epsfxsize=70mm
      \epsfbox{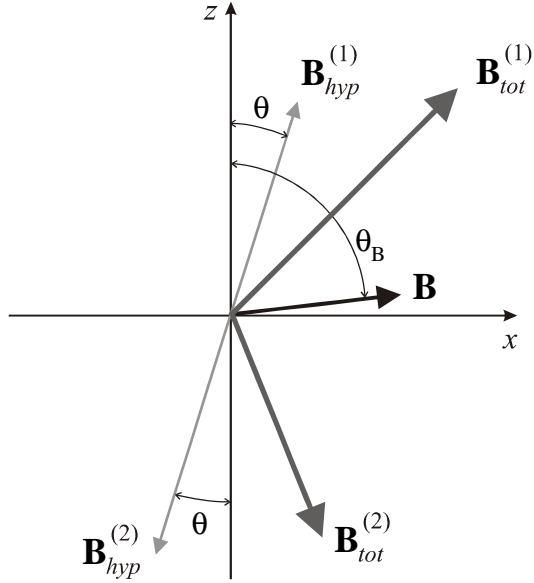}
\end{center}
\caption{\label{angles} Sketch of the orientation of hyperfine
fields in Mn$^{4+}$ ($\vec{B}_{\mathrm{hyp}}^{(1)}$) and Mn$^{3+}$
($\vec{B}_{\mathrm{hyp}}^{(2)}$) ions, relative to the crystal
axes $x$ and $z$ of the molecule, in the presence of an external
magnetic field $\vec{B}$ at an angle $\theta_B$ from the $z$ axis.
The total field at the nuclear site, $B_{\mathrm{tot}}$, which
determines the NMR frequency, is obtained as vector sum of
$\vec{B}_{\mathrm{hyp}}$ and $\vec{B}$.}
\end{figure}

\begin{figure}[t]
\begin{center}
      \leavevmode
      \epsfxsize=100mm
      \epsfbox{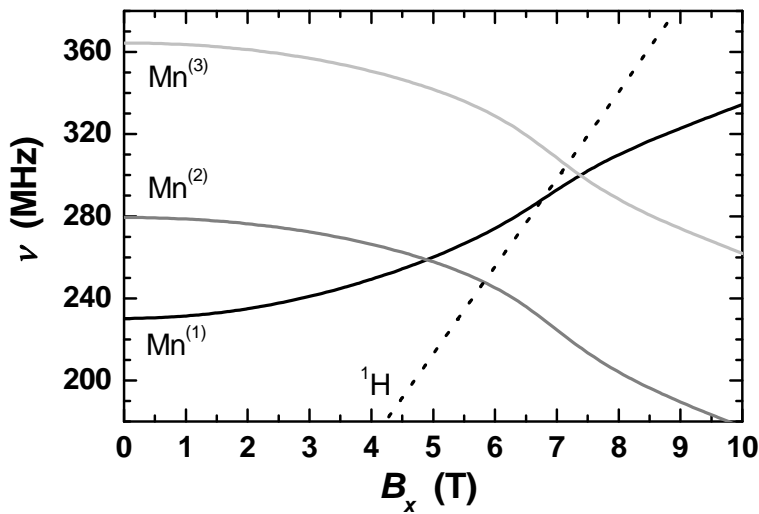}
\end{center}
\caption{\label{spectraBx} Perpendicular field dependence of the
$^{55}$Mn resonance frequencies, calculated according to Eqs.
(\ref{Btot4}) and (\ref{Btot3}) assuming $\theta_B = 90^{\circ}$
exactly. For comparison, we also plot the Larmor frequency of
$^1$H.}
\end{figure}

If the magnetic field is applied perpendicular to the anisotropy
axis, the situation is complicated by the fact that the electron
spins, and as a consequence also the hyperfine fields, tend to
cant towards the direction of the external field
\cite{chudnovsky88PRL}. Recently, Furukawa \textit{et al.}
\cite{furukawa03PRB} have analyzed the $^{55}$Mn NMR spectrum of
Mn$_{12}$-ac in perpendicular fields by assuming that the canting
angle $\theta$ between the direction of the electron spins and the
$c$-axis of the molecules is given by $\sin\theta =
M_{\perp}/M_s$, where $M_s$ is the saturation magnetization of the
$S=10$ ground state, and $M_{\perp}$ is the component of the
magnetization along the field direction. At the very low
temperatures involved in the present work the contribution of
excited electron spin states to the total magnetization is
negligible, and the above approach is equivalent to expressing the
canting angle as\footnote{Here and in the following we assume that
the field is always applied in the $xz$ plane.}:
\begin{eqnarray}
\sin\theta = \frac{\langle S_x \rangle}{10}, \label{canting}
\end{eqnarray}
where $\langle S_x \rangle = \langle \psi_{\mathrm{G}}| S_x
|\psi_{\mathrm{G}}\rangle$ is the expectation value of the
$x$-component of the spin in the ground state
$|\psi_{\mathrm{G}}\rangle$, to be obtained by numerical
diagonalization of the Hamiltonian (\ref{hamiltonianMn12}).
There's no difficulty in allowing the field $\vec{B}$ to lie at an
arbitrary angle $\theta_B$ with respect to $\vec{z}$, which helps
accounting for the possibility of misalignments between external
field an crystal axes. Assuming that the hyperfine field
$\vec{B}_{\mathrm{hyp}}$ has always the same strength as deduced
from the zero-field spectra and a direction given by the canting
angle calculated above, the total field at the nuclear site is
obtained as:
\begin{eqnarray}
B_{\mathrm{tot}}= \sqrt{(B_{\mathrm{hyp}}\sin\theta +
B\sin\theta_B)^2+(B_{\mathrm{hyp}}\cos\theta+B\cos\theta_B)^2}
\label{Btot4}
\end{eqnarray}
for Mn$^{4+}$ ions and
\begin{eqnarray}
B_{\mathrm{tot}}=\sqrt{(B_{\mathrm{hyp}}\sin\theta -
B\sin\theta_B)^2 - (B_{\mathrm{hyp}}\cos\theta - B\cos\theta_B)^2}
\label{Btot3}
\end{eqnarray}
for the Mn$^{3+}$ ions, for which the direction of
$\vec{B}_{\mathrm{hyp}}$ is opposite to the total magnetic moment
of the cluster (Fig. \ref{angles}). The $^{55}$Mn Larmor frequency
can be directly calculated as:
\begin{eqnarray}
\nu(\vec{B}) = \frac{\gamma_N}{2\pi} B_{\mathrm{tot}},
\end{eqnarray}
yielding the field dependence of the Larmor frequencies shown in
Fig. \ref{spectraBx}.

The condition $B_{\mathrm{hyp}}(\vec{B}) = B_{\mathrm{hyp}}(0)$
for any value of $\vec{B}$, which is contained in Eq.
(\ref{canting}) and used to derive Eqs. (\ref{Btot4}) and
(\ref{Btot3}), is actually equivalent to assuming that the total
spin of the molecule $S = \sqrt{\langle S_x \rangle ^2 + \langle
S_z \rangle ^2}$ remains equal to 10 at all fields. By
diagonalizing the Hamiltonian (\ref{hamiltonianMn12}) one finds
that this is the case only if the external field is not exactly
perpendicular to $\vec{z}$ plane but has (e.g. due to a small
misalignment with respect to the crystal axes) a longitudinal
component $B_z$. This yields a bias $\xi = 2g \mu_B S_z B_z$ which
forces the eigenstates of (\ref{hamiltonianMn12}) to localize at
one side or the other of the anisotropy barrier, whereas with a
perfect perpendicular field ($\theta_B=90^{\circ}$) the
eigenstates of (\ref{hamiltonianMn12}) are delocalized over both
sides of the barrier and yield $\langle S_z \rangle = 0$, thus $S
= \langle S_x \rangle$. As a rule of thumb, one obtains $S=10$
when $\xi > \Delta$, and $\langle S_z \rangle = 0$, $S \approx
\langle S_x \rangle$ otherwise. In fact, in zero external field
($\xi = 0$) all the eigenstates of (\ref{hamiltonianMn12}) have
$S=0$, which would lead to $B_{\mathrm{hyp}}(0) = 0$, contrary to
the experimental findings. The solution to this paradox is in the
fact that the Hamiltonian (\ref{hamiltonianMn12}) refers to a
single, isolated molecule. For a realistic description, we must
take into account the fact that the clusters are embedded in a
crystalline structure and interact with each other by intercluster
dipolar coupling, producing bias fields that act on the electron
spin levels.

\section{Dipolar fields} \label{sec:dipolar}

\begin{figure}[t]
\begin{center}
      \leavevmode
      \epsfxsize=100mm
      \epsfbox{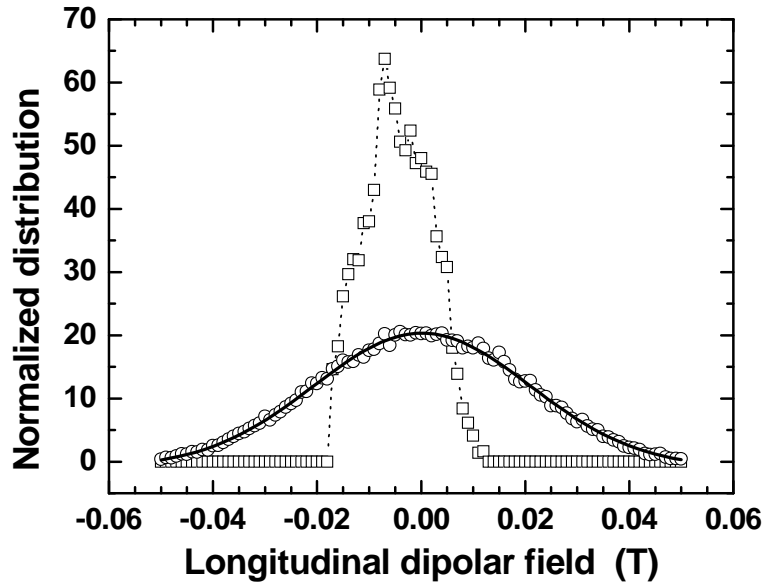}
\end{center}
\caption{\label{dipolardistr} Calculated distribution of the
longitudinal component $B_{\mathrm{dip}}^{(z)}$ of the dipolar
fields in a ZFC (circles) and FC (squares) sample in zero external
field. The solid line is a gaussian fit with width
$2\sigma_{\mathrm{dip}} \simeq 0.042$ T, whereas the dotted line
is just a guide for the eye (courtesy of I. S. Tupitsyn).}
\end{figure}

\begin{figure}[t]
\begin{center}
      \leavevmode
      \epsfxsize=100mm
      \epsfbox{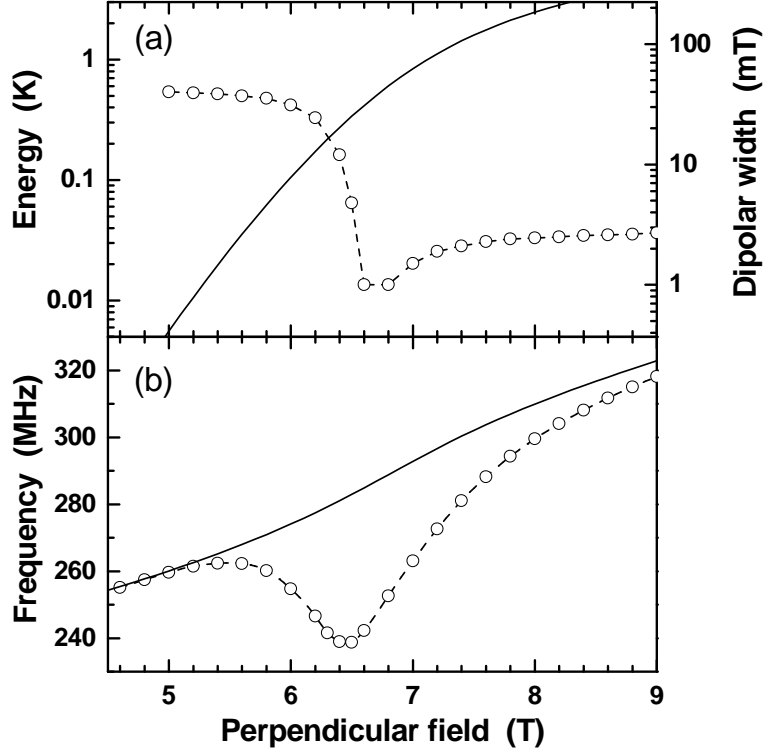}
\end{center}
\caption{\label{WdipvsDelta} (a) Perpendicular field dependence of
the full width $2\sigma_{\mathrm{dip}}$ of the longitudinal
dipolar field distribution (circles), compared to that of the
tunneling splitting $\Delta$ (solid line). (b) Calculated NMR
spectra for the Mn$^{(1)}$ site as a function of perpendicular
field. Solid line: calculated from (\ref{Btot4}, taking $S=10$ and
independent of $B_{\perp}$ . Circles: calculated from
(\ref{Bfree}), with $S(B_{\perp})$ as obtained by averaging over
the dipolar fields distribution (courtesy of I. S. Tupitsyn).}
\end{figure}

In a real sample consisting of a lattice of identical clusters,
each molecule is also subject to the dipolar field
$\vec{B}_{\mathrm{dip}}$ created by its neighbors, which adds a
term $-g \mu_B \vec{S}\cdot \vec{B}_{\mathrm{dip}}$ in the
Hamiltonian. A numerical calculation of the statistical
distribution of such dipolar fields has been performed by I. S.
Tupitsyn (Kurchatov Institute Moscow, Russia) by taking a
``real''\footnote{That is, not treating the cluster spin as a
point dipole but considering the single atomic moments.} $40
\times 40 \times 40$ lattice of Mn$_{12}$-ac molecules, and
assuming a certain value of the total magnetization. Fig.
\ref{dipolardistr} shows the distribution of the longitudinal
component $B_{\mathrm{dip}}^{(z)}$ of the dipolar fields in the
case of a fully polarized, field-cooled (FC) sample and a
demagnetized, zero-field cooled (ZFC) sample. Whereas the FC
distribution is asymmetric and obviously depends on the size and
shape of the crystal, the ZFC distribution can be well
approximated by a gaussian with width $2\sigma_{\mathrm{dip}}
\simeq 0.042$ T. In practice, this means that the vast majority of
clusters is subject to a longitudinal bias $\xi \sim 0.1$ K,
several orders of magnitude greater than the tunneling splitting;
diagonalizing the Hamiltonian (\ref{hamiltonianMn12}) including
$\vec{B}_{\mathrm{dip}}$ yields indeed $S=10$, as argued on basis
of the observed hyperfine fields. Nevertheless, there remains the
possibility to reduce the total spin to less than 10 by applying a
perpendicular field strong enough to produce a tunneling splitting
larger than the dipolar bias. The exact way in which $\Delta$
becomes greater than $\xi$ is a delicate problem that must be
solved by self-consistent numerical iteration, because as soon as
some molecules have a small enough bias to cause the spin to
reduce from 10 to $\langle S_x \rangle$, also the dipolar field
they produce on their neighbors will suddenly be reduced, causing
a drop in the width of the distribution of dipolar biases. Fig.
\ref{WdipvsDelta}(a) shows the full width $2\sigma_{\mathrm{dip}}$
of the distribution of longitudinal dipolar fields, calculated
self-consistently on the real Mn$_{12}$-ac lattice and expressed
in millitesla\footnote{Notice that we may convert from field to
energy units by multiplying by the factor $g \mu_B S$, but
$S(B_{\perp})$ is not a constant! The vertical scales in Fig.
\ref{WdipvsDelta}(a) are such that the field and energy units
match in the low-field region ($B_{\perp} \lesssim 5.5$ T) where
$S=10$, but at higher $B_{\perp}$ only the field units scale
(right-hand side) should be used for $2\sigma_{\mathrm{dip}}$.}.

Since we are interested in the consequences of such a phenomenon
for the $^{55}$Mn NMR spectrum, we first calculated numerically
the values of $\langle S_z \rangle$ and $\langle S_x \rangle$ as
function of $B_x$ by averaging over the self-consistently
calculated distribution of dipolar biases. We then recalculated
the total hyperfine field in Mn$^{(1)}$ sites by substituting
\begin{eqnarray}
\tan\theta = \frac{\langle S_x \rangle}{\langle S_z \rangle} \nonumber \\
B_\mathrm{hyp}(B_x) = B_\mathrm{hyp}(0) \frac{\sqrt{\langle S_x
\rangle ^2 + \langle S_z \rangle ^2}}{10} \label{Bfree}
\end{eqnarray}
into Eq. (\ref{Btot4}). The resulting Larmor frequency is plotted
in Fig. \ref{WdipvsDelta}(b) and compared with the frequency that
would be obtained assuming a constant value of $B_{\mathrm{hyp}}$,
i.e. a constant total spin $S=10$. It clearly appears that the
deviation from the $S=10$ spectrum occurs when $\Delta$ approaches
the width of the dipolar field distribution. An actual observation
of such a peculiar spectrum would imply that the total electron
spin of the cluster is occupying a ground state consisting of a
superposition of macroscopically distinct states for a time longer
than the intrinsic NMR timescale, and that such a superposition of
states is robust enough to survive the interaction with the
environment and the measurement performed via the nuclear spins.
To be completely correct, one should include the hyperfine
interaction $\sum_i  \vec{I}_i \cdot \mathbf{A}_i\cdot \vec{s}_i$
in the Hamiltonian (\ref{hamiltonianMn12}) and solve the complete
problem, but the size of the matrices would become intractable. We
notice anyway that the hyperfine interactions are of the same
order of magnitude as the dipolar fields, thus we can empirically
assume that their effect on the localization of the electron spins
should not be much larger than what we calculated so far by
considering only $B_{\mathrm{dip}}$.

\section{Nuclear spin dynamics}

\subsection{Nuclear spin-lattice relaxation (NSLR)} \label{NSLR}

The general theory of nuclear relaxation, which expresses the
spin-lattice relaxation rate as a function of the time correlation
of a perturbing field\cite{abragam61}, has been successfully
applied to interpret the temperature dependence of the NSLR rate
in the thermally activated regime of molecular magnets, first with
proton NMR in
Fe$_8$\cite{lascialfari98PRL,furukawa01PRBFe8,ueda02PRB} and more
recently with $^{55}$Mn in
Mn$_{12}$-ac\cite{furukawa01PRB,goto03PRB,morello03POLY}. We shall
briefly review it here, mainly to point out why it breaks down
upon approaching the pure quantum regime for the electron spin
fluctuations. The implicit assumption that has always been
made\cite{lascialfari98PRL,furukawa01PRBFe8,ueda02PRB,furukawa01PRB,goto03PRB,morello03POLY}
is that the nuclear spins have a certain Zeeman splitting, thus a
certain Larmor frequency $\omega_N$, given by a static magnetic
field, and that the transitions between the Zeeman levels of the
static Hamiltonian can be computed within perturbation theory,
assuming the presence of fluctuations in the hyperfine field which
are small compared to the static field. In most cases, as in the
$^1$H experiments on Mn$_{12}$-ac
\cite{lascialfari98PRL,furukawa01PRBFe8} (with the exception of
\cite{ueda02PRB}), the static field is an externally applied
field, much larger than the hyperfine fields. For $^{55}$Mn in
Mn$_{12}$-ac however, the reverse is true and one may carry out an
NMR experiment exploiting the local hyperfine field alone, which
is made ``quasi-static'' by the high anisotropy barrier for the
reversal of the electron spins. Thus, in the absence of an
external field, the static field is just the value of
$B_{\mathrm{hyp}}$ obtained when the electron spin is in its
ground state. Fluctuations of the local field may then arise when
the cluster spin is thermally excited to a higher level. Calling
$b(t)$ the fluctuating component of $B_{\mathrm{hyp}}$, the NSLR
rate $W$ can still be computed within the framework of
perturbation theory if $b \ll B_{\mathrm{hyp}}$. $W$ is
proportional to the spectral density at the Larmor frequency of
the perpendicular component $b_{\perp}(t)$ of the fluctuating
field, i.e. the Fourier transform of the time correlation
function, $\langle b_{\perp}(t)b_{\perp}(0)\rangle$:
\begin{eqnarray}
W = \frac{\gamma_N^2}{4}\int \langle
b_{\perp}(t)b_{\perp}(0)\rangle \exp(i\omega_N t) dt.
\label{Wfirst}
\end{eqnarray}
We may further assume that the correlation function of
$b_{\perp}(t)$ is related to the phonon-induced fluctuation rate
$\tau_{\mathrm{s-ph}}^{-1}$ of the cluster electron spin as:
\begin{eqnarray}
\langle b_{\perp}(t)b_{\perp}(0)\rangle = \langle \Delta
b_{\perp}^2 \rangle \exp(-t/\tau_{\mathrm{s-ph}}).
\end{eqnarray}
Here $\langle \Delta b_{\perp}^2 \rangle$ is the square of the
average change in the perpendicular component of the hyperfine
field when the electron spin state is excited from $m = \pm 10$ to
$m = \pm 9$. The spin-phonon transition rate
$\tau_{\mathrm{s-ph}}^{-1}$ depends on the lifetime, $\tau_9$, and
the activation energy, $\Delta E = E_9 - E_{10}$, of the $m = \pm
9$ states as $\tau_{\mathrm{s-ph}}^{-1} = \tau_9^{-1}\exp(-\Delta
E/k_B T)$, where:
\begin{eqnarray}
\frac{1}{\tau_{9}} = \frac{C_{\mathrm{s-ph}} \Delta E^3}{1 -
\exp(-\Delta E/k_B T)}.
\end{eqnarray}
$\tau_9$ incorporates therefore the spin-phonon coupling constant
$C_{\mathrm{s-ph}}$ \cite{hartmann96IJMPB,leuenberger00PRB}. For
the purpose of the present discussion we may neglect the higher
lying levels, $|m| \geq 8$, which is a reasonable approximation
below $T \sim 3$ K since $\Delta E > 10$ K. Inserting into Eq.
(\ref{Wfirst}) we find:
\begin{eqnarray}
W = \frac{\gamma_N^2}{4} \langle \Delta b_{\perp}^2 \rangle
\frac{\tau_{s-ph}}{1+\omega_N^2 \tau_{s-ph}^2} \approx
\frac{\langle \Delta b_{\perp}^2 \rangle}{4 B_{\mathrm{tot}}^2}
\tau_9^{-1}\exp\left(-\frac{\Delta E}{k_B T}\right),
 \label{Wlast}
\end{eqnarray}
which shows clearly the expected exponential temperature
dependence of the NSLR rate, since $\tau_9 \simeq const.$ for $k_B
T \ll \Delta E$.

Eq. (\ref{Wlast}) has the same form as the expression used by Goto
\textit{et al.}\cite{goto03PRB}, except that the latter authors
insert the $T$-dependence into an effective fluctuating field,
expressed as:
\begin{eqnarray}
b_{\mathrm{eff}} = \Delta
b_{\perp}\frac{\tau_9}{\tau_{10}},\nonumber
\\
\frac{1}{\tau_{10}} = \frac{C_{\mathrm{s-ph}} \Delta
E^3}{\exp(\Delta E/k_B T) - 1},
\end{eqnarray}
where $\tau_{10}$ is the lifetime of the $m = \pm 10$ states,
which is exponentially $T$-dependent. Goto \textit{et al.} also
pointed out that, in order to obtain a transverse component of the
fluctuating hyperfine field upon excitation of the electron spin
from $m=\pm 10$ to $m=\pm 9$, the hyperfine coupling tensor
$\mathbf{A}$ (\ref{Bhypgen}) must contain non-vanishing
off-diagonal terms. Then we may write $b_{\perp} = -\delta s_z
A_{xz}/(\gamma_N \hbar)$, where $\delta s_z$ is the change in the
$z$ component of the electron spin \emph{in each single ion}. As
mentioned in \S\ref{hyperfineMn12}, only $^{55}$Mn nuclei in
Mn$^{3+}$ have $A_{xz} \neq 0$ \cite{kubo02PRB,goto03PRB}, while
$\mathbf{A}$ in Mn$^{4+}$ sites is a diagonal tensor. We shall
come back to the implications of this remark in \S\ref{zerofield}
and \S\ref{mn3+}.

As will become clear from the experiments presented in
\S\ref{lowfield}, although Eq. (\ref{Wlast}) properly fits the
data in the thermally activated regime, it fails completely in the
quantum regime ($T < 0.8$ K), where the NSLR is found to become
nearly independent of temperature. This phenomenon is observed in
both the proton-NMR experiments in Fe$_8$ \cite{ueda02PRB} and in
our $^{55}$Mn-NMR results, and clearly calls for a different
theoretical approach. As will be further argued in
\S\ref{lowfield}, the NSLR in the quantum regime can be ascribed
to the hyperfine field fluctuations arising from incoherent
quantum tunneling of the electron spin between the $|+10\rangle$
and $|-10\rangle$ states. In this context, the crucial point is
that the complete hyperfine field acting on the $^{55}$Mn nuclei
in a certain molecule will be suddenly inverted when the electron
spin of that molecule tunnels, and will stay like that until the
next tunneling event occurs. This means that a formalism such as
that leading to Eq. (\ref{Wlast}) is unsuitable for describing the
$^{55}$Mn relaxation produced by tunneling events, because the
effect of tunneling cannot be treated as a perturbation on a
static Hamiltonian. As discussed in \S\ref{coupling}, the
inversion of the hyperfine field takes place in a time
$\tau_{\mathrm{tr}} \sim 10^{-12}$ s, i.e. instantaneous compared
to $\omega_N^{-1} \sim 10^{-9}$ s, whereas the interval
$\tau_{\mathrm{T}}$ between tunneling events is, as we shall show
below, very long compared to the NMR window. The description of
the nuclear spin dynamics in the presence of incoherent tunneling
requires therefore a completely different approach, for which the
basis has been laid by the Prokof'ev-Stamp
theory\cite{prokof'ev96JLTP}, which we shall review in
\S\ref{PStheory}.

\subsection{Transverse spin-spin relaxation (TSSR)} \label{TSSR}

Another essential aspect of the nuclear spin dynamics is the
transverse spin-spin relaxation (TSSR). Under the same hypotheses
as for the NSLR, Goto \textit{et al.}\cite{goto03PRB} have derived
an expression for the TSSR rate $T_2^{-1}$ for the $^{55}$Mn, that
is applicable in the thermally activated regime of Mn$_{12}$-ac:
\begin{eqnarray}
T_2^{-1} = \frac{1}{\tau_{10}} \times \frac{(\gamma_N \Delta b_z
\tau_9)^2}{1+(\gamma_N \Delta b_z \tau_9)^2} , \label{invT2gen}
\end{eqnarray}
where $\Delta b_z$ is the change in longitudinal hyperfine field
whenever the electron spin makes a transition between $m = \pm 10$
and $m = \pm 9$. In this approach the temperature dependence is
incorporated in the lifetime $\tau_{10}$ of the $m = \pm 10$
states. Based on the analysis of the field-dependence of
$T_2^{-1}$, they also conclude that the regime $(\gamma_N \Delta
b_z \tau_9) \gg 1$ holds, so that (\ref{invT2gen}) becomes:
\begin{eqnarray}
T_2^{-1} \approx \tau_{10}^{-1} = \frac{C_{\mathrm{s-ph}} \Delta
E^3}{(\Delta E/k_B T) - 1}. \label{invT2}
\end{eqnarray}

In this model the decay of the transverse magnetization is
basically due to a phase shift that occurs locally and randomly as
a consequence of the sudden changes in the longitudinal component
of the hyperfine hyperfine field (and thus in the Larmor
frequency), for the nuclei in a thermally-excited cluster. In
fact, this mechanism describes the dephasing of nuclear precession
without considering any interaction between the nuclear spins.

The nuclear dipole-dipole interaction would add an extra
contribution, whose magnitude can be calculated by taking into
account the ``flip-flop'' term $\mathcal{H}_{f-f}$ in the dipolar
Hamiltonian for two like spins $I^{(j)}$ and $I^{(k)}$ placed at a
distance $r_{jk}$ and an angle $\theta_{jk}$ from each other
\cite{abragam61}:
\begin{eqnarray}
\mathcal{H}_{f-f} = -\frac{\hbar^2 \gamma_N^2}{4r^3} (1-3\cos^2
\theta_{jk})(I_+^{(j)} I_-^{(k)} + I_-^{(j)} I_+^{(k)}).
\label{flipflop}
\end{eqnarray}
Using the Van Vleck formula one then obtains the contribution of
the flip-flop term to the second moment $M_2$ of the resonance
line, and the TSSR rate $T_2^{-1}$ as its square root:
\begin{eqnarray}
M_{2} = \frac{1}{12}\gamma_N^{4}\hbar
^{2}I(I+1)\sum_{k}\frac{(1-3\cos
^{2}\theta _{jk})^{2}}{r_{jk}^{6}}, \nonumber\\
T_{2}^{-1} = \sqrt{M_{2}}. \label{vanvleck}
\end{eqnarray}
Applying this formula to, for instance, the nuclei in Mn$^{(1)}$
sites, gives $T_2^{-1} = 1790$ s$^{-1}$ if the Mn$^{(1)}$ nuclei
in the same cluster are included, whereas restricting the
summation to nuclei in neighboring clusters leads to $T_2^{-1} =
111$ s$^{-1}$. The latter value gives an order of magnitude of the
rate of intercluster nuclear spin diffusion which, as we shall
discuss below, is an essential addition to the Prokof'ev-Stamp
theory needed to understand our experiments.

\section{Prokof'ev-Stamp theory} \label{PStheory}

\begin{figure}[p]
\begin{center}
      \leavevmode
      \epsfxsize=120mm
      \epsfbox{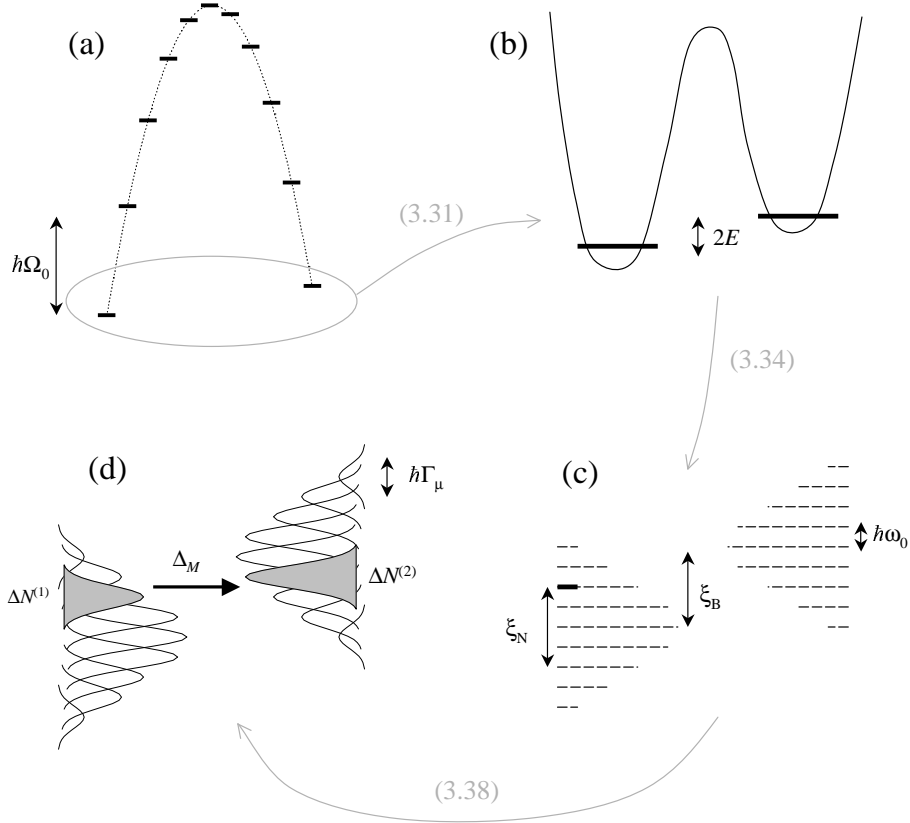}
\end{center}
\caption{\label{Figprokstamp} ``Top-down'' sketch of the model
used by Prokof'ev and Stamp to describe the coupled system
``central spin + spin bath''. The numbers in brackets indicate the
equations that describe each further step. (a) Giant spin with
crystal field anisotropy and external bias. (b) Truncation to the
ground doublet, with static bias $\xi_B$. (c) Coupling with
nuclear spins: this adds an extra bias $\xi_N$ which depends on
the polarization state $\Delta N = N^{\uparrow} - N^{\downarrow}$
of the nuclei (cf. Fig. \ref{PSenergies}). (d) Nuclear spin
diffusion: each polarization group is broadened by $\hbar
\Gamma_{\mu}$, so that groups on either side of the barrier may
overlap and create a tunneling window. Calling $\Delta N^{(1)}$
and $\Delta N^{(2)}$ the nuclear polarizations before and after
the flip, the tunneling probability is governed by the effective
tunneling splitting $\Delta_M$, with $2M = \Delta N^{(2)} - \Delta
N^{(1)}$.}
\end{figure}

The Prokof'ev- Stamp (PS) theory of the spin bath
\cite{prokof'ev95CM,prokof'ev96JLTP,prokof'ev00RPP} presents a
major step forward in understanding the quantum dynamics of a
giant spin (in our case the cluster spin) coupled to a bath of
environmental spins (here the nuclear spins). The starting point
for its development was to recognize that, in several physical
systems, the ``oscillator bath'' theory \cite{leggett87RMP} cannot
be applied, because the environmental degrees of freedom cannot be
described as a set of non-interacting oscillators, and the
couplings to the central spin are not weak.
Initially\cite{prokof'ev95CM,prokof'ev96JLTP}, the PS theory was
developed for a single central spin. The short-time $\sqrt{t}$ law
for the quantum relaxation of an ensemble of spins (namely SMMs)
has been obtained by including the effect of intercluster dipolar
coupling in the form of an effective (bias) field
\cite{prokof'ev98PRL}. This section contains a review the
essential aspects of the PS theory, with emphasis on the issues
that will be necessary to interpret the experimental results
presented chapter IV. Fig. \ref{Figprokstamp} can be used as a
visual guide through the ingredients necessary to arrive at the
complete description of the problem.

\subsection{Central spin + spin bath} \label{centralspinbath}

The first step for describing the coupled system of ``central spin
+ spin bath'' is to consider a single giant spin subject to
crystal fields and, eventually, a magnetic field, and to truncate
its total Hamiltonian to the ground doublet, neglecting all the
physics at energies $> \Omega_0$ (see \S\ref{coupling}): in this
way the central spin can be treated as an effective spin 1/2
described by Pauli matrices $\hat{\tau}$, where $|\Uparrow\rangle$
and $|\Downarrow\rangle$ are the eigenstates of $\hat{\tau}_z$. In
particular, for an isolated spin having a tunneling splitting
$\Delta$ in the ground doublet, the truncated Hamiltonian is:
\begin{eqnarray}
\mathcal{H}_{\mathrm{trunc}} = \frac{\Delta}{2} \hat{\tau}_x
\cos(\pi S) - \frac{\xi}{2} \hat{\tau}_z, \label{truncated}
\end{eqnarray}
where $\cos(\pi S)$ expresses the fact that in half-integer spins
there is a destructive interference of tunneling paths
\cite{loss92PRL,garg93EPL}, and $\xi$ is a longitudinal
bias\footnote{In the whole Prokof'ev-Stamp literature, $\Delta$
and $\xi$ denote the matrix elements in (\ref{truncated}), whereas
throughout this thesis we always use those symbols for ``real''
energy differences. Therefore it always holds
$\Delta(\mathrm{this\ thesis}) = 2\Delta(\mathrm{PS})$ and
$\xi(\mathrm{this\ thesis}) = 2\xi(\mathrm{PS})$, and all
equations in this section are consistently adapted to the present
notation.}. If such a bias is produced by an external static
field, it is labelled $\xi_B = -2g \mu_B \vec{S} \cdot \vec{B}$.
As is well known from textbook quantum mechanics \cite{cohen}, in
the absence of any other couplings the probability $P_{\Uparrow
\Uparrow}(t)$ for $\vec{S}$ to remain in the state
$|\Uparrow\rangle$ at time $t$ when prepared in $|\Uparrow\rangle$
at $t=0$ is:
\begin{eqnarray}
P_{\Uparrow \Uparrow}(t) = 1 - \frac{(\Delta/2)^2}{E^2} \sin^2(E
t), \label{Psimple}
\end{eqnarray}
where $E = \sqrt{(\xi/2)^2 + (\Delta/2)^2}$.

The central spin is further assumed to be coupled to $N$
environmental (e.g. nuclear) spins $\hat{\vec{\sigma}}_k$, $k = 1
\ldots N$, assumed for simplicity to have spin 1/2, described by
vector Pauli matrices $\hat{\vec{\sigma}}_k$. In the notation of
Prokof'ev and Stamp, the single hyperfine couplings have strength
$\hbar \omega_k = \vec{\gamma}_k \cdot \vec{\sigma}_k$, where
$\vec{\gamma}_k$ is the hyperfine field expressed in energy
units\footnote{With the notation used in the rest of this thesis,
$\vec{\gamma}_k = -\hbar \gamma_N \vec{B}_{\mathrm{hyp}}$.}, and
they are centered around an average value $\hbar \omega_0$. The
energy levels of the central spin are now represented by two
hyperfine-split manifolds, each containing $2^N$ states
distributed over a Gaussian with half-width $\sim
\sqrt{N}\omega_0$. Degenerate states within the manifold can be
labelled by their ``polarization group'', i.e. according to the
total nuclear polarization $\Delta N = N^{\uparrow} -
N^{\downarrow}$ (see Fig. \ref{PSenergies}).

\begin{figure}[t]
\begin{center}
      \leavevmode
      \epsfxsize=90mm
      \epsfbox{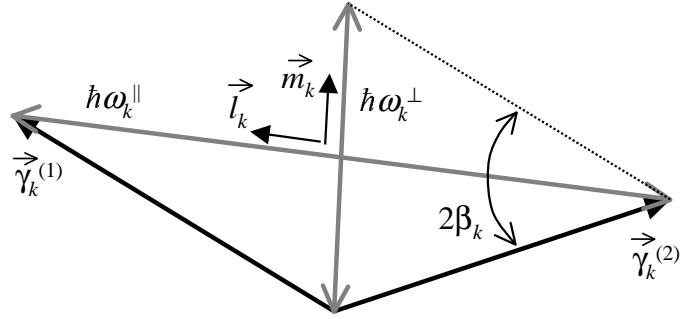}
\end{center}
\caption{\label{PSangles} Scheme of the relative orientations of
the hyperfine fields before ($\vec{\gamma}_k^{(1)}$) and after
($\vec{\gamma}_k^{(2)}$) the electron spin flip, and the
consequent sum ($\hbar \omega_k^{\parallel}$) and difference
($\hbar \omega_k^{\perp}$) terms in the hyperfine energies, as
defined in Eq. (\ref{PSsumdiff}). The angle $\beta_k$ is typically
very small since $\omega_k^{\parallel} \gg \omega_k^{\perp}$.}
\end{figure}

By calling $\vec{\gamma}_k^{(1)}$ and $\vec{\gamma}_k^{(2)}$ the
hyperfine fields before and after the tunneling of the central
spin, respectively, we can define the sum and difference terms:
\begin{eqnarray}
\hbar \omega_k^{\parallel} \vec{l}_k = \vec{\gamma}_k^{(1)} -
\vec{\gamma}_k^{(2)} \nonumber \\
\hbar \omega_k^{\perp} \vec{m}_k = \vec{\gamma}_k^{(1)} +
\vec{\gamma}_k^{(2)} \label{PSsumdiff}
\end{eqnarray}
where $\vec{l}_k$ and $\vec{m}_k$ are unit vectors (cf. Fig.
\ref{PSangles}). $\hbar \omega_k^{\perp}$ represents the component
of the hyperfine coupling energy that does not change during a
flip of the central spin, whereas $\hbar \omega_k^{\parallel}$ is
the variable part. For instance, in the case of $^{55}$Mn nuclei
in Mn$_{12}$-ac, one would find $\omega_k^{\perp} \approx 0$ since
the hyperfine fields acting on $\vec{\sigma}_k$ before and after
the flip are just antiparallel and equal in magnitude\footnote{In
fact, as long as we strictly consider \emph{only one} electron
spin, $\vec{\gamma}_k^{(1)} = - \vec{\gamma}_k^{(2)}$ always. In
most realistic cases, the nuclei would be subject to the field of
different electron spins (or to an externally applied field), so
that $\omega_k^{\perp} \neq 0$ is often true, although usually
$\omega_k^{\parallel} \gg \omega_k^{\perp}$.}. These terms add a
static (Zeeman) contribution to the Hamiltonian of the coupled
system:
\begin{eqnarray}
\mathcal{H}_{\mathrm{static}} = \frac{\hbar}{2} \left(
\hat{\tau}_z \sum_k \omega_k^{\parallel} \vec{l}_k \cdot
\hat{\vec{\sigma}}_k + \sum_k \omega_k^{\perp} \vec{m}_k \cdot
\hat{\vec{\sigma}}_k \right).  \label{Hstatic}
\end{eqnarray}
The first term is a longitudinal coupling that yields an internal
field bias acting on $\vec{\tau}_z$ that depends on the nuclear
polarization, yielding a hyperfine bias\footnote{Let us recall
that, throughout this thesis, we \emph{always} express the
hyperfine field $\vec{B}_{\mathrm{hyp}}$ as the field created by
the electron spin and acting on the nuclear spin, which produces a
nuclear Zeeman splitting $\hbar \omega = \hbar \gamma
B_{\mathrm{hyp}}$. Obviously, this energy is exactly the same as
the bias $\xi_N$ produced by the nuclear spin on the electron spin
levels, which some readers may like to express as $\xi_N = -2 g
\mu_B \vec{B}_{\mathrm{hyp}}^{(e)} \cdot \vec{S}$, where
$B_{\mathrm{hyp}}^{(e)}$ is the field produced by the nuclear spin
and acting on the electron spin. Therefore, there is no
inconsistency in writing, for instance, the Hamiltonian
(\ref{Hstatic}) for the central spin in terms of the hyperfine
field acting on the nuclei! Moreover, it should be clear from Fig.
\ref{PSenergies} that the total bias on the electron spin levels
can be obtained by summing up the Zeeman splittings of the nuclei,
provided we account (by means of $\Delta N$) for their
polarization state. The only subtlety is that, in general, the
nuclear spins cannot be treated as a classical source of external
static field acting on the electron spin, since there is the
possibility that some nuclei may coflip with $\vec{S}$, thereby
changing $\Delta N$.} $\xi_N = \hbar \omega_0 \Delta N$ (see Fig.
\ref{PSenergies}). Formally it plays the same role as $\xi_B$
discussed above, with the crucial difference that $\xi_N$ may vary
in time, as we shall see below. The second term causes the
so-called \emph{``orthogonality blocking''}, since when
$\vec{\gamma}_k^{(1)} \neq - \vec{\gamma}_k^{(2)}$, the basis for
the eigenstates of $\vec{\sigma_k}$ (defined taking the
quantization axis parallel to the instantaneous local field
direction) after the flip of $\vec{S}$ is not exactly orthogonal
to the basis before the flip. This means that after the flip
$\vec{\sigma_k}$ will start to precess around a different axis,
which quantum mechanically is equivalent to a transition between
different nuclear Zeeman levels. The number of nuclear spins that
would coflip by this mechanism is given by the parameter $\kappa$,
defined as (cf. Fig. \ref{PSangles} for $\beta_k$):
\begin{eqnarray}
e^{- \kappa} = \prod_k \cos \beta_k \approx e^{-\frac{1}{2}\sum_k \beta_k^2}, \nonumber \\
\cos(2 \beta_k) = \frac{-\vec{\gamma}_k^{(1)} \cdot
\vec{\gamma}_k^{(2)}}{|\gamma_k^{(1)}||\gamma_k^{(2)}|} .
\end{eqnarray}
Clearly, $\kappa$ depends only on the direction of the hyperfine
fields, and not on the timescale of the central spin flip.

\begin{figure}[t]
\begin{center}
      \leavevmode
      \epsfxsize=120mm
      \epsfbox{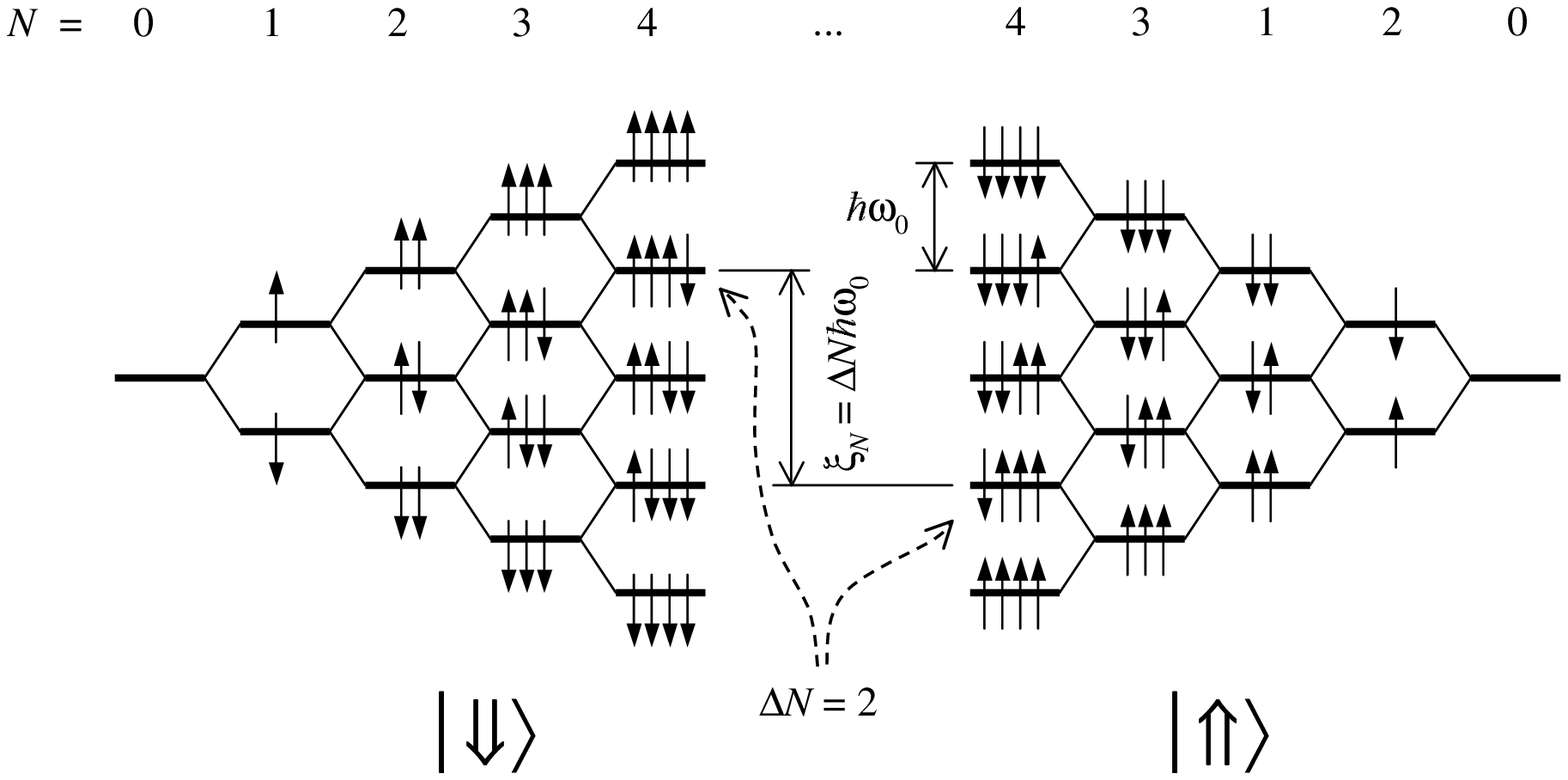}
\end{center}
\caption{\label{PSenergies} Sketch of the hyperfine-split manifold
obtained by coupling the central spin to $N$ nuclear spins, with
$N=1,2,\ldots$, assuming that all spins have $I=1/2$ and Zeeman
splitting $\hbar \omega_0$. Notice how the bias $\xi_N = \hbar
\omega_0 \Delta N$ on the electron spin levels depends on the
nuclear polarization $\Delta N$. This picture is valid only when
all the nuclei have the same $\omega_0$; in a realistic case, the
spread in hyperfine coupling makes the energy level scheme much
more intricate.}
\end{figure}

Furthermore, the presence of environmental spins adds extra Berry
phase terms to the $\cos(\pi S)$ discussed above, leading to the
so-called \emph{``topological decoherence''}. The detailed
derivation of the total phase is not essential here, but we
mention that it contains terms like $\alpha_k = (\pi / 2)\omega_k
/ \Omega_0$, which depend on the ratio between the nuclear Larmor
frequencies and the bounce frequency of the central spin (cf.
\S\ref{coupling}). $\alpha_k^2$ is the probability for
$\vec{\sigma}_k$ to coflip with $\vec{S}$ (see also
\cite{garg95PRL}), but now for a different reason than the
orthogonality blocking. What matters here is only the timescale of
the flip compared to the period of Larmor precession, i.e. whether
or not the nuclei can adiabatically follow the rotation of
$\vec{S}$. A parameter
\begin{eqnarray}
\lambda = 1/2 \sum_k \alpha_k^2
\end{eqnarray}
is introduced, that represents that number of nuclear spins that
coflip with $\vec{S}$ because of the topological decoherence
mechanism alone.

An additional aspect that must be included for a realistic
description is the \emph{``degeneracy blocking''}, which consists
in allowing the hyperfine couplings to have a spread $\hbar \delta
\omega_k$ around the central value $\hbar \omega_0$. This spread
can have different origins (quadrupolar splitting, nuclear
dipole-dipole interaction, transfer hyperfine interactions,
Nakamura-Suhl interactions \cite{suhl58PR}, etc.) and could
potentially block the tunneling dynamics of the electron spin,
since it breaks the degeneracy of the hyperfine-split electron
spin states within the same polarization group. Because of the
spread of nuclear frequencies, each polarization group is no
longer represented by a sharp line, but rather by a Gaussian peak
of width
\begin{eqnarray}
\Gamma_{\mu} = \sqrt{N} \delta \omega_k.
\end{eqnarray}
In most cases, the gap between unperturbed polarization groups is
much smaller than $\Gamma_{\mu}$, so that different polarization
groups are in fact largely overlapping.

Because of the mutual dipolar coupling between nuclear spins:
\begin{eqnarray}
\mathcal{H}_{\mathrm{dip}}(\{ \sigma_k \}) = \sum_{k \neq k'}
V_{kk'}^{\alpha \beta} \hat{\sigma}_k^{\alpha}
\hat{\sigma}_{k'}^{\beta}, \label{Hdipnucl}
\end{eqnarray}
there is an intrinsic broadening $\delta \omega_k^{(\mathrm{dip})}
\sim T_2^{-1}$ of the nuclear Zeeman levels. The crucial point is
to recognize that this broadening is \emph{dynamic}, due to
nuclear spin diffusion. In practice, \emph{the hyperfine bias can
fluctuate and cover the whole range of each polarization group},
i.e. an amplitude $\hbar\Gamma_{\mu}$, on a timescale $\sim T_2$.
The effect of spin diffusion is therefore to eliminate the
degeneracy blocking. Notice that this sort of fluctuations arises
from what one would call ``intracluster'' spin diffusion, since
the model is written for one single central spin coupled to his
nuclear spins. The transition between different polarization
groups would therefore require the exchange of energy with an
external reservoir, on a timescale $T_1$; in the PS theory, this
mechanism is assumed to take very long times. On the other hand,
by taking many identical ``central spin + spin bath'' units, one
may allow for ``intercluster'' spin diffusion as well, i.e.
between equivalent nuclei belonging to different clusters. This
possibility is not explicitly discussed in the early PS papers,
but will play an essential role in the interpretation of our
experimental results.

\subsection{Tunneling rate} \label{PStunnelrate}

The strategy adopted by Prokof'ev and Stamp to derive the
incoherent tunneling rate of the central spin coupled to a spin
bath, is to estimate it from the evolution of the probability
$P_{\Uparrow \Uparrow}(t)$ for $\vec{S}$ to remain in the same
state $|\Uparrow\rangle$ until time $t$. Its trivial expression
for an isolated spin is Eq. (\ref{Psimple}); including the spin
bath one has to average over the different values $P_{M}(t)$ that
$P_{\Uparrow \Uparrow}(t)$ may take, depending on the change $2M$
of the polarization state $\Delta N = N^{\uparrow} -
N^{\downarrow}$ of the environmental spins caused by the flip of
$\vec{S}$. $P_{M}(t)$ is itself obtained by averaging, over the
``topological'' and ``orthogonality blocked'' processes, the
quantity $P_{M}^{(0)}(t)$, defined as:
\begin{eqnarray}
P_{M}^{(0)}(t) = 1 - \frac{(\Delta_M/2)^2}{E_M^2} \sin^2(E_M t) \\
E_M^2 = (\xi/2)^2 + (\Delta_M/2)^2 \label{EM} \\
\Delta_M \sim \Delta_0 \frac{(\lambda - \lambda')^{M/2}}{M!},
\label{PM}
\end{eqnarray}
where $\lambda' = \lambda$ in the case of pure topological
decoherence and $\lambda' = 0$ for pure orthogonality blocking,
and for $\xi$ we take the total bias $\xi = \xi_N + \xi_B$ which
may include the effect of an externally applied field. Roughly
speaking, this means that the effect of the spin bath is
incorporated in a renormalization of the tunneling splitting,
assigning a vanishingly small value of $\Delta_M$ to those
tunneling transitions that would require too many nuclear coflips,
as compared to the ``natural values'' of $\kappa$ and $\lambda$.
The really essential point is that $\xi$ in Eq. (\ref{EM}) is no
longer a static bias, as soon as nuclear spin diffusion is
allowed. From (\ref{Hdipnucl}) one finds that $\xi_N$ contains a
time-dependent part which fluctuates on a timescale $T_2^{-1}$
over a range $\sim \hbar \Gamma_{\mu}$, which is the change in
longitudinal hyperfine bias when $N$ pairwise $|\uparrow
\downarrow \rangle \rightarrow |\downarrow \uparrow \rangle$
nuclear flip-flops occur. With the assumption that $T_2$ is much
shorter than the typical tunneling interval $\tau_T$, the average
of $P_{M}^{(0)}(t)$ over the fluctuating bias can be written as:
\begin{eqnarray}
\langle P_{M}^{(0)}(t,\xi) \rangle_{\xi} - \frac{1}{2}
= \frac{1}{2} e^{-t/\tau_M}\\
\tau_M^{-1} = \frac{\Delta_M^2}{2\sqrt{\pi}\hbar^2 \Gamma_{\mu}}.
\label{tauM}
\end{eqnarray}
The parameter $\tau_M^{-1}$ represents therefore the rate for
$\vec{S}$ to tunnel accompanied by the coflip of $M$ nuclear
spins. In the presence of an external static bias $\xi_B$,
$\tau_M^{-1}$ starts to be suppressed as soon as $\xi_B$ is larger
than the typical hyperfine field spread. Combining this with the
fact that the largest tunneling splitting is obtained for $M=0$,
one finds that the leading term in the tunneling rate takes the
form:
\begin{eqnarray}
\tau^{-1} = \frac{
\Delta_0^2}{2\sqrt{\pi}\hbar^2\Gamma_{\mu}}e^{-|\xi_B|/\xi_0}.
\label{taufinal}
\end{eqnarray}
where $\xi_0$ is the width of the ``tunneling window''. When some
nuclear spins may coflip with $\vec{S}$ by topological decoherence
($\lambda$) or orthogonality blocking ($\kappa$), then $\xi_0$ is
given by the energy that the central spin can exchange with the
nuclear spin bath, $\xi_0 \sim (\lambda + \kappa) \hbar \omega_0$.
Otherwise, in the absence of nuclear coflips the tunneling window
takes the width $\xi_0 \sim \hbar \Gamma_{\mu}$ due to nuclear
spin diffusion only. From this discussion it is clear that the
major role in creating a useful tunneling window for the central
spin is played by those nuclei that more easily coflip with
$\vec{S}$. We anticipate that this is an experimentally testable
prediction (\S\ref{deuterated}).

The global relaxation, i.e. the precise form of $P_{\Uparrow
\Uparrow}(t)$, is obtained by combining all the results discussed
above. This leads to very interesting phenomena to be observed in
the quantum relaxation, but goes beyond the scope of our research.
We shall mainly make use of the tunneling rates
(\ref{tauM}),(\ref{taufinal}) and the concepts of topological
decoherence and orthogonality blocking to evaluate the numbers of
coflipping nuclei $\lambda$ and $\kappa$.

\subsection{Comparison with the Landau-Zener formalism}
\label{sec:PSLZ}

\begin{figure}[t]
\begin{center}
      \leavevmode
      \epsfxsize=100mm
      \epsfbox{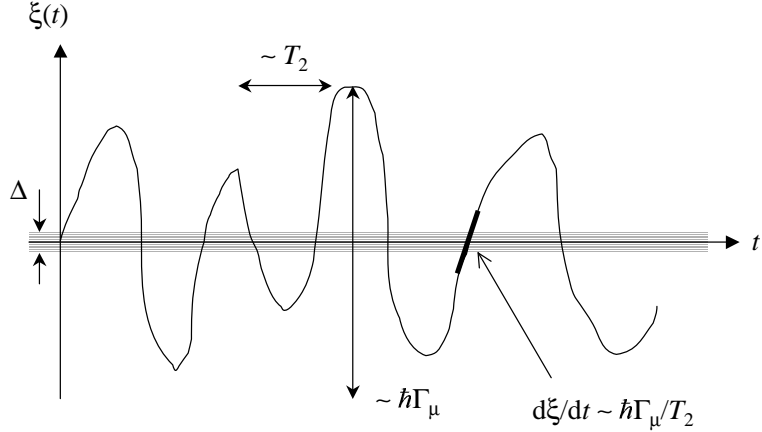}
\end{center}
\caption{\label{FigPSLZ} Time evolution of the bias $\xi$ on the
central spin levels as a consequence of nuclear spin diffusion:
the crossing through the tunneling resonance happens with slope
$\mathrm{d}\xi/\mathrm{d}t \sim \hbar\Gamma_{\mu}/T_2$.}
\end{figure}

It is rather instructive to compare the Prokof'ev-Stamp theory
with the Landau-Zener formula (\ref{PLZ}) for the calculation of
an incoherent tunneling rate. In the presence of nuclear spin
diffusion, assuming for simplicity the external bias $\xi_B = 0$,
the bias $\xi$ fluctuates over a range $\hbar \Gamma_{\mu}$ on a
timescale $T_2$, thereby forcing the energy levels of the central
spin $\vec{S}$ to cross back and forth through the tunneling
resonance: a sketch of the situation is given in Fig.
\ref{FigPSLZ}. We could therefore think of applying Eq.
(\ref{PLZ}), simplified by noting that $\Delta^2$ is a very small
number and expanding the exponential:
\begin{eqnarray}
P_{LZ} \approx \frac{\pi
\Delta^2}{2\hbar}\left(\frac{\mathrm{d}\xi}{\mathrm{d}t}\right)^{-1}.
\end{eqnarray}
Clearly, in our case $\mathrm{d}\xi/\mathrm{d}t \sim
\hbar\Gamma_{\mu}/T_2$, yielding $P_{LZ} \sim
(\Delta^2/\hbar)[(T_2/(\hbar\Gamma_{\mu})]$. To obtain the
tunneling rate, we multiply the single tunneling probability by
the number of crossings per unit time, i.e. $T_2^{-1}$, and obtain
\begin{eqnarray}
\tau_{LZ}^{-1} \sim \frac{\Delta^2}{\hbar^2\Gamma_{\mu}},
\end{eqnarray}
which has indeed the same form as Eq. (\ref{tauM}). Deriving it
via the LZ formalism clarifies why the tunneling rate does not
contain the timescale $T_2$ of the bias fluctuations. On the other
hand, by this we have completely lost any information about the
effect of the central spin dynamics on its nuclei, which in Eq.
(\ref{tauM}) is incorporated in the renormalized tunneling
splitting $\Delta_M$, cf. (\ref{PM}). An essential feature of the
PS theory is indeed that the fluctuating bias is not added
artificially as an ``environmental noise'', uncorrelated with the
dynamics of $\vec{S}$, but is built in the model in a
self-consistent way, including the back-action of $\vec{S}$ on the
nuclear spins \cite{stamp02CM}.

\subsection{Application to SMMs}   \label{PSSMMS}

When applying the PS theory to the combined electron-nuclear spin
dynamics in SMMs, it is useful to distinguish between nuclei
coupled directly to one single ion, like $^{55}$Mn in
Mn$_{12}$-ac, and nuclei located at the ligand molecules, like
$^1$H (and a small fraction of $^{13}$C). In the latter case, the
strength of the coupling is of course much lower than for
$^{55}$Mn, since it is due mainly to dipolar fields, but the total
field is the sum of the dipolar fields from different molecules.
If one molecule flips, it is only one component of the total field
at the $^1$H site that flips, which may yield a situation where
$\beta_k$, and therefore the chance of nuclear coflip, is rather
large (cf. Fig. \ref{angles}). This means that the protons can be
the main contributors to the width of the tunneling window
$\xi_0$, despite the relatively weak coupling.

On the contrary, $\beta_k \simeq 0$ in $^{55}$Mn, since the whole
hyperfine field is inverted by $180^{\circ}$ upon tunneling. This
yields $\kappa \simeq 0$ and, because of the short tunneling
traversal time (cf. \S\ref{coupling}), $\lambda \simeq 0$ as well.
Clearly, we can expect the change of polarization group for
$^{55}$Mn nuclei to be very difficult, unless nuclei in different
clusters are linked with each other. This possibility is not
explicitly discussed in the PS theory, but we shall see in the
next chapter that it is both experimentally observed
(\S\ref{zerofield}), and necessary to explain the observed
$^{55}$Mn NSLR rate in Mn$_{12}$-ac.

\def\baselinestretch{1}
\chapter{Nuclear spin dynamics in Mn$_{12}$-ac}

The nuclear spin dynamics in a single-molecule magnet is one of
the most crucial aspects influencing its quantum behavior,
nonetheless the experimental knowledge was so far almost totally
lacking, in particular in the low-$T$ quantum regime for the
cluster spin fluctuations. We dedicate therefore this chapter to a
detailed survey of our results on the $^{55}$Mn NMR in
Mn$_{12}$-ac, starting with the temperature dependence of the
nuclear spin-lattice relaxation (NSLR) and transverse spin-spin
relaxation rates (TSSR) in zero external field. The data indicate
that quantum tunneling of the electron spin is the mechanism
responsible for the observed features in the nuclear spin
dynamics. The dependencies on small longitudinal fields,
magnetization state of the electron spins, isotope substitution,
and crystallographic Mn site, yield precious additional
information. We then discuss the influence on the NSLR of an
external perpendicular field, which can be used to tune the
tunneling splitting. Finally, an essential and original
information is provided by the study of the nuclear spin
temperature. The results are very surprising and call for an
extension of the existing theory of quantum tunneling in magnetic
molecules. The chapter is concluded by a suggestion for an
approach that should unify all the experimental evidences into a
new and more complete theoretical description.

\section{Measurements and data analysis} \label{sec:measanal}

About 100 mg of Mn$_{12}$-ac crystallites were mixed with Stycast
1266 epoxy and introduced into a $\varnothing$ $6$ mm capsule,
then inserted in the room-temperature bore of a 9.4 T magnet and
allowed to orient for 24 hours. A two-turns NMR coil was wound
around the capsule, the whole assembly was inserted inside the
mixing chamber of the dilution refrigerator (\S\ref{sec:dilution})
and connected to the NMR electronics as described in
\S\ref{NMRcircuit}.

The $^{55}$Mn nuclear precession was detected by the spin-echo
technique. A typical pulse sequence includes a first $\pi /
2$-pulse with duration $t_{\pi / 2} = 12$ $\mu$s, a waiting
interval of 45 $\mu$s, and a 24 $\mu$s $\pi$-pulse for refocusing.
At the lowest temperatures provided by the dilution refrigerator,
such a pulse sequence was found to cause an increase in
temperature of about 5 mK in the $^3$He/$^4$He mixture. With a
waiting time of 600 s between subsequent pulse trains we could
easily keep the operating temperature around $15-20$ mK (cf. Fig.
\ref{thermaleff}). On the other hand, at such low temperature the
signal intensity is so high that we could obtain an excellent S/N
ratio without need of averaging, so that a typical measurement
sequence (e.g. an inversion recovery) would last less than 12
hours. Above 100 mK it proved convenient to take a few averages,
but then the heating due to the \textit{rf}-pulses becomes
negligible, and the waiting time can be reduced to $\sim 100$ s.

\begin{figure}[t]
\begin{center}
      \leavevmode
      \epsfxsize=100mm
      \epsfbox{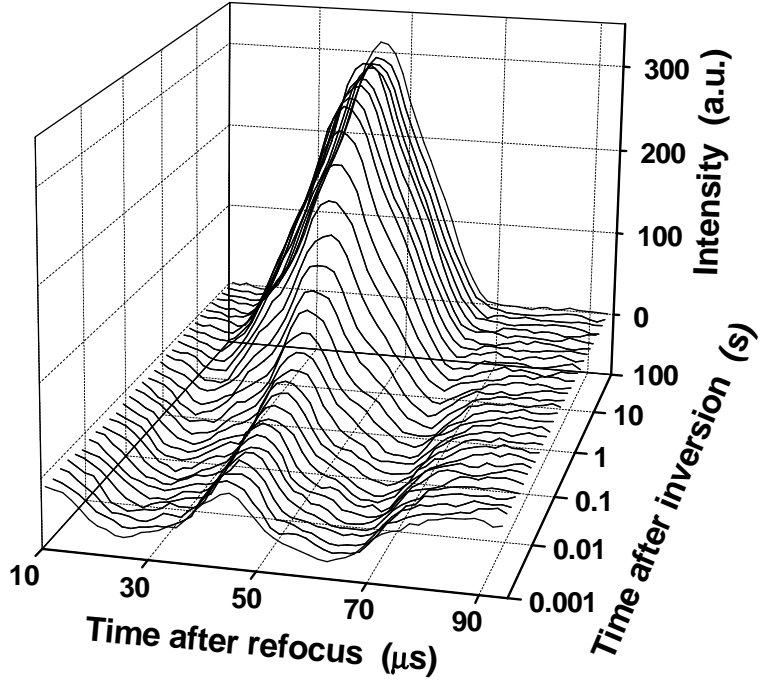}
\end{center}
\caption{\label{echo3D} An example of ``real time'' echo signals
recorded during an inversion recovery, i.e. measuring the echo
intensity at increasing delays after an inversion pulse. In
particular, these are single-shot (no averaging) raw data taken at
$B=0$ and $T=20$ mK in the Mn$^{(1)}$ site. The (normalized)
integral of the echoes is reported in Fig. \ref{recoveries}, open
squares.}
\end{figure}

\begin{figure}[t]
\begin{center}
      \leavevmode
      \epsfxsize=100mm
      \epsfbox{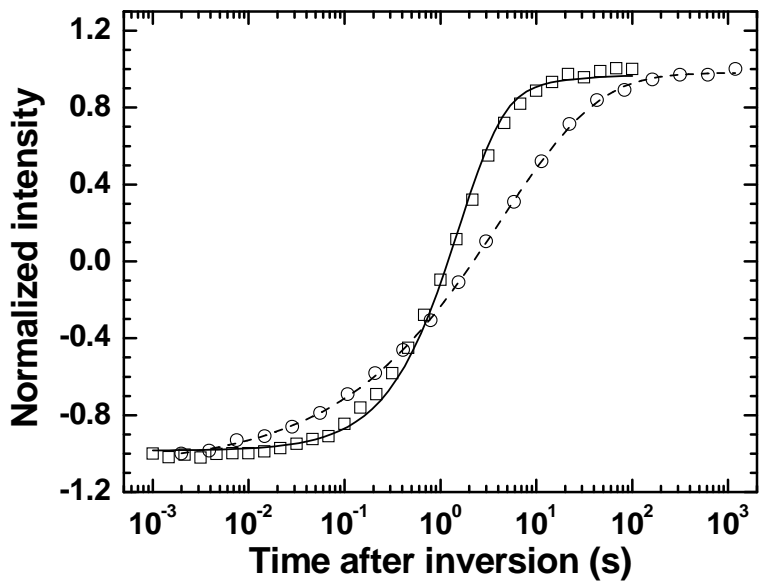}
\end{center}
\caption{\label{recoveries} Examples of nuclear inversion
recoveries in zero field (squares) and strong perpendicular field
(circles). The NSLR rate $W$ is extracted by fitting the data with
Eq. (\ref{recovery}) or Eq. (\ref{strexp}), respectively.}
\end{figure}

The NSLR was studied by measuring the recovery of the nuclear
magnetization after an inversion pulse (cf. Fig.
\ref{measureT1T2}). We preferred this technique to the more
widely-used saturation recovery
\cite{furukawa01PRB,kubo02PRB,goto03PRB} because it avoids the
heating effects of the saturation pulse train, but we checked at
intermediate temperatures that the two techniques indeed lead to
the same value of NSLR rate. An example of echo signals obtained
as a function of the waiting time after the inversion pulse is
shown in Fig. \ref{echo3D}. By integrating the echo intensity we
obtained recovery curves such as those shown in Fig.
\ref{recoveries}. For the ease of comparison between different
curves, we renormalize the vertical scale such that
$M(0)/M(\infty)=-1$ and $M(t \gg T_1)/M(\infty)=1$, even though
usually $|M(0)|<|M(\infty)|$, as could be deduced from Fig.
\ref{echo3D}\footnote{This is just an artifact that occurs when
NMR line is much broader than the spectrum of the inversion pulse,
and does not mean that the length of the $\pi$-pulse is
incorrect.}. Since the $^{55}$Mn nuclei have spin $I = 5/2$, the
recovery of the nuclear relaxation for the central line in the
quadrupolar split manifold is described by \cite{suter98JPCM}:
\begin{eqnarray}
    \frac{M(t)}{M(\infty)} = 1 - \left[ \frac{100}{63} e^{-30 W t} + \frac{16}{45} e^{-12 W t}
    + \frac{2}{35} e^{-2 W t} \right]
\label{recovery}
\end{eqnarray}
where $W$ is the nuclear spin-lattice relaxation rate\footnote{In
the simple case of a spin 1/2, where the NSLR is described by a
single exponential, $W$ is related to the relaxation time $T_1$ by
$2W=T_1^{-1}$.}. All data in zero field and moderate parallel
fields could be accurately fitted by Eq. (\ref{recovery}), which
indicates that the relaxation mechanism can be characterized by a
single rate $W$. Conversely, a proper analysis of the inversion
recoveries in strong perpendicular fields requires the
introduction of a stretching exponent $\alpha$:
\begin{eqnarray}
    \frac{M(t)}{M(\infty)} = 1 - \left[ \frac{100}{63} e^{-(30 W t)^{\alpha}} + \frac{16}{45} e^{-(12 W t)^{\alpha}}
    + \frac{2}{35} e^{-(2 W t)^{\alpha}} \right]
\label{strexp}
\end{eqnarray}
This can be attributed to a small distribution of orientations of
the crystallites in the sample, which causes the NMR signal in
high fields to consist of a mixture of lines from different
crystallites, each of them characterized by a different NSLR rate.
A comparison between an inversion recovery at zero field, fitted
by Eq. (\ref{recovery}), and a similar recovery in strong
perpendicular field fitted by Eq. (\ref{strexp}), is shown in Fig.
\ref{recoveries}.\\

The TSSR rate $T_2^{-1}$ was studied by measuring the decay of
echo intensity upon increasing the waiting time $\tau$ between the
$\pi/2$- and the $\pi$-pulses. The decay of transverse
magnetization $M_{\perp}(\tau)$ can be fitted by a single
exponential
\begin{eqnarray}
\frac{M_{\perp}(2\tau)}{M_{\perp}(0)}=\exp\left(-\frac{2\tau}{T_2}\right)
\label{T2}
\end{eqnarray}
except at the lowest temperatures ($T \lesssim 0.2$ K), where also
a gaussian component $T_{2G}^{-1}$ needs to be included:
\begin{eqnarray}
\frac{M_{\perp}(2\tau)}{M_{\perp}(0)}=\exp \left(
-\frac{2\tau}{T_{2}}\right)  \exp\left(
-\frac{(2\tau)^2}{2T_{2G}^2}\right) \label{T2G}
\end{eqnarray}

As regards the experiments to determine the nuclear spin
temperature, the measurements were performed by monitoring the
echo intensity at regular intervals while changing the temperature
$T_{\mathrm{bath}}$ of the $^3$He/$^4$He bath in which the sample
is immersed. Recalling (\S\ref{NMRbasics}) that nuclear
magnetization is related to the nuclear spin temperature
$T_{\mathrm{nucl}}$ by the Curie law:
\begin{eqnarray}
M(T_{\mathrm{nucl}})=N \frac{\gamma_N^2 \hbar^2
I(I+1)}{3k_{\mathrm{B}}T_{\mathrm{nucl}}}, \label{curie}
\end{eqnarray}
and assuming that $T_{\mathrm{bath}} = T_{\mathrm{nucl}}$ at a
certain temperature $T_0$ (e.g. 0.8 K), we can define a
calibration factor $K$ such that $M(T_0) =
K/T_{\mathrm{nucl}}(T_0)$ and use that definition to derive the
time evolution of the nuclear spin temperature as
$T_{\mathrm{nucl}}(t) = K/M(t)$ while the bath temperature is
changed.

\section{Nuclear relaxation in moderate parallel fields}
\label{lowfield}

In this section we present and discuss the experimental results on
the $^{55}$Mn spin-lattice and spin-spin relaxation rates in
Mn$_{12}$-ac, starting with the temperature dependencies in zero
field, and then discussing the effect of small longitudinal fields
and isotope substitution. All data refer to Mn$^{(1)}$ sites,
except for \S\ref{mn3+}, where a comparison with the Mn$^{(2)}$
site is presented.

\subsection{NSLR and TSSR in zero field} \label{zerofield}

\begin{figure}[p]
\begin{center}
      \leavevmode
      \epsfxsize=100mm
      \epsfbox{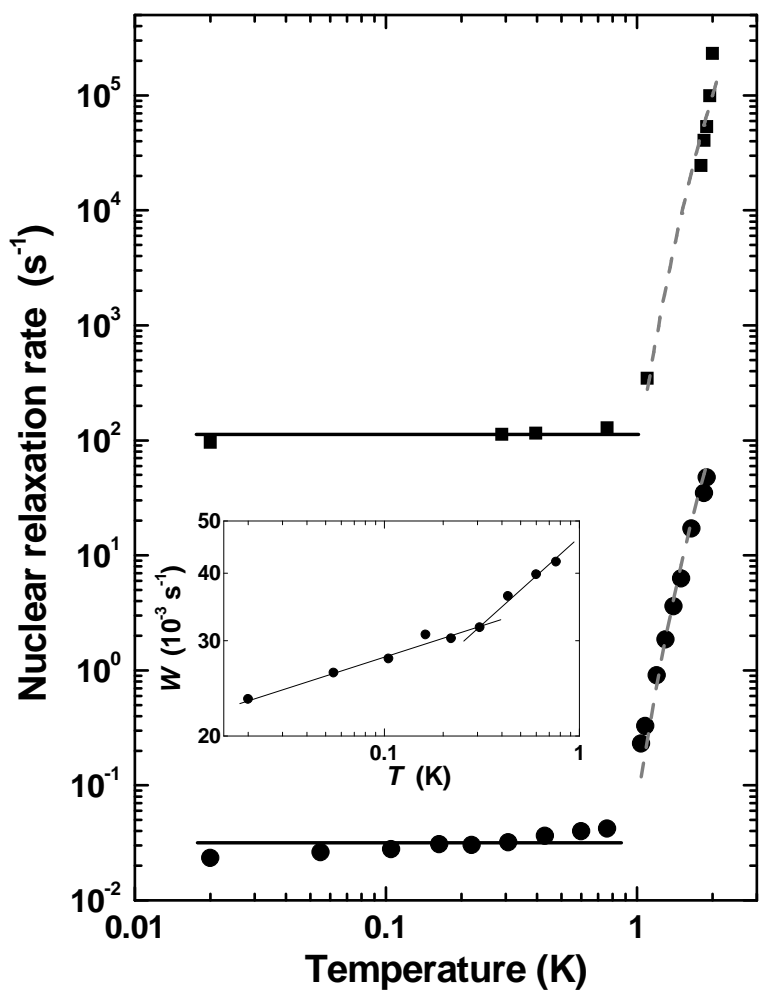}
\end{center}
\caption{\label{W1W2vsT} Temperature-dependence of the NSLR
(circles) and TSSR (squares) rates for a ZFC sample in zero field
and $\nu = 231$ MHz. The dashed lines are fits to Eqs.
(\ref{Wlast}) and (\ref{invT2}), with parameters discussed in the
text. The inset is an enlargement of $W(T)$ for $T<0.8$ K, with
lines as guides for the eye.}
\end{figure}

As mentioned in \S\ref{NSLR}, the NSLR and TSSR in Mn$_{12}$-ac
have already been studied in quite some detail in the thermally
activated regime by other groups
\cite{furukawa01PRB,goto03PRB,morello03POLY}, whereas the purpose
of our work has been to extend the analysis deep into the pure
quantum regime \cite{morello03POLY}. We begin by discussing the
temperature dependencies of the NSLR and TSSR rates in a
demagnetized, zero-field-cooled (ZFC) sample, shown in Fig.
\ref{W1W2vsT}.

What clearly appears is a sharp crossover at $T \simeq 0.8$ K
between a roughly exponential $T$-dependence and an almost
$T$-independent plateau. We attribute the $T$-independent nuclear
relaxation to the effect of ground-state tunneling fluctuations of
the cluster spins, and we shall dedicate most of this chapter to
discuss our further results supporting this statement. It should
be noted that the crossover from thermally activated to
ground-state tunneling has been also observed by analyzing the
$T$-dependence of the steps in the magnetization loops
\cite{chiorescu00PRL,bokacheva00PRL}. The important advantage of
our NMR measurements is that the nuclear dynamics is sensitive to
\emph{fluctuations} of the cluster electron spins without even
requiring a change in the macroscopic magnetization of the sample.
Clearly, no macroscopic probe (except perhaps an extremely
sensitive magnetic noise detector) would be able to detect the
presence of tunneling fluctuations in a zero-field cooled sample
in zero external field, since the total magnetization is zero and
remains so. In fact, below $T \sim 1.5$ K the steps in the
hysteresis loops of Mn$_{12}$-ac can be observed only at
relatively high values of external field
\cite{chiorescu00PRL,bokacheva00PRL}, which is indeed predicted to
lead to a less sharp transition between thermally activated and
quantum regimes \cite{garanin98PRB}. We fitted the high-$T$ part
of the NSLR rate by Eq. (\ref{Wlast}), as shown by the dotted line
in Fig. \ref{W1W2vsT}: the curve depicted is obtained by fixing
$\Delta E = D(10^2 - 9^2) + B_4(10^4 - 9^4) = 14.4$ K which yields
$\langle \Delta b_{\perp}^2 \rangle / \tau_9 \simeq 1.1 \times
10^7$ T$^2$ s$^{-1}$. The fit, in particular its slope, may be
improved by leaving also $\Delta E$ as a free parameter, which
then yields $\Delta E \simeq 12.2$ K and $\langle \Delta
b_{\perp}^2 \rangle / \tau_9 \simeq 2.2 \times 10^6$ T$^2$
s$^{-1}$. Despite the reasonable quality of the fit, Eq.
(\ref{Wlast}) is based on a model which is seriously questionable
for two reasons: (i) it considers only hyperfine field fluctuation
due to ``intrawell'' excitation, i.e. it does not account for
thermally-assisted tunneling through excited doublets, and (ii) it
does not explain what is the source of transverse fluctuating
field $b_{\perp}$, since the hyperfine coupling tensor in
Mn$^{4+}$ does not contain non-diagonal elements (cf.
\S\ref{NSLR}). We may already notice that including ``interwell''
fluctuations due to thermally assisted tunneling could solve the
paradox of the transverse hyperfine field component, provided we
can demonstrate that quantum tunneling of the electron spin is a
possible source of nuclear relaxation. We shall come back to this
issue in \S\ref{mn3+}.

The roughly $T$-independent plateau in the NSLR rate below $T
\simeq 0.8$ K is characterized by a value of $W \simeq 0.03$
s$^{-1}$ that is surprisingly high, which at first sight may
appear like an argument against the interpretation in terms of
tunneling fluctuations of the electron spin. Experimentally it is
indeed well known \cite{thomas99PRL} that the relaxation of the
magnetization in Mn$_{12}$-ac in zero field may take years at low
$T$, which means that the tunneling events are in fact extremely
rare. Based on this, we are forced to assume that the tunneling
events take place only in a small minority of the clusters, and
that some additional mechanism takes care of the relaxation of the
nuclei in molecules that do not tunnel. This is a very realistic
assumption indeed, since it is well known that all samples of
Mn$_{12}$-ac contain a fraction of fast-relaxing molecules (FRMs)
\cite{sun99CC,wernsdorfer99EPL}, as anticipated in
\S\ref{paramMn12}. Moreover, that the tunneling dynamics has to be
ascribed to FRMs is supported by the experimental fact that, when
measuring the NSLR in zero field in a saturated field-cooled
sample (FC), we observe that the sample magnetization does not
change during the experiment. This can be done by applying a small
longitudinal field $B_z \sim 0.5$ T and measuring the NMR signal
intensity at the shifted frequencies $\nu_{\uparrow} = \nu(0) +
\gamma_N B_z$ and $\nu_{\downarrow} = \nu(0) - \gamma_N B_z$,
corresponding to nuclei in clusters whose spin is respectively
parallel or antiparallel to the external field
\cite{jang00PRL,kubo02PRB}. In a fully saturated sample, the NMR
signal at $\nu_{\downarrow}$ is indeed zero, and we verified that
it remains that way even after several weeks of experiments as
long as the sample is kept constantly at low temperature. In other
words, the macroscopic magnetization \emph{as seen by the nuclei
that contribute to the NMR signal} does not change in time,
implying that the observable NMR signal comes from nuclei that
belong to ``frozen'' clusters\footnote{We cannot tell for sure
whether also nuclei in FRMs participate to the signal, but the
chance is indeed small since their $T_2$ would be very short, due
to the quickly changing hyperfine field they experience.},
therefore the relaxation channel must involve the tunneling in
FRMs.

The question remains what is the mechanism that links the
tunneling in FRMs to the nuclear spins in frozen molecules. One
possibility is to ascribe it to the fluctuating dipolar field
produced by a tunneling FRM at the nuclear sites of neighboring
frozen molecules. In that case we may give an estimate of $W$
using an expression of the form:
\begin{eqnarray}
W \approx \frac{\gamma_N^2}{4} b_{\mathrm{dip}}^2
\frac{\tau_{\mathrm{T}}}{1 + \omega_N^2 \tau_{\mathrm{T}}^2}
\approx
\frac{b_{\mathrm{dip}}^2}{4B_{\mathrm{tot}}}\tau_{\mathrm{T}}^{-1},
\label{Wdipolar}
\end{eqnarray}
where $b_{\mathrm{dip}}$ is the perpendicular component of the
fluctuating dipolar field produced by a tunneling molecule on its
neighbors and $\tau_{\mathrm{T}}^{-1}$ is the tunneling rate. Even
for nearest neighbors $b_{\mathrm{dip}} < 3$ mT, which leads to
the condition $W \simeq 0.03$ s$^{-1}$ $\Rightarrow
\tau_{\mathrm{T}} \gg 10^6$ s$^{-1}$. This is completely
unrealistic, even for the FRMs with two flipped Jahn-Teller axes.
It is therefore unavoidable to consider the effect of a tunneling
molecule on the nuclei that \emph{belong} to the molecule itself,
and to look for an additional mechanism that links nuclei in FRMs
with equivalent nuclei in frozen clusters. It is natural to seek
the origin of such a mechanism in the intercluster nuclear spin
diffusion, and we shall provide strong experimental evidences to
support such an interpretation.

By magnifying the NSLR rate in the quantum regime (inset of Fig.
\ref{W1W2vsT} we notice that the plateau is in fact not perfectly
horizontal but seems to have two slopes, with most pronounced one
for $T \gtrsim 0.3$ K. We may attribute this slope to the
thermally-assisted intrawell fluctuations in the FRMs, i.e. the
same phenomenon as observed for $T>0.8$ in the whole sample. Since
the FRMs have a lower anisotropy and therefore lower $\Delta E$,
we may expect their crossover from quantum to thermally activated
regime to take place at a lower temperature. Magnetization
measurements \cite{wernsdorfer99EPL} have indicated that the
crossover takes place at $T\simeq 0.6$ K in the FRMs with just one
flipped Jahn-Teller axis, but this value could be of course lower
for the FRMs with two flipped axis and 15 K barrier.

Fig. \ref{W1W2vsT} also shows the $T$-dependence of the TSSR rate
$T_2^{-1}(T)$. We observe that above 0.8 K $T_2^{-1}(T)$ has the
same $T$-dependence as the NSLR rate, and can be fitted by Eq.
(\ref{invT2}) with $\Delta E = 14.4$ K. This indicates that the
fluctuations arising from thermal activation of the electron spin
are responsible for the TSSR as well, at least in this temperature
regime.

Below 0.8 K the TSSR also saturates to a nearly $T$-independent
plateau. By fitting the decay of transverse magnetization with Eq.
(\ref{T2}) at all temperatures we obtain $T_2^{-1} \simeq 100$
s$^{-1}$. This is just a rough estimate since, as mentioned in
\S\ref{sec:measanal}, the transverse relaxation is not well
described by a single exponential decay below $T \lesssim 0.2$ K.
Nevertheless, we find this estimate to be comparable with the
value $T_2^{-1} = 111$ s$^{-1}$ calculated on basis of the
dipole-dipole coupling of nuclei in neighboring clusters
(\S\ref{TSSR}). It is of great interest to point out that in a ZFC
sample, to which the data in Fig. \ref{W1W2vsT} refer, only half
of the nuclei in neighboring molecules should be taken into
account as possible partners for flip-flop transitions. Since half
of the cluster spins point up and the other half point down, the
nuclear spins in equivalent sites should be split in two groups
having Larmor frequencies $+\omega_{N}$ and $-\omega_N$,
respectively. For the purpose of calculating the TSSR rate that
results from intercluster nuclear spin diffusion, this is
equivalent to having diluted the system by a factor 2. Carrying
out the sum (\ref{vanvleck}) over half of the spins only, reduces
the calculated TSSR rate by a factor $\sqrt{2}$, yielding
$T_2^{-1} = 78$ s$^{-1}$ for a ZFC sample, whereas the
above-mentioned $T_2^{-1} = 111$ s$^{-1}$ remains the correct
estimate for a FC sample. Moreover, the dipole-dipole coupling
should lead to a roughly gaussian decay of the transverse nuclear
magnetization \cite{abragam61}.

\begin{figure}[t]
\begin{center}
      \leavevmode
      \epsfxsize=100mm
      \epsfbox{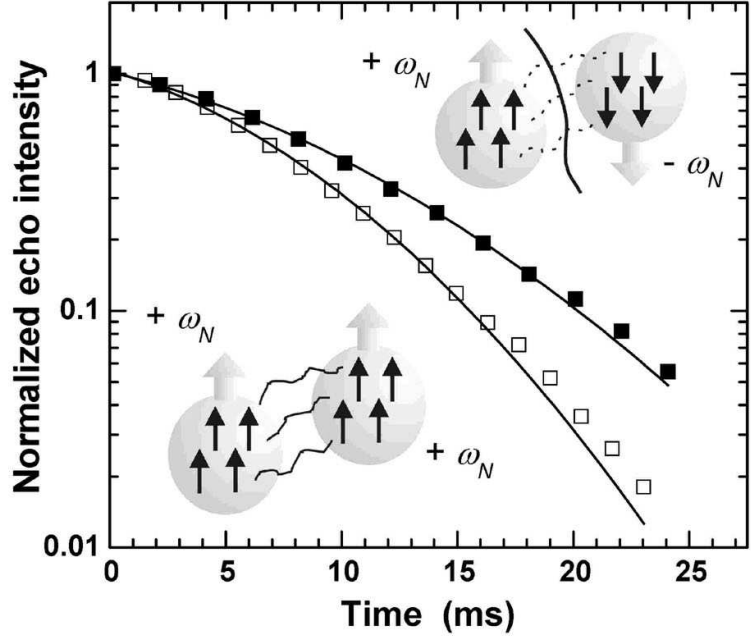}
\end{center}
\caption{\label{T2FCZFC} Decay of transverse nuclear magnetization
in FC (open circles) and ZFC (full circles) sample at $T=20$ mK,
zero applied field and $\nu=231$ MHz. The lines are fits to Eq.
(\ref{T2G}) yielding the ratio
$T_{2G}^{-1}(\mathrm{FC})/T_{2G}^{-1}(\mathrm{ZFC}) = 1.35 \simeq
\sqrt{2}$. The inset sketches represent pictorially the fact that
intercluster spin diffusion is possible in a FC sample since all
the nuclei have the sam Larmor frequency, contrary to the case of
a ZFC sample.}
\end{figure}

We have indeed been able to verify both these predictions by
measuring the TSSR at $T=20$ mK in a FC and a ZFC sample, as shown
in Fig. \ref{T2FCZFC}. The decay of the transverse magnetization
is indeed best fitted by Eq. (\ref{T2G}), whereby the Gaussian
component, $T_{2G}^{-1}$, is separated from the Lorentzian one,
$T_{2L}^{-1}$. From the Gaussian component of the decay we can
extract directly the effect of the nuclear dipole-dipole
interaction, whereas the other mechanisms of dephasing (e.g.
random changes in the local field, cf. \S\ref{TSSR}) contribute
mainly to the Lorentzian part. The fit yields
$T_{2G}^{-1}(\mathrm{FC}) = 104 \pm 3$ s$^{-1}$ and
$T_{2G}^{-1}(\mathrm{ZFC}) = 77 \pm 3$ s$^{-1}$, respectively; the
ratio $T_{2G}^{-1}(\mathrm{FC})/T_{2G}^{-1}(\mathrm{ZFC}) = 1.35$
of the TSSR rates in FC and ZFC samples agrees indeed with the
prediction of a $\sqrt{2}$ reduction of the intercluster flip-flop
rate in a ZFC sample\footnote{In reality it is likely that, even
in the ZFC sample, some clusters may have locally the same spin
orientation, thereby reducing the ratio to something less than
$\sqrt{2}$, as observed.} and constitutes solid evidence for the
presence of intercluster nuclear spin diffusion. This experimental
finding points to the need of an extension of the basic assumption
of the Prokof'ev-Stamp theory \cite{prokof'ev96JLTP} to the case
of an assembly of interacting ``central spin - spin bath'' units.
It also supports our previous interpretation of the mechanism of
nuclear spin-lattice relaxation in terms of the combined effect of
tunneling in FRMs plus nuclear spin diffusion. We stress that, in
view of our quantitative analysis of the TSSR, the fact that the
NSLR and the TSSR are both roughly $T$-independent below 0.8 K
does not mean that they have the same origin in this regime.
Rather, we attribute them to two different mechanisms, both
$T$-independent: the quantum tunneling of the electron spin (for
the NSLR) and the nuclear spin diffusion (for the TSSR). It is
only for $T>0.8$ that the same thermal fluctuations are
responsible for NSLR and TSSR at the same time.

We point out that, in spite of the even quantitative agreement
with the simple picture of intercluster nuclear spin diffusion as
described by Eq. (\ref{vanvleck}), a puzzling point remains. Due
to the dipolar field produced by neighboring clusters, $^{55}$Mn
nuclei in different molecules will have different Zeeman energies,
and the typical magnitude of the energy spread involved is
actually much greater than the intercluster nuclear dipole-dipole
coupling, and should lead to a strong suppression of the flip-flop
probability. We shall suggest a solution to this problem in the
next subsection, when discussing the field dependence of the TSSR
rate.

\subsection{Field dependencies} \label{fielddep}

Further insight in the interplay between quantum tunneling and
nuclear spin dynamics is provided by the study of the dependence
of the NSLR and TSSR rates on a magnetic field $B_z$ applied along
the anisotropy axis. It is clear from the Hamiltonian
(\ref{hamiltonianMn12}) that, in the absence of other
perturbations, such a longitudinal field destroys the resonance
condition for corresponding electron spin states on both sides of
the barrier and therefore inhibits the quantum tunneling. In the
presence of static dipolar fields, $B_{\mathrm{dip}}$, by studying
the tunneling rate as a function of $B_z$ one may in principle
obtain information about the distribution of longitudinal
$B_{\mathrm{dip}}$, since at a given value of $B_z$ there will be
a fraction of molecules for which $B_{\mathrm{dip}} = -B_z$ and
will therefore be allowed to tunnel just by the application of the
external bias. We measured the longitudinal field dependence of
the NSLR rate $W(B_z)$ while shifting the measuring frequency $\nu
(B_z) = \nu (0) + \gamma_N B_z$ with $\nu(0)=230$ MHz, in order to
stay on the center of the NMR line that corresponds to the
molecules that are aligned exactly parallel with the applied
field.

\begin{figure}[t]
\begin{center}
      \leavevmode
      \epsfxsize=100mm
      \epsfbox{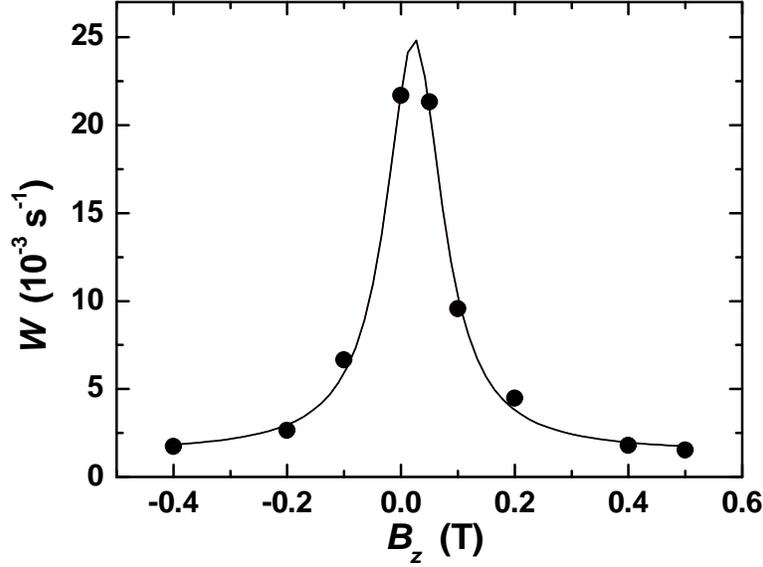}
\end{center}
\caption{\label{WZFCvsB} Longitudinal field dependence of the NSLR
rate $W$ in ZFC sample at $T=20$ mK. The measuring frequencies are
$\nu(B_z) = 230 + \gamma_N B_z$ MHz. The solid line is a
Lorentzian fit with full width at half maximum $\Delta B_z = 0.12$
T.}
\end{figure}

\begin{figure}[t]
\begin{center}
      \leavevmode
      \epsfxsize=130mm
      \epsfbox{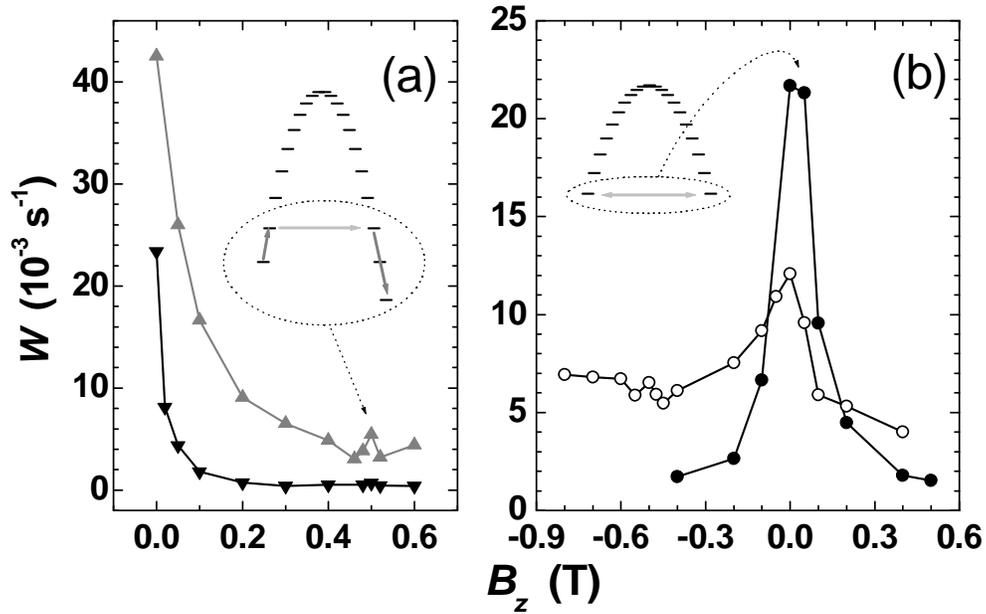}
\end{center}
\caption{\label{WvsBall} (a) Longitudinal field dependence of the
NSLR rate in ZFC sample at $T=20$ mK (down triangles) and $T=720$
mK (up triangles). The measuring frequency in these datasets is
$\nu = 231 + \gamma_N B_z$ MHz. (b) Longitudinal field dependence
of the NSLR rate in FC (open circles) and ZFC (full circles)
sample at $T=20$ mK, with measuring frequency $\nu = 230 +
\gamma_N B_z$ MHz. The insets are sketches of the electron spin
transitions responsible for the observed features.}
\end{figure}

The results for the ZFC sample at $T=20$ mK are shown in Fig.
\ref{WZFCvsB}. Since for a ZFC sample the magnetization is zero,
the field dependence should be the same when $B_z$ is applied in
opposite directions, as is observed. The data can be fitted by a
Lorentzian with a full width at half maximum $\Delta B_z = 0.12$
T, thus different both in shape and in width from the calculated
dipolar bias distribution (Fig. \ref{dipolardistr}). A possible
reason for the extra broadening observed in $W(B_z)$ could be the
presence of misaligned molecules in the sample, combined with
quadrupolar broadening of the NMR line. The manifold of
quadrupolar-split Mn$^{(1)}$ lines can be closely described as a
gaussian with total width $2 \sigma_{\nu} \simeq 2.4$ MHz, i.e.
$\simeq 0.23$ T in field units, thus already broader than
$W(B_z)$. Although we follow the center of the line for the
well-aligned molecules, the misaligned ones (which experience a
smaller longitudinal field) could still be in condition to tunnel
and contribute to the NMR signal by means of their quadrupolar
satellite lines. An obvious source of misalignment is the fact
that our sample consists of an ensemble of crystallites, thus
there is the chance for part of them to be misoriented. In fact,
the misalignment is in any case unavoidable for fast-relaxing
molecules, which we consider to be the actual source of quantum
tunneling events. It is known (cf. \S\ref{paramMn12}) that the
FRMs with one flipped Jahn-Teller axis also have a local
anisotropy axis for the cluster spin which deviates $\sim
10^{\circ}$ from the crystallographic $\hat{c}$-axis, and the
misalignment could be even greater for the FRMs with two flipped
J-T axes. A confirmation for the role of misaligned molecules in
the width of $W(B_z)$ comes from the fact that, if we repeat the
experiment taking $\nu(0) = 231$ MHz, thus slightly higher than
the center of the line, $W(B_z)$ appears indeed much narrower
[Fig. \ref{WvsBall}(a)] since the tails of the NMR lines in
misaligned molecules are now farther away from the measuring
frequency. The central frequency for the data in Fig.
\ref{WZFCvsB} is $\nu(0) = 230$ MHz, which is just slightly below
the center of the line, and this is sufficient to cause the
maximum of $W(B_z)$ to be at $B_z \neq 0$

Fig. \ref{WvsBall}(a) also shows a comparison between $W(B_z)$ at
$T=20$ and 720 mK, thus just at the edge of the thermally
activated regime. It appears that both $W(0)$ and the width of the
resonance increase by raising the temperature. In addition, at
$T=720$ mK an extra small peak appears at $B_z \simeq 0.5$ T,
which value coincides with the first level crossing for
slow-relaxing molecules. Since the peak is not visible at 20 mK,
we may attribute it to tunneling of slow molecules through higher
excited doublets, e.g. $S_z = -9 \rightarrow +8$, which becomes
favorable as soon as the thermally activated regime is approached
(\S\ref{coupling}). Moreover, the fact that $W(B_z)$ at $T=720$ mK
and fields $0 < B_z < 0.5$ T is largely increased as compared to
the data at 20 mK, would confirm that the thermal activation in
FRMs becomes an important source of relaxation already much below
0.8 K, as already argued from the analysis of the slope of $W(T)$
in the quantum regime.

In Fig. \ref{WvsBall}(b) we compare the ZFC data of Fig.
\ref{WZFCvsB} with a similar field scan of $W(B_z)$ at $T=20$ mK
for an FC sample. We notice that in the FC case it is possible to
define both positive and negative longitudinal fields, i.e.
corresponding to whether $B_z$ is applied parallel or opposite to
the magnetization of the sample. One may observe in Fig.
\ref{WvsBall}(b) that the shape of the resulting field-dependence
is asymmetric, in a way that resembles the asymmetry of the
dipolar field distribution (Fig.\ref{dipolardistr}). On the other
hand, the width in the FC is much larger than in the ZFC sample,
contrary to the prediction for $B_{\mathrm{dip}}$. Moreover, as
mentioned before, the dipolar field distribution is one order of
magnitude narrower than the peaks in $W(B_z)$. At increasingly
large negative fields (where negative means opposite to the
magnetization the sample) $W(B_z)$ stabilizes to a rather high
plateau with a small anomaly around $B_z \simeq -0.5$ T. We shall
suggest a possible explanation for the high values of $W(B_z)$ in
FC sample at negative fields when discussing Fig. \ref{spinTsweep}
in \S\ref{superradiance}.

The peak at $B_z \simeq 0.5$ T in ZFC sample at 720 mK and the
anomaly at $B_z \simeq -0.5$ T in FC sample suggest that tunneling
in slow-relaxing molecules may also give a contribution under
special conditions like high $T$ or inverted magnetization. An
important and interesting observation is that, contrary to the
TSSR (cf. Fig. \ref{T2FCZFC}), at zero applied field the NSLR is
faster in the ZFC than in the FC sample \cite{morello03JMMM}.
Since we argued above that intercluster nuclear spin diffusion
should be faster for the FC than for the ZFC sample, this would
indicate that there are more tunneling events in a ZFC sample. As
mentioned before, such an observation is only possible thanks to
the fact that NMR is sensitive to magnetic fluctuations rather
than to macroscopic changes in the magnetization.

\begin{figure}[t]
\begin{center}
      \leavevmode
      \epsfxsize=90mm
      \epsfbox{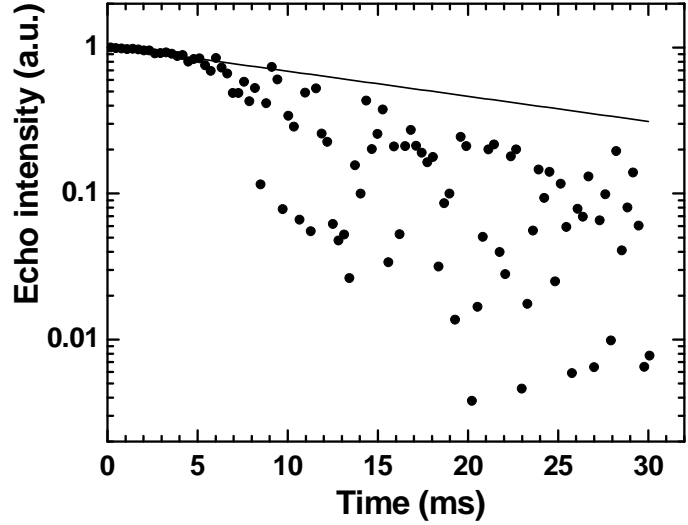}
\end{center}
\caption{\label{T2decayB02} Decay of transverse magnetization at
$T=20$ mK in ZFC sample, with an applied field $B_z = 0.2$ T.
Notice the chaotic behavior after about 5 ms. The solid line is a
fit to Eq. (\ref{T2}) in the first part of the decay.}
\end{figure}

We also investigated the field dependence of the TSSR rate, but we
encountered an unexpected and rather intriguing phenomenon:
already for $\left| B_z \right| \gtrsim 0.05$ T, the decay of
transverse magnetization is characterized by a ``chaotic region''
that appears after a few milliseconds, as shown in Fig.
\ref{T2decayB02}. In other words, after an initially well-behaving
decay, the echo intensity fluctuates at random, although it always
remains below the envelope that can be extrapolated from the first
part of the decay. In the same way, by measuring several times the
echo intensity with the same delay between pulses, one would
randomly find any intensity between zero and the height of the
extrapolated envelope. Also, the extension of the initial
well-behaving decay tends to reduce with increasing field. Such a
chaotic behavior is observed both in FC and ZFC samples, and both
in Mn$^{4+}$ and Mn$^{3+}$ sites (see \S\ref{mn3+}). We stress
that this has nothing to do with instrumental noise: at any point
we see a clean echo signal, just the amplitude and the phase are
totally random. We don't have an explanation for this phenomenon,
but it's intriguing to remark that 0.05 T is just the maximum
spread of the dipolar bias distribution (Fig. \ref{dipolardistr}).

\begin{figure}[t]
\begin{center}
      \leavevmode
      \epsfxsize=130mm
      \epsfbox{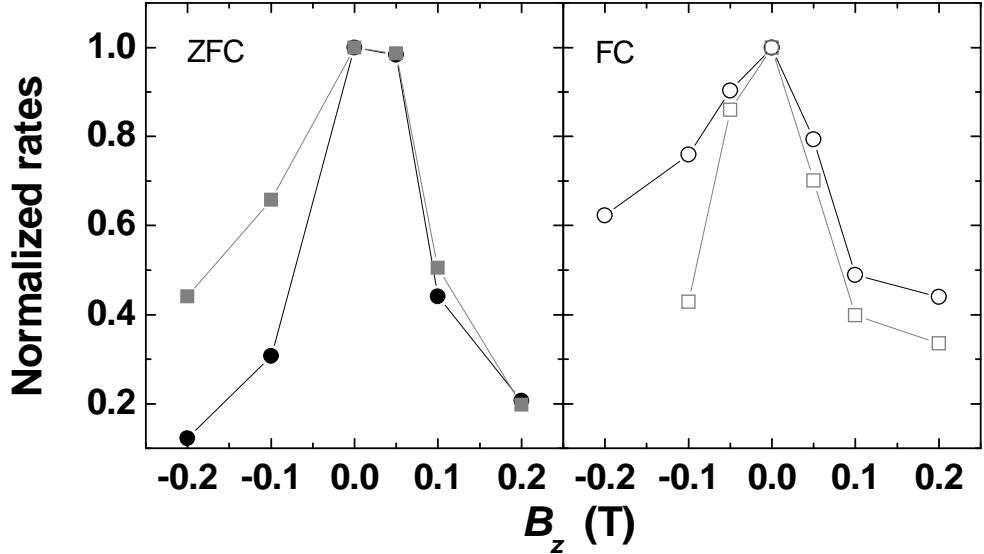}
\end{center}
\caption{\label{W1W2vsBren} Comparison between the field
dependencies of the NSLR (circles) and TSSR (squares) rates,
normalized to the zero-field value, in ZFC (left panel) and FC
(right panel) sample, at $T=20$ mK.}
\end{figure}

By fitting just the initial part of the decay we can still
tentatively extract the field dependence of the TSSR rate
$T_2^{-1}(B_z)$. Unfortunately, on such a short interval it is not
possible to separate the gaussian and exponential contributions to
the decay, thus we fitted all the data to Eq. (\ref{T2}) which for
the zero-field dataset gives just an average value. Given the
circumstances it's not easy to evaluate the error bars, but we may
assume them to be rather large. Fig. \ref{W1W2vsBren} shows a
comparison between $W(B_z)$ and $T_2^{-1}(B_z)$, both normalized
to the zero-field value. As it appears, the field dependencies of
the NSLR and TSSR rates are quite similar, although curiously the
similarity seems closer for the positive field direction, relative
to the FC sample. Moreover, the observation that the TSSR is
faster in a FC than in ZFC sample remains valid when $B_z \neq 0$.

We suggest that there is a possibility to explain at the same time
the resemblance of the field dependencies of NSLR and TSSR rates
and the efficiency of the intercluster nuclear spin diffusion. The
tunneling in fast-relaxing molecules produces a fluctuating
dipolar field, such that the Zeeman energy of $^{55}$Mn nuclei in
equivalent sites of neighboring molecules also fluctuates in time;
in this way, there is a chance that at some instant the nuclei in
different molecules will have precisely the same Zeeman splitting
and will therefore easily undergo a flip-flop transition. If we
accept this mechanism as responsible for the efficiency of the
intercluster spin diffusion, then it's clear that by applying a
longitudinal field the drastic reduction of tunneling rate should
also decrease the probability that two nuclei in neighboring
clusters would come at exact resonance, resulting in a reduction
of the TSSR rate with the same field dependence as the NSLR rate,
as indeed observed.

\subsection{Deuterated sample} \label{deuterated}

The role of the fluctuating hyperfine bias on the incoherent
tunneling dynamics of SMMs, predicted by Prokof'ev and Stamp
(\S\ref{PStheory}), has been very clearly demonstrated by
measuring the quantum relaxation in Fe$_8$ crystals where the
hyperfine couplings were artificially modified by substituting
$^{56}$Fe by $^{57}$Fe or $^1$H by $^2$H \cite{wernsdorfer00PRL}.
For instance, the time necessary to relax 1\% of the saturation
magnetization below 0.2 K was found to increase from 800 s to 4000
s by substituting protons by deuterium, whereas it decreased to
300 s in the $^{57}$Fe enriched sample. We already mentioned in
\S\ref{PStunnelrate} that the ``tunneling window'' for the cluster
spin is most effectively enlarged by those nuclei that easily
coflip with the cluster spin. In SMMs, this is indeed the case for
the nuclei located at ligand molecules (like $^1$H) because their
local field is the sum of the dipolar fields from several
clusters, whereas the $^{55}$Mn or the $^{57}$Fe nuclei are less
likely to coflip since the hyperfine fields they experience before
and after the flip are almost exactly antiparallel.

\begin{figure}[t]
\begin{center}
      \leavevmode
      \epsfxsize=130mm
      \epsfbox{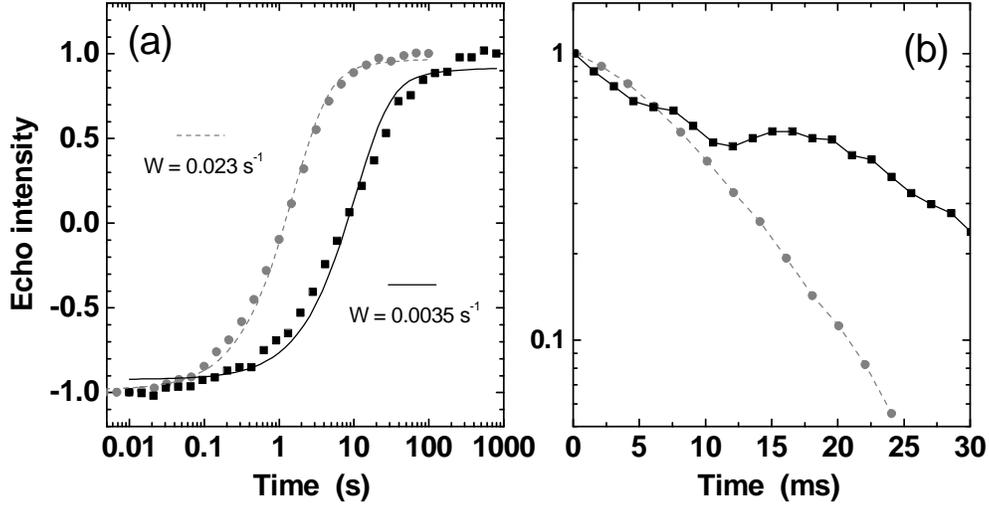}
\end{center}
\caption{\label{T1T2D} Comparison between the nuclear inversion
recoveries (a) and the decays of transverse magnetization (b) in
the "natural" Mn$_{12}$-ac (circles) and in the deuterated sample
(squares), at $T=20$ mK in zero field and ZFC sample, for the
Mn$^{(1)}$ site. The solid lines in (a) are fits to Eq.
(\ref{recovery}).}
\end{figure}

In this sense, it is understandable that the largest effect on the
tunneling rate is seen by isotope substitution of $^1$H by $^2$H.
The deuterium nuclei have spin $I = 1$ but their gyromagnetic
ratio is 6.5 times lower than for protons. This is the reason why,
in the framework of the PS theory, one expects indeed a reduction
of the incoherent tunneling rate due to the weaker (fluctuating)
hyperfine bias. Since in Mn$_{12}$-ac the only possible isotope
substitution is $^1$H $\rightarrow ^2$H, we performed a short set
of measurements on a deuterated sample. The sample consists of
much smaller crystallites than the ``natural'' ones used in all
other experiments reported in this chapter. Despite having been
prepared following the same procedure as described in
\S\ref{sec:measanal}, the orientation of the deuterated sample
turned out to be almost completely random, probably due to the too
small shape anisotropy of the crystallites\footnote{At room
temperature, the shape anisotropy is indeed the only mechanism by
which the crystallites may orient in a magnetic field.}. We report
therefore only experiments in zero external field, where the
orientation is in principle irrelevant.

The results are shown in Fig. \ref{T1T2D}: the $^{55}$Mn NSLR rate
at $T=20$ mK in zero field and ZFC sample is indeed reduced to
$W_{\mathrm{deut}} \simeq 0.0035$ s$^{-1}$, i.e. precisely 6.5
times lower than in the ``natural'' sample! This clearly points to
a reduction of the electron spin tunneling rate of the same type
as observed in the deuterated Fe$_8$ \cite{wernsdorfer00PRL}, and
confirms once more that the NSLR is determined precisely by such
tunneling fluctuations. As regards the TSSR, the results is very
intriguing: slow but rather ample oscillations are superimposed to
the decay of transverse magnetization, somehow reminiscent of what
is observed in the natural sample upon application of a small
field (Fig. \ref{T2decayB02}), but without the chaotic features.
Fitting the average decay constant [Eq. (\ref{T2})] yields
$T_2^{-1} \simeq 38$ s$^{-1}$, i.e. a TSSR $\sim 2.5$ times slower
than in the natural sample. Such reduction is consistent with the
hypothesis that the intercluster nuclear spin diffusion is helped
by the fluctuation of the bias on the nuclear Zeeman levels when
tunneling events occurs (\S\ref{fielddep}). As a matter of fact,
both the NSLR and the TSSR in the deuterated sample are rather
similar to the data in the ``natural'' sample with an applied
field $B_z \simeq 0.05$ T. The oscillations in the decay of
transverse magnetization remain an obscure phenomenon, but
observing them even in this zero-field experiment seems to rule
out instrumental artifacts, and suggests that something bizarre
happens in the nuclear spin-spin interactions as soon as the
tunneling fluctuations slow down below a certain threshold.

\subsection{Comparison with a Mn$^{3+}$ site}  \label{mn3+}

Some rather intersting results come from the analysis of extra
measurements performed on the NMR line of the Mn$^{(2)}$ site,
i.e. a Mn$^{3+}$ ion, which at low temperature is centered at
$\nu^{(2)} \simeq 283.7$ MHz. We didn't follow the whole
temperature evolution of the spectrum, but comparing our data with
the previous experiments\cite{furukawa01PRB,kubo02PRB} at higher
$T$ it seems that the resonance line has slightly shifted to
higher frequencies upon cooling.

\begin{figure}[t]
\begin{center}
      \leavevmode
      \epsfxsize=130mm
      \epsfbox{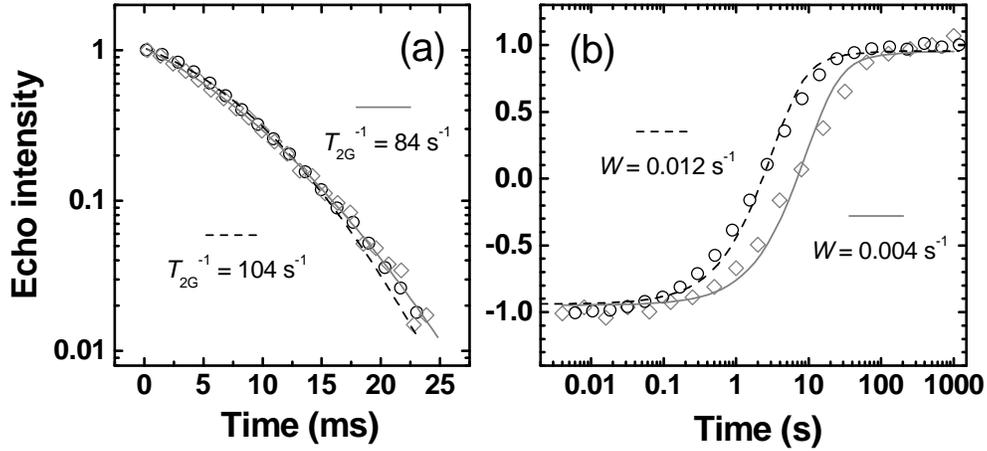}
\end{center}
\caption{\label{T1T2Mn34}  Comparison between the decay of
transverse magnetization (a) and the recovery of longitudinal
magnetization (b) in Mn$^{(1)}$ (circles) and Mn$^{(2)}$
(diamonds) sites, at $T=20$ mK in FC sample and zero external
field. The solid (Mn$^{(2)}$) and dashed (Mn$^{(1)}$) lines are
fits to Eq.(\ref{T2G}) in panel (a) and Eq.(\ref{recovery}) in
panel (b).}
\end{figure}

\begin{figure}[t]
\begin{center}
      \leavevmode
      \epsfxsize=100mm
      \epsfbox{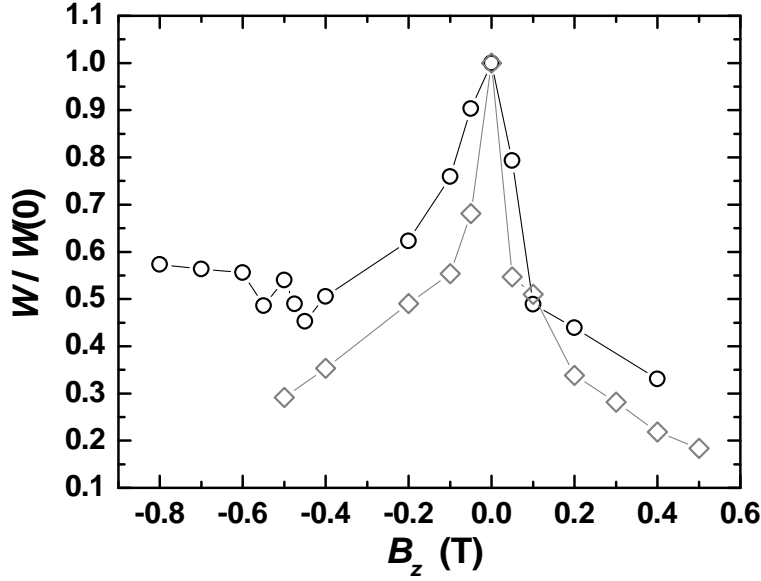}
\end{center}
\caption{\label{WvsB34}  Longitudinal field dependencies of the
NSLR rates in Mn$^{(1)}$ (circles) and Mn$^{(2)}$ (diamonds)
sites, normalized at the zero-field value. The data are taken at
$T=20$ mK in FC sample with central measuring frequencies
$\nu^{(1)}(0) = 230$ MHz and $\nu^{(2)}(0) = 283.7$ MHz.}
\end{figure}

Fig.\ref{T1T2Mn34} shows a comparison between the decay of
transverse magnetization and the recovery of the longitudinal
magnetization in Mn$^{(1)}$ and Mn$^{(2)}$ sites, at $T=20$ mK in
FC sample and zero external field. The TSSR is very similar in
both sites, although a closer inspection evidences that the
gaussian nature of the decay is less pronounced in the Mn$^{(2)}$
sites, which leads to $T_{2G}^{-1} = 84$ s$^{-1}$ instead of the
$T_{2G}^{-1} = 104$ s$^{-1}$ found in Mn$^{(1)}$. More
importantly, the NSLR is three times slower in the Mn$^{(2)}$
site, as seen in Fig. \ref{T1T2Mn34}(b). This is opposite to the
high-$T$ regime\cite{furukawa01PRB,goto03PRB}, where the Mn$^{3+}$
sites were found to have much faster relaxation. Furthermore, the
field dependence of the NSLR rate appears sharper in the
Mn$^{(2)}$ site, as shown in Fig.\ref{WvsB34}. The asymmetry in
$W(B_z)$ for a FC sample is still present, but less evident than
in the Mn$^{(1)}$ site due to the more pronounced decrease of $W$
already for small applied fields. These results suggest that the
tunneling in fast-relaxing molecules is less effective in relaxing
the nuclei in a Mn$^{3+}$ site than in a Mn$^{4+}$ site, although
the intercluster nuclear spin diffusion appears to work comparably
well. In fact, the measured TSSR rate is mainly determined by
intercluster spin diffusion between nuclei in slow molecules,
since the nuclear spins in FRMs are just a small minority. On the
contrary, the NSLR rate is determined by the combination of
tunneling in FRMs and nuclear spin diffusion \emph{between fast
and slow molecules}. Since the FRMs differ from the majority
clusters precisely in the electron spin arrangement of the
Mn$^{3+}$ sites, it is conceivable that this would be an obstacle
for the coupling between the nuclei in Mn$^{3+}$ sites of fast and
slow molecules, which are no longer equivalent, whereas the
Mn$^{4+}$ sites remain relatively unaffected by the distortion in
FRMs. This would explain both the much slower NSLR because of the
``broken link'' between nuclei in fast and slow molecules, and the
slightly slower TSSR where we just notice the ``absence'' of the
small fraction of nuclei located at the FRMs from the
participation to intercluster spin diffusion.

When the temperature is increased such that also the slow-relaxing
molecules start to have considerably fast tunneling fluctuations
via higher excited levels ($T > 0.8$ K), the problem of bad
nuclear spin diffusion at Mn$^{3+}$ sites between fast and slow
molecules becomes irrelevant. We may then expect that the NSLR in
Mn$^{3+}$ sites is produced by the combined action of the
fluctuating transverse hyperfine field due to intrawell
excitations [the model on which Eq. (\ref{Wlast}) is based], plus
the effect of thermally-assisted tunneling. In Mn$^{4+}$ sites the
first contribution should be absent, which could explain why the
NSLR in the thermally-activated regime is lower than in Mn$^{3+}$
sites.

We may therefore conclude that most if not all the different data
presented in \S\ref{lowfield} can be consistently explained within
our model of tunneling in FRMs plus intercluster nuclear spin
diffusion.

\section{Effects of strong perpendicular fields}

In this section we describe the effects on the NSLR rate of a
strong magnetic field $B_{\perp}$, applied perpendicular to the
anisotropy axis. The interpretation of the NSLR is tightly related
to the NMR spectra, which provide useful information about the
orientation of the crystallites and the crossover between
$^{55}$Mn and $^1$H Larmor frequencies.

\subsection{NMR spectra}

In a single crystal having the $\hat{c}$-axis exactly
perpendicular to the external field, the three $^{55}$Mn NMR lines
should evolve as in Fig. \ref{spectraBx}, neglecting the
possibility of delocalization of the spin wave functions
(\S\ref{sec:spectraBx}). Our sample consists of a large number of
crystallites, oriented (within a certain spread) along an axis
which is orthogonal to both the external field and the axis of the
NMR coil. In this way, whatever is the canting angle of the
electron spins towards the external field, the total field at the
nuclei is always perpendicular to the axis of the NMR coil.

\begin{figure}[p]
\begin{center}
      \leavevmode
      \epsfxsize=110mm
      \epsfbox{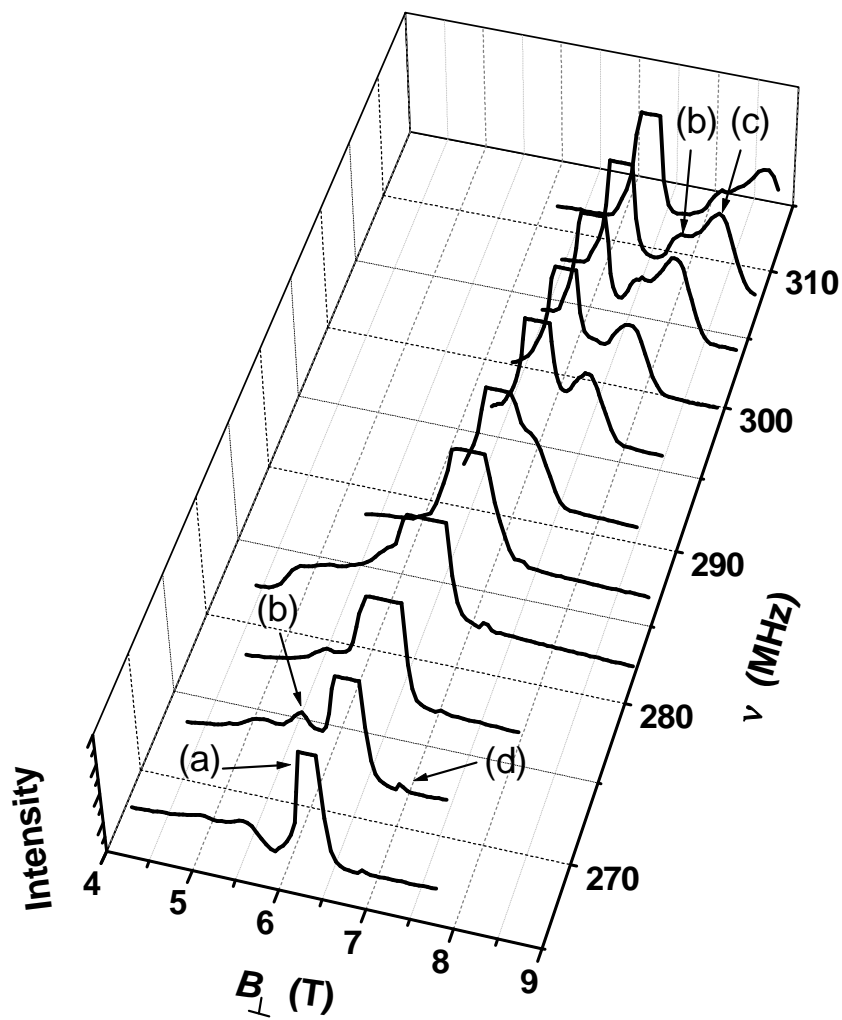}
\end{center}
\caption{\label{plot3D} Field-sweep NMR spectra taken at $T=20$
mK. Several features, marked by letters, are visible and discussed
in the text. The high and chopped peak observed in the range $6-7$
T is due to the protons in the sample.}
\end{figure}

Fig. \ref{plot3D} shows field-sweep spectra taken at $T=20$ mK.
The sample can be considered as saturated, since we applied 9 T
before starting the sweeps. By this procedure we expect most of
the cluster spins to be aligned in the same direction within each
crystallite, except eventually for those crystallites that are
oriented perpendicular to the field within a fraction of a degree.
The signal belonging to the protons (a) is recognizable by the
chopped top; in fact, the $^1$H line is so intense that it
saturates the receiver on this scale, and to observe it properly
we had to attenuate the input by at least 30 dB. Focusing on the
Mn$^{(1)}$ line, we observe that at low frequencies and fields the
signal is characterized by a peak (b), plus a more spread-out
background that we attribute to misoriented crystallites. The peak
fits indeed with the calculated position for crystallites with
easy axis oriented at an angle $\theta_B = 88^{\circ} -
90^{\circ}$ with the field (see below), as explained in
\S\ref{sec:spectraBx}. By increasing the field, line (b) crosses
and is retrieved on the right hand side of line (a), but now two
peaks are clearly visible. The peak on the right, labelled (c),
tends to increase in intensity at the expenses of (b) when moving
to higher fields and frequencies. For instance, the ratio between
the amplitudes of the (b) and (c) peaks is 0.73 at $\nu = 303.2$
MHz and 0.67 at $\nu = 311$ MHz. An extra series of small and
sharp peaks (d) is also visible on the right hand side of the
$^1$H line for $\nu \lesssim 290$ MHz.

\begin{figure}[p]
\begin{center}
      \leavevmode
      \epsfxsize=120mm
      \epsfbox{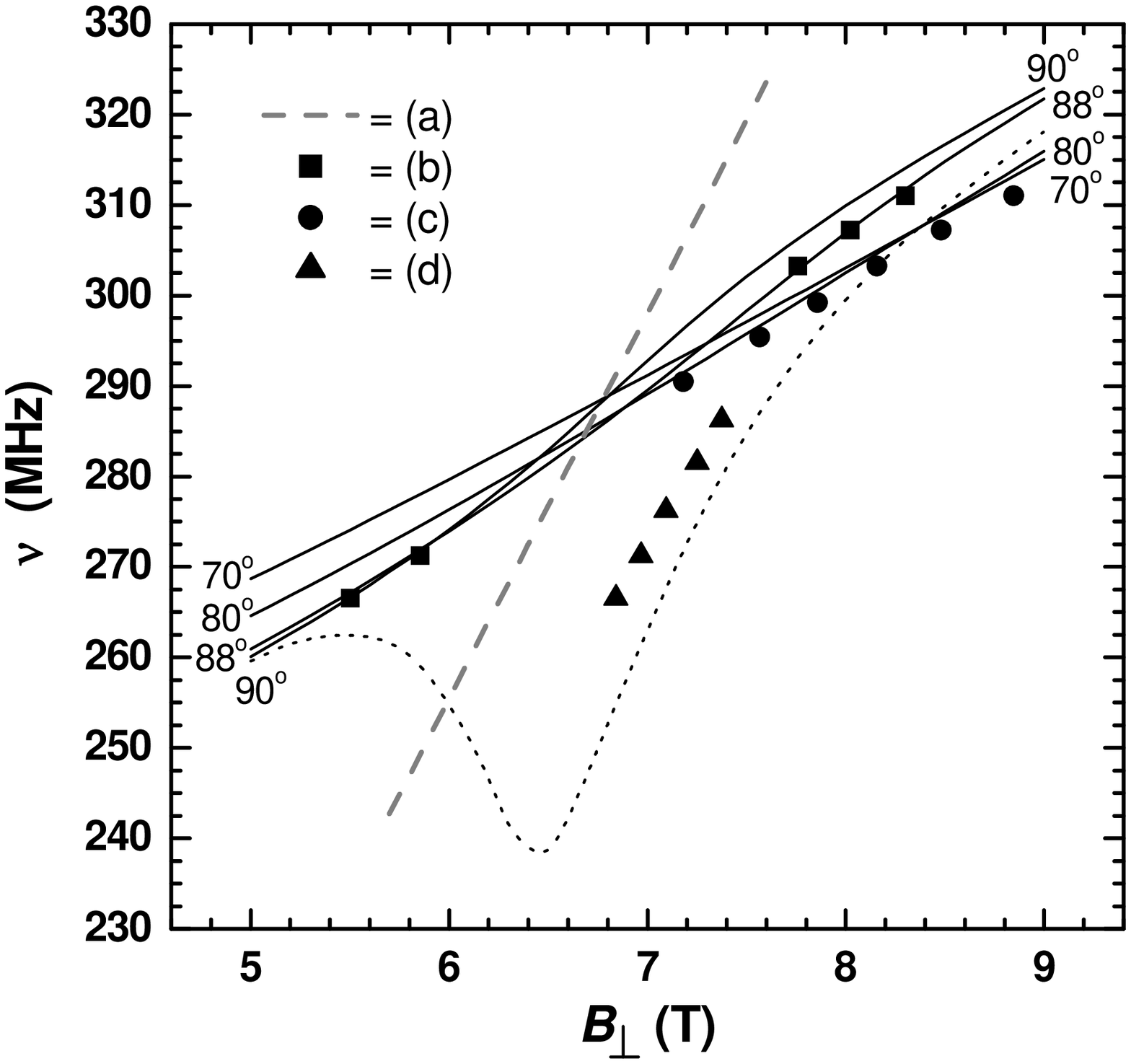}
\end{center}
\caption{\label{peaks} Positions of the peaks in the field-sweep
NMR spectra: the symbols correspond to lines (b), (c) and (d) as
specified in the inset. For the protons (a) only the calculated
central value is given, plotted as a dashed gray line. Solid lines
are calculated spectra according to Eq. (\ref{Btot4}), with the
indicated orientation of the $z$ axis relative to the magnetic
field. The dotted line is the calculated spectrum for $\theta_B =
90^{\circ}$, taking into account the reduction of the total spin
by quantum superposition [Eq. (\ref{Bfree})]}
\end{figure}

In Fig. \ref{peaks} we report the positions of all these peaks and
compare them with the calculations. It is then clear that, of the
two broad peaks at $B_{\perp} > 7$ T and $\nu > 295$ MHz, the left
must indeed be the continuation of line (b) corresponding to
nuclei in crystallites oriented at $88^{\circ}$ from the field.
Remarkably, if line (b) at low fields contained contributions from
crystallites at $\theta_B \simeq 90^{\circ}$ (it's hard to believe
that this would not be the case), they seem to have disappeared at
higher fields. The right peak (c), of increasing height, could
have two origins. On the one hand, the calculated spectra for
$\theta_B = 70^{\circ}- 80^{\circ}$ tend to accumulate precisely
around the position of (c), thus the signal intensity that is
found in the region 5 T - 270 MHz should in principle pile up just
around (c) for $7.5 < B_{\perp} < 8.5$ T. On the other hand, for
$B_{\perp} > 8$ T the tunneling splitting becomes so large (Fig.
\ref{FigDeltavsBx}) as to overcome the bias $\xi$ originating from
the longitudinal component of the dipolar fields, but also that
arising from the misalignment of the crystallites as long as
$(90^{\circ} - \theta_B)$ is not too large. One may expect
(\S\ref{sec:spectraBx}) the NMR signal intensity to start being
displaced towards the calculated line for delocalized electron
spins, beginning with the best aligned crystallites and involving
more and more amounts of sample as $B_{\perp}$ increases. This may
explain why, upon increasing $B_{\perp}$, line (c) seems to grow
at the expenses of line (b), and why there is no trace of signal
at $\theta_B \simeq 90^{\circ}$. Most likely, a detailed
description of the NMR spectra involves a combination of the
phenomena mentioned above.

One may ask to what extent can we trust such fine-detailed
comparisons between measured and calculated spectra, in view of
the fact that the parameters of the spin Hamiltonian are affected
by some uncertainty. We found out that the calculations are in
fact very robust, since the effect of varying the spin Hamiltonian
parameters, even over a range much wider that the accepted error
margins for the commonly used values, is hardly visible on the
scale of Fig. \ref{peaks}.

The extra small line (d) is rather fascinating: its position is in
fact close to the calculated spectrum for a perfectly oriented
crystallite, where the total spin falls into a symmetric
superposition of $|\Uparrow\rangle$ and $|\Downarrow\rangle$ as
soon as $\Delta > \xi_D$ (\S\ref{sec:dipolar}, Fig
\ref{WdipvsDelta}). This calculation, contrary to those performed
assuming $S=10$ (solid lines in Fig. \ref{peaks}), is extremely
sensitive to the values of the parameters of the spin Hamiltonian.
One could easily allow the dotted line in Fig. \ref{peaks} to be
drawn just on top of the data, by varying the parameters in the
calculation within a reasonable margin. To further investigate
this issue we measured field-sweep spectra on another sample,
consisting of deuterated Mn$_{12}$-ac crystallites having an
almost random orientations (\S\ref{deuterated}). Line (d) is still
visible but its intensity is even smaller than shown in Fig.
\ref{plot3D}. Moreover, it continues linearly down to at least 230
MHz, i.e. much lower than expected from the calculation (cf. Fig.
\ref{WdipvsDelta}), without showing any turning point. The most
plausible origin of line (d) is therefore a set of $^1$H nuclei
very close to the magnetic core and experiencing an extra dipolar
field $\simeq 0.6$ T; in the deuterated sample, the intensity is
further reduced because of the substitution of most of the $^1$H
by $^2$H. The $^{55}$Mn signal in the deuterated sample does not
yield well-resolved peaks because of the random orientation of the
crystallites.

The NMR lines corresponding to nuclei in Mn$^{3+}$ sites, whose
calculated frequencies for $\theta_B = 90^{\circ}$ are shown in
Fig. \ref{spectraBx}, are broader and less intense than the
Mn$^{(1)}$ line \cite{furukawa01PRB,kubo02PRB}. The distribution
of $\theta_B$ broadens their spectrum even further, such that
their contribution to the measured signal is basically just a
uniform background.

Finally, we remark that all the discussion made so far is relative
to $^{55}$Mn nuclei in Mn$^{(1)}$ sites of standard, slow relaxing
molecules. Nuclei in fast-relaxing molecules have their resonance
frequency shifting much more quickly to higher values, because the
lower anisotropy implies that the electron spin can cant more
easily towards the field. Moreover, the direction of the local
hyperfine field in FRMs is likely to deviate from the
crystallographic $\hat{c}$-axis even n zero applied field
(\S\ref{paramMn12}), yielding an extra contribution to the spread
of the resonance line.

\subsection{NSLR in perpendicular fields}

In view of the results described in \S\ref{lowfield}, it appeared
very interesting to check whether the possibility to increase the
tunneling splitting by means of a perpendicular field can give
visible features in the NSLR rate $W(B_{\perp})$. One would expect
an acceleration of the tunneling rate and therefore an increase of
$W$; in reality, the problem is made more complicated by the
misalignment of the crystallites with respect to the field, and by
the fact that, at low fields, the tunneling dynamics is provided
only by FRMs. Whatever the orientation of a crystallite may be, it
is very unlikely that many FRMs will be exactly perpendicular to
the field, since their local anisotropy axis does not coincide
with $\hat{c}$. As a consequence, one would expect any external
field to effectively destroy the tunneling dynamics of most of the
FRMs.

\begin{figure}[p]
\begin{center}
      \leavevmode
      \epsfxsize=120mm
      \epsfbox{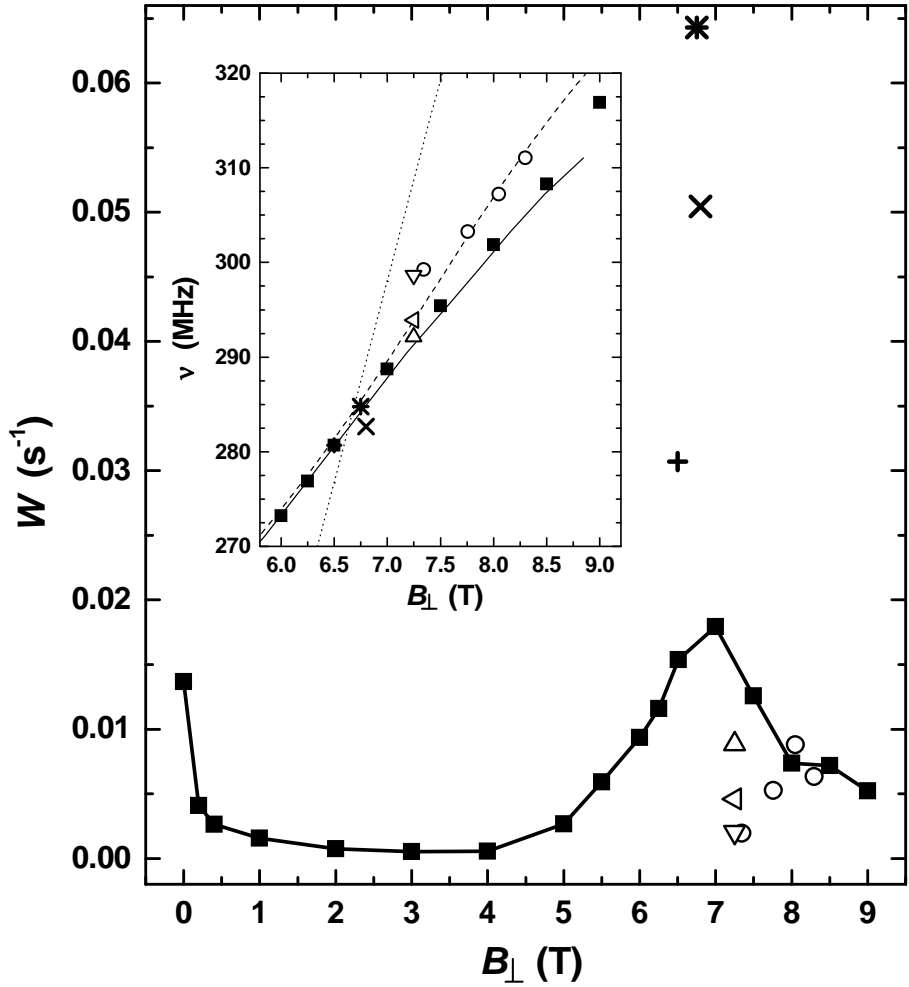}
\end{center}
\caption{\label{WvsBx} NSLR rate as function of perpendicular
field at $T=20$ mK, following the highest signal intensity
(squares), at the crossing between $^1$H and $^{55}$Mn (crosses),
at $B_{\perp} > 7.25$ T as function of frequency (triangles), and
along line (b) (circles). The inset shows the measuring frequency
for each set of datapoints (same symbols as in the main panel),
compared with line (c) (solid), line (b) (dashed) and line (a)
(dotted).}
\end{figure}

Fig. \ref{WvsBx} shows $W(B_{\perp})$ at $T=20$ mK; the values of
$W$ are obtained by fitting the inversion recoveries to Eq.
(\ref{strexp}), i.e. by a stretched exponential. As expected, as
soon as $B_{\perp} > 0.2$ T we observe a strong reduction of $W$,
due to the suppression of tunneling in FRMs. For $B_{\perp} > 4$
T, the NSLR starts to accelerate again: at this point $\Delta \sim
10^{-5}$ K, already larger than in most FRMs, but the effect of
misalignment can still be very important. At higher fields it
becomes essential to specify at which position we take the data.
The main results (square symbols) obey the criterion of maximal
signal intensity at a given field: in practice, line (b) for
$B_{\perp} < 6.5$ T and line (c) for $B_{\perp} > 6.5$ T.

At $B_{\perp} \simeq 6.7$ T, the Mn$^{(1)}$ line crosses the
protons line, which is 30 dB more intense that the $^{55}$Mn line
because of the much larger number of protons present in each
molecule. This fact makes it is very easy to recognize when
protons are involved, but it also means that the Mn$^{(1)}$ signal
is in fact completely masked by the $^1$H signal when the crossing
of the Larmor frequencies occurs. The measured NSLR (crosses in
Fig. \ref{WvsBx}) must therefore be interpreted as the NSLR of the
protons.

\begin{figure}[t]
\begin{center}
      \leavevmode
      \epsfxsize=100mm
      \epsfbox{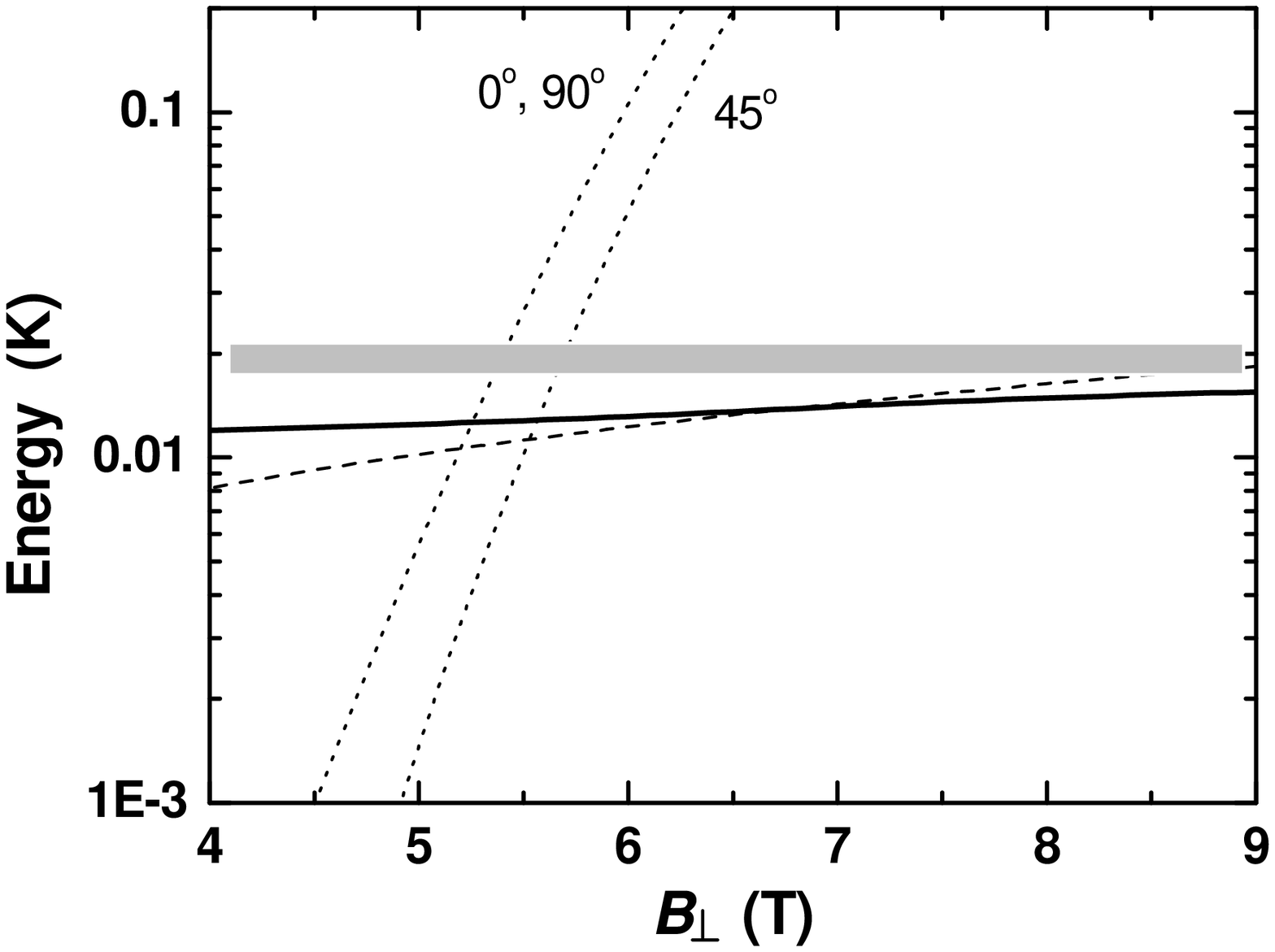}
\end{center}
\caption{\label{deltavsnu} Perpendicular field dependencies of the
Zeeman energies in Mn$^{(1)}$ (solid) and $^1$H (dashed), the
tunneling splitting (dotted) for the indicated orientations in the
$xy$ plane, and the thermal energy (grey shadow).}
\end{figure}

By increasing $B_{\perp}$ above 7 T, $W(B_{\perp})$ starts to
decrease again. As shown in Fig. \ref{deltavsnu}, the tunneling
splitting is at this point much higher than both the thermal and
the nuclear Zeeman energies. Under these conditions, due to
depopulation of the upper level of the tunnel-split ground
doublet, the electron spin fluctuations should indeed slow down,
consistent with the observation of a Schottky anomaly in the
field-dependent specific heat $C_m(B_{\perp})$
\cite{luis00PRL,mettes01PRB} which occurs at the crossing between
thermal energy and tunneling splitting. Although the transition
rate between the symmetric ground state and the antisymmetric
excited state is here measured less directly than in a specific
heat experiment, the largely similar field dependencies of
$W(B_{\perp})$ and $C_m(B_{\perp})$ for 5 T $< B_{\perp} < 9$ T
points to a common physical origin for the two phenomena. The
crossing between Mn$^{(1)}$ and $^1$H lines also gives rise to an
apparent increase in the NSLR, but this is simply due to the fact
that the protons line is itself characterized by a higher NSLR
rate, and completely masks the Mn$^{(1)}$ line. On the other hand,
as seen in Fig. \ref{plot3D}, the crossing range should be limited
to the $6.5 - 7$ T range and is therefore definitely not wide
enough to justify the whole peak in $W(B_{\perp})$.

We remark that the mentioned possibility to observe the field
dependence of the NSLR from time-dependent specific heat
experiments has interesting implications. In order to measure the
hyperfine contribution to $C_m$, the nuclei must be able to absorb
heat quickly enough as compared to the experimental time window of
the specific heat experiment ($\sim 100$ s). Indeed, this
condition was observed to hold only for $B_{\perp} > 5$ T
\cite{mettes01PRB}.

\begin{figure}[t]
\begin{center}
      \leavevmode
      \epsfxsize=100mm
      \epsfbox{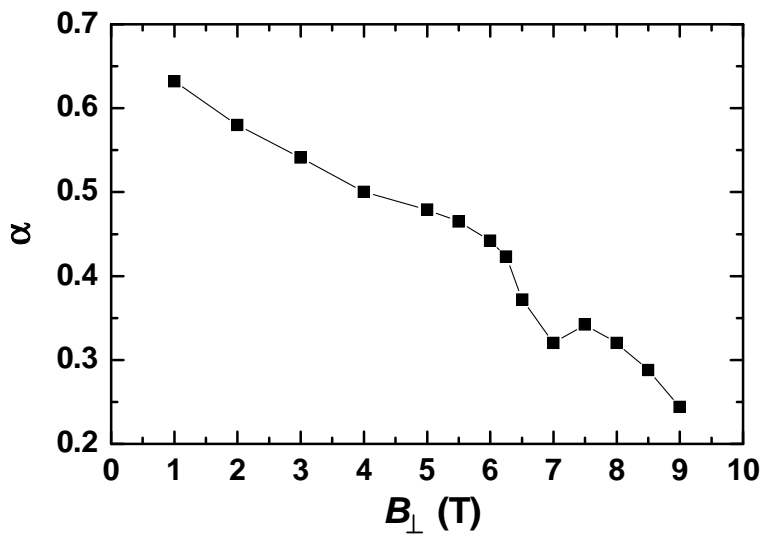}
\end{center}
\caption{\label{alpha} Perpendicular field dependence of the
stretching exponent $\alpha$ for the inversion recovery
(\ref{strexp}), at $T=20$ mK.}
\end{figure}

Two extra sets of data points have been measured. One set (circles
in Fig. \ref{WvsBx}) is taken along line (b) on the high-field
side, i.e. where crystallites with $\theta_B \simeq 88^{\circ}$
are expected to show up, and shows an apparently opposite behavior
compared to line (c), since $W(B_{\perp})$ strongly decreases just
where it would go towards a maximum on line (c). The other dataset
(triangles) is taken at constant $B_{\perp} = 7.25$ T while
shifting the frequency towards lower values, i.e. moving from line
(b) to line (c). Indeed $W$ grows along that direction. These
observations are difficult to understand if we interpret line (c)
as the bunching of the signal coming from crystallites oriented at
$70^{\circ} - 80^{\circ}$ from the field, since with such
misalignment in the field range $7 - 9$ T the two lowest energy
levels are separated by at least 10 K, due to the large
longitudinal bias. One would then expect the electron spin
fluctuations to be totally frozen at $T=20$ mK. As mentioned when
discussing Fig. \ref{plot3D}, it is also possible that line (c)
contains (at least for a part of its total intensity) a signal
from a growing fraction of molecules that are very well oriented
perpendicularly and whose electron spin occupies stably the
symmetric ground state. The excitations to the antisymmetric state
would then represent tunneling oscillations of the cluster spin,
and, given the magnitude of $\Delta$ in this regime, we may even
expect such oscillations to be coherent for fairly long times
\cite{stamp03CM}. If we adopt this picture, (c) should be a highly
inhomogeneous line, consisting of a mixture of signals from very
well oriented ($\theta_B = 88^{\circ} - 90^{\circ}$) and badly
misoriented ($\theta_B <80^{\circ}$) crystallites. In this case
one expects the inhomogeneity of the line to show up in the
stretching exponent $\alpha$ of the inversion recovery
(\S\ref{sec:measanal}). Fig. \ref{alpha} shows that
$\alpha(B_{\perp})$ is indeed a good indicator for the
inhomogeneity of the signal: it first decreases smoothly up to
$B_{\perp} \sim 6$ T, then shows an abrupt dip where also the
protons participate to the signal, finally it keeps decreasing
quite steeply along line (c), possibly because of the increasing
contribution of nuclei from well aligned molecules in their
symmetric ground state.

\section{Nuclear spin temperature} \label{spinT}

\begin{figure}[t]
\begin{center}
      \leavevmode
      \epsfxsize=120mm
      \epsfbox{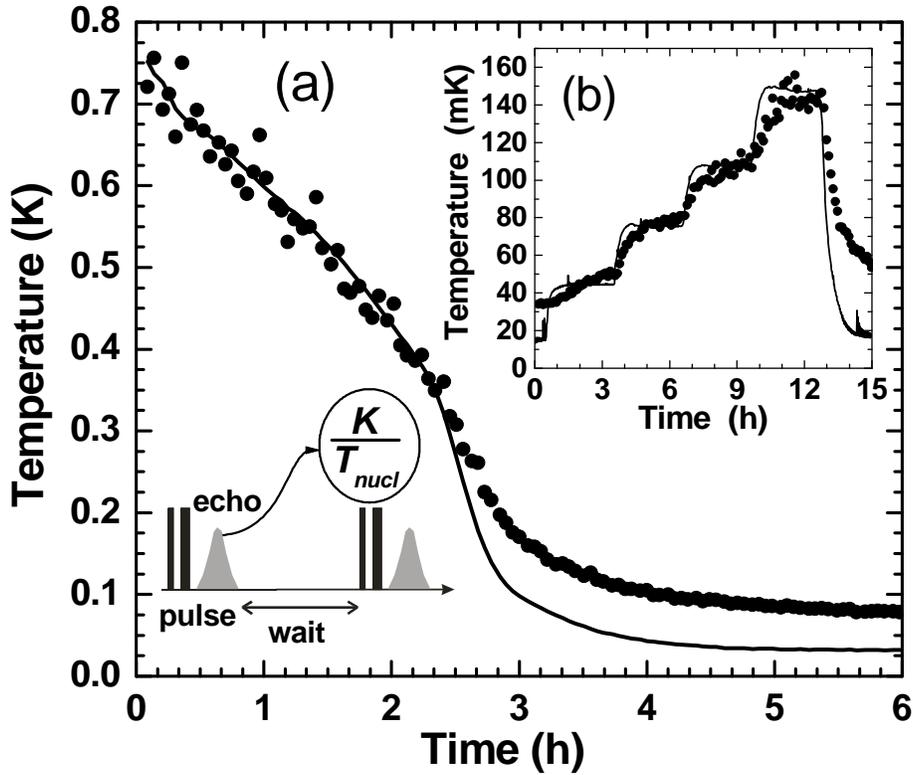}
\end{center}
\caption{\label{Tnuclcooldown} Comparison between bath temperature
$T_{\mathrm{bath}}$(solid lines) and nuclear spin temperature
$T_{\mathrm{nucl}}$ (circles), while cooling down the system (a)
and while applying step-like heat loads (b). The inset illustrates
how $T_{\mathrm{nucl}}$ is obtained from the NMR data. The waiting
time between NMR pulses was 60 s in (a) and 180 s in (b). Both
datasets are at zero field in ZFC sample.}
\end{figure}

A crucial issue in the nuclear spin dynamics of single-molecule
magnets is to which extent the nuclear spin temperature
\cite{goldman70} is able to follow the lattice temperature (as
determined by the phonons in the molecular crystal), and what is
the mechanism that links the two systems. To our knowledge there
is neither experimental nor theoretical work published on this
topic so far.

\subsection{Time-evolution of the nuclear spin temperature}

We have investigated this problem by cooling down the refrigerator
from 800 mK to 20 mK while monitoring simultaneously the
temperature $T_{\mathrm{bath}}$ of the $^3$He bath in the mixing
chamber (just next to the sample) and the NMR signal intensity of
the Mn$^{(1)}$ line, checked by an echo pulse sequence every 60 s,
in zero external field and on a ZFC sample. The nuclear spin
temperature $T_{\mathrm{nucl}}$ is obtained as described in
\S\ref{sec:measanal}, and plotted in Fig. \ref{Tnuclcooldown}
together with $T_{\mathrm{bath}}$. We find that the nuclear spin
temperature indeed strictly follows the bath temperature, with a
small deviation below $\sim 200$ mK that we attribute to the
heating effects of the NMR pulses. In fact, the pulse rate used in
this experiment is ten times higher than in all the measurements
shown in \S\ref{lowfield}, so it is likely that not all the energy
provided to the nuclear spin system can dissipate to the $^3$He
stream at such a rate. We therefore repeated the experiment at the
lowest temperatures applying a pulse train every 180 s and
step-like heat loads. As shown in the inset of Fig.
\ref{Tnuclcooldown}, $T_{\mathrm{nucl}}$ indeed follows quite
accurately $T_{\mathrm{bath}}$ except for a remaining discrepancy
at the lowest $T$, but now much less evident than in the previous
experiment with high pulse rate.

\begin{figure}[p]
\begin{center}
      \leavevmode
      \epsfxsize=120mm
      \epsfbox{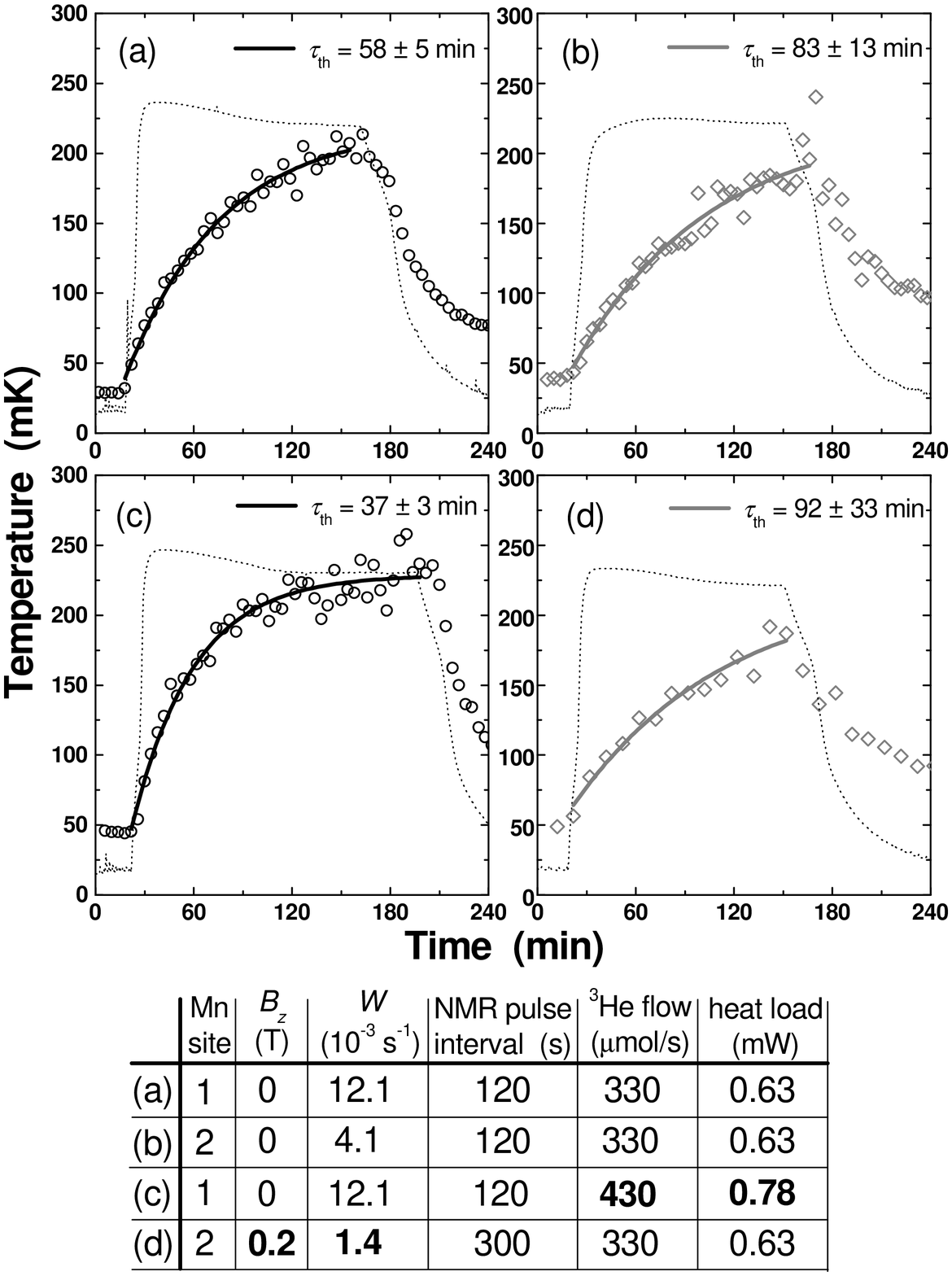}
\end{center}
\caption{\label{spinTcompare}  Time evolution of the nuclear spin
temperature (open symbols) and the bath temperature (dotted lines)
upon application of a step-like heat load. The solid lines are
fits to Eq. (\ref{Tspinexp}), yielding the thermal time constants
$\tau_{\mathrm{th}}$ reported in each panel. All data are for an
FC sample, with the Mn site, external magnetic field $B_z$, NSLR
rate $W$, NMR pulse waiting time, $^{3}He$ flow rate and applied
heat load as given in the annexed table.}
\end{figure}

The application of step-like heat loads is a useful method to
investigate the effective time constant $\tau_{\mathrm{th}}$ for
the thermalization of the nuclear spin system with the helium
bath, in particular the relationship between $\tau_{\mathrm{th}}$
and the NSLR rate $W$ as obtained from the inversion recovery
technique and discussed in \S\ref{lowfield}. This can be done by
monitoring $T_{\mathrm{nucl}}$ in different longitudinal fields
and different Mn sites, by which we can easily change the NSLR
rate. Since also the NMR signal intensity changes under each
different condition, we must redefine every time the conversion
factor $K$ between signal intensity and $T_{\mathrm{nucl}}$.
Moreover the monitoring pulse rate must be adjusted as well, and
we found that it is indeed the combination of pulse rate and NSLR
rate that determines the discrepancy between the observed lowest
spin temperature and the base temperature of the refrigerator.
Upon heating up to, say, 200 mK, this effect becomes irrelevant as
seen in the inset of Fig. \ref{Tnuclcooldown}. It is therefore a
natural choice to define $K$ in such a way that the asymptotic
value of $T_{\mathrm{nucl}}$ matches the measured
$T_{\mathrm{bath}}$ at the end of the heat step. Fig.
\ref{spinTcompare} shows four examples of the time evolution of
$T_{\mathrm{nucl}}$ under the application of a heat load for $\sim
2$ hours, in Mn$^{(1)}$ and Mn$^{(2)}$ sites, with or without an
applied field, and with an increased $^{3}$He flow rate. We fitted
the data to the phenomenological function:

\begin{eqnarray}
T_{\mathrm{nucl}}(t) = T_{\mathrm{nucl}}(0) +
[T_{\mathrm{nucl}}(\infty) - T_{\mathrm{nucl}}(0)] \left[1 - \exp
\left(-\frac{t - t_0}{\tau_{\mathrm{th}}}\right)\right],
\label{Tspinexp}
\end{eqnarray}

where $T_{\mathrm{nucl}}(\infty)$ is set by definition equal to
$T_{\mathrm{bath}}$ at the end of the step, $T_{\mathrm{nucl}}(0)$
follows automatically from the above constraint, and $t_0$ is the
time when the heat pulse is applied. We find that
$\tau_{\mathrm{th}}$ is always much longer than the nuclear
relaxation time $T_1 \sim 1/2W$, and that the dependence on Mn
site and applied field is not proportional to the change in $T_1$
under different conditions. What really seems to make the
difference is the $^{3}$He flow rate: within the errors, the ratio
of heat transfer from the $^{3}$He stream to the nuclear spins is
proportional to the $^{3}$He flow rate, given the same conditions
of nuclear site and external field. This confirms that the thermal
contact between nuclear spins and lattice phonons is quite fast,
as we demonstrated by the values of the NSLR rate, and that the
main bottleneck is at the interface between the crystallites, the
embedding epoxy and the $^{3}$He stream.

\subsection{Effect of large magnetic fields and superradiance}
\label{superradiance}

\begin{figure}[t]
\begin{center}
      \leavevmode
      \epsfxsize=100mm
      \epsfbox{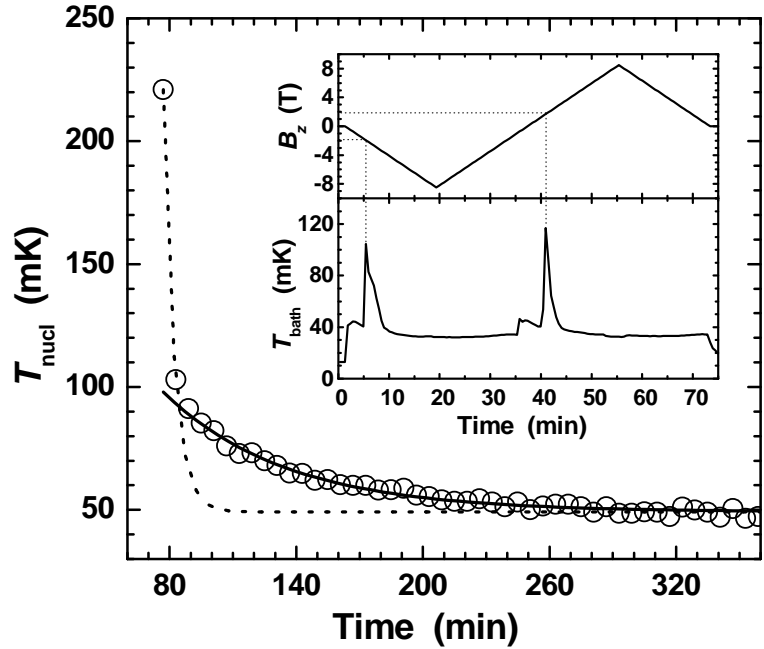}
\end{center}
\caption{\label{spinTsweep} Time evolution of $T_{\mathrm{nucl}}$
(open symbols) measured with NMR pulse intervals of 180 s on the
Mn$^{(1)}$ line just after a magnetic hysteresis loop performed
with a field sweep rate $|dB_z / dt| = 0.5$ T/min. The sample is
FC with initial positive magnetization. The dashed and solid lines
are fits to Eq. (\ref{Tspinexp}) for the fast (initial) and slow
part of the relaxation, respectively, having imposed
$T_{\mathrm{nucl}}(\infty) = 49$ mK for both fits. The inset shows
$B_z$ and $T_{\mathrm{bath}}$ during the field sweep: notice that
the time scale in the main panel is the continuation of the one in
the inset.}
\end{figure}

To conclude this section we discuss an issue which may be relevant
to most of the experiments on quantum tunneling studied by
magnetization loops or fast field-sweeps. While sweeping the
magnetic field in such a way that many cluster spins are forced to
flip, also the hyperfine fields at the nuclear sites are being
inverted at an abnormal rate. One may then ask whether the nuclear
spin system remains in thermal equilibrium, or the macroscopic
inversion of the electron spin polarization leads to a remarkable
increase of $T_{\mathrm{nucl}}$. To answer this question we
performed a full magnetization loop $0 \rightarrow -8.5
\rightarrow +8.5 \rightarrow 0$ T at a sweep rate $|\mathrm{d}B_z
/ \mathrm{d}t| = 0.5$ T/min on a sample with initial positive
magnetization and measured $T_{\mathrm{nucl}}$ as soon as the loop
was finished, with the field back to zero. Because of the
continuous shift of the resonance frequency it would have
obviously been impossible to monitor $T_{\mathrm{nucl}}$ during
the field sweep. As shown in Fig. \ref{spinTsweep}, we find that
$T_{\mathrm{nucl}}$ at the end of the field sweep is much higher
than $T_{\mathrm{bath}}$. Interestingly, the evolution of
$T_{\mathrm{nucl}}$ is characterized by an initially very fast
decay, whereas after a few minutes $T_{\mathrm{nucl}}$ relaxes to
its equilibrium value with precisely the same time constant
$\tau_{\mathrm{th}} \simeq 57$ min as already found in the heat
transient experiment in Fig. \ref{spinTcompare}(a). By extracting
$T_{\mathrm{nucl}}(\infty)$ from the fit to Eq. (\ref{Tspinexp})
and imposing that it should be the same also for the initial fast
decay, we refitted just the very first datapoints obtaining
$T_{\mathrm{nucl}}(0) \sim 350$ mK and
$\tau_{\mathrm{th}}^{\mathrm{(initial)}} \simeq 300$ s. The latter
value is still a bit longer than $T_1 = (2W)^{-1} \simeq 40$ s in
FC sample at $T=20$ mK [see Fig. \ref{T1T2Mn34}(b), open circles]
but not incompatible with the idea that the very first part of the
relaxation is the real spin-lattice relaxation, whereas afterwards
the equilibrated lattice and spins relax to the bath temperature
at the usual rate mainly determined by the $^3$He circulation.

In the inset of Fig. \ref{spinTsweep} we also report $B_z$ and
$T_{\mathrm{bath}}$ as a function of time during the field sweep.
$T_{\mathrm{bath}}$ remains rather stable around 35 mK except at
two special points, $-1.9$ T sweeping down and $+1.9$ T sweeping
up, where it suddenly jumps to $\sim 100$ mK. We notice that this
occurs only when $B_z$ is opposite to the instantaneous
magnetization, and that in fact $|B_z| = 1.9$ T is a level
crossing field for the electron spins in slow-relaxing molecules,
which suggest that the phenomenon may be due to magnetic
avalanches. A similar heat emission was already observed by
specific heat experiments \cite{fominaya97PRL}, but at $T \sim 2 -
3$ K. On the other hand, it is rather surprising to find that the
main heat emission takes place already at $B_z = \pm 1.9$ T, i.e.
the fourth level crossing, whereas it is
known\cite{chiorescu00PRL} that small Mn$_{12}$-ac crystals at
very low $T$ start to appreciably invert their magnetization only
at $B_z \sim \pm 4$ T. Moreover we know from experience that,
because of the large volume of our refrigerator, it is virtually
impossible to heat the $^3$He bath so abruptly as it appears in
the inset of Fig. \ref{spinTsweep}. In fact, such a sudden
temperature increase resembles very much the effect we find when
applying a NMR pulse (cf. Fig. \ref{thermaleff}): we first observe
a sharp peak in the measured $T$, which is due to the direct
heating of the thermometer by the \textit{rf}-pulse, and only in
the subsequent minutes we observe a smooth temperature increase,
when the thermometer indeed senses the heating of the $^3$He bath.
In a recent work \cite{tejada03CM}, Tejada \textit{et al.} report
the observation of radiation emitted by a large amount of
Mn$_{12}$-ac crystals when measuring magnetic hysteresis loops. In
their experiment the radiation bursts are observed at $B_z = \pm
1.4$ T, thus the third level crossing, but the measurements are
performed at $T = 1.8$ K, much higher than in our case. The
proposed explanation is based on the phenomenon of superradiance
\cite{dicke54PR}: when the wavelength of the radiation emitted by
each electron spin flip is greater than the size of the sample,
large numbers of spins may tunnel simultaneously because they
interact with a common radiation field, thereby producing
radiation avalanches. A prerequisite for this phenomenon to occur
is indeed a population inversion, in this case a total electron
spin magnetization opposite to the direction of the applied field,
which is precisely what we observe.

In fact, even though the thermometer detects only the radiation
bursts produced by large avalanches, we may expect that weak hints
of superradiance could occur even at small values of $B_z$,
provided that the magnetization is opposite to the field. An
example is given by the data in Fig. \ref{WvsBall}(b), where we
investigated the NSLR rate in a FC sample with ``positive''
magnetization, and observed indeed that $W(B_z)$ shows a
surprisingly high plateau for $B_z < 0$. An argument against the
attribution of that plateau to weak hints of superradiance could
be that we didn't notice any substantial changes in the sample
magnetization while performing those measurements. On the other
hand, we are still speaking about relatively low values of $W$,
which would not require an extremely high number of tunneling
events. A pragmatic solution to this puzzle would be given only by
continuously repeating the same nuclear relaxation experiment
(e.g. at $B_z = -0.5$ T) and checking whether $W$ changes in time,
in the same way as the total magnetization.

\section{Proposal for a unifying theory} \label{uniftheory}

We are now at a stage to gather all the experimental results
discussed so far and use them as basic elements to build a
complete description of the quantum dynamics of the coupled system
``cluster spins + nuclei''. The starting framework is the
Prokof'ev-Stamp theory (\S\ref{PStheory}), but we shall supplement
it with some extra assumptions in order to account for the
observed intercluster nuclear spin diffusion and the thermal
equilibrium between nuclear spins and lattice phonons.

\subsection{Basic assumptions} \label{basicassumptions}

The model we are going to set up is based upon the following
(experimentally justified) starting points:

(i) the tunneling dynamics takes place only in fast-relaxing
molecules, whereas the neighboring slow molecules can be safely
considered as frozen during the timescale of interest. For our
purpose, the slow molecules serve simply as a ``reservoir of
nuclear polarization'';

(ii) for the ease of discussion, we assume that every cluster is
coupled to $N$ nuclear spins $I = 1/2$, and that the hyperfine
field $\vec{B}_{\mathrm{hyp}}$ always lies along the $\vec{z}$
axis; upon tunneling, the direction of $\vec{B}_{\mathrm{hyp}}$ is
turned by $180^{\circ}$. The latter assumption simulates indeed
the real situation for $^{55}$Mn nuclei in Mn$_{12}$-ac. We also
define a \emph{local} nuclear polarization for the $i$-th cluster
$\Delta N(i) = N_i^{\uparrow} - N_i^{\downarrow}$ along the
$+\vec{z}$ axis, and the associated \emph{local} hyperfine bias
$\xi_N(i) = \hbar \gamma_N \Delta N(i) \vec{B}_{\mathrm{hyp}}
\cdot \vec{z}$. Notice that the sign of $\xi_N(i)$ depends on the
instantaneous direction of $\vec{B}_{\mathrm{hyp}}$, whereas the
sign of $\Delta N(i)$ does not.

(iii) nuclei in neighboring clusters are coupled by dipole-dipole
interactions that lead to intercluster nuclear spin diffusion.
Experimentally, the spin diffusion rate is quite fast: we may
therefore assume that any sudden change in local hyperfine bias is
quickly redistributed among all clusters in order to attain the
equilibrium within the nuclear spin system.

For all other aspects we rely on the Prokof'ev-Stamp model for the
description of the coupling between central spin and nuclei. We
assume therefore that the hyperfine interactions split the
electron spin levels into a manifold of states, indexed by $\Delta
N$ and broadened by intracluster nuclear spin diffusion, spanning
a total width $\Xi_N = \xi_N^{(\mathrm{max})} -
\xi_N^{(\mathrm{min})} = N \hbar \gamma_N B_{\mathrm{hyp}}$. From
assumption (ii) it follows that, since $\vec{B}_{\mathrm{hyp}}$
only takes two antiparallel directions ($\Rightarrow
\omega_k^{\perp} = 0$), the hyperfine-split manifolds on either
sides of the anisotropy barrier are simply the mirror of each
other. In the absence of external bias, to any state with $\Delta
N^{(1)}$ on the left side of the barrier corresponds a state with
$-\Delta N^{(2)}$ on the right side\footnote{In
\S\ref{centralspinbath} we used the superscripts $^{(1)}$ and
$^{(2)}$ for the hyperfine fields and the nuclear polarizations
before and after the flip, respectively. The notation used here is
consistent with the above as long as we assume that the the
electron spin is in the left well at $t=0$ and subsequently
tunnels to the right.}. Also, $\omega_k^{\perp} = 0$ implies that
the tunneling of the central spin does not produce any coflipping
of nuclear spins by the ``orthogonality blocking'' mechanism
(\S\ref{centralspinbath}), i.e. $\kappa \simeq 0$. Furthermore,
the ``topological decoherence'' is negligible as well; the typical
nuclear Larmor frequencies are $\omega_N \sim 10^9$ s$^{-1}$,
whereas the tunneling traversal time (\S\ref{coupling}) is
$\tau_{\mathrm{tr}} \sim \Omega_0^{-1} \sim 10^{-12}$ s. In the PS
language, this leads to $\lambda \sim N(\omega_N / \Omega_0)^2
\simeq 0$. All this means that the only allowed tunneling
transitions are those that do not require any coflip of nuclear
spins, i.e. $\Delta N = const.$; the relevant tunneling splitting
is therefore $\Delta_0$. On basis of the assumption (iii) about
the internal equilibrium in the nuclear spin system we may write
the nuclear spin temperature $T_{\mathrm{nucl}}$ in terms of the
populations of nuclear Zeeman levels. For this purpose it is
convenient to introduce the populations of the nuclear Zeeman
levels \emph{relative to the local direction of
$B_{\mathrm{hyp}}$}, $N_+$ and $N_-$, where $N_+$ is the number of
nuclei per cluster that are in the Zeeman ground state, and $N_-$
counts the nuclei in the excited state:
\begin{eqnarray}
\frac{\langle N_- \rangle}{\langle N_+ \rangle} =
\exp\left(-\frac{\hbar \omega_N}{k_B T_{\mathrm{nucl}}}\right),
\label{theorTspin}
\end{eqnarray}
where $\langle \ \rangle$ indicates the average over all the
clusters in the sample. Calling $n = N_+ - N_-$,\footnote{In other
words, $n = \Delta N$ if $\vec{B}_{\mathrm{hyp}}$ is aligned along
$+\vec{z}$, and $n = -\Delta N$ otherwise.} $\langle N_+ \rangle =
(N + \langle n \rangle)/2$, $\langle N_- \rangle = (N - \langle n
\rangle)/2$ and $\langle \xi_N \rangle = -\hbar \omega_N \langle n
\rangle$, (\ref{theorTspin}) becomes:

\begin{eqnarray}
\frac{\Xi_N + \langle \xi_N \rangle}{\Xi_N - \langle \xi_N
\rangle} = \exp\left(- \frac{\hbar \omega_N}{k_B
T_{\mathrm{nucl}}}\right).
\end{eqnarray}

\subsection{Tunneling mechanism}

\begin{figure}[t]
\begin{center}
      \leavevmode
      \epsfxsize=120mm
      \epsfbox{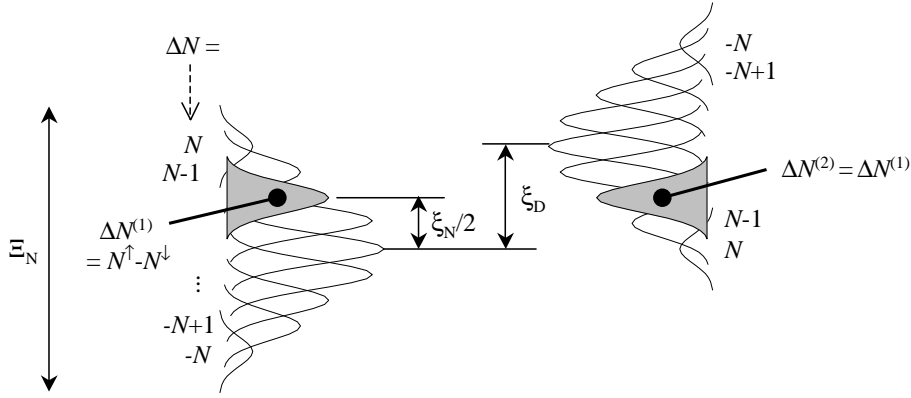}
\end{center}
\caption{\label{tunnelbias} Sketch of a possible configuration
where tunneling may occur in the presence of a dipolar bias
$\xi_D$. The gray areas represent the nuclear polarization groups
on opposite side of the barrier having the same $\Delta N$, such
that tunneling does not require any nuclear coflips. In order to
populate one of those groups, an initial hyperfine bias $\xi_N =
\xi_D$ must be achieved via intercluster nuclear spin diffusion.}
\end{figure}

In the absence of an external bias on the cluster electron spin,
tunneling would take place only when $\xi_N = 0$ as well, since
the polarization groups with $\Delta N = 0$ are the only ones
having equal energy and equal nuclear polarization. In the more
general case where we include the quasi-static dipolar bias
$\xi_D$ arising from neighboring clusters, the tunneling
transition with $\Delta N = const.$ can still take place, but
requires now an initial hyperfine bias $\xi_N = \xi_D$. After the
tunneling, since $\vec{B}_{\mathrm{hyp}}$ is now reversed, the
hyperfine bias becomes $-\xi_D$; in other words, the effect of
tunneling is in fact to convert the hyperfine energy into dipolar
energy or vice versa. A sketch of the situation is depicted in
Fig. \ref{tunnelbias}: starting from a cluster where the electron
spin is favorably aligned with respect to the dipolar field
produced by the neighbors, a tunneling transition requires
``charging'' the local nuclei up to polarization group where
$\xi_N = \xi_D$, such that on the opposite side of the barrier
there is another polarization group having identical $\Delta N$.
After tunneling, the cluster spin is oriented opposite to the
local dipolar field, and the required energy has been taken from
the nuclear reservoir, which is now in a favorable polarization
state with respect to the new direction $\vec{B}_{\mathrm{hyp}}$.

It is clear that intercluster nuclear spin diffusion is an
essential ingredient to attain the condition $\xi_N = \xi_D$ at
all. Flip-flop processes between nuclei in different clusters
conserve the \emph{total} energy of the nuclear system, but allow
the \emph{local} hyperfine bias to wander through the whole
hyperfine-split manifold until the tunneling resonance is found. A
tunneling event like in Fig. \ref{tunnelbias} would have the
effect of ``locally extracting'' an energy $\xi_D$ from the
nuclear spin system; by the same mechanism of intercluster spin
diffusion, this energy can then be redistributed among the
neighbors, thereby attaining an effective reduction of the nuclear
spin temperature. On the other hand, the tunneling process can
work in the opposite direction, reducing the dipolar energy at the
expenses of an increase of $T_{\mathrm{nucl}}$.

Let us recall that this picture, in particular as a consequence of
assumption (ii), accounts only for nuclei like $^{55}$Mn in
Mn$_{12}$-ac, i.e. nuclei where the hyperfine fields flips by
$180^{\circ}$ when a tunneling event occurs. We have shown in
\S\ref{deuterated} that actually the most important role in
triggering the incoherent tunneling dynamics is played by nuclei
such as protons, precisely because they have a large coflipping
probability due to the non-antiparallel directions of
$\vec{B}_{\mathrm{hyp}}^{(1)}$ and $\vec{B}_{\mathrm{hyp}}^{(2)}$.
In that case, the condition $\Delta N = const.$ may be relaxed to
$|\Delta N^{(1)} - \Delta N^{(2)}| \leq 2M$, where $M$ is the
number of coflipping nuclei. Although such an extension is
essential to properly account for the tunneling rate of the
electron spin, we shall not include it in our discussion as long
as we focus on the tunneling-driven nuclear relaxation of
$^{55}$Mn (which is the main subject of our experiments) rather
than the hyperfine-driven tunneling rate of the electron spin.

\subsection{Coupling with lattice phonons}

So far we have not introduced any thermostat to set the
temperature to which the equilibrium $T_{\mathrm{nucl}}$ should
relax. The only natural way to obtain a certain equilibrium
nuclear spin temperature from the tunneling mechanism described
above, is to assume detailed balance between the rate
$w^{\downarrow}$ of the transitions that reduce
$T_{\mathrm{nucl}}$ at the expenses of dipolar energy (like in
Fig. \ref{tunnelbias}), and the rate $w^{\uparrow}$ for
transitions in the opposite direction:
\begin{eqnarray}
\frac{w^{\uparrow}}{w^{\downarrow}} = e^{-\xi_N/k_B T}.
\end{eqnarray}
On the other hand, as long as the tunneling rates are derived from
the standard Prokof'ev-Stamp expression (\ref{taufinal}), there's
no way to distinguish between $w^{\downarrow}$ and $w^{\uparrow}$.

We propose therefore that, in order to account for the observed
equilibrium between nuclear spin and lattice temperatures, the PS
theory must be extended to include the possibility of inelastic
tunneling events, accompanied by creation or annihilation of
lattice phonons. In particular, we argue that an essential role
could be played by the time-dependence of the dipolar bias $\xi_D$
that arises when the distance between neighboring clusters is
modulated by a phonon. Due to the peculiar structure of
single-molecule magnets, containing very rigid clusters bound by
soft organic ligands, we may expect a clear separation between the
intracluster, high-energy phonon modes, and the intercluster
low-energy modes, corresponding to the relative displacement of
the clusters considered as rigid objects. This approach has been
already successfully used to account for the M\"{o}ssbauer
recoil-free fraction in metal cluster molecules
\cite{paulus01PRB}, which are bound by similarly soft ligands,
yielding a Debye temperature $\Theta_D \sim 10-20$ K for
intercluster phonon modes. In some sense, this idea is a
``molecular version'' of the well-known Waller mechanism
\cite{waller32ZP} for the spin-lattice relaxation of paramagnetic
ions. The problem is that, if used to estimate the nuclear
spin-lattice relaxation rate in the standard way, i.e. by
calculating the spectral density of the fluctuating dipolar field
\emph{directly at the nuclear sites} \cite{abragam70}, the Waller
mechanism would lead to extremely long relaxation times. What we
suggest is that the Waller mechanism can be regarded as a source
of fluctuating bias producing a similar effect as the nuclear spin
diffusion in the PS theory (\S\ref{centralspinbath}); similar
conclusions on the effect of the Waller mechanism could be drawn
by using the Landau-Zener formalism (see \S\ref{coupling} and
\S\ref{sec:PSLZ}) to calculate the probability of incoherent
tunneling while the dipolar bias sweeps through the tunneling
resonance. We already remarked in \S\ref{sec:PSLZ} that the
tunneling rate $\tau_T^{-1}$ obtained in this way is independent
of the bias sweep rate, in this case the frequency of the phonon
modulating $\xi_D$. $\tau_T^{-1}$ does depend on the range of
fluctuation $\hbar \Gamma_{\mu}$, whose role would here be taken
by the amplitude $\Delta\xi_D$ of the change in dipolar bias
arising from lattice strain associated with the phonon modes. By
inspecting Eq. (\ref{taufinal}) one finds that the prefactor
$\propto \Delta^2 / \Gamma_{\mu} \rightarrow \Delta^2 / \Delta
\xi_D$ in $\tau_T^{-1}$ is inversely proportional to the amplitude
of fluctuation. On the other hand, the extra factor $e^{-|\xi| /
\xi_0}$ may lead to at least a partial cancellation effect if
$\Delta \xi_D$ contributes to $\xi_0$ as well, which could justify
the approximately $T$-independent NSLR rate found below 0.8 K.
Since the fluctuation of dipolar bias has to be combined with the
fluctuating hyperfine bias, deriving the exact expression for the
inelastic tunneling rate constitutes a very challenging but
formidable theoretical task, that largely exceeds the scope of
this thesis.

\subsection{Nuclear spin-lattice relaxation rate}

Irrespective of the precise expressions for $w^{\uparrow}$ and
$w^{\downarrow}$, their deviation from the average value
$\tau_T^{-1} = (w^{\uparrow} + w^{\downarrow})/2$ will be very
small, unless the temperature becomes much lower than the typical
range of the dipolar interaction $\xi_{D} \sim 0.1$ K. The reason
why the relevant energy scale is $\xi_{D}$ and not $\hbar
\omega_N$ is that, in our model, the nuclear spin system can
change its global polarization only by converting hyperfine energy
into dipolar energy ($\xi_N = \xi_D$). In the extreme case where
$\xi_D = 0$ in all clusters, all tunneling events would take place
at $\Delta N = 0$, with no effects on the nuclear
spins\footnote{This argument obviously does not hold for nuclear
spins like protons, which have a non-negligible coflipping
probability.}. In fact, it would be very intriguing to find a
system where $\hbar \omega_N \gg \xi_{D}$ (e.g. by diluting the
magnetic clusters) and check if the thermalization of the nuclear
spins works as efficiently as we find here.

\begin{figure}[t]
\begin{center}
      \leavevmode
      \epsfxsize=100mm
      \epsfbox{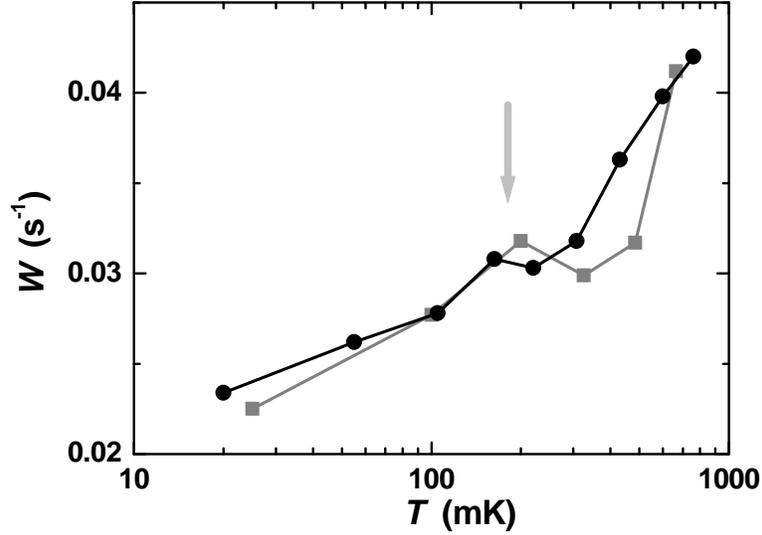}
\end{center}
\caption{\label{maxWvsT} NSLR rate in the quantum regime, for
Mn$^{(1)}$ site in ZFC sample and zero external field. Circles are
the same data as in Fig. \ref{W1W2vsT}, squares are from an
experiment carried out in a different run several months before:
the arrow indicates the position of the weak maximum in $W(T)$
around 180 mK.}
\end{figure}

Furthermore, by focusing on the NSLR rate $W(T)$ only in the
quantum regime, we have noticed that the ``roughly''
$T$-independent plateau has not only a slope, but also a slight
and broad maximum around $T \simeq 0.18$ K, below which $W(T)$
tends to decrease somewhat more steeply. As shown in Fig.
\ref{maxWvsT}, by measuring $W(T)$ in the quantum regime in two
independent runs, we got the impression that the maximum around
0.18 K is a weak albeit reproducible feature. The value of 0.18 K
may not be a coincidence in view of the fact that, as we discuss
in chapter V, a system like Mn$_6$ orders ferromagnetically at
$T_c = 0.16$ K purely by the effect of dipolar interactions. It
would be very interesting to see whether a fully developed
theoretical description can account for this peak in $W(T)$, and
if this is indeed related to the dipolar interaction energy.

An estimate of the NSLR rate in the quantum regime can be obtained
by writing a master equation for the populations of the nuclear
Zeeman levels \emph{relative to the local hyperfine field
direction}, $N_+$ and $N_-$ (cf. \S\ref{basicassumptions}). For
simplicity, since the internal equilibrium is reestablished within
a time $T_2 \ll \tau_T$ after each tunneling event, we assume that
just before tunneling all clusters have the same values of $N_+$
and $N_-$, neglecting fluctuations around the mean values. If a
fast-relaxing molecule tunnels at time $t$, the polarization of
its own nuclei is abruptly inverted. Each time a tunneling
transition lowers the energy of the local nuclei, which occurs at
a rate $w_{\downarrow}$, then $N_-$ nuclei have been added to the
total number of nuclei in the Zeeman ground state. After a time
$T_2$ this decrease in local hyperfine bias has been redistributed
over the sample: calling $x$ the fraction of FRMs over the total,
then the tunneling event has increased $N_+$ to $N_+ + xN_-$. The
same reasoning holds for transitions that increase the hyperfine
bias. The master equation is therefore:
\begin{eqnarray}
\frac{\mathrm{d}N_+}{\mathrm{d}t} = xN_- w_{\downarrow} - xN_+
w_{\uparrow}.  \label{mastereq}
\end{eqnarray}
From here the NSLR rate can be obtained by standard textbook
calculations \cite{slichterB}. Writing $N_+=(N+n)/2$ and
$N_-=(N-n)/2$, (\ref{mastereq}) becomes:
\begin{eqnarray}
\frac{\mathrm{d}n}{\mathrm{d}t} = xN(w_{\downarrow} -
w_{\uparrow}) - xn(w_{\downarrow} + w_{\uparrow}),
\end{eqnarray}
which can be rewritten as:
\begin{eqnarray}
\frac{\mathrm{d}n}{\mathrm{d}t} = 2W(n_0 - n),\label{masterfinal}\\
n_0 = N \frac{w_{\downarrow} - w_{\uparrow}}{w_{\downarrow} +
w_{\uparrow}},\\
W = x \frac{w_{\downarrow} + w_{\uparrow}}{2} = x \tau_T^{-1}
\label{Wmodel},
\end{eqnarray}
where $n_0$ is the equilibrium nuclear polarization and $W$ is the
desired NSLR rate, since the solution of (\ref{masterfinal}) is
precisely of the form $n(t) = n(0) - [n(0)-n_0][1-\exp(-2Wt)]$.

Comparing Eq. (\ref{Wmodel}) with the experimental value of $W
\simeq 0.03$ s$^{-1}$ reported in \S\ref{zerofield}, one would
deduce that, for instance, 1\% of FRMs tunneling at a rate
$\tau_T^{-1} \sim 3$ s$^{-1}$ is sufficient to explain the data.
Such a tunneling rate is high but not unreasonably high,
especially considering the FRMs having two flipped Jahn-Teller
axes and a barrier $\sim 15$ K (\S\ref{paramMn12}). We recall that
for an easy-axis spin $S \gg 1$, described by a Hamiltonian
$\mathcal{H} = -DS_z^2 + ES_x^2$ with $D \gg E$, the tunneling
rate is proportional to $D(E/D)^S$ \cite{chudnovskyB}. By simply
reducing the height of the barrier by a factor 4, we may already
expect an increase of five orders of magnitude in the tunneling
rate with respect to a normal molecule! FRMs are also likely to
have higher values of $E$, thereby increasing the tunneling rate
even further.


Let us remark that the interplay between electron spin (tunneling)
fluctuations and nuclear spin diffusion is completely different
from the standard picture of relaxation by paramagnetic impurities
\cite{lowe68PR}. In the presence of an impurity spin, the closest
nuclei are relaxing very fast because of the strong fluctuations,
but the static component of the field produced by the impurity
drives them out of resonance with respect to the nuclei that are
farther away: one finds that nuclear spin diffusion starts to work
only outside a sphere whose radius is determined by the strength
of the nuclear dipolar coupling compared with the field produced
by the impurity. In this sense, intercluster nuclear spin
diffusion in SMMs is a very peculiar phenomenon, since all the
nuclei are equivalent and there is no minimum radius for the spin
diffusion.

\section{Conclusions}

We now briefly summarize the results presented in this chapter, in
order to place them in a complete framework of observations and
interpretations, and in particular to make clear what the
breakthroughs of our research are.

(i) Incoherent quantum tunneling of the electron spin provides a
relaxation channel for the nuclear spin system, and its effects
are much more pronounced than expected \cite{prokof'ev96JLTP}.
This can be deduced from the roughly $T$-independent plateau in
the $^{55}$Mn NSLR rate below 0.8 K, and from the coincidence
between maxima in the NSLR rate and tunneling resonances for the
electron spins. The observation of the effect of tunneling on the
nuclear spins can be done even in zero field and demagnetized
sample, thus without any changes in the macroscopic magnetization
of the cluster spins. We also suggest that the correct
interpretation of the observed NSLR for $T>0.8$ K should include
the effect of thermally-assisted tunneling as well.

(ii) The intercluster nuclear spin diffusion is fast on the
timescale of the NSLR and of the electron spin tunneling rate. Its
existence can be inferred from the $T$-independent value of the
TSSR rate below 0.8 K, and from the $\simeq \sqrt{2}$ ratio
between the TSSR rates in a fully polarized and a demagnetized
sample. This is the first direct experimental observation of such
a phenomenon in SMMs. Moreover, the field dependencies of NSLR and
TSSR suggest that the fluctuating dipolar field produced by
tunneling events may play a role in facilitating the intercluster
nuclear spin diffusion.

(iii) The tunneling dynamics in zero external field is ascribed to
the fast-relaxing molecules present in any real sample of
Mn$_{12}$-ac. This follows from both a quantitative analysis of
the NSLR rate, and from the experimental observation that the
macroscopic magnetization of the sample does not change during the
measurements. Furthermore, the comparison between NSLR and TSSR in
Mn$^{4+}$ and Mn$^{3+}$ sites indicates that the intercluster
nuclear spin diffusion between FRMs and normal molecules is less
efficient between nuclei in Mn$^{3+}$ sites, which are precisely
those where the FRMs are different.

(iv) The isotope substitution of $^1$H by $^2$H leads, as
expected, to a significant decrease of the incoherent tunneling
rate, which manifests itself in a congruent reduction of the NSLR
and TSSR rates.

(v) The application of strong perpendicular fields produces a
broad peak in the NSLR rate which is consistent with the observed
increase of excitations within the tunneling-split ground doublet,
as previously observed by field-dependent specific heat
experiments \cite{luis00PRL}.

(vi) Even in the regime ($T<0.8$ K) where the nuclear spins are
relaxed only by ground state quantum tunneling of the electron
spins, the nuclear spin temperature remains in equilibrium with
the bath temperature. This is a new and unexpect result, and its
implication have never been included in any theoretical
discussion. We attempt to account for this observation by a
suggesting a model that combines the Prokof'ev-Stamp theory with
the effect of modulation of the intercluster dipolar fields
(Waller mechanism), in order to elucidate how the cluster spin
tunneling can relax the nuclear spins, and why it is necessary to
attribute different rates to tunneling transitions that increase
rather than decrease the nuclear spin energy.

(vii) Experiments where a longitudinal magnetic field is applied
opposite to the direction of the sample magnetization could be
interpreted by assuming the presence of magnetization avalanches,
which also manifest themselves in the form of electromagnetic
radiation \cite{tejada03CM}.

\def\baselinestretch{1}
\chapter{Long-range dipolar ordering in Mn$_{6}$}

Few examples of long-range magnetic order induced by purely
dipolar interactions are known as yet
\cite{white93PRL,bitko96PRL}. Therefore, the possibility to study
such phase transitions and the associated long-time relaxation
phenomena in detail in high-spin molecular cluster compounds, with
varying crystalline packing symmetries and different types of
anisotropy, presents an attractive subject
\cite{fernandez00PRB,fernandez02PRB}. However, for the most
extensively studied molecular clusters sofar such as Mn$_{12}$,
Fe$_8$ and Mn$_4$
\cite{sangregorio97PRL,aubin98JACS,thomas99PRL,mettes01PRB} the
uniaxial anisotropy experienced by the cluster spins is very
strong. Consequently, the electronic spin-lattice relaxation time
$T_{1}^{el}$ becomes very long at low temperatures and the cluster
spins become frozen at temperatures of the order of 1 K, i.e. much
higher than the ordering temperatures $T_c \sim 0.1$ K expected on
basis of the intercluster dipolar couplings
\cite{fernandez00PRB,fernandez02PRB,martinez01EPL}. Although
quantum tunneling of these cluster spins has been observed
\cite{sangregorio97PRL,aubin98JACS,thomas99PRL,mettes01PRB}, and
could in principle provide a relaxation path towards the
magnetically ordered equilibrium state
\cite{fernandez00PRB,fernandez02PRB}, the associated rates in zero
field are extremely small ($<100$ s$^{-1}$). For these systems,
tunneling only becomes effective when strong transverse fields
$B_{\perp}$ are applied to increase the tunneling rate. Although
the tunability of this rate and thus of $T_{1}^{el}$ by
$B_{\perp}$ could recently be demonstrated for Mn$_{12}$, Fe$_8$
and Mn$_4$ \cite{mettes01PRB}, no ordering has yet been observed.

The obvious way to obtain a dipolar molecular magnet is thus to
look for a high-spin molecule having sufficiently weak magnetic
anisotropy and negligible {\em inter}-cluster super-exchange
interactions. Here we report data for\\
Mn$_{6}$O$_{4}$Br$_{4}$(Et$_{2}$dbm)$_{6}$, abbreviated as
Mn$_{6}$ \cite{aromi99JACS}, whose net magnetocrystalline
anisotropy (cf. \S\ref{paramMn6}) proves to be sufficiently small
to enable measurements of its equilibrium magnetic susceptibility
and specific heat down to our lowest temperatures ($15$ mK).

In this chapter we report the observation that the magnetic
clusters do undergo a transition to a long-range ferromagnetically
ordered state at $T_{c}=0.161(2)$ K. The transition can be
observed both by \textit{ac}-susceptibility and by specific heat
experiments, thanks to the fact that the magnetic relaxation
remains very fast, even at very low temperatures, as compared to
the intrinsic experimental times ($0.1 - 5$ ms for
\textit{ac}-susceptibility and $1-100$ s for specific heat). The
magnetic relaxation has been investigated by field-dependent
specific heat and $^{55}$Mn NMR. By combining the two techniques
we are able to study the interplay between electron- and nuclear-
spin-lattice relaxation.

\section{Long-range ferromagnetic ordering}

\subsection{\textit{ac}-susceptibility}

\begin{figure}[p]
\begin{center}
      \leavevmode
      \epsfxsize=120mm
      \epsfbox{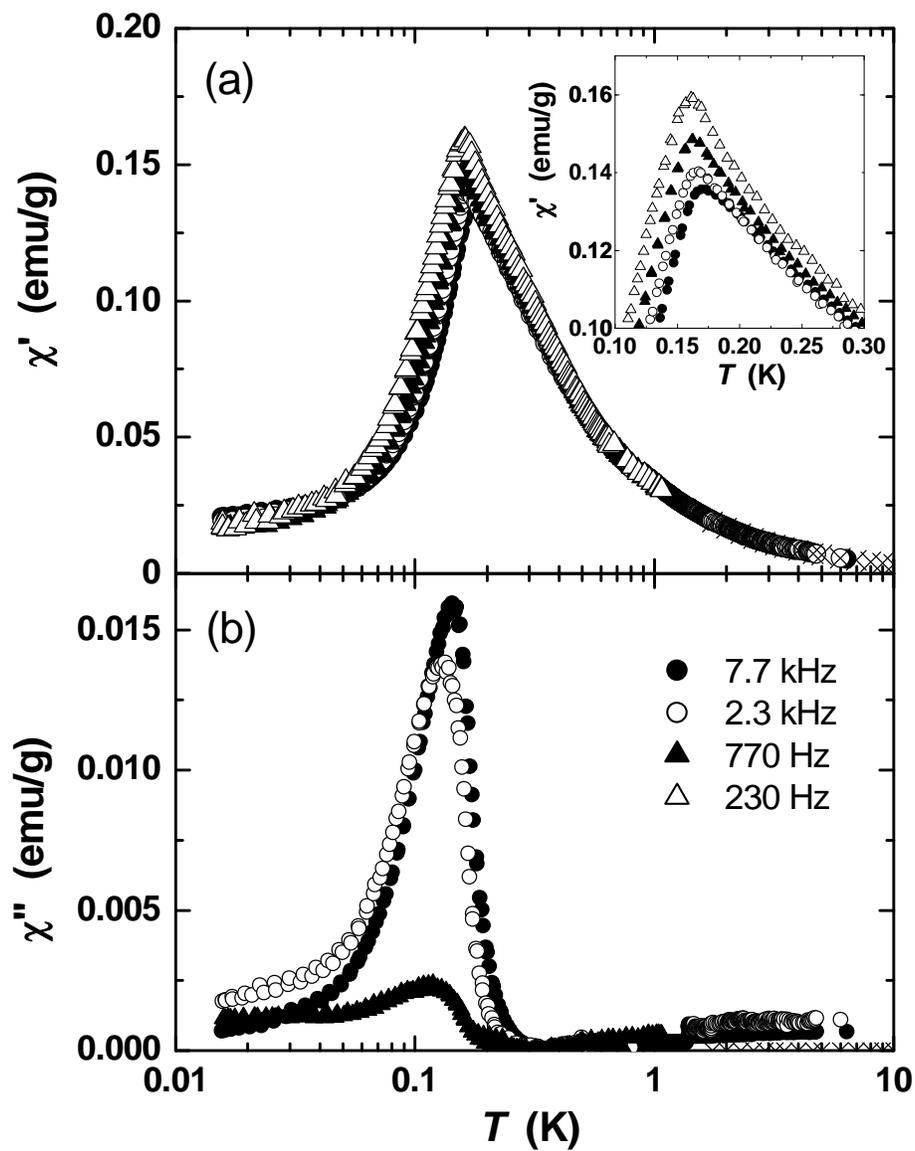}
\end{center}
\caption{\label{X12vsT} Real (a) and imaginary (b) parts of the
\textit{ac}-susceptibility at the indicated frequencies. Inset:
magnification of $\chi'(T)$ to evidence the frequency dependence
of the peak.}
\end{figure}

The \textit{ac}-susceptibility data in Fig. \ref{X12vsT} were
taken using the susceptometer described in \S\ref{acsusc}, varying
the frequency $\nu$ between $230$ to $7700$ Hz and recording the
data while slowly cooling down the refrigerator. Additional
measurements at $T > 1.8$ K were performed in a commercial SQUID
susceptometer. For this zero-field experiment, the susceptometer
was mounted in the upper Araldite pot of the dilution refrigerator
(cf. Fig. \ref{dilution}), after having replaced the Kapton tail
by a closed plug. This has the advantage of allowing a higher
$^3$He circulation rate, and assuring a the perfect coincidence of
the sample and thermometer temperatures even for $T>0.8$ K, when
some rather sudden temperature jumps may occur.

The real part $\chi^{\prime}$ shows a sharp maximum at $T_{c} =
0.161(2)$ K. We found that $\chi^{\prime}$ at $T_{c}$ is close to
the estimated limit for a ferromagnetic powder sample, $1/(\rho
N_{s}+\rho_{\mathrm{sam}} N_{c}) \simeq 0.14 \pm 0.02$ emu/g,
where $\rho = 1.45$ g/cm$^{3}$ and $\rho_{\mathrm{sam}} \simeq
0.45$ g/cm$^{3}$ are respectively the densities of bulk Mn$_{6}$
and of the powder sample, $N_{s}=4 \pi/3$ is the demagnetizing
factor of a crystallite, approximated by a sphere, and
$N_{c}\simeq 2.51$ is the demagnetizing factor of the sample
holder. This indicates that Mn$_{6}$ is ferromagnetically ordered
below $T_{c}$.

The temperature $T_{\mathrm{peak}}$ at which the maximum value of
$\chi^{\prime}$ is found, as shown in detail in the inset of Fig.
\ref{X12vsT}, is seen to vary only weakly with $\nu$, which we
attribute to the anisotropy. The total activation energy of
Mn$_{6}$ amounts to $DS^{2} \simeq 1.5$ K, i.e. about $45$ times
smaller than for Mn$_{12}$. Accordingly, one expects the
superparamagnetic blocking of the Mn$_{6}$ spins to occur when $T
\simeq T_{B}($Mn$_{12})/45$, that is below $\simeq 0.12$ K. Since
this value is close to the actual $T_c$, one may expect that for
$T \rightarrow T_c$ the approach to equilibrium begins to be
hindered by the anisotropy of the individual molecular spins. We
stress however, that the frequency dependence of $\chi^{\prime}$
observed here is quite different from that of the well known
anisotropic superparamagnetic clusters. A way to quantify the
frequency dependence of the peak in $\chi '$ is by means of the
parameter $\Delta T_{\mathrm{peak}}/[T_{\mathrm{peak}} \Delta
(\log \nu)]$, which gives the variation of $T_{\mathrm{peak}}$ per
decade of frequency. We find here $\Delta
T_{\mathrm{peak}}/[T_{\mathrm{peak}} \Delta (\log \nu)] \simeq
0.04$, to be compared with the typical values of $\sim 0.20$ for
superparamagnetic blocking. In fact it is closer to the value
$\sim 0.06$ found for certain types of spin glasses
\cite{mydoshB}, but the peak found here is much higher and
sharper.

\begin{figure}[t]
\begin{center}
      \leavevmode
      \epsfxsize=120mm
      \epsfbox{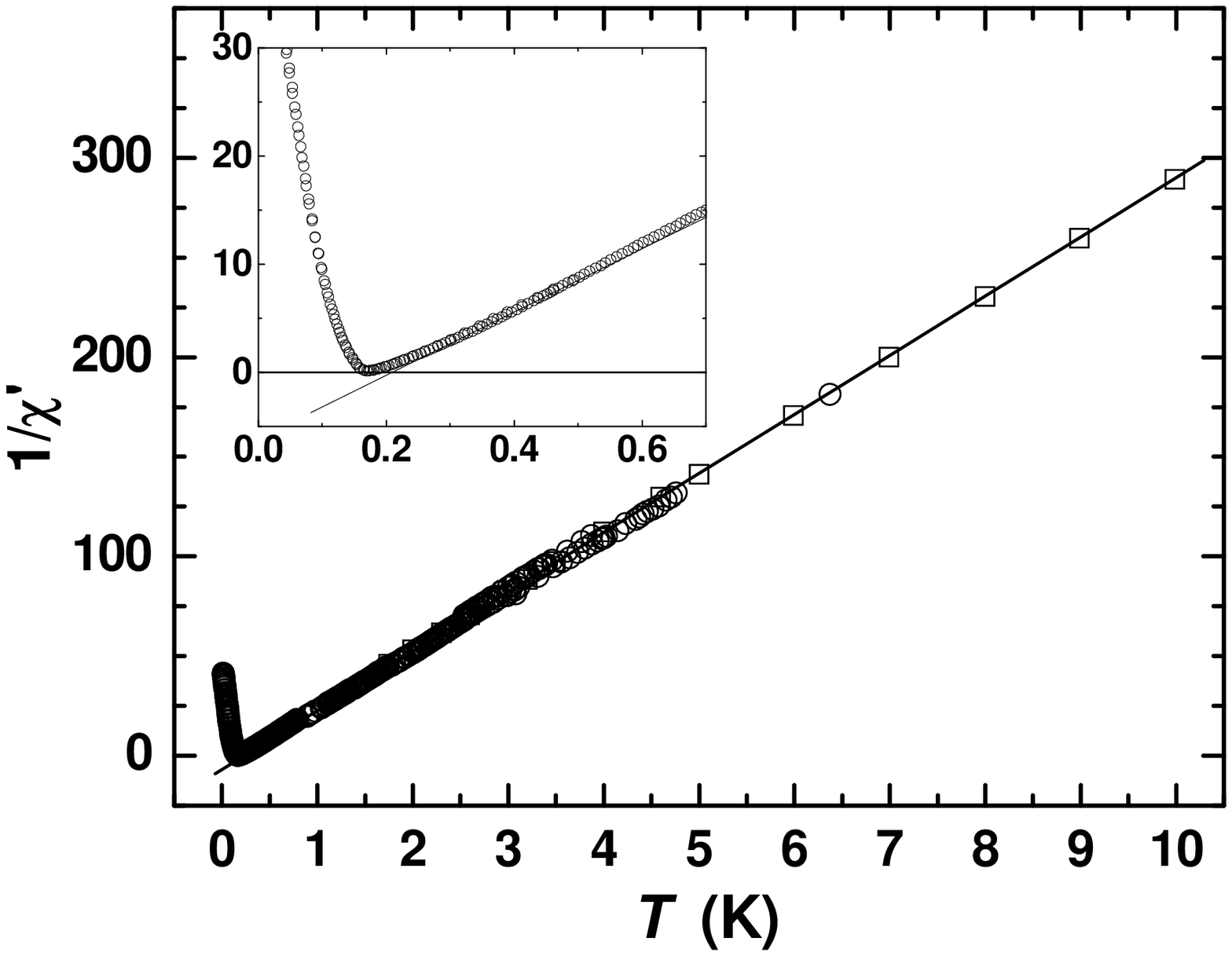}
\end{center}
\caption{\label{invXvsT.eps} $T$-dependence of the inverse of the
real part of the \textit{ac}-susceptibility. Circles: low-$T$
susceptometer, $\nu = 7700$ Hz. Squares: SQUID susceptometer. The
inset shows a magnification of the temperature range around
$T_c$.}
\end{figure}

Below $T_{c}$, $\chi^{\prime}$ decreases rapidly, as expected for
an anisotropic ferromagnet in which the domain-wall motions become
progressively pinned. The associated domain-wall losses should
then lead to a frequency dependent maximum around $T_c$ in the
imaginary part, $\chi^{\prime \prime}$, as seen experimentally
indeed [cf. Fig. \ref{X12vsT}(b)]. In fact, although the Mn$_6$
spins can be considered as nearly isotropic at high temperatures,
the anisotropy energy is still large compared with the dipolar
interaction energy $\mu^2 / r^3 \simeq 0.1$ K between nearest
neighbor molecules. Thus the ordering should be that of an Ising
dipolar ferromagnet.

Susceptibility data $\chi_{i}$ corrected for the demagnetizing
field ($\chi_{i} = \chi^{\prime}/[1 - (\rho N_{s} +
\rho_{\mathrm{sam}} N_{c}) \chi^{\prime}]$) follow the Curie-Weiss
law $\chi = C/(T - \theta)$ down to approximately $0.3$ K, with $C
= 0.034(1)$ cm$^{3}$K/g and $\theta= 0.20(3)$ K. The constant $C$
equals, within the experimental errors, the theoretical value for
randomly oriented crystals with Ising-like anisotropy
$N_{A}g^{2}\mu_{B}^{2}S(S+1)/3k_{B}P_{m} = 0.0332$ cm$^{3}$K/g,
where $S = 12$, $g = 2$, and the molecular weight $P_{m} =
2347.06$. The positive $\theta$ confirms the ferromagnetic nature
of the ordered phase. From the mean-field equation $\theta =
2zJ_{eff}S(S+1)/3k_{B}$, we estimate the effective inter-cluster
magnetic interaction $J_{eff} \simeq 1.6 \times 10^{-4}$ K, and
the associated effective field $H_{eff} = 2zJ_{eff}S/g\mu_{B} =
3.5 \times 10^{2}$ Oe coming from the $z=12$ nearest neighbors.

\subsection{Specific heat}

\begin{figure}[t]
\begin{center}
      \leavevmode
      \epsfxsize=120mm
      \epsfbox{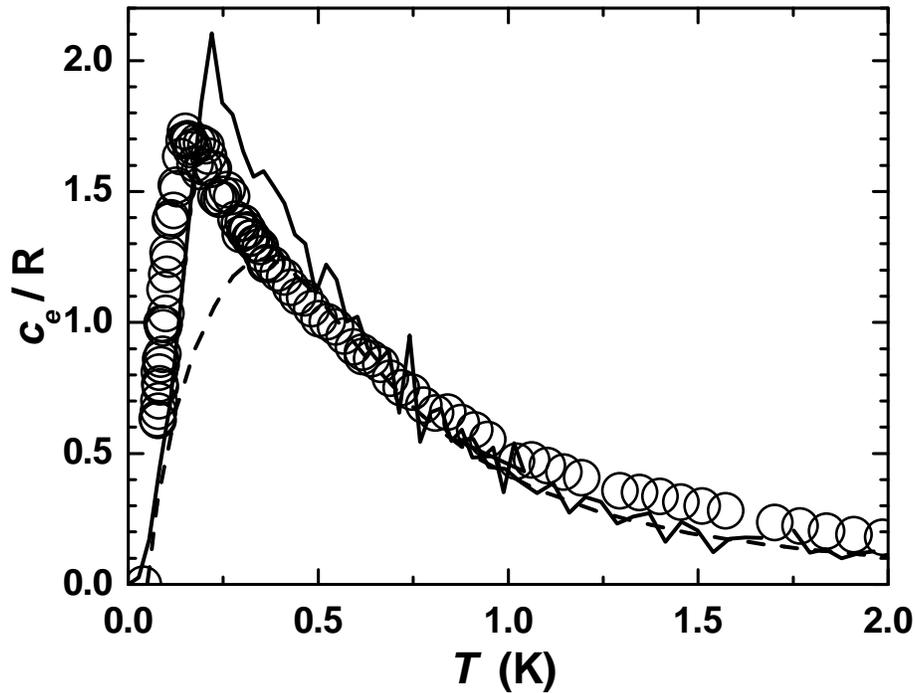}
\end{center}
\caption{\label{CevsT} $T$-dependence of the electronic specific
heat at zero applied field (circles). The dotted line is the
Schottky anomaly calculated with $D=0.013$ K. The full line is the
Monte Carlo (MC) calculation for an orthorhombic lattice of $1024$
Ising spins with periodic boundary conditions. For each point, $2
\times 10^{4}$ MC steps per spin were performed.}
\end{figure}

Additional evidence for the ferromagnetic transition at $T_{c}$ is
provided by the electronic specific heat $c_{e}$, shown in Fig.
\ref{CevsT} \cite{mettesT}. The specific heat of a few milligrams
of sample, mixed with Apiezon grease, was measured at low-T in a
home-made calorimeter \cite{mettesT,mettes01PRB} that makes use of
the thermal relaxation method. An important advantage of this
method is that the characteristic time $\tau_{e}$ of the
experiment (typically, $\tau_{e} \simeq 1 - 100$ seconds at low-T)
can be varied by changing the dimensions (and therefore the
thermal resistance) of the Au wire that acts as a thermal link
between the calorimeter and the mixing chamber of the dilution
refrigerator.

The electronic specific heat was obtained by subtracting from the
total specific heat the contribution of the lattice, which follows
the well-known Debye approximation for low temperatures
$c_{\mathrm{lattice}} \propto (T/\Theta_{D})^{3}$, where
$\Theta_{D} \simeq 48$ K, as well as the contribution
$c_{\mathrm{nucl}}$ arising from the nuclear spins, as discussed
in \S\ref{sec:CmvsB} below. $c_e (T)$ reveals a sharp peak at
$0.15(2)$ K, indeed very close to the peak in the real part of the
\textit{ac}-susceptibility. Numerical integration of $c_{e}/T$
between $0.08$ K and $4$ K gives a total entropy change of about
$3.4 k_{B}$ per molecule, which is indeed very close to the value
$k_{B} \ln(2S+1)=3.22k_{B}$ for a fully-split $S=12$ spin
multiplet. We may therefore attribute the peak to the long-range
order of the molecular spins. We note, however, that at $T_c$ the
entropy amounts to about $1 k_B$ per spin, showing that only the
lowest energy spin states take part in the magnetic ordering.

\subsection{Monte-Carlo Simulations}

In order to simulate the zero-field specific heat data,
Monte-Carlo (MC) calculations were performed by J. F.
Fern\'{a}ndez (univ. of Zaragoza, Spain) for an $S=12$ Ising model
of magnetic dipoles on an orthorhombic lattice with axes $a_x =
15.7$ \AA, $a_y=23.33$ \AA, and $a_z=16.7$ \AA, which approximates
the crystal structure of Mn$_{6}$. The Hamiltonian includes
dipolar interaction term as well as the anisotropy term $-DS_z^2$
given in Eq. (\ref{hamiltonianMn6}). When dipolar interactions are
neglected, the zero field splittings of the $S=12$ multiplet
produced by this crystal field term lead to the Schottky anomaly
in $c_{e}$ shown by the dotted curve in Fig. 3. The fit in the
range above 0.5 K yields $D/k_B \simeq 0.013$ K, in good agreement
with our above estimates. For $T < 0.5$ K the intermolecular
dipolar interactions come into play and remove the remaining
degeneracy of the lowest-lying $|\pm m\rangle$ spin doublets. The
MC simulations show that the ground state is ferromagnetically
ordered, as observed, and predict a shape for $c_{e}$ that is in
very good agreement with the experiment. In Fig. \ref{CevsT}, we
show $c_{e}$ calculated assuming all molecular easy ($z$) axes to
point along $a_z$, i.e. one of the two nearly equivalent short
axes of the actual lattice. Similar results were obtained for
other orientations chosen for the anisotropy ($z$) axis. We note
that the Ising simulations give $T_{c} = 0.22$ K, which is
slightly higher than the experimental $T_c=0.161(2)$ K. This
difference may be due to the Ising approximation taken for the
intercluster dipolar interaction.

\begin{figure}[t]
\begin{center}
      \leavevmode
      \epsfxsize=100mm
      \epsfbox{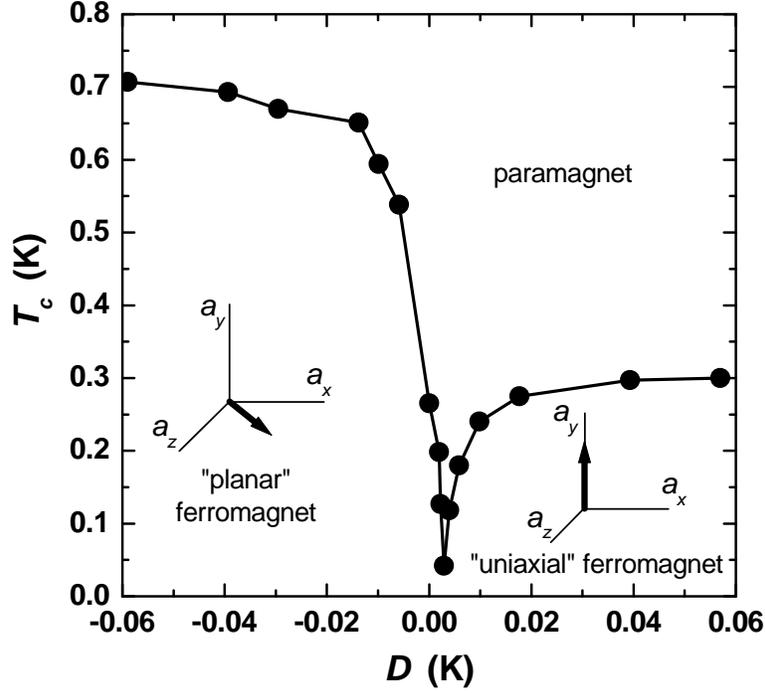}
\end{center}
\caption{\label{TcvsD} Calculated critical temperature $T_c$ as a
function of the anisotropy parameter $D$, for classical Heisenberg
spins.}
\end{figure}

In order to pursue this point further, additional MC simulations
were performed for the same crystal lattice, but with classical
Heisenberg spins replacing the ($S=12$) Ising spins. To
investigate the sensitivity to the type of anisotropy, the sign
and magnitude of $D$ were varied. These calculations resulted in
the phase diagram shown in Fig. \ref{TcvsD}, which we include here
since it illustrates the complicated way in which the nature of
the actual ground state and the value of $T_c$ may depend on the
combination of the long-range dipolar interaction and the
anisotropy parameter. Although the ground state for this lattice
is always ferromagnetic, it can be either ``uniaxial'', with
strong preference for the $a_y$ axis, or ``planar'', in the sense
that the $a_x - a_z$ plane becomes an easy plane with a weak
preference for a given direction in the plane, as indicated in the
figure. Interestingly, the switching point between these
orientations is not at $D=0$ but at $D=3$ mK. The reason for this
is as follows. Because the crystal lattice under consideration is
far from being cubic, the dipolar interaction energy is rather
anisotropic. The energy is minimized when the magnetization
$\vec{M}$ points in the direction (in the $a_x - a_z$ plane) shown
in the inset, on the left-hand side of Fig. \ref{TcvsD}.
Therefore, $\vec{M}$ points along this direction for $D<0$. On the
other hand, the energy minimization for $D>0$ is a competing
process. Clearly, dipolar interaction must become dominant for
sufficiently small values of $D$. The numerical results show that
this occurs if $0 < D \lesssim 3$ mK. The numerical datapoints in
Fig. \ref{TcvsD} also show that $T_c$ varies sharply within the
$-0.01 < D < 0.01$ K range. Because of the discreteness of the MC
simulations we can not tell whether $T_c$ vanishes at $D \simeq 3$
mK. The lowest numerical value obtained is as small $T_c \simeq
0.03$ K at $D \simeq 3$ mK. Outside this narrow range of $D$,
$T_c$ is already almost equal to the limiting values of $\simeq
0.7$ K and $\simeq 0.3$ K, reached for infinite negative and
positive $D$, respectively. Such a variation of $T_c$ with
anisotropy, as well as the form of the calculated and observed
specific heat ordering anomaly, appear to be specific for dipolar
interactions. They differ widely from the corresponding behavior
known for 3-dimensional (Heisenberg, Ising, XY) ferromagnetic
lattices with nearest-neighbor interaction only. For instance, in
those models the variation of $T_c$ with anisotropy is restricted
to about 20\% and the $T_c$ value is highest for the Ising case.

\section{Spin-lattice relaxation}

The electron- and nuclear- spin-lattice relaxation in Mn$_6$ is an
interesting topic to investigate and compare with the typical
behavior of the highly anisotropic SMMs like Mn$_{12}$-ac. As we
shall see, provided that a Zeeman splitting for the electron spin
is created by the application of an external field, the nuclear
spin dynamics in Mn$_6$ closely resembles the behavior of the
nuclei in Mn$_{12}$-ac in the thermally activated regime, where
the electron spin energy levels are split by crystal-field effects
instead.

\subsection{Field-dependent specific heat}  \label{sec:CmvsB}

\begin{figure}[t]
\begin{center}
      \leavevmode
      \epsfxsize=110mm
      \epsfbox{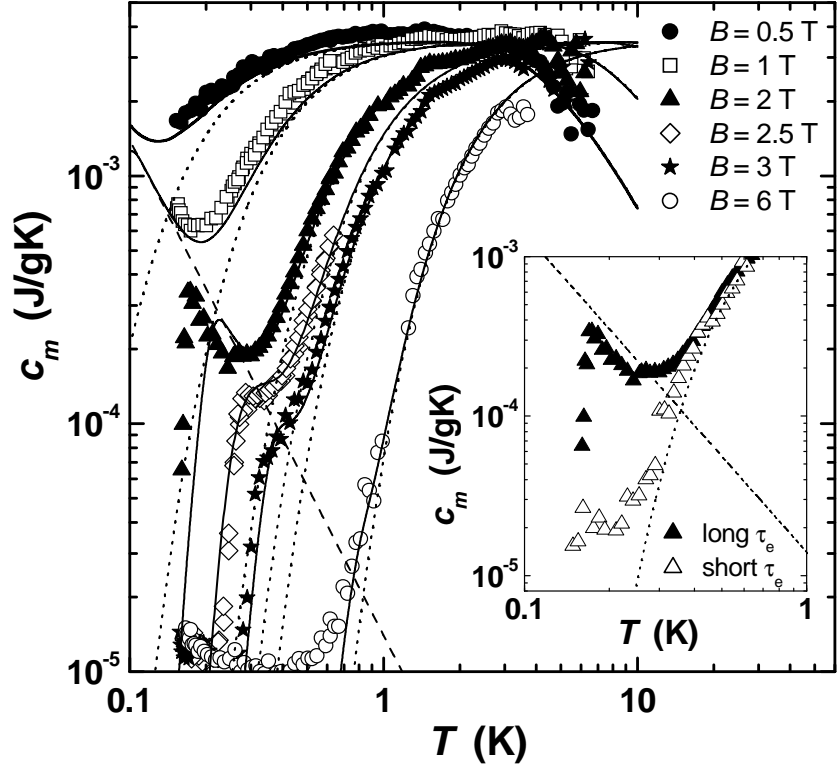}
\end{center}
\caption{\label{CmvsT-B} $T$-dependence of the magnetic specific
heat at the specified temperatures. The lines are the calculated
electronic (dotted) and nuclear (dashed) contributions. The solid
lines are the total (electronic + nuclear) $c_m$, calculated
accounting for the nuclear $T_1$ (see text). Inset: detail of
$c_m(T)$ at $B=2$ T, for long ($\sim 100$ s) and short ($\sim 1$
s) experimental times.}
\end{figure}

We begin by discussing the specific heat data obtained in varying
magnetic field $B$ \cite{mettesT}, plotted in Fig. \ref{CmvsT-B}.
Even for the lowest $B$ value, the ordering anomaly is fully
suppressed, as expected for a ferromagnet \cite{dejongh01AP}.
Accordingly, we may account for these data with the Hamiltonian
(\ref{hamiltonianMn6}) neglecting dipolar interactions. The Zeeman
term splits the otherwise degenerate $| \pm m \rangle$ doublets,
and already for $B \sim 0.5$ T the level splittings become
predominantly determined by $B$, so that the anisotropy term can
also be neglected. As seen in Fig. \ref{CmvsT-B}, the calculations
performed with $D=0$ reproduce the data quite satisfactorily at
higher temperatures (dotted curves).

However, when the maxima of the Schottky anomalies are shifted to
higher $T$ by increasing $B$, an additional contribution at low
$T$ is revealed. It is most clearly visible in the curves for $1$
T$ < B < 2.5$ T, and varies with temperature as $cT^{2}/R = 4
\times 10^{-3}$. We attribute this to the high-temperature tail of
the specific heat contribution $c_{\mathrm{nucl}}$ arising from
the Mn nuclear spins ($I=5/2$), whose energy levels are split by
the hyperfine interaction with the Mn$^{3+}$ electronic spins $s$.
This interaction can be approximated by $\mathcal{H}_{hf}=A
\vec{I} \cdot \vec{s}$, where $A$ is the hyperfine coupling
constant (cf. \S\ref{sec:hyperfine}). At high temperatures (i.e.,
when $A \, s \ll k_{B}T$) $c_{\mathrm{nucl}}/R \simeq
\frac{1}{3}A^{2}s^{2}I(I+1)T^{-2}$ \cite{abragam70}. Taking
$A=7.6$ mK as used previously to simulate ESR spectra measured on
a Mn$_{4}$ cluster \cite{zheng96IC}, we obtain the dashed line
shown in \ref{CmvsT-B}. This contribution was subtracted from the
zero-field data shown in Fig. \ref{CevsT}.

A remarkable feature of the experimental data that is not
reproduced by these calculations is that, at the lowest $T$, the
nuclear specific heat drops abruptly to about $10^{-5}$ J/gK. The
temperature $T_n$ where the drop occurs depends on $B$ but also on
the characteristic time constant $\tau_{e}$ of our
(time-dependent) specific heat experiment: the deviation from the
(calculated) equilibrium specific heat is found at a lower $T$
when the system is given more time to relax, as shown in the inset
of Fig. \ref{CmvsT-B}. We conclude that the drop indicates that
the $^{55}$Mn nuclear spins can no longer reach thermal
equilibrium within time $\tau_{e}$. We may write:
\begin{eqnarray}
c_{nucl}(\tau_{e})= c_{nucl}^{(\mathrm{eq})}
[1-\exp(-\tau_{e}/T_{1})],  \label{CnuclT1}
\end{eqnarray}
showing that the transition should occur when the nuclear
spin-lattice relaxation time $T_{1}$ becomes of the order of
$\tau_{e}$. These transitions to non-equilibrium provide therefore
direct information on the temperature and field dependence of
$T_{1}$.

As already discussed in \S\ref{NSLR}, $T_1(T,B)$ can be related to
the fluctuation of the transverse hyperfine field, as produced by
the phonon-induced transitions (at a rate $\tau_{\mathrm{s-ph}}$)
between different levels of the electronic spin. In the case of
Mn$_6$ these are simply the Zeeman levels of the $S=12$ cluster
spin in the magnetic field $B$, split by an energy $\Delta E = g
\mu_B B$.

\begin{figure}[t]
\begin{center}
      \leavevmode
      \epsfxsize=110mm
      \epsfbox{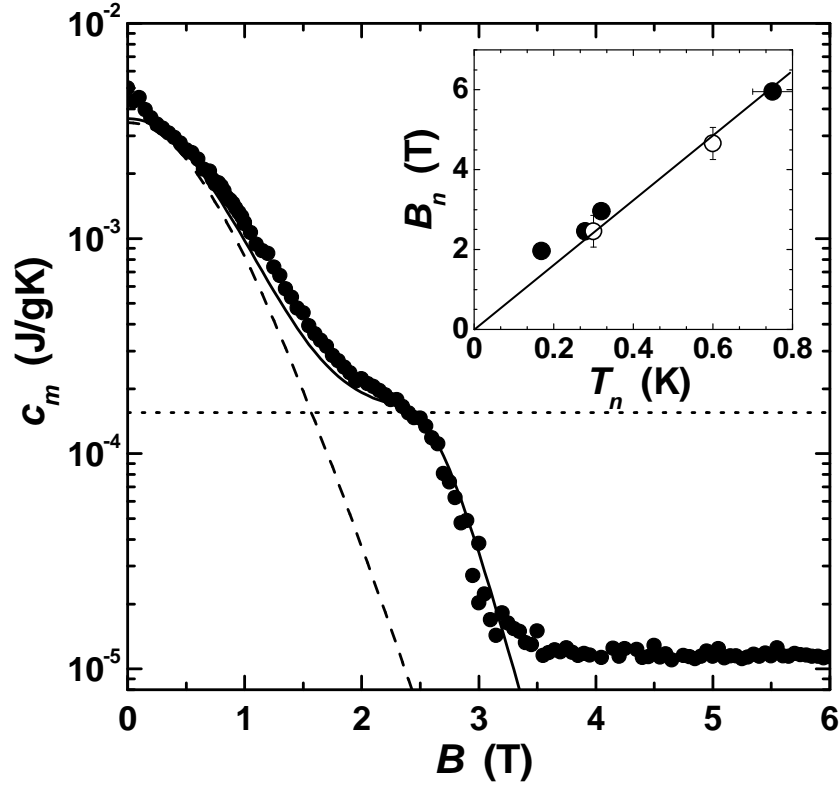}
\end{center}
\caption{\label{CmvsB} Field-dependence of the magnetic specific
heat at $T=0.3$ K. The dotted line is the calculated electronic
specific heat, the dashed line is the equilibrium nuclear specific
heat, and the solid line is the calculated total $c_m$, accounting
for the nuclear $T_1$. Inset: $B_n(T_n)$ obtained from $T$-sweeps,
as in Fig. \ref{CmvsT-B} (full circles), or from $B$-sweeps, as in
the present figure (open circles). From the linear fit (solid
line) we extract the average $\tau_0 \simeq 3 \times 10^{-4}$ s.}
\end{figure}

At low $T$ and high $B$, only the ground and the first excited
states, $m=+12$ and $m=+11$, need to be considered. We may
therefore rewrite Eq. (\ref{Wlast}) as\footnote{Recall that $1/T_1
= 2W$.}:
\begin{eqnarray}
\frac{1}{T_{1}} = \frac{\gamma_N^2}{2} \langle \Delta b_{\perp}^2
\rangle \frac{\tau_{s-ph}}{1+\omega_N^2 \tau_{s-ph}^2} \approx
\tau_{0}^{-1}\exp\left(-\frac{g \mu_B B}{k_B T}\right),
\label{T1Mn6}
\end{eqnarray}
where $\tau_{0}^{-1} = \langle \Delta b_{\perp}^2 \rangle / (2
B_{\mathrm{tot}}^2 \tau_{11})$ incorporates the lifetime
\begin{eqnarray}
\tau_{11} = \frac{1 - \exp(-g \mu_B B/k_B T)}{C_{\mathrm{s-ph}} (g
\mu_B B)^3}
\end{eqnarray}
of the first excited state $m=+11$, and the ratio between the
average value of the transverse component of the fluctuating
hyperfine field $\Delta b_{\perp}$ and the square of the total
field at the nuclei $B_{\mathrm{tot}}$. The approximated
expression of (\ref{T1Mn6}) is valid when $\omega_N \tau_{s-ph}
\gg 1$, which is practically always the case, except at very low
field.

It is easy to see from Eq. (\ref{T1Mn6}) that the nuclear spins
can be taken out of of equilibrium either by decreasing $T$ down
to $T_n$ at constant field, as in Fig. \ref{CmvsT-B}), or by
increasing $B$ up to a given value $B_n$ at constant $T$. This is
indeed observed experimentally, as shown in Fig. \ref{CmvsB}. The
effect of the field is just to polarize the electron spins, which
reduces the fluctuations of the hyperfine field, thus effectively
disconnecting nuclear spins from the lattice. It also follows
that, for a given $\tau_{e}$, $B_{n}$ is linear in $T_{n}$ as long
as the field-dependence of $\tau_{11}$ can be neglected as
compared to the exponential dependence of $T_1^{-1}$ on $T$ and
$B$. The inset of Fig. \ref{CmvsB} shows $B_n(T_n)$ obtained
either from $T$-sweeps at constant $B$ (as in Fig. \ref{CmvsT-B})
or from $B$-sweeps at constant $T$ (as in Fig. \ref{CmvsB}). The
two methods prove to be indeed perfectly consistent with each
other. The slope of the linear fit of $B_n(T_n)$ gives an average
value of $\tau_{0} \simeq 3 \times 10^{-4}$ s in the high-field
regime. We observe that at lower fields $B_n(T_n)$ tends to
deviate from the linear fit, which we attribute to the decrease of
$\tau_{11}$. Using the average value of $\tau_0$ we have
calculated the time-dependent $c_{\mathrm{nucl}}$ from Eq.
(\ref{CnuclT1}), which has been added to the calculated electronic
specific heat to yield the solid lines in Figs. \ref{CmvsT-B} and
\ref{CmvsB}, and seen to be in good agreement with the
experimental data.

\subsection{$^{55}$Mn NMR}

\begin{figure}[t]
\begin{center}
      \leavevmode
      \epsfxsize=130mm
      \epsfbox{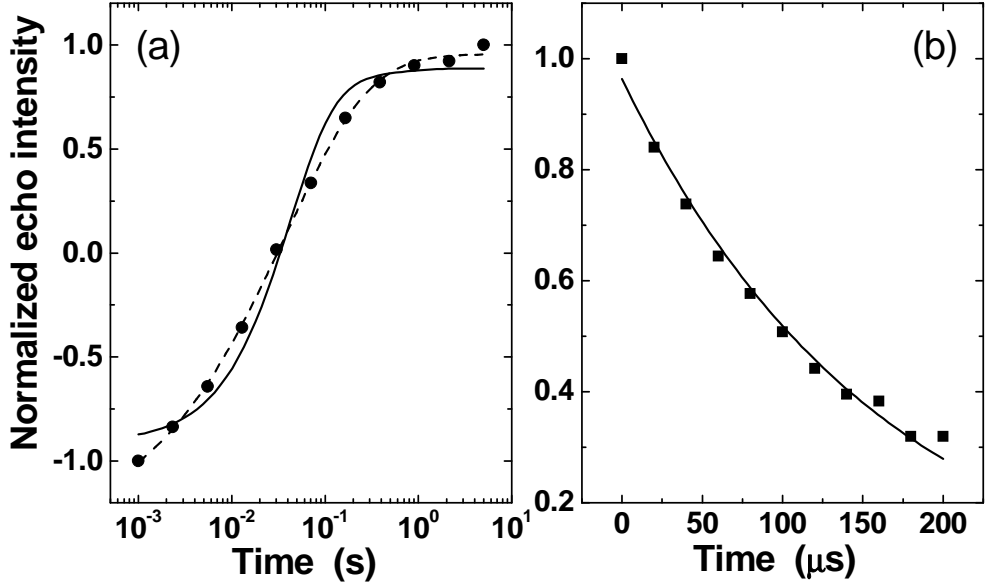}
\end{center}
\caption{\label{T1T2Mn6} Inversion recovery (a) and decay of
transverse magnetization (b) for the $^{55}$Mn nuclei at $T=0.9$
K, $B = 5$ T and $\nu = 251.5$ MHz. The lines in (a) are fits to
Eq. (\ref{recovery}) (solid), and Eq. (\ref{strexp}) with $\alpha
\simeq 0.5$ (dashed). The solid line in (b) is a fit to Eq.
(\ref{T2}).}
\end{figure}

We have now the possibility to compare the NSLR rate as inferred
from the specific heat experiments, with direct measurements of
this quantity by NMR experiments. The $^{55}$Mn NSLR and TSSR
rates were obtained following the same procedures and analysis as
described in \S\ref{sec:measanal} (Fig. \ref{T1T2Mn6}), but using
a standard $^3$He system instead of the dilution refrigerator. The
advantage of the $^3$He cryostat is that, because of its
construction, we do not need to insert a $\lambda$-cable in the
NMR resonant circuit (cf. \S\ref{NMRcircuit}). This increases the
quality factor of the resonator and allows to measure large enough
signals without need to reach millikelvin temperatures, where the
very slow NSLR would make the measurements almost unfeasible.
Unfortunately, the useful frequency range becomes accordingly much
narrower.

\begin{figure}[t]
\begin{center}
      \leavevmode
      \epsfxsize=100mm
      \epsfbox{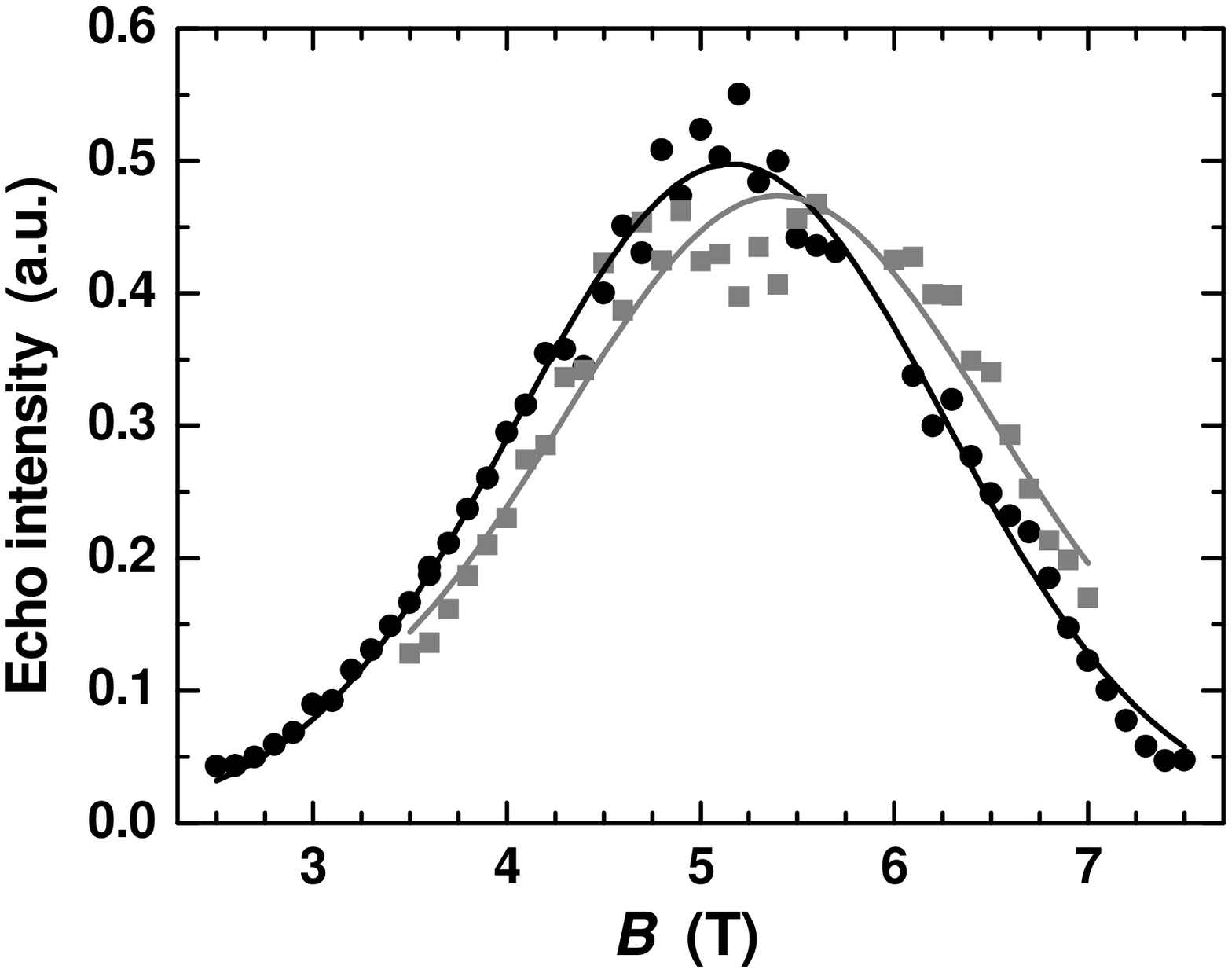}
\end{center}
\caption{\label{spectrumMn6} $^{55}$Mn NMR spectra at $T=0.9$ K
and measuring frequencies $\nu=256.5$ MHz (black circles) and
$\nu=261.5$ MHz (gray squares). The gap in the data around $B
\simeq 5.8$ T is due to the presence of the $^1$H line. The lines
are Gaussian fits with total width $2 \sigma_B \simeq 2.2$ T.}
\end{figure}

The inversion recoveries [Fig. \ref{T1T2Mn6}(a)] are best fitted
by a stretched exponential [Eq. (\ref{strexp})] with $\alpha \sim
0.5$, although the choice of the stretching exponent does not
strongly influence the extracted value of $W$. The field-sweep NMR
spectra in Fig. \ref{spectrumMn6} clearly show that it is
impossible to determine whether there are inequivalent sites in
the molecule, as regards the hyperfine coupling (compare with the
case of Mn$_{12}$-ac in Fig. \ref{Mnlines}). This may be due to
the large quadrupolar splitting expected in Mn$^{3+}$ sites, plus
the fact that our sample is an unoriented powder. As expected from
the internal ferromagnetic structure of the cluster electron spins
(cf. \S\ref{paramMn6}), the $^{55}$Mn spectrum shifts to higher
fields by lowering the frequency. The spectra can be fitted by a
Gaussian shape with total width $2 \sigma_B \simeq 2.2$ T; if this
width were due to quadrupolar splitting only, it would imply
$\Delta \nu_Q \sim 7$ MHz\footnote{This estimate is obtained by
comparing to the Mn$^{(1)}$ line in Mn$_{12}$-ac, where $\Delta
\nu_Q = 0.72$ MHz yields $2 \sigma_{\nu} = 2.4$ MHz.}, even larger
than the highest $\Delta \nu_Q \simeq 4.3$ MHz in the Mn$^{(2)}$
sites of the less symmetric Mn$_{12}$-ac cluster. We expect
therefore that the random orientation of the crystallites and,
eventually, the presence of inequivalent Mn sites as regards the
hyperfine coupling, should contribute as well to the observed
broadening. Indeed, when decreasing the frequency by 5 MHz the
maximum of the spectrum shifts only by 0.24 T, instead of the 0.47
T that would be expected when all the local hyperfine fields are
antiparallel to $\vec{B}$. We conclude that the observed spectrum,
as well as the NSLR and TSSR data, should be considered as
obtained from a mixture of nuclear signals arising from randomly
oriented crystallites with largely overlapping and
quadrupolar-split NMR lines from all the Mn sites in the cluster.
Extrapolating to $B=0$ the field-dependence of the peak of the
spectrum, one would obtain $\nu(0) \simeq 360$ MHz $\Rightarrow
B_{\mathrm{hyp}} \simeq 34$ T, very similar to the value found in
the Mn$^{(3)}$ site of Mn$_{12}$-ac.

\begin{figure}[t]
\begin{center}
      \leavevmode
      \epsfxsize=100mm
      \epsfbox{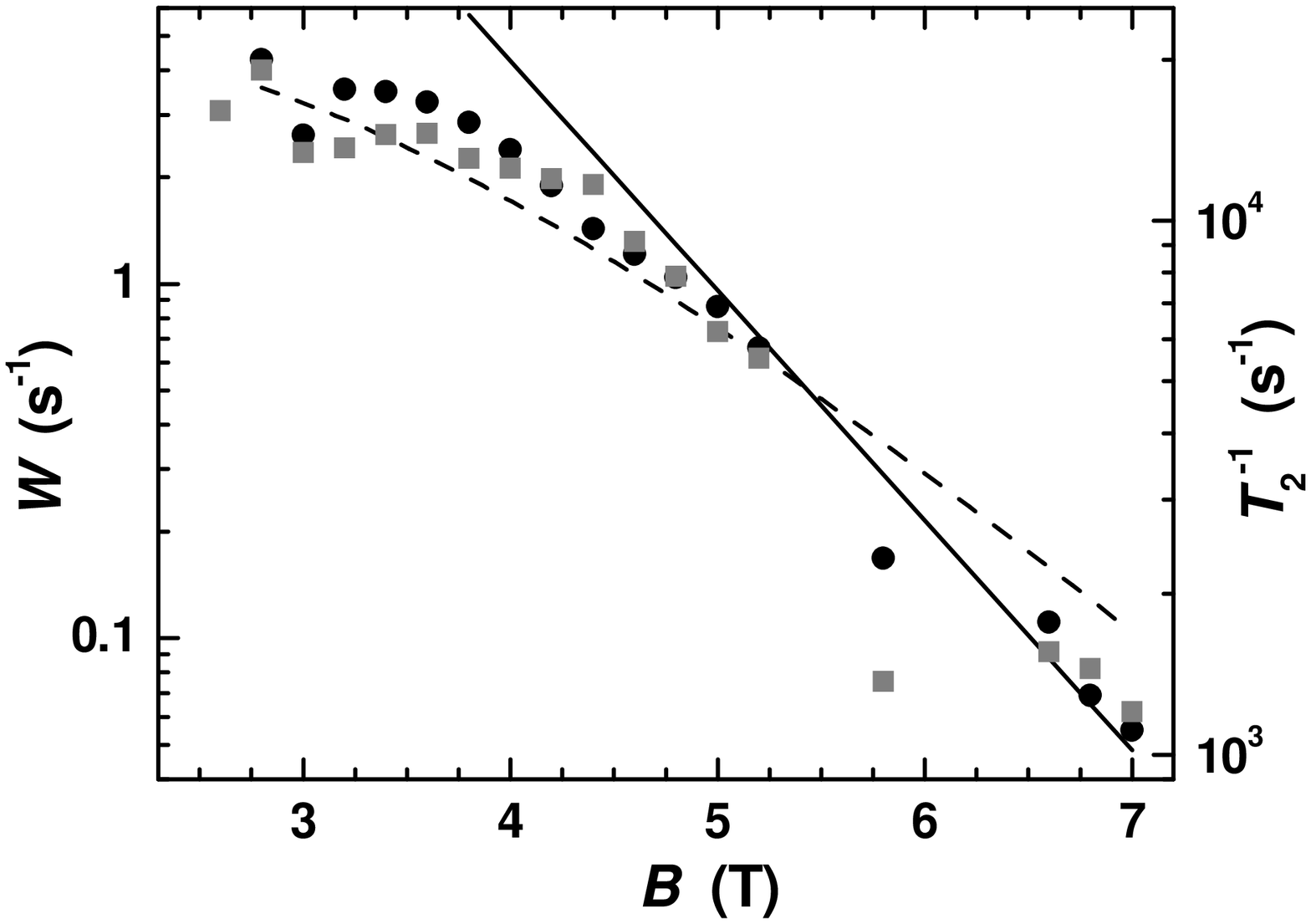}
\end{center}
\caption{\label{W1W2Mn6} Field dependencies of the $^{55}$Mn NSLR
(black circles, left scale) and TSSR (grey squares, right scale)
at $T=0.9$ K and $\nu=251.5$ MHz. The vertical scales are adjusted
to evidence the similar field dependencies. Solid line: calculated
$W(B)$ according to Eq. (\ref{T1Mn6}) with $\tau_0 = 3 \times
10^{-4}$ s. Dashed line: fit to Eq. (\ref{WMn6}).}
\end{figure}

Fig. \ref{W1W2Mn6} shows the field-dependencies of the NSLR rate
$W$ and the TSSR rate $T_2^{-1}$, measured at constant frequency
$\nu = 251.5$ MHz and temperature $T=0.9$ K. From the discussion
above it is clear that these data must be interpreted with a
certain caution, since shifting $B$ at constant $\nu$ means that
we are sampling each time a different portion of the NMR signal,
which means different quadrupolar satellites, different
orientation of the crystallites, etc. Nevertheless, the agreement
with the estimate of $T_1$ obtained by specific heat data (inset
of Fig. \ref{CmvsT-B}) is quite satisfactory. We can directly
compare the NSLR rate $W(B)$ as obtained by NMR with the estimated
$T_1$ from specific heat data by plotting the line $W(B) = 1/(2T_1
(B))$ with $T_1(B)$ calculated from Eq. \ref{T1Mn6}, fixing
$\tau_{0} = 3 \times 10^{-4}$ s and $T=0.9$ K (solid line in Fig.
\ref{W1W2Mn6}). The agreement is very good on the high-field side
of the data, whereas at low fields the NSLR rate is seen to level
off instead of increasing exponentially. This is actually the same
phenomenon that we already observed in the low-field part of
$B_n(T_n)$ in the inset of Fig. \ref{CmvsT-B}, due to the fact
that $\tau_0$ is not constant. We fitted therefore $W(B)$
including the field-dependence of $\tau_0$:
\begin{eqnarray}
W(B) \approx \frac{\langle \Delta b_{\perp}^2 \rangle}{4
B_{\mathrm{tot}}^2} \times \frac{C_{\mathrm{s-ph}} (g \mu_B B)^3}
{1 - \exp(-g \mu_B B/k_B T)} \exp\left(-\frac{g \mu_B B}{k_B T}
\right).  \label{WMn6}
\end{eqnarray}
As shown by the dashed line in Fig \ref{W1W2Mn6}, this function
indeed fits much better the low-field part of the data.

In Fig. \ref{W1W2Mn6} we also show the TSSR rate $T_2^{-1}(B)$,
rescaled in order to evidence the practically identical
$B$-dependence as found for the NSLR rate. This is indeed to be
expected (cf. \S\ref{NSLR} and \S\ref{TSSR}) if the TSSR is
induced by the random change of local hyperfine field when the
electron spin makes thermally-assisted transitions between the
ground- and the first excited states. Eq. (\ref{invT2}) can be
immediately readapted for Mn$_6$ by substituting the energy gap
$\Delta E = g \mu_B B$ and the lifetime of the ground state
$\tau_{12} = C_{\mathrm{s-ph}}\Delta E^3 / [(\Delta E /k_B T) -
1]$, which yields:
\begin{eqnarray}
T_2^{-1} \approx \tau_{12}^{-1} = \frac{C_{\mathrm{s-ph}} (g \mu_B
B)^3}{\exp(g \mu_B B/k_B T)-1} \approx C_{\mathrm{s-ph}}(g \mu_B
B)^3 \exp\left( -\frac{g \mu_B B}{k_B T} \right), \label{invT2Mn6}
\end{eqnarray}
i.e. the same dependence as $W(B)$\footnote{Notice that in Eq.
(\ref{WMn6}) the factor $1 - \exp(-g \mu_B B/k_B T) \simeq 1$ at
the denominator is a constant except at very small fields.}.

The similarity between NSLR and TSSR rates produced by thermal
fluctuations of the electron spin was already observed in the
$T$-dependence of $W$ and $T_2^{-1}$ in Mn$_{12}$-ac for $T>0.8$ K
(cf. \S\ref{zerofield}). It is interesting to observe that in
Mn$_6$ we find $T_2^{-1} / T_1^{-1} \simeq 3800$, i.e. larger than
the value $T_2^{-1} / T_1^{-1} \simeq 1000$ found for the
Mn$^{4+}$ site of Mn$_{12}$-ac and \emph{much larger} that
$T_2^{-1} / T_1^{-1} \sim 200$ for its Mn$^{3+}$ sites
\cite{goto03PRB}. This in fact confirms the hypothesis that a
complete description of the NSLR in the thermally-assisted regime
of Mn$_{12}$-ac should include the effect of thermally-assisted
tunneling, yielding a contribution to the NSLR which is obviously
absent in Mn$_6$.

\section{Conclusions}

In conclusion, our experiments on Mn$_{6}$ demonstrate that
dipole-dipole interactions between molecular magnetic clusters may
indeed induce long-range magnetic order at low temperatures,
provided that the anisotropy is sufficiently small. The
spin-lattice relaxation then becomes fast enough to produce
equilibrium conditions down to the low temperatures needed. We
should add that similar conditions could in principle also be
reached in the highly anisotropic cluster systems, for which it
was shown that by applying magnetic fields perpendicular to the
anisotropy axis, the spin lattice relaxation can be tuned and made
similarly fast through the process of magnetic quantum tunneling.
However, magnetic ordering phenomena have not been seen in those
systems so far. This is most probably due to the fact that, given
the magnitude of the fields needed to have a considerable increase
of the relaxation rate ($B_{\perp} \gg 1$ T), any longitudinal
component of the field would create a Zeeman splitting that is
much larger than the energy involved in the magnetic dipolar
ordering. We found indeed that in Mn$_6$ the ordering transition
is removed already for relatively small fields ($\sim 0.5$ T).

We have also studied the nuclear spin dynamics of Mn$_6$, either
directly by NMR experiments, or through the hyperfine contribution
to the field-dependent specific heat. The agreement between the
two techniques is very good, and provides an interesting
comparison with the nuclear spin dynamics in the anisotropic SMM
Mn$_{12}$-ac. We find that the splitting of the electron spin
levels in Mn$_6$, as produced by the application of an external
field, yields similar features in the nuclear spin dynamics as the
crystal field splitting in Mn$_{12}$-ac. On the other hand, since
the quantum tunneling fluctuations are strictly absent in Mn$_6$,
the quantitative comparison of the data confirms that, even in the
thermally activated regime, a correct description of the $^{55}$Mn
NSLR of Mn$_{12}$-ac should include the effects of
thermally-assisted tunneling.

\cleardoublepage

\begin{small}

\end{small}

\nonumchapter{Samenvatting}

\hyphenation{belang-rijke} \hyphenation{be-staan}
\hyphenation{na-tuur-kun-di-gen} \hyphenation{ver-krij-ging}
\hyphenation{mag-ne-ti-sche} \hyphenation{bij-voor-beeld}
\hyphenation{mi-nia-tu-ri-se-ring}
\hyphenation{kern-spin-po-la-ri-sa-tie} \hyphenation{an-der-zijds}
\hyphenation{pro-du-ce-ren}

\begin{flushleft}

\emph{van het proefschrift met de titel:}

\large{\textbf{``Kwantum spin dynamica in Enkel-Molecuul
Magneten''}}

\end{flushleft}

\noindent Terwijl de fundamentele wetenschap steeds meer bezig is
met onderzoek aan complexe systemen, blijft de industrie
investeren in de voortdurende miniaturisering van, bijvoorbeeld,
elektronische circuits en geheugens. Deze twee
ontwikkelingsrichtingen gaan elkaar ontmoeten op de
nanometer-schaal, met gevolg dat de nanowetenschap een zeer
belangrijke onderzoeksgebied eordt voor zowel de fundamentele als
de toegepaste fysica.

Een van de meest belangwekkende verschijnselen die op de
nanometer-schaal kunnen worden bestudeerd is de overgang tussen de
klassieke- en de kwantum- fysica. Natuurkundigen zijn gewend aan
de toepassing van de kwantummechanica op het atomaire niveau, maar
zodra het te beschrijven systeem groter wordt, kunnen sommige
aspecten van de kwantummechanica (met name het meet-postulaat) tot
paradoxale conclusies leiden (de ``Schr\"{o}dinger kat'' paradox).
In dit opzicht is de theorie nog niet afgerond, het is dus
essentieel om systemen te vinden die geschikt zijn om
experimenteel te testen welke factoren van belang zijn voor de
overgang tussen kwantum en klassiek gedrag.

Dit proefschrift beschrijft een onderzoek aan het
kwantummechanisch gedrag van Enkel-Molecuul Magneten (EMMen). Dit
zijn chemische verbindingen die bestaan uit een centrale cluster
van magentische ionen, omringd door organische liganden. De
magnetische wisselwerking tussen de elektronspins van de ionen in
eenzelfde cluster resulteert in een ``gigantisch'' magnetisch
moment (gigantisch t.o.v. de typische waarden voor atomaire
momenten) voor de cluster als geheel. De moleculaire magnetische
clusters zijn geordend in een kristalstructuur, waarin de
moleculaire magnetische momenten (spins) slechts een zwakke
onderlinge wisselwerking vertonen, zodat het systeem beschouwd mag
worden als een ensemble van identieke nagenoeg onafhakelijke
moleculaire spins. Hiermee wordt het mogelijk om de bekende
experimentele technieken voor vaste stof fysica toe te passen op
de studie van het gedrag van de EMMen. Hoofdstuk II beschrijft de
eigenschappen en het ontwerp van de experimentele opstellingen die
voor onze onderzoek zijn gebruikt.

Hoofdstuk III is besteed aan de theoretische beschrijving van de
fysische eigenschappen van de EMMen. Een essentieel aspect is de
magnetische anisotropie, d.v.z. de voorkeur van de cluster spin om
zich te orienteren in een van de twee tegenovergestelde richtingen
langs een bepaalde voorkeursas. Om de spin om te klappen moet er
een bepaalde energie worden toegevoerd, bijvoorbeeld via
thermische excitatie. Als het systeem tot een temperatuur veel
lager dan de anisotropie energie wordt afgekoeld, dan zijn de
cluster spins ``bevroren'', wat suggereert dat het mogelijk zou
zijn om EMMen te gebruiken als magnetische geheugens, op
voorwaarde dat er geen andere mechanisme is dat de richting van de
spins kan inverteren.

Dat extra mechanisme bestaat feitelijk wel, als een compleet
kwantummechanisch model van de cluster spin wordt toegepast. Omdat
twee tegenovergestelde richtingen van de spin dezelfde energie
hebben, is er een mogelijkheid voor de spin om te tunnelen door de
barri\`{e}re, d.v.z. zonder dat extra energie wordt toegevoegd.
Dit fenomeen bestaat niet in de klassieke fysica, en het is vrij
wonderlijk om het te kunnen waarnemen in een zo groot systeem als
een EMM. De Prokof'ev-Stamp theorie van het ``spin-bad'' is een
antwoord op de vraag hoe kwantumtunneling in een EMM kan
plaatsvinden ondanks zijn koppeling met de omgeving, die normaal
gesproken de mogelijkheid voor tunnelen geheel uitsluit. De totale
elektronspin van een EMM is magnetisch gekoppeld aan de kernspins
in zijn omgeving, maar deze koppeling is sterk be\"{i}nvloed door
de \emph{dynamica} van de kernspins. Bovendien, zijn de dipolaire
koppeling tussen clusterspins en de invloed van roostertrillingen
(fononen) ook van belang voor een nauwkeurige beschrijving van het
kwamtumgedrag van EMMen.

Een van de meest aantrekkelijke eigenschappen van de EMMen voor de
studie van kwantummechanische verschijnselelen op moleculaire
schaal, is dat het mogelijk is om de tunnelwaarschijnlijkheid te
veranderen door een extern magnetisch veld aan te leggen,
loodrecht op de anisotropie as. Deze ``instelbaarheid'' is niet
alleen handig voor het fundamentele onderzoek, maar geeft ook de
hoop om coherente kwantummechanische oscillaties van de cluster
spin te kunnen produceren, waardoor de EMMen als basis element
voor een kwantumcomputer zouden kunnen worden gebruikt.

We hebben de bovengenoemden aspecten van de interactie tussen
EMMen en hun omgeving bestudeerd door middel van experimenten aan
de kernspin dynamica van $^{55}$Mn kernen in de Mn$_{12}$-ac
molecuul, beschreven in hoofdstuk IV. De Mn$_{12}$-ac moleculen
bevatten clusters van 12 mangaan atomen, magnetisch gekoppeld tot
een cluster met totale spin $S =10$. Dankzij de hoge
anisotropiebarriere, is het bij temperaturen lager dan $T \sim 2$
K mogelijk om kernspinresonantie (NMR) experiementen uit te voeren
zonder dat een extern magnetisch veld aangelegd hoeft te worden,
d.w.z met behulp van het bijna-statische hyperfijne veld dat door
de electronspins geproduceerd is. De metingen van de kern
spin-rooster relaxatie duiden dat de magnetische fluctuaties
veroorzaakt worden door kwantumtunnelen van de cluster spin, dat
zich manifesteert in een temperatuur-onafhakelijk gedrag van de
relaxatietijden voor $T < 0.8$ K. Als extra bewijs voor de
anwezigheid van kwantumtunnel verschijnselen, vinden we dat er een
maximum van de kernspinrelaxatie bij veld nul is, wat klopt met de
resonantievoorwaarde voor de cluster spin. Verder, bewijzen we dat
het mogelijk is om ``flip-flop'' overgangen tussen kernspins in
verschillende clusters te hebben. Dit betekent dat de
kernspinpolarisatie via spin-diffusie over het hele sample
verdeeld kan worden, wat van belang is voor het interne evenwicht
van het kernspinsysteem. In feite, is de gemeten spin-rooster
relaxatietijd verrassend snel, in aanmerking nemend hoe klein de
kans op tunnelen wel is. We hebben daarom voorgesteld dat het
kwantumtunnelen in een kleine fractie zogenaamde ``snelle
moleculen'' plaatsvindt: dat zijn moleculen waar, vanwege een
lokale verstoring, de anisotropie barriere veel lager is dan
normaal, zodat de cluster spin een veel hogere
tunnelwaarschijnlijkheid heeft. De \emph{eigen} kernspins van de
snelle moleculen worden door het kwantumtunnelen gerelaxeerd,
waarna de kernspin diffusie zorgt voor de overdracht van energie
naar de kernen in de langzame moleculen. Deze interpretatie van
het mechanisme van kernspinrelaxatie wordt bevestigd door de
vergelijking tussen het gedrag van de kernspins in verschillende
niet-equivalente Mn ionen in het molecuul. Anderzijds, zoals
voorspeld door Prokof'ev en Stamp, \emph{stiumuleert} de dynamica
van de kernspins het kwantumtunnelen van de cluster spin. We
hebben NMR metingen uitgevoerd op een sample waarin de waterstof
atomen gedeeltelijk vervangen waren door deuterium atomen,
waardooe de kernen een kleiner magnetisch moment hebben. Inderdaad
vonden wij daarop een veel langzamere kernspin-rooster en
spin-spin relaxatie, waarmee is aangetoond dat de
tunnelwaarschijnlijkheid kleiner wordt als de elektron-kern
koppelingen worden verzwakt.

Additionele metingen aan de afhankelijkheid van de kern
spin-rooster relaxatie van een veld aangelegd loodrecht op de
voorkeursrichting bevestigen vroegere resultaten van soortelijke
warmte experimenten in onze groep.

Ten slotte hebben we een aspect van de kernspindynamica
onderzocht, dat nooit eerder in beschouwing was genomen, namelijk
het verband tussen de kernspintemperatuur en de kwantumtunnel
fluctuaties van de cluster spins. Het is inderdaad helemaal niet
triviaal om te voorspellen wat de evenwichtspolarisatie van het
kernspinsysteem zal zijn, als de dynamica gedreven wordt door
\emph{temperatuur-onafhankelijke} kwantum verschijnselen. Wij
vonden tot onze verrassing uit onze experimenten dat de
kernspintemperatuur in evenwicht blijft met de roostertemperatuur.
Dit resultaat vraagt om een uitbreiding van de bestaande
theorie\"{e}n van het gekoppeld systeem ``kwantum spin +
spin-bad'' met een koppeling naar het rooster, en is van groot
belang voor een nauwkeurige berekening van de decoherentie
snelheid. We sluiten hoodfstuk IV dan ook af met een voorstel voor
een model dat al deze observaties in beschouwing neemt.

In hoofdstuk V beschrijven we een aantal experimenten aan een vrij
bijzondere EMM, de moleculaire cluster verbinding Mn$_6$, die
clusters bestaande uit 6 mangaan ionen in een zeer symmetrische
structuur bevat, zozeer dat de totale anisotropie van de cluster
bijna verwaarloosbaar is. Hier verwachten we dus geen
kwantummechanisch tunnel verschijnsel, maar wel dat de cluster
spins in evenwicht met het rooster blijven tot zeer lage
temperaturen. Dit wordt inderdaad bevestigd door onze metingen van
de \textit{ac}-susceptibiliteit en de soortelijke warmte. Dankzij
de snelle relaxatie krijgen de cluster spins de mogelijkheid om de
thermodynamische evenwichtstoestand te vinden, die bij voldeoende
lage temperatuur (beneden 0.16 K) correspondeert met een
ferromagnetische lange-afstands-ordening. Het bijzondere van deze
lange afstands ordening is dat zij uitsluitend wordt veroorzaakt
door dipolaire interacties tussen de cluster spins, zoals
bevestigd wordt door numerieke berekeningen m.b.v. Monte-Carlo
simulaties. De metingen van de veld-afhankelijkheid van de
soortelijke warmte laat zien hoe de kernspins buiten thermsiche
evenwicht kunnen worden gebracht als de cluster elektronspins
geprolariseerd zijn door een sterke aangelegde veld. We hebben dit
verder kunnen bevestigen en onderzoeken door NMR metingen aan de
kernspin-rooster relaxatie van ditzelfde systeem.

Samenvattend kunnen we stellen dat het proefschrift een grondig
onderzoek beschrijft naar de kwantumdynamica van moleculaire
spins, in het bijzonder naar de interactie tussen de kwantumspin
en zijn omgeving. Dit is van groot fundamenteel belang voor de
kwantummechanische beschrijving van mesoscopische systemen, en zou
zelfs essentieel kunnen zijn voor toekomstige toepassingen van de
EMMen op het gebied van magnetische geheugens of kwantumcomputers.

\nonumchapter{List of Publications}

\begin{enumerate}

\item A. Morello, O.N. Bakharev, H.B. Brom, N.J. Zelders, I.S. Tupitsyn, R. Sessoli, and L.J. de
Jongh, \emph{Nuclear spin dynamics and quantum tunneling of
magnetization in Mn$_{12}$-ac single-molecule magnets}, in
preparation.

\item  A. Morello, F.L. Mettes, F. Luis, J.F. Fern\'{a}ndez, J.
Krzystek, G. Arom\'{i}, and L.J. de Jongh, \emph{Long-range
ferromagnetic dipolar ordering and spin-lattice relaxation of the
isotropic Mn$_6$ single-molecule magnets}, in preparation.

\item  A. Morello, W.G.J. Angenent, G. Frossati, and L.J. de Jongh,
\emph{An automated and versatile ultra-low temperature SQUID
magnetometer}, in preparation.

\item  A. Morello, O.N. Bakharev, H.B. Brom, R. Sessoli, and L.J. de
Jongh, \emph{Nuclear spin dynamics in the quantum regime of a
single-molecule magnet}, submitted to Phys. Rev. Lett.

\item  F. Luis, F.L. Mettes, M. Evangelisti, A. Morello, and L.J. de
Jongh, \emph{Approach of single-molecule magnets to thermal
equilibrium}, to appear in J. Phys. Chem. Solids.

\item  A. Morello, O.N. Bakharev, H.B. Brom, and L.J. de Jongh,
\emph{Low-temperature NMR study of quantum tunneling of
magnetization in the molecular magnet Mn$_{12}$-ac}, to appear in
J. Mag. Mag. Mat.

\item  A. Morello, O.N. Bakharev, H.B. Brom, and L.J. de Jongh,
\emph{Quantum tunneling of magnetization in Mn$_{12}$-ac studied
by $^{55}$Mn NMR}, Polyhedron \textbf{22}, 1745 (2003).

\item  A. Morello, F.L. Mettes, F. Luis, J.F. Fern\'{a}ndez, J.
Krzystek, G. Arom\'{i}, G. Christou, and L.J. de Jongh,
\emph{Long-range ferromagnetic dipolar ordering of high-spin
molecular clusters}, Phys. Rev. Lett. \textbf{90}, 017206 (2003).

\item  P. Szab\'{o}, P. Samuely, J. Kacmarcik, T. Klein, A.G.M. Jansen,
A. Morello, and J. Marcus, \emph{Vortex glass transition versus
irreversibility line in superconducting BKBO}, Int. J. Mod. Phys.
B \textbf{16}, 3221 (2002).

\item R.S. Gonnelli, A. Morello, G.A. Ummarino, V.A. Stepanov, G.
Behr, G. Graw, S.V. Shulga, and S.-L. Drechsler,
\emph{Electron-phonon coupling origin of the resistivity in
YNi$_{2}$B$_2$C single crystals}, Int. J. Mod. Phys. B
\textbf{14}, 2840 (2000).

\item  R.S. Gonnelli, V.A. Stepanov, A. Morello, G.A. Ummarino, G.
Behr, G. Graw, S.V. Shulga, and S.-L. Drechsler,
\emph{Resisitivity and electron-phonon coupling in YNi$_{2}$B$_2$C
single crystals}, Physica C \textbf{341-348}, 1957 (2000).

\item  A. Morello, A.G.M. Jansen, R.S. Gonnelli, and S.I. Vedeneev,
\emph{3D-melting features of the irreversibility line in overdoped
Bi$_2$Sr$_2$CuO$_6$ at ultra-low temperature and high magnetic
field}, Physica C \textbf{341-348}, 1321 (2000).

\item  A. Morello, A.G.M. Jansen, R.S. Gonnelli, and S.I. Vedeneev,
\emph{Irreversibility line of overdoped Bi$_2$Sr$_2$CuO$_6$ at
ultralow temperatures and high magnetic fields}, Phys. Rev. B
\textbf{61}, 9113 (2000).

\end{enumerate}

\nonumchapter{Curriculum Vitae}

of Andrea Morello, born in Pinerolo (Italy), on June 26, 1972.

\ \\

In the summer of 1991 I obtained the diploma in Electronics at the
Technical High School ``VIII I.T.I.S.'' in Torino, after which I
joined the Faculty of Engineering at the Politecnico di Torino,
where I followed the graduation course in Electronics, with
specialization in Electromagnetism and Solid State Physics. In
1997 I worked at the High Magnetic Field Laboratory of the
Max-Planck Institut f\'{u}r Festk\"{o}rperforschung in Grenoble,
under the guidance of dr. A. G. M. Jansen, for a research project
on the vortex phase of Bi$_2$Sr$_2$CuO$_6$ and
Ba$_{1-x}$K$_x$BiO$_3$ superconductors at ultra-low temperatures
and high magnetic fields. This one-year stage resulted in my
graduation thesis. In 1998 I was back in Torino to accomplish the
Substitutional Civil Service in an organization for psychiatric
patients. At the end of that year I graduated ``cum laude'' in
Electronics Engineering at the Politecnico di Torino.

In 1999 I worked in the group of prof. R. S. Gonnelli at the
Physics Department of the Politecnico di Torino, for a research on
the transport properties of La$_{2-x}$Sr$_x$CuO$_4$ and
YNi$_{2}$B$_2$C superconductors. At the same time, I was
collaborating with dr. M. Rajteri at the Metrological Institute
``G. Ferraris'' in Torino, to set up a low-temperature facility
for single-photon counters based on Josephson junctions.

In October 1999 I started a Ph.D. project in the Kamerlingh Onnes
Laboratory of the University of Leiden, under the supervision of
prof. dr. L. J. de Jongh, as an OIO (Onderzoeker in Opleideing) of
the Stichting voor Fundamenteel Onderzoek der Materie (FOM). This
research, concerning the low-temperature quantum properties of
nanometer-scale molecular magnets, resulted in the work presented
in this thesis. In July 2000 I followed the Summer School ``Atomic
Clusters and Nanoparticles'' in Les Houches. During my stay at the
University of Leiden, I was Assistant at the Laboratory Training
for the second-year students of the Faculty of Physics.

\nonumchapter{Acknowledgements}


The work presented in this thesis is the result of the efforts of
several people who, to a larger or smaller extent, were involved
in making it a successful scientific project. I would like to
start by expressing a special gratitude to those people who
contributed ineffaceably to my scientific growth, in addition to
their immediate and tangible help. A special thank is therefore
deserved by O. N. Bakharev for his help in introducing me to the
beauty and power of the NMR technique, and by P. C. E. Stamp, I.
S. Tupitsyn and B. V. Fine, who have been a precious source of
illuminating discussions and clarifications about the theoretical
background of my research. Thanks also to F. Luis, for his
guidance in my first contact with molecular magnetism and his
constant availability to discuss newer or subtler details. I am
deeply indebted with the technicians of the \emph{fine}-mechanical
workshop, A. Kuijt, R. van Egmond and E. de Kuyper, and of the
electronics development center, R. Hulstman and M. Pohlkamp, for
their patience and their exceptional skills in turning my
experimental hopes into reality, but also for what I learned while
collaborating with them.

This research would not have been possible without the samples
provided by R. Sessoli, A. Caneschi (Mn$_{12}$-ac), G. Arom\'{i}
(Mn$_6$) and, although the results do not appear in this thesis,
by J. van Slageren, A. Cornia and W. T. Fu. Because of the several
discussions about our projects, their contribution goes in fact
much beyond the simple supply of materials.

Thanks to J. Sese, K. Siemensmeyer and L. Gottardi for their help
and suggestions about the design of the SQUID magnetometer, T. J.
Gortenmulder for the collaboration on the construction of the
torque magnetometer, R. J. van Kuyk and W. van der Geest for the
(large) supply of cryogenic liquids, and F. J. Kranenburg and R.
Zweistra for the computer management. I am grateful to M.
Evangelisti and T. G. Sorop for their participation in the tests
and calibration of the torquemeter and the
\textit{ac}-susceptometer, for their support with magnetic
measurements above 1 K, and for the uncountable discussions.
Contributions from F. L. Mettes, J. F. Fern\'{a}ndez, J. Krzystek
and W. Wernsdorfer have often completed the picture arising from
my experiments and are gratefully acknowledged.

A special thank to the students that joined my research and
preciously contributed to its success: W. G. J. Angenent for the
automation of the experimental setups and the constant support on
any computer-related issue, and N. J. Zelders for the numerical
calculations and the thorough discussions about quantum mechanics.

Finally I would like to mention, for the suggestions or the
discussions, F. Meier, S. I. Mukhin, Y. Furukawa, F. Borsa, A.
Lascialfari, I. Chiorescu, G. Aeppli and J. van den Brink.

\ \\
\ \\
\ \\
\ \\
\ \\
\ \\
\ \\
\ \\
\ \\

\clearpage

\thispagestyle{empty}
\begin{figure}[ht]
\begin{center}
      \leavevmode
      \epsfxsize=105mm
      \epsfbox{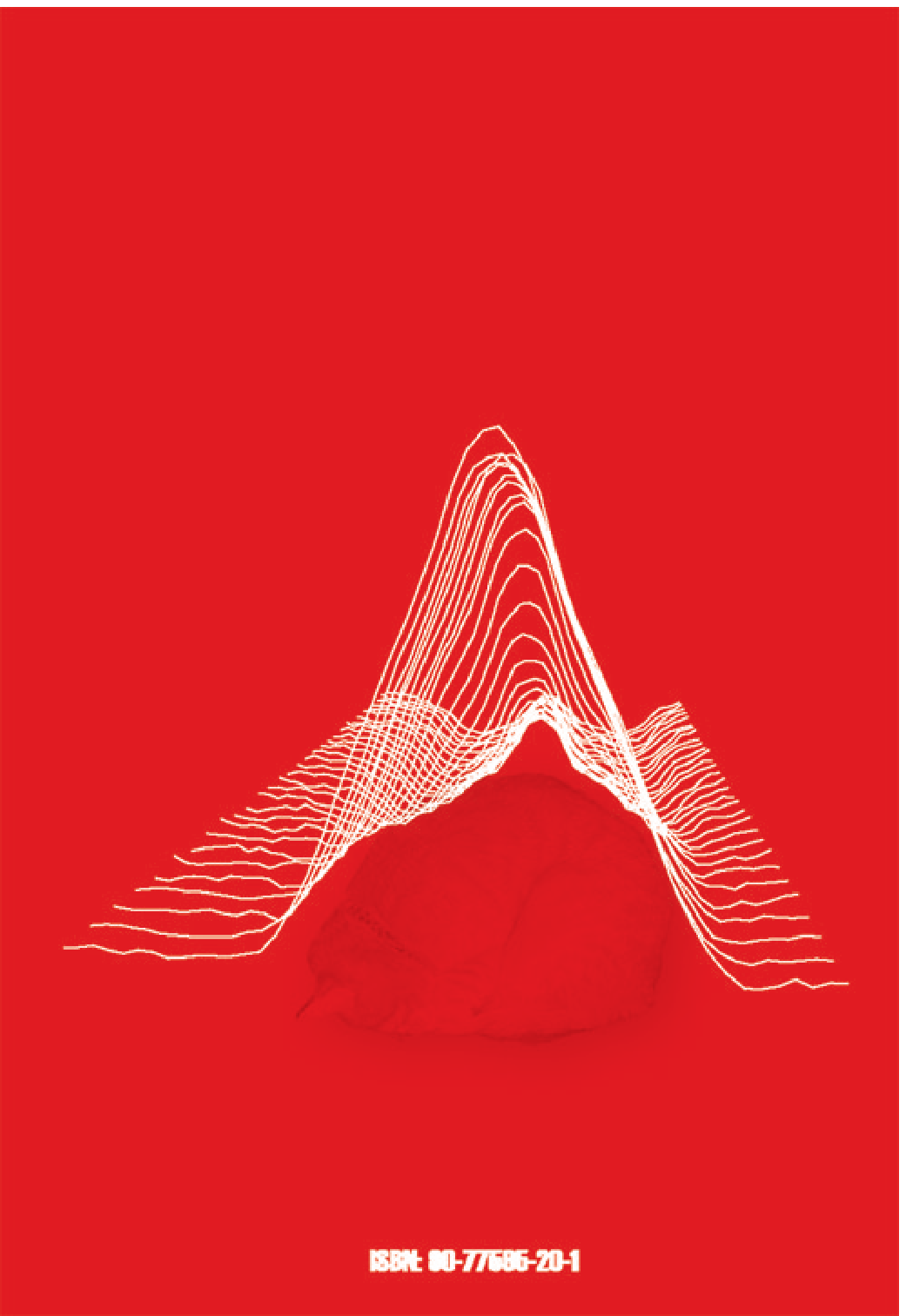}
\end{center}
\end{figure}

\end{document}